%% file: Lambda_Regge_model.tex
\documentclass[prd, reprint, showpacs, nofootinbib, longbibliography, floatfix, raggedbottom]{revtex4-1}
\usepackage{epsf}

\usepackage{amsmath, mathtools, graphicx, color, latexsym, array, hyperref, amssymb}
\usepackage[inline]{enumitem}
\usepackage[retainorgcmds]{IEEEtrantools}
\usepackage[explicit]{titlesec}
\usepackage[caption=false]{subfig}
\usepackage[english]{babel}

\allowdisplaybreaks

\newcommand*{\Figuref}[1]{Figure \ref{#1}}
\newcommand*{\FigRef}[1]{Fig.\ \ref{#1}}
\newcommand*{\tabref}[1]{Table \ref{#1}}
\newcommand*{\secref}[1]{Section \ref{#1}}
\newcommand*{\appendref}[1]{Appendix \ref{#1}}

\renewcommand*{\imath}{\iota}

\newcommand*{\Action}{\mathcal{S}}

\newcommand*{\metric}{g}
\newcommand*{\RicScal}{R}
\newcommand*{\RicTens}{\RicScal}
\newcommand*{\Riemann}{\RicScal}

\newcommand*{\EinsTens}{G}
\newcommand*{\StressEnergy}{T}

\newcommand*{\firstform}{h}
\newcommand*{\secondform}{\chi}
\newcommand*{\firstprojection}[2]{\firstform^{#1}_{\phantom{#1}{#2}}}

\newcommand{\Tensb}[1]{\mathbf{#1}}

\newcommand*{\FriedScale}{a}
\newcommand*{\CosmoConst}{\Lambda}

\newcommand*{\Cauchy}{\Sigma}
\newcommand*{\Cauchyt}[1]{\Cauchy_{#1}}

\newcommand*{\CWstrutlength}{m}
\newcommand*{\CWstrut}[1]{\CWstrutlength_{#1}}
\newcommand*{\CWdiag}[1]{d_{#1}}
\newcommand*{\TrapArea}[1]{\Area{trap}{#1}}
\newcommand*{\AreaA}[1]{\Area{A}{#1}}
\newcommand*{\AreaB}[1]{\Area{B}{#1}}
\newcommand*{\TriArea}[1]{\Area{tri}{#1}}
\newcommand*{\CWdef}{\delta}
\newcommand*{\CWdeficitbar}[1]{\bar{\CWdef}_{#1}}
\newcommand*{\CWdeficit}[1]{\CWdef_{#1}}
\newcommand*{\TrapDeficit}[1]{\CWdeficit{#1}^{\,\text{trap}}}
\newcommand*{\DeficitA}[1]{\DeficitS{\text{A}}{#1}}
\newcommand*{\DeficitB}[1]{\DeficitS{\text{B}}{#1}}
\newcommand*{\TriDeficit}[1]{\CWdeficit{#1}^{\,\text{tri}}}
\newcommand*{\FourVol}[1]{\Vol^{(4)}_{#1}}
\newcommand*{\Num}{N}
\newcommand*{\N}[1]{\Num_{#1}}
\newcommand*{\Ns}[1]{\Num^{#1}}
\newcommand*{\Nvert}{{\N{0}}}
\newcommand*{\Nedge}{{\N{1}}}
\newcommand*{\Ntri}{{\N{2}}}
\newcommand*{\Ntet}{{\N{3}}}

\newcommand*{\Odt}[1]{O \! \left(dt^{#1} \right)}
\newcommand*{\Odtone}{O \! \left(dt \right)}
\newcommand*{\Oconst}{O \! \left(1 \right)}

\newcommand*{\axisl}{h}
\newcommand*{\axislen}[1]{\axisl_{#1}}
\newcommand*{\axislentilde}[1]{\tilde{\axisl}_{#1}}

\newcommand*{\dihedral}[2]{\theta^{(#1)}_{#2}}
\newcommand*{\deficit}[1]{\epsilon_{#1}}

\newcommand*{\RegTime}[1]{t_{#1}}
\newcommand*{\CWstrutl}[1]{\CWstrut{i}^\ell}
\newcommand*{\CWstrutg}[1]{\CWstrut{i}^g}
\newcommand*{\CWstrutdot}[1]{\dot{\CWstrutlength}_{#1}}
\newcommand*{\Area}[2]{A^\text{{#1}}_{#2}}

\newcommand*{\AreaAi}[2]{\Area{A{#1}}{#2}}
\newcommand*{\AreaBi}[2]{\Area{B{#1}}{#2}}
\newcommand*{\CWstrutA}[1]{\CWstrut{#1}^\text{A}}
\newcommand*{\CWstrutB}[1]{\CWstrut{#1}^\text{B}}

\newcommand*{\DeficitS}[2]{\CWdef^{#1}_{#2}}
\newcommand*{\DeficitSub}[2]{\DeficitS{({#1})}{#2}}

\newcommand*{\CWTrapDihe}{\theta}
\newcommand*{\CWTrapdihedral}[1]{\CWTrapDihe_{#1}}
\newcommand*{\CWTrapdihedraldot}[1]{\dot{\CWTrapDihe}_{#1}}
\newcommand*{\CWTrapdiS}[2]{\CWTrapDihe^{#1}_{#2}}
\newcommand*{\CWTrapdiSdot}[2]{\dot{\CWTrapDihe}^{#1}_{#2}}
\newcommand*{\CWTrapdiSbar}[1]{\bar{\CWTrapDihe}_{#1}}
\newcommand*{\CWSubdivTrapdi}[2]{\CWTrapdiS{({#1})}{#2}}
\newcommand*{\CWSubdivTrapdiSbar}[2]{\bar{\CWTrapDihe}^{({#1})}_{#2}}

\newcommand*{\CWTridihe}{\phi}
\newcommand*{\CWTridihedral}[2]{\CWTridihe^{#1}_{#2}}
\newcommand*{\CWTridihedraldot}[2]{\dot{\CWTridihe}^{#1}_{#2}}

\newcommand*{\Vol}{V}
\newcommand*{\VolSub}[2]{Vol^{#1}_{#2}}
\newcommand*{\FourVolbar}[1]{\bar{\Vol}^{(4)}_{#1}}

\newcommand*{\BrewinBase}[1]{\BrewinB_{#1}}
\newcommand*{\BrewinParent}[1]{\BrewinP_{#1}}
\newcommand*{\BrewinCentral}[1]{\BrewinC_{#1}}
\newcommand*{\BrewinBasedot}[1]{\dot{\BrewinB}_{#1}}

\newcommand*{\BrewinBaseddot}[1]{\ddot{\BrewinB}_{#1}}

\newcommand*{\Rd}{R}
\newcommand*{\Rad}[1]{\Rd_{#1}}
\newcommand*{\RadS}[2]{\Rd^{({#1})}_{#2}}
\newcommand*{\Radhat}[1]{\hat{\Rd}_{#1}}
\newcommand*{\Radbar}[1]{\bar{\Rd}_{#1}}
\newcommand*{\Radtilde}[1]{\tilde{\Rd}_{#1}}

\newcommand*{\Rtio}{Z}
\newcommand*{\SphRatio}[1]{\Rtio_{#1}}
\newcommand*{\SphRatioS}[1]{\Rtio^{({#1})}}
\newcommand*{\SphRatiohat}[1]{\hat{\Rtio}_{#1}}

\newcommand*{\Lamblength}{l}
\newcommand*{\Lamblen}[1]{\Lamblength_{#1}}
\newcommand*{\Lamblenl}[1]{\Lamblen{#1}^\ell}
\newcommand*{\Lamblendot}[1]{\dot{\Lamblength}_{#1}}
\newcommand*{\Lamblenddot}[1]{\ddot{\Lamblength}_{#1}}
\newcommand*{\Lambstrut}[2]{\CWstrut{#2}^{({#1})}}
\newcommand*{\Lambdiag}[2]{\CWdiag{#2}^{({#1})}}

\newcommand*{\midpt}[1]{({#1})}
\newcommand*{\central}{\midpt{ABCD}}
\newcommand*{\BrewinB}{v}
\newcommand*{\BrewinP}{u}
\newcommand*{\BrewinC}{p}

\newcommand*{\LinEl}{s}

\renewcommand{\thesection}{\Roman{section}}
\titleformat{\section}[block]
  {\normalfont\bfseries}{\thesection.}{1em}{\centering \MakeUppercase{#1}}
\renewcommand{\thesubsection}{\Alph{subsection}}
\titleformat{\subsection}[block]
  {\normalfont\bfseries}{\thesubsection.}{1em}{\centering {#1}}

\begin{document}

\title{\bf Regge calculus models of the closed vacuum $\CosmoConst$-FLRW universe}
\author{Rex G \surname{Liu}}
\email{R.Liu@damtp.cam.ac.uk}
\affiliation{Trinity College, Cambridge CB2 1TQ, UK,}
\affiliation{DAMTP, CMS, Wilberforce Road, Cambridge CB3 0WA, UK.}
\author{Ruth M \surname{Williams}}
\email{R.M.Williams@damtp.cam.ac.uk}
\affiliation{Girton College, Cambridge CB3 0JG, UK,}
\affiliation{DAMTP, CMS, Wilberforce Road, Cambridge CB3 0WA, UK.}

\begin{abstract}
The Collins-Williams Regge calculus models of FLRW space-times and Brewin's subdivided models are applied to closed vacuum $\CosmoConst$-FLRW universes.  In each case, we embed the Regge Cauchy surfaces into 3-spheres in $\mathbf{E}^4$ and consider possible measures of Cauchy surface radius that can be derived from the embedding.  Regge equations are obtained from both global variation, where entire sets of identical edges get varied simultaneously, and local variation, where each edge gets varied individually.  We explore the relationship between the two sets of solutions, the conditions under which the Regge Hamiltonian constraint would be a first integral of the evolution equation, the initial value equation for each model at its moment of time symmetry, and the performance of the various models.  It is revealed that local variation does not generally lead to a viable Regge model.  It is also demonstrated that the various models do satisfy their respective initial value equations.  Finally, it is shown that the models reproduce the correct qualitative dynamics of the space-time.  Furthermore, the approximation's accuracy is highest when the universe is small but improves overall as we increase the number of tetrahedra used to construct the Regge Cauchy surface.  Eventually though, all models gradually fail to keep up with the continuum FLRW model's expansion, with the models with lower numbers of tetrahedra falling away more quickly.  We believe this failure to keep up is due to the finite resolution of the Regge Cauchy surfaces trying to approximate an ever expanding continuum Cauchy surface; each Regge surface has a fixed number of tetrahedra and as the surface being approximated gets larger, the resolution would degrade.  Finally, we note that all Regge models end abruptly at a point when the time-like struts of the skeleton become null, though this end-point appears to get delayed as the number of tetrahedra is increased.
\end{abstract}

\pacs{04.25.-g, 98.80.Jk, 98.80.-k}

\maketitle

\section{Introduction}

Since its initial formulation in 1961, Regge calculus \cite{Regge}, a discretisation of general relativity, has remained an enduring subject of interest, particularly because of its applications to numerical relativity and to path-integral formulations of quantum gravity \cite{WilliamsTuckey, ReggeWilliams, Hamber, DRS, Perez, AJGL}.  The focus of this paper is on its numerical applications.  Regge calculus can, in principle, approximate any solution to the Einstein field equations of general relativity, and indeed, it has been shown that solutions of Regge calculus converge at second order in the discretisation edge-lengths to solutions of the Einstein field equations \cite{BrewinGentle}.  Regge calculus has been applied to approximate a wide range of space-times \cite{GHMW, GentleMiller1998, Gentle1999, Gentle2004, Gentle2013, Brewin2015}, and comparison of the Regge approximation with the continuum solution, when known, has demonstrated good agreement, indicating that Regge calculus offers a viable approximation to general relativity.

The main value of Regge calculus though is that it can be used to approximate space-times where exact solutions are difficult to obtain.  In particular, it offers a non-perturbative approach to approximating space-times, potentially elucidating certain physics that, owing to the non-linear nature of the Einstein field equations, would not be evident using perturbative approaches.  For example, it has been argued that to unambiguously understand the effects of inhomogeneities on cosmological observables, a non-perturbative model of inhomogeneities is necessary \cite{KMR, ClarksonMaartens, ClarksonUmeh}; Regge calculus may assist in this respect.  Regge calculus can also complement other approaches to numerical relativity: when no exact solution is available for comparison, consistency of results between different numerical approaches would strengthen confidence in any conclusions.

The focus of this paper is on further exploring a Regge approximation to the \emph{Friedmann-Lema\^itre-Robertson-Walker} (FLRW) space-times.  We shall focus on a Regge formalism first devised by Collins and Williams (CW) \cite{CollinsWilliams} and subsequently expanded by Brewin \cite{Brewin}.  This formalism was first applied to closed dust-filled FLRW cosmologies, and the resulting approximation was able to reproduce the behaviour of the continuum space-time rather well.  In this paper, we shall test the CW formalism against another cosmological space-time for which the exact solution is well-known, the closed FLRW universe with non-zero cosmological constant $\CosmoConst$.  A key difference between this universe and that studied by Collins, Williams, and Brewin is that the dust-filled universe eventually collapses back in on itself whereas the $\CosmoConst$-FLRW universe expands indefinitely.  By comparing the approximation against the exact solution, we can learn more about the CW formalism's performance and range of applicability in general.  As this paper will show, we have also uncovered several interesting properties of the CW formalism that were not explored in the original papers.  With a better understanding of this formalism, we should then be better-equipped to adopt it, for instance, to modelling inhomogeneous cosmologies where exact solutions are not known.  As an example, drawing on the lessons learned here, we have recently applied this formalism to model `lattice universes' wherein the matter content consists of point particles distributed into a regular lattice \cite{RGL-Williams}. \footnote{There has been great interest recently in studying such toy universes to try and understand whether dark energy is indeed needed to explain the observed acceleration of the universe or whether this acceleration can be explained away as an apparent effect arising from the influence of the late universe's inhomogeneous matter distribution on cosmological observables \cite{EllisBuchert, *Wiltshire2007, *Mattsson, *Ellis2011, *CELU}.}

The FLRW metric is founded upon the Copernican principle which posits that the universe admits a foliation of constant-time Cauchy surfaces such that each surface is perfectly homogeneous and isotropic; this leads to a family of metrics that can be expressed in the form
\begin{equation}
d\LinEl^2 = - dt^2 + \FriedScale^2(t) \left[\frac{dr^2}{1- k r^2} + r^2 \big(d\theta^2 + \sin^2\theta\, d\phi^2 \big) \right],
\label{FLRW}
\end{equation}
where $a(t)$ is a time-dependent function known as the \emph{scale factor}, $t$ is the time parameter, and $k$ is a curvature constant.  The sign of $k$ determines whether the constant-$t$ Cauchy surfaces are open, flat, or closed, with $k<0$ being open, $k=0$ being flat, and $k>0$ being closed.

By inserting this metric into the Einstein field equations, one obtains a pair of differential equations for $\FriedScale(t)$ known as the Friedmann equations; these are
\begin{IEEEeqnarray}{rCl}
\left( \frac{\dot{\FriedScale}}{\FriedScale} \right)^2 &=& \frac{1}{3} \Big(8\pi\rho + \CosmoConst \Big) - \frac{k}{\FriedScale^2},
\label{Friedmann1}\\
\frac{\ddot{\FriedScale}}{\FriedScale} \hphantom{\left. \vphantom{\frac{\dot{\FriedScale}}{\FriedScale}}\right)^2 \!} &=& -\frac{4\pi}{3} \Big(\rho + 3p \Big) + \frac{\CosmoConst}{3},
\label{Friedmann2}
\end{IEEEeqnarray}
where $\rho$ and $p$ are the energy density and pressure of any fluid filling the space, and $\CosmoConst$ is the cosmological constant.

For closed vacuum universes with a non-zero cosmological constant, we have that $k>0$, $\rho=p=0$, and $\CosmoConst \neq 0$; solving the Friedmann equations then yields
\begin{equation}
\FriedScale(t) = \frac{1}{2}\sqrt{\frac{3}{\CosmoConst}} \left(e^{-\sqrt{\frac{\CosmoConst}{3}}t} + e^{\sqrt{\frac{\CosmoConst}{3}}t} \right),
\label{continuum_a}
\end{equation}
where the integration constant has been chosen so that $\dot{\FriedScale}(t)=0$ when $t=0$, and where we have chosen a scaling of $\FriedScale(t)$ such that it corresponds to the radius of curvature of constant-$t$ Cauchy surfaces; for such a scaling, the curvature constant $k$ would in turn be re-scaled to $k=1$.

Constant-$t$ Cauchy surfaces of any closed FLRW universe can always be embedded as 3-spheres in 4-dimensional Euclidean space $\mathbf{E}^4$.  Such an embedding requires scaling $k$ to be unity so that $\FriedScale(t)$ equals the 3-sphere radius.  The embedding is then given by
\begin{equation}
\begin{aligned}
r &= \sin \chi,\\
x^1 &= \FriedScale(t)\cos\chi,\\
x^2 &= \FriedScale(t)\sin\chi\cos\theta,\\
x^3 &= \FriedScale(t)\sin\chi\sin\theta\cos\phi,\\
x^4 &= \FriedScale(t)\sin\chi\sin\theta\sin\phi,
\end{aligned}
\label{FLRW:closedE4}
\end{equation}
for $0 \leq \chi, \theta \leq \pi$ and $0 \leq \phi < 2\pi$.  Such a 3-sphere would have a volume of
\begin{equation}
U_\text{FLRW}(t) = 2\pi^2 \FriedScale(t)^3, \label{FLRW:closedU}
\end{equation}
and an expansion rate of
\begin{equation}
\dot{U}_\text{FLRW}(t) = 6\pi^2 \FriedScale(t)^2 \, \dot{\FriedScale}(t).
\label{FLRW:closeddUdt}
\end{equation}
The FLRW metric can then be expressed as
\begin{equation}
d\LinEl^2 = - dt^2 + \FriedScale^2(t) \left[d\chi^2 + \sin^2\chi\, \left( d\theta^2 + \sin^2\theta\, d\phi^2  \right) \right].\label{FLRW:closedmetric}
\end{equation}

Regge calculus \cite{Regge} approximates any curved space-time using a piece-wise linear manifold constructed out of flat 4-blocks: the blocks are `glued' together such that neighbouring blocks share an entire 3-face, and as the blocks are flat, the metric inside is the Minkowski metric.  Regge space-times are generally referred to as \emph{skeletons}.  Curvature in a skeleton manifests itself as conical singularities concentrated on the sub-faces of co-dimension 2; these are known as \emph{hinges}.  If a hinge were flat, then the dihedral angles between all 3-faces meeting at it would sum to $2\pi$; any deviation from $2\pi$ provides a measure of the curvature and is known as the \emph{deficit angle}.  The edge-lengths serve as the Regge analogue of the metric and are determined by the \emph{Regge field equations}, a set of equations analogous to the Einstein field equations of general relativity; and just as the Einstein field equations can be obtained by varying the Einstein-Hilbert action with respect to the metric, so can the Regge field equations be obtained by varying a \emph{Regge action} with respect to the edges.

One of the central features of the CW formalism is its skeleton, designed to strongly mirror the structure and symmetries of the continuum FLRW space-time it is approximating \cite{CollinsWilliams}.  In analogy with FLRW space-times, CW skeletons are also foliated by a one-parameter family of space-like Cauchy surfaces.  But now, each surface is essentially a triangulation of an FLRW 3-sphere using equilateral tetrahedra such that all vertices, edges, and triangles are identical to each other; in this way, the surfaces would mimic as closely as possible the Copernican symmetries.  According to Coxeter \cite{Coxeter}, such a triangulation of the 3-sphere is only possible with 5, 16, and 600 tetrahedra; \tabref{tab:primary} summarises the numbers of vertices, edges, triangles, and tetrahedra for each case.
\begin{table} [htb]
\renewcommand{\arraystretch}{1.2}
\caption{\label{tab:primary}The number of simplices in each of the three triangulations of the 3-sphere with equilateral tetrahedra as well as the number of triangles meeting at any edge.  We introduce $\Ntet$, $\Ntri$, $\Nedge$, and $\Nvert$ to denote the numbers of parent tetrahedra, triangles, edges, and vertices in the Cauchy surface.}
\begin{tabular*}{8.6cm}{>{\centering\arraybackslash}m{1.7cm} >{\centering\arraybackslash}m{1.7cm} >{\centering\arraybackslash}m{1.2cm} >{\centering\arraybackslash}m{1.7cm} >{\centering\arraybackslash}m{1.7cm}}
\hline\hline
Tetrahedra $(\Ntet)$ & Triangles $(\Ntri)$ & Edges $(\Nedge)$ & Vertices $(\Nvert)$ & Triangles per edge \\
\hline
5 & 10 & 10 & 5 & 3\\
16 & 32 & 24 & 8 & 4\\
600 & 1200 & 720 & 120 & 5\\
\hline\hline
\end{tabular*}
\end{table}
As with FLRW Cauchy surfaces, all CW Cauchy surfaces are required to be identical to each other apart from an overall scaling, represented by the length $\Lamblen{}(\RegTime{i})$ of the tetrahedral edge, $\RegTime{i}$ being a discrete time parameter labelling the foliation; thus $\Lamblen{}(\RegTime{i})$ would be a Regge analogue of the FLRW scale factor $\FriedScale(t)$.  To complete the skeleton's construction, the CW Cauchy surfaces are glued together by a series of time-like edges connecting vertices in one surface to their time-evolved images in the next; these edges are known as \emph{struts}.  The world-tubes of the tetrahedra between two consecutive Cauchy surfaces would then correspond to the 4-blocks for this skeleton.  To ensure homogeneity of all vertices in any Cauchy surface, all struts between pairs of consecutive Cauchy surfaces are required to be identical to each other.  Finally, by taking the limit where the separation between surfaces goes to zero, one can generate a continuum time formulation of Regge calculus.  Collins and Williams first applied their construction to model closed dust-filled FLRW universes, and the continuum time function $\Lamblen{}(\RegTime{})$ for the edge-lengths behaved very similarly to the equivalent FLRW scale-factor $\FriedScale(t)$, with models with a greater number of tetrahedra yielding better accuracy.

The CW formulation was further explored and extended by Brewin \cite{Brewin}.  In particular, Brewin devised an algorithm to triangulate each tetrahedron into smaller tetrahedra thereby generating secondary models.  The virtue of this algorithm is that it can in principle be repeated indefinitely, thereby yielding even finer approximations to the underlying FLRW surfaces.  However, subdivision comes at the expense of some of the symmetries inherent in the original CW surfaces: the new tetrahedra would no longer be identical nor necessarily equilateral.  For instance, after the first generation of subdivision, each Cauchy surface would instead have three sets of vertices, three sets of edge-lengths, three sets of triangles, and three sets of tetrahedra.  All members of any set would be identical to each other, and in this sense, there was still some Copernican symmetry.  We shall refer to CW's original models as the \emph{parent} models and any subdivided ones as \emph{children} models.  The children models were also applied to closed dust-filled FLRW universes and were found to better approximate the continuum universe compared to their parent counterparts, again with accuracy commensurate with the number of tetrahedra.

Both parent and children Cauchy surfaces can also be embedded in 3-spheres in $\mathbf{E}^4$, effectively embedding them in FLRW Cauchy surfaces like those described by \eqref{FLRW:closedE4}.  Such an embedding was first mentioned by Collins and Williams and more extensively developed by Brewin.  Since the CW Cauchy surfaces are approximations of 3-spheres, the embedding would map all vertices from a parent Cauchy surface into a single 3-sphere, and it would, in general, map each of the three sets of vertices in a child Cauchy surface into its own 3-sphere.  However in Brewin's embedding of the child Cauchy surface, he considered only the case where the three sets of vertices had been constrained to lie on the same 3-sphere.  He then proposed that, for both the parent and children models, the corresponding 3-sphere radius $\Rad{}(t)$ could provide a more accurate analogue to the FLRW scale factor $\FriedScale(t)$ than the tetrahedral edge-length.  In this paper, we shall generalise Brewin's investigation in two ways: we shall look at alternative radii that can also serve as analogues to $\FriedScale(t)$, and we shall also consider models where the three sets of children vertices do not necessarily lie on the same 3-sphere.

Brewin has also pointed out certain analogies between the ADM formalism and the CW formalism.  He has likened the tetrahedral edge-lengths to the 3-metric of an ADM foliation and the Regge equations obtained from varying the tetrahedral edges to the evolution equations.  He has also likened the struts and diagonals to the ADM lapse and shift functions, respectively, and the Regge equations obtained from their variation to the Hamiltonian and momentum constraints, respectively.  Thus in this paper, we shall refer to the Regge equations obtained from the tetrahedral edges as the evolution equations, from the struts as the Hamiltonian constraints, and from the diagonals as the momentum constraints.

This paper is organised as follows.  The second section provides a brief review of Regge calculus in general and the CW formalism specifically.  The third section applies the parent CW formalism to the $\CosmoConst$-FLRW universes.  We begin with an exposition of the 4-block's geometry.  We follow this by presenting possible embeddings of parent CW Cauchy surfaces into 3-spheres and discussing possible measures for the Cauchy surfaces' radii that can serve as analogues to $\FriedScale(t)$.  We then vary the Regge action in two different manners to arrive at the Regge field equations.  In the first approach, we impose the symmetry constraints on the edges first before varying the action.  The skeleton must then continue satisfying the constraints even under variation; so if we vary one edge, then all edges constrained to share the same length must get varied at the same time: we would simultaneously vary either all tetrahedral edges in a Cauchy surface or all struts between a pair of consecutive surfaces.  This was the approach followed by Collins and Williams, and we shall call it \emph{global variation}.  From this approach, we find that the CW Hamiltonian constraint for our model is actually a first integral of the evolution equation; thus, the Hamiltonian constraint is sufficient to determine our model's evolution.  In the second approach, we instead vary each edge individually first and afterwards impose the symmetry constraints on the resulting Regge equations; this is the more standard way of varying the action in Regge calculus, and we shall call this \emph{local variation}.  In this case, we find that the Hamiltonian constraints are again first integrals of the evolution equations provided we also satisfy the momentum constraints.  However, these momentum constraints impose rather unphysical constraints, and we therefore dismiss the local model as unviable.  The global and local Regge equations can also be related through a chain rule; when we do this, we again arrive at the same conclusion about the local model's unviability.  We next consider the initial value equation in the context of Regge calculus and demonstrate that the global models do satisfy this equation at its moment of time symmetry.  We conclude our investigation of the parent models with a brief speculation on the reasons for the local models' breakdown before finally examining the evolution of the global models.

In the final section, we turn our attention to the secondary models obtained by subdividing the tetrahedra of the parent models.  We begin by presenting Brewin's scheme for subdividing the CW models.  This is followed by the embedding of children Cauchy surfaces into 3-spheres.  We next present the geometric quantities needed to compute the varied Regge action.  After briefly discussing the local variation of the action, we determine the global Regge equations obtained by varying with respect to both the struts and the tetrahedral edge-lengths, and we briefly discuss the conditions under which the Hamiltonian constraint would be a first integral of the evolution equation.  We then consider the initial value equation for the children models and demonstrate that the models do satisfy this equation.  Whereas our method for applying the equation to the parent models is based on a proposal by Wheeler \cite{Houches}, we have devised a different form of the equation for the children models; this is because Wheeler's proposal can only be applied to Regge Cauchy surfaces where there is a well-defined `volume per vertex', which is the case for the parent models but not for the children models.  To the best of our knowledge, our alternative form of the initial value equation is completely novel and can be applied to any time-symmetric Regge Cauchy surface in general.  Finally, we examine the evolution of the children models, comparing their performance against those of the parent models, and we speculate briefly on how the children models might be extended.

In this paper, we shall use geometric units where $G = c = 1$.

\section{Regge calculus and the CW skeletons}

In general relativity, the Einstein field equations can be derived by varying the Einstein-Hilbert action
\begin{equation}
\Action_{EH} = \displaystyle{\frac{1}{16\pi}\int \left( \RicScal - 2\, \CosmoConst \right) \sqrt{-\metric} \; d^4x},
\label{Einstein-Hilbert}
\end{equation}
with respect to the metric tensor $\metric_{\mu\nu}$, where $\CosmoConst$ is the cosmological constant, $\RicScal$ is the Ricci scalar, and $\metric = \det (\metric_{\mu\nu})$.

When applied to a piece-wise linear manifold, as found in Regge calculus, this reduces to the Regge action \cite{Regge}
\begin{equation}
\Action_{Regge} = \displaystyle{\frac{1}{8\pi} \left( \sum_{i \,\in\, \left\{ \text{hinges}\right\}} \mkern-9mu A_i\, \CWdeficit{i} \; - \mkern-26mu \sum_{i \, \in \, \left\{ \text{4-blocks} \right\} } \mkern-21mu \CosmoConst\, \FourVol{i} \right) },
\label{ReggeAction}
\end{equation}
where $A_i$ is the area of a hinge in the Regge skeleton, $\CWdeficit{i}$ its corresponding deficit angle, and $\FourVol{i}$ the volume of a 4-block.  The first summation is over all hinges in the skeleton while the second is over all 4-blocks.  The deficit angle $\CWdeficit{i}$ at hinge $i$ is given by
\begin{equation}
\CWdeficit{i} = 2\pi - \sum_j \dihedral{i}{j},
\label{def_ang}
\end{equation}
where $\dihedral{i}{j}$ is the dihedral angle between the two faces of block $j$ meeting at the hinge and the summation is over all blocks meeting at the hinge.

Since the skeletal edge-lengths are the Regge analogue of the metric, the Regge action is varied with respect to an edge-length $\ell_j$ to get the Regge field equation
\begin{equation}
0 = \frac{1}{8\pi} \left( \sum_i \frac{\partial A_i}{\partial \ell_j} \CWdeficit{i} - \CosmoConst \sum_i \frac{\partial \FourVol{i}}{\partial \ell_j} \right),
\label{ReggeEqn}
\end{equation}
where the variation of the deficit angles has cancelled out owing to the well-known Schl\"afli identity \cite{Regge},
$$
\sum_i A_i \frac{\partial \dihedral{i}{k}}{\partial \ell_j} = 0;
$$
this identity holds for any individual block $k$, with the summation being over all hinges in the block and $\dihedral{i}{k}$ being the block's dihedral angle at hinge $A_i$.\footnote{In the standard formulation of Regge calculus, one actually uses a simplicial manifold where every block is a 4-simplex, and the Schl\"afli identity is usually formulated in terms of simplices rather than arbitrary blocks.  However, any block can always be triangulated into simplices, and one can then apply the simplicial form of the Schl\"afli identity to the triangulated block to obtain the form of the identity we have above, using the chain rule if necessary to satisfy any constraints on the block's geometry.}

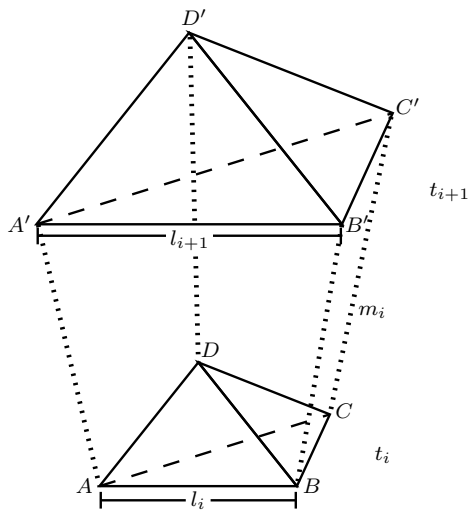
\begin{figure}[tbh]
\input{4-block.pspdftex}
\caption{\label{fig:4block}An equilateral tetrahedron of edge-length $\Lamblen{i}$ at time $\RegTime{i}$ evolves to a tetrahedron of edge-length $\Lamblen{i+1}$ at time $\RegTime{i+1}$, tracing out a 4-dimensional world-tube.  The struts are all of equal length.}
\end{figure}

Regge calculus customarily uses simplicial manifolds where the skeleton's fundamental building block is the 4-simplex.  The geometry of a single $n$-simplex can always be completely determined by specifying the lengths of its $C(n+1,2)$ edges; therefore, the geometry of an entire simplicial skeleton can be completely determined by specifying the lengths of all its edges.  However, the fundamental building blocks of CW skeletons are instead 4-blocks corresponding to the truncated world-tubes of the tetrahedra as they evolve from one Cauchy surface to the next, as illustrated in \FigRef{fig:4block}.  Without extra constraints on the 4-block's internal geometry, the skeleton's geometry would not be completely determined.  As an analogous 2-dimensional example, consider the geometry of a quadrilateral where only its four external edge-lengths are known: there is a wide range of possible quadrilaterals that would have these four edge-lengths, ranging from trapezia to irregular quadrilaterals; without extra constraints, it would be impossible to determine a unique quadrilateral.

The CW 4-block's geometry is therefore constrained by requiring the tetrahedron to simply expand or contract uniformly about its centre as it evolves from one end of the 4-block to the other.  Brewin has indicated that such a geometry is equivalent to imposing the following two requirements on the 4-block:
\begin{enumerate*}[label=(\roman*)]
\item that all struts have the same length; and
\item that there be no twist or shear along the 4-block.
\end{enumerate*}
He has likened these requirements to a choice of lapse and shift function in the ADM formalism.  Indeed, the standard form of the FLRW metric \eqref{FLRW} also implies a certain foliation of FLRW space-time, and Collins and Williams' choice seems closest to the lapse and shift implicit in this foliation.

The 4-block geometry can be described by introducing a Cartesian co-ordinate system into the 4-block.  We shall denote by $\Cauchyt{i}$ the Cauchy surface at time $t_i$.  Let the tetrahedron in $\Cauchyt{i}$ have length $\Lamblen{i}:=\Lamblen{}(\RegTime{i})$, and label its vertices by $A$, $B$, $C$, and $D$.  The origin is then set to be at the tetrahedron's centre; the $x$-axis is taken to lie parallel to $AB$, the $y$-axis to pass through vertex $C$, and the $z$-axis to pass through vertex $D$.  The vertices' co-ordinates are therefore\footnote{We could more generally have used a time co-ordinate $\imath T_i:=\imath T(\RegTime{i})$ instead for the co-ordinates in \eqref{vertices}, but this would still lead to the same set of final equations.  So for simplicity, we choose to use $\RegTime{i}$ as the vertices' time co-ordinates.}
\begin{equation}
\renewcommand{\arraystretch}{2}
\begin{aligned}
A &= \displaystyle{\left(-\frac{\Lamblen{i}}{2}, -\frac{\Lamblen{i}}{2\sqrt{3}}, -\frac{\Lamblen{i}}{2\sqrt{6}}, \, \imath \RegTime{i} \right)},\\
B &=\displaystyle{\left(\frac{\Lamblen{i}}{2}, -\frac{\Lamblen{i}}{2\sqrt{3}}, -\frac{\Lamblen{i}}{2\sqrt{6}}, \, \imath \RegTime{i} \right)},\\
C &= \displaystyle{\left(0, \frac{\Lamblen{i}}{\sqrt{3}}, -\frac{\Lamblen{i}}{2\sqrt{6}}, \, \imath \RegTime{i} \right)},\\
D &= \displaystyle{\left(0, 0, \frac{\sqrt{3}\, \Lamblen{i}}{2\sqrt{2}}, \, \imath \RegTime{i} \right)}.
\end{aligned}
\label{vertices}
\end{equation}
To help simplify future calculations with this geometry, we have chosen to use a Euclidean metric; we have therefore introduced the imaginary unit $\imath$ in our time co-ordinate so that inner products would effectively yield a signature of $(+,+,+,-)$.

This tetrahedron will evolve to another of edge-length $\Lamblen{i+1}:=\Lamblen{}(\RegTime{i+1})$ in surface $\Cauchyt{i+1}$.  We shall denote the image of vertices $A$, $B$, $C$, $D$ in $\Cauchyt{i+1}$ by $A^\prime$, $B^\prime$, $C^\prime$, $D^\prime$, respectively.  Their co-ordinates are given by an analogous expression to \eqref{vertices}, where each vertex is replaced by its primed counterpart and each subscript $i$ by $i+1$.  This ensures the struts all have the same length.  Additionally, in this co-ordinate system, the tetrahedron would simply expand or contract uniformly about its centre in the spatial dimensions, as required.  \Figuref{fig:4block} illustrates this particular 4-block.  For simplicity, we shall sometimes refer to the tetrahedron in $\Cauchyt{i+1}$ as the upper tetrahedron and the one in $\Cauchyt{i}$ as the lower tetrahedron, as that is how they appear in the figure.

The CW constraints on the 4-block geometry are imposed in different manners depending on whether we are globally or locally varying the skeleton.  Under local variation, we must first completely triangulate the skeleton, thereby generating a completely simplicial manifold.  All edges in this fully triangulated skeleton are then considered independent of all others, including the newly introduced edges; when one edge is locally varied, all others are held constant.  After the Regge equations are obtained, the constraints on the geometry are then imposed; this is done by setting the various edges to have the appropriate lengths consistent with the CW 4-block geometry: that is, all tetrahedral edge-lengths in a Cauchy surface would be set equal; all strut-lengths between a pair of consecutive surfaces would be set equal; and all diagonal-lengths between a pair of consecutive surfaces would be set equal.

\begin{figure}[htb]
\input{allhinges.pspdftex}
\caption{\label{fig:allhinges}The world-sheets generated by the six tetrahedral edges and their triangulation into triangular time-like hinges.}
\end{figure}
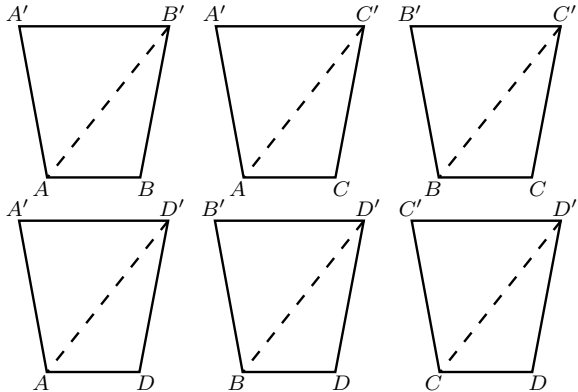

The specific 4-block represented by \FigRef{fig:4block} and co-ordinate system \eqref{vertices} is triangulated by introducing diagonals $AD^\prime$, $BD^\prime$, $CD^\prime$, $AC^\prime$, $BC^\prime$, and $AB^\prime$; this corresponds to one diagonal above each of the tetrahedral edges, as illustrated in \FigRef{fig:allhinges}, and these diagonals divide the 4-block into four distinct 4-simplices, $ABCDD^\prime$, $ABCC^\prime D^\prime$, $ABB^\prime C^\prime D^\prime$, and $AA^\prime B^\prime C^\prime D^\prime$.  One can generate a consistent triangulation of the entire skeleton by triangulating each of its 4-blocks in this manner.
The CW geometry is then imposed on a 4-block by requiring its lower tetrahedral edges to have length $\Lamblen{i}$, its upper tetrahedral edges to have length $\Lamblen{i+1}$, its struts to have length $\CWstrut{i}$ given by
\begin{align}
\CWstrut{i}^2 &= \left(\frac{3}{8}\,\Lamblendot{i}^{\,2} - 1 \right) \delta \RegTime{i}^2,\label{strut}\\
\intertext{where we have introduced the notation}
\Lamblendot{i} &:= \frac{\Lamblen{i+1}-\Lamblen{i}}{\RegTime{i+1}-\RegTime{i}},\nonumber
\end{align}
and its diagonals to have length $\CWdiag{i}$ given by
\begin{equation}
\CWdiag{i}^{\,2} = \frac{1}{3}\Lamblen{i}^{\,2} + \frac{1}{24}\left(3\, \Lamblen{i+1} + \Lamblen{i} \right)^2 - \delta \RegTime{i}^{\,2}.
\label{diag}
\end{equation}
These lengths are all derived from the co-ordinates \eqref{vertices} and its $\Cauchyt{i+1}$ counterpart.

Under global variation, the 4-block geometry is instead constrained to have the CW geometry before the skeleton gets varied, and this is done without introducing any new edges.  Global variation preserves the CW geometry because when a tetrahedral edge-length gets varied, all tetrahedral edges in the same Cauchy surface gets varied; in each 4-block, the varied tetrahedron simply gets re-scaled uniformly about its centre.  Similarly, when a strut-length gets varied, all struts between the same pair of surfaces gets varied.  In either case, each 4-block simply becomes a different CW 4-block.  We note that under global variation, each Cauchy surface would consist of only two independent edges, the tetrahedral edge and the strut.

As we mentioned in the Introduction, local variation is the standard way of doing Regge calculus; it is more similar to how standard general relativity is done.  In general relativity, global variation would be analogous to requiring the metric in the Einstein-Hilbert action to be of FLRW form \eqref{FLRW}, and then varying the action with respect to $\FriedScale(t)$; this imposes the Copernican symmetries prior to varying the Einstein-Hilbert action.  The standard approach would be to vary the Einstein-Hilbert action first, yielding the Einstein field equations, and then setting the metric to be of FLRW form.

Brewin \cite{Brewin} has explored to greater depth the relationship between solutions of the global Regge equations and solutions of the local Regge equations.  He has shown that, in general, global and local equations would not necessarily lead to the same set of solutions; rather the local solutions would, under certain circumstances, form a subset of the global solutions.  We shall demonstrate this relationship explicitly for the parent model later on.

\section{Parent models of closed vacuum $\CosmoConst$-FLRW universes}

\subsection{Embedding Cauchy surfaces into a 3-sphere}

As Collins and Williams first noted and Brewin fully explored, a CW Cauchy surface $\Cauchyt{i}$ can be embedded into $\mathbf{E}^4$ such that all vertices lie on a 3-sphere of radius $\Rad{i}$.  It is most natural to parametrise this 3-sphere using a set of polar co-ordinates $(\chi, \theta, \phi)$.  If $(x^1, x^2, x^3, x^4)$ is a set of Cartesian co-ordinates in $\mathbf{E}^4$, then points on the 3-sphere are can be parametrised by
\begin{equation}
\begin{aligned}
x^1 &= \Rad{i} \cos\chi,\\
x^2 &= \Rad{i} \sin\chi\cos\theta,\\
x^3 &= \Rad{i} \sin\chi\sin\theta\cos\phi,\\
x^4 &= \Rad{i} \sin\chi\sin\theta\sin\phi,
\end{aligned}
\label{3sphere}
\end{equation}
with $\chi, \theta \in [0,\pi]$ and $\phi \in [0,2\pi)$.  This embedding provides a natural framework within which to study and elucidate the underlying geometry of the CW Cauchy surface.  Most importantly, it makes clearer the relationship between the CW Cauchy surfaces and the FLRW 3-spheres they approximate.
\begin{figure}[htb]
\input{3sphere.pspdftex}
\caption{\label{fig:ParentEmbedding}A schematic diagram of an equilateral tetrahedron of edge-length $\Lamblen{i}$ embedded into a 3-sphere of radius $\Rad{i}$.  One dimension has been projected out.}
\end{figure}
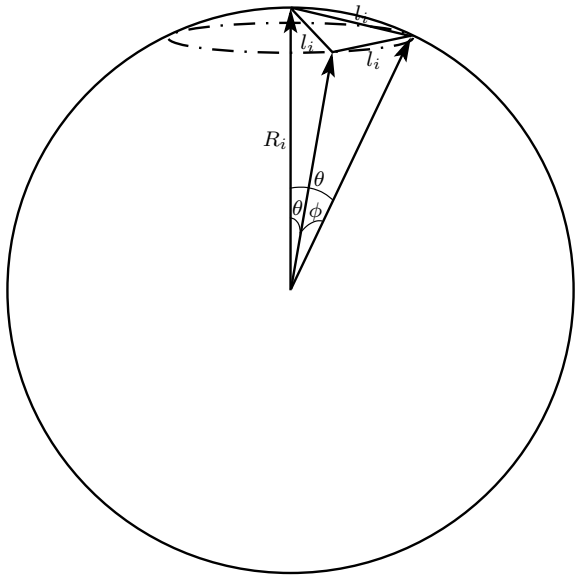

\begin{table} [tbh]
\renewcommand{\arraystretch}{1.2}
\caption{The polar co-ordinates of $n+2$ neighbouring vertices.}
\begin{tabular}{>{\centering\arraybackslash}m{2cm} >{\centering\arraybackslash}m{2cm} >{\centering\arraybackslash}m{2cm} >{\centering\arraybackslash}m{2cm}}
\hline\hline
Vertex & $\chi$ & $\theta$ & $\phi$\\
\hline
1 & 0 & 0 & 0\\
2 & $\chi_0$ & 0 & 0\\
3 & $\chi_1$ & $\theta_0$ & 0\\
4 & $\chi_1$ & $\theta_0$ & $\frac{2\pi}{n}$\\
$\vdots$ & $\vdots$ & $\vdots$ & $\vdots$\\
$n+2$ & $\chi_1$ & $\theta_0$ & $\frac{2(n-1)\pi}{n}$\\
\hline\hline
\end{tabular}
\label{tab:3sphvertices}
\end{table}

We begin by embedding a subset of the vertices into the 3-sphere; a schematic diagram of our embedding is given in \FigRef{fig:ParentEmbedding}.  By symmetry of the 3-sphere, we can always choose polar co-ordinates such that one vertex is located at $(\chi, \theta, \phi)=(0,0,0)$.  Then, we also have freedom to choose $\theta$ and $\phi$ co-ordinates such that one of the neighbouring vertices is at $(\chi, \theta, \phi)=(\chi_0,0,0)$ for some $\chi_0$.  If $n$ is the number of triangles meeting at an edge, then there will be $n$ vertices surrounding the edge formed by the first two vertices.  To see this, we can consider the first two vertices as forming a common base for the $n$ triangles and each of the $n$ vertices as forming the apex of each of the $n$ triangles.  The values for $n$ corresponding to the different models are listed in the final column of \tabref{tab:primary}.  We again have freedom to choose the $\phi$ co-ordinate such that one of these $n$ vertices is located at $(\chi, \theta, \phi)=(\chi_1,\theta_0,0)$ for some $\chi_1$ and $\theta_0$.  Once this choice is made, the remaining $n - 1$ vertices would be located at $\left(\chi_1,\theta_0, 2\pi / n \right)$, $\left(\chi_1,\theta_0, 4\pi / n \right)$, $\dots$, $\left(\chi_1,\theta_0, 2(n-1)\pi / n\right)$.  The co-ordinates of these vertices have been summarised in \tabref{tab:3sphvertices}.  By requiring all distances to be $\Lamblen{i}$ between any pair of neighbouring vertices, we obtain the equations
\begin{equation}
\begin{aligned}
(L_{12})^2 &= \Lamblen{i}^2 = 2\Rad{i}^2(1-\cos\chi_0)\\
(L_{13})^2 &= \Lamblen{i}^2 = 2\Rad{i}^2(1-\cos\chi_1)\\
(L_{23})^2 &= \Lamblen{i}^2 = 2\Rad{i}^2(1-\cos\chi_0\cos\chi_1 - \sin\chi_0\sin\chi_1\cos\theta_0)\\
(L_{34})^2 &= \Lamblen{i}^2 = 2\Rad{i}^2\sin^2\chi_1\sin^2\theta_0\left(1-\cos\frac{2\pi}{n}\right),
\end{aligned}
\end{equation}
where $L_{ij}$ denotes the distance between vertices $i$ and $j$.  We then solve these to obtain
\begin{IEEEeqnarray}{rCl}
\chi_0 &=& \chi_1,\\
\cos\chi_0 &=& \frac{\cos\frac{2\pi}{n}}{1-2\cos\frac{2\pi}{n}},\\
\cos\theta_0 &=& \frac{\cos\frac{2\pi}{n}}{1-\cos\frac{2\pi}{n}},\\
\SphRatio{0}:= \frac{\Lamblen{i}}{\Rad{i}} &=& \sqrt{2}\left(\frac{1-3\cos\frac{2\pi}{n}}{1-2\cos\frac{2\pi}{n}}\right)^{1/2}.
\label{Z0}
\end{IEEEeqnarray}
These are the relations Brewin \cite{Brewin} obtained for the parent models' embedding, although we have derived them here independently of him.

From this embedding, we see that as the edge-lengths $l(t_i)$ expand and contract, the 3-sphere simply expands and contracts about its centre, and the vertices simply move radially inwards or outwards accordingly.  The angular positions of these vertices remain constant.

Two interesting features of the Cauchy surface geometry come to light from this embedding.  First, we can now see that $\Lamblen{i}$ and $\Rad{i}$ are related by the constant ratio $\SphRatio{0}$.  Most notably, this ratio is independent of the label $i$ and hence of time $\RegTime{i}$.  Thus, we can define a radius $\Rad{}(\RegTime{i}) = \Lamblen{}(\RegTime{i})/\SphRatio{0}$ for our CW Cauchy surfaces, and this serves as a natural analogue to the FLRW scale factor $\FriedScale(t)$.  Secondly, we see that as the edge-lengths $\Lamblen{}(\RegTime{i})$ expand and contract, the 3-sphere simply expands and contracts about its centre, and the vertices simply move radially inwards or outwards accordingly; their angular positions remain constant.

Our embedding above has yielded one possible definition of radius for the CW Cauchy surface, namely the vertices' embedding radius.  This was the definition Brewin chose as his Regge analogue to the FLRW scale factor of $\FriedScale(t)$.  However, there are other equally plausible definitions of radius for the Cauchy surface: we could just as well have chosen the radius of any other point in the tetrahedra, as there will always be a set of points in the Cauchy surface sharing that same radius.  Some possibilities are the radius $\Rad{1}$ to the centres of edges
\begin{align}
\Rad{1}(\RegTime{i}) &= \frac{1}{\sqrt{2}}\left(\frac{1-\cos\frac{2\pi}{n}}{1-2\cos\frac{2\pi}{n}}\right)^{1/2} \Rad{}(\RegTime{i}),\label{PEdgeRad}\\
\SphRatio{1} := \frac{\Lamblen{}(\RegTime{i})}{\Rad{1}(\RegTime{i})} &= 2 \left(\frac{1-3\cos\frac{2\pi}{n}}{1-\cos\frac{2\pi}{n}}\right)^{1/2},
\intertext{the radius $\Rad{2}$ to the centres of triangles}
\Rad{2}(\RegTime{i}) &= \frac{1}{\sqrt{3}}\left(\frac{1}{1-2\cos\frac{2\pi}{n}}\right)^{1/2} \Rad{}(\RegTime{i}),\label{PFaceRad}\\
\SphRatio{2} := \frac{\Lamblen{}(\RegTime{i})}{\Rad{2}(\RegTime{i})} &= \sqrt{6} \left(1-3\cos\frac{2\pi}{n}\right)^{1/2},
\intertext{or the radius $\Rad{3}$ to the centres of tetrahedra}
\Rad{3}(\RegTime{i}) &= \frac{1}{2}\left(\frac{1+\cos\frac{2\pi}{n}}{1-2\cos\frac{2\pi}{n}}\right)^{1/2} \Rad{}(\RegTime{i}),\label{PCentrRad}\\
\SphRatio{3} := \frac{\Lamblen{}(\RegTime{i})}{\Rad{3}(\RegTime{i})} &= 2\sqrt{2} \left(\frac{1-3\cos\frac{2\pi}{n}}{1+\cos\frac{2\pi}{n}}\right)^{1/2}.
\end{align}
The main point to notice is that regardless of which radius we choose, the ratio between that radius and $\Lamblen{}(\RegTime{i})$ is always a constant independent of $\RegTime{i}$.

Since no particular choice of radius seems preferred, one natural choice would be to average over all radii across the entire Cauchy surface.  We have numerically computed this average radius $\Radbar{}$ in terms of $\Rad{}(\RegTime{i})$ to be
\begin{equation}
\Radbar{}(\RegTime{i}) = \begin{dcases*}
0.484066\,\Rad{}(\RegTime{i}) & 5 tetrahedra model,\\
0.627392\,\Rad{}(\RegTime{i}) & 16 tetrahedra model,\\
0.940901\,\Rad{}(\RegTime{i}) & 600 tetrahedra model.
\end{dcases*}
\label{AvgParentRad}
\end{equation}
The derivation of these numbers has been explained in \appendref{AvgRad}.

We can also consider the effective radius $\Radtilde{}(\RegTime{i})$ obtained by treating the volume of the Cauchy surface as if it were the volume of a 3-sphere.  The volume of a 3-sphere of radius $\Radtilde{}(\RegTime{})$ is
\begin{equation}
U(\RegTime{}) = 2\pi^2 \Radtilde{}(\RegTime{})^3,\label{3sphvol}
\end{equation}
while the volume of an $\Ntet$-tetrahedra CW universe is
\begin{equation}
U_\Ntet(\RegTime{i}) = \frac{\Ntet}{6\sqrt{2}} \, \Lamblen{}(\RegTime{i})^3.
\label{ParentVol}
\end{equation}
If we equate the two expressions, we find that the effective radius is
\begin{equation}
\Radtilde{}(\RegTime{i}) = \left(\frac{\Ntet}{12\sqrt{2}\pi^2}\right)^{1/3}\Lamblen{}(\RegTime{i}).\label{EffParentRad}
\end{equation}

For comparison with the average radii in \eqref{AvgParentRad}, the numerical values for $\Radtilde{}(t)$ in terms of the vertex radius $\Rad{}(\RegTime{i})$ are
\begin{equation}
\Radtilde{}(\RegTime{i}) = \begin{dcases*}
0.490488\,\Rad{}(\RegTime{i}) & 5 tetrahedra model,\\
0.646482\,\Rad{}(\RegTime{i}) & 16 tetrahedra model,\\
0.945651\,\Rad{}(\RegTime{i}) & 600 tetrahedra model,
\end{dcases*}
\end{equation}
and the fractional difference between these numerical factors and those in \eqref{AvgParentRad} are
\begin{equation}
\frac{\Radbar{}(\RegTime{i})-\Radtilde{}(\RegTime{i})}{\Radbar{}(\RegTime{i})} = \begin{dcases*}
-0.0132668 & 5 tetrahedra model,\\
-0.0304271 & 16 tetrahedra model,\\
-0.00504837 & 600 tetrahedra model.
\end{dcases*}
\end{equation}
The two radii are very close to each other, but with $\Radtilde{}(\RegTime{i})$ consistently greater than $\Radbar{}(\RegTime{i})$ by a slight amount.  

Finally, we shall consider one more possible definition of the radius when we have obtained the equations for $\Lamblen{}(t)$ and $\Lamblendot{}(t)$ in the continuum time limit.  Like all other radii, this radius $\Radhat{}(t)$ is related to $\Lamblen{}(t)$ by a constant $\SphRatiohat{}$,
\begin{equation}
\Radhat{}(t) = \SphRatiohat{}\, \Lamblen{}(t),
\label{R-hat}
\end{equation}
and we define $\SphRatiohat{}$ by requiring $\Radhat{} = \FriedScale(t)$ when both $d\Radhat{} / dt=0$ and $\dot{\FriedScale}=0$.

\subsection{Global variation of the parent models}

Under the global constraints, the CW skeleton simplifies significantly.  Most importantly, there are now only two distinct sets of hinges associated with any Cauchy surface.  The first corresponds to the world-sheets of the tetrahedral edges between surfaces $\Cauchyt{i}$ and $\Cauchyt{i+1}$.  These hinges are time-like and trapezoidal; an example would be hinge $ABA^\prime B^\prime$ in \FigRef{fig:4block}.  The second corresponds to the equilateral triangular faces of the tetrahedra.  These hinges are space-like and have edges of length $\Lamblen{i}$; an example would be the hinge $ABC$ in \FigRef{fig:4block}.

Thus for a skeleton satisfying the global constraints, the Regge action can be written as
\begin{equation}
8\pi \Action_{global} = \mkern-35mu \sum_{i \in \left\{\substack{\text{trapezoidal}\\\text{hinges}}\right\}} \mkern-37mu \TrapArea{i} \TrapDeficit{i} + \mkern-30mu \sum_{i \in \left\{\substack{\text{triangular}\\\text{hinges}}\right\}} \mkern-35mu \TriArea{i} \TriDeficit{i} - \CosmoConst \mkern-30mu \sum_{i \in \left\{\vphantom{\substack{\text{trapezoidal}\\\text{hinges}}}\text{4-blocks}\right\}} \mkern-20mu \FourVol{i}.
\label{GlobAction}
\end{equation}

We shall now derive the Regge equations by global variation.  As mentioned previously, there are only two distinct types of edges characterising a global skeleton, the struts and the tetrahedral edges.  If we vary the action with respect to a strut, we obtain the Hamiltonian constraint
\begin{equation}
0 = \Nedge \frac{\partial \TrapArea{i}}{\partial \CWstrut{i}} \TrapDeficit{i} - \Ntet \, \CosmoConst \frac{\partial \FourVol{i}}{\partial \CWstrut{i}},
\label{GlobConstraint0}
\end{equation}
where $\Nedge$ and $\Ntet$ are the numbers of edges and tetrahedra in a Cauchy surface and equal the numbers of trapezoidal hinges and 4-blocks, respectively, between any two consecutive Cauchy surfaces $\Cauchyt{i}$ and $\Cauchyt{i+1}$.  If we vary with respect to a tetrahedral edge, we obtain the evolution equation
\begin{equation}
\begin{split}
& \Nedge \left(\frac{\partial \TrapArea{i}}{\partial \Lamblen{i}} \TrapDeficit{i} + \frac{\partial \TrapArea{i-1}}{\partial \Lamblen{i}} \TrapDeficit{i-1} \right) + \Ntri \frac{\partial \TriArea{i}}{\partial \Lamblen{i}} \TriDeficit{i}\\
& \qquad {}= \Ntet \, \CosmoConst \left(\frac{\partial \FourVol{i}}{\partial \Lamblen{i}} + \frac{\partial \FourVol{i-1}}{\partial \Lamblen{i}} \right),
\end{split}
\label{GlobalEvol}
\end{equation}
where $\Ntri$ is the number of triangles in a Cauchy surface; all three numbers $\Nedge$, $\Ntri$, and $\Ntet$ are given in \tabref{tab:primary}.  From these equations, we see that there are only three types of geometric quantities relevant to the global Regge equations: the varied hinge areas, the corresponding deficit angles, and the varied 4-volumes.  We shall now derive each in turn.

The area of any trapezoidal hinge between $\Cauchyt{i}$ and $\Cauchyt{i+1}$ is
\begin{equation}
\TrapArea{i} = \frac{\imath}{2}(\Lamblen{i+1}+\Lamblen{i})\left[\frac{1}{4}(\Lamblen{i+1}-\Lamblen{i})^2 - \CWstrut{i}^2 \right]^{1/2},
\label{TrapArea}
\end{equation}
while the area of any triangular hinge in $\Cauchyt{i}$ is
\begin{equation}
\TriArea{i} = \frac{\sqrt{3}}{4}\, \Lamblen{i}^{\,2}.
\label{TriArea}
\end{equation}
If the two hinge areas are varied with respect to $\CWstrut{j}$, only the variation of $\TrapArea{i}$ will be non-zero, yielding
\begin{equation}
\frac{\partial \TrapArea{i}}{\partial \CWstrut{j}} = -\frac{\imath}{2}\CWstrut{i}(\Lamblen{i+1}+\Lamblen{i})\left[\frac{1}{4}(\Lamblen{i+1}-\Lamblen{i})^2 - \CWstrut{i}^2 \right]^{-1/2} \delta_{ij}.
\label{TrapAreaVaried}
\end{equation}
If the space-like triangular hinges are varied with respect to $\Lamblen{i}$, we obtain
\begin{equation}
\frac{\partial \TriArea{i}}{\partial \Lamblen{i}} = \frac{\sqrt{3}}{2}\Lamblen{i}.
\end{equation}
If the trapezoidal hinges are varied, there will actually be two sets of hinges that get affected because each edge $\Lamblen{i}$ is attached to two trapezoidal hinges, one between surfaces $\Cauchyt{i}$ and $\Cauchyt{i+1}$ and the other between $\Cauchyt{i}$ and $\Cauchyt{i-1}$.  Varying a `future' hinge with respect to $\Lamblen{i}$ yields
\begin{equation}
\frac{\partial \TrapArea{i}}{\partial \Lamblen{i}} = -\frac{\imath}{2} \frac{\frac{1}{2} \Lamblen{i}(\Lamblen{i+1}-\Lamblen{i})+\CWstrut{i}^2}{\sqrt{\frac{1}{4} (\Lamblen{i+1}-\Lamblen{i})^2-\CWstrut{i}^2}},
\end{equation}
and varying the corresponding `past' hinge yields
\begin{equation}
\frac{\partial \TrapArea{i-1}}{\partial \Lamblen{i}} = \frac{\imath}{2} \frac{\frac{1}{2} \Lamblen{i}(\Lamblen{i}-\Lamblen{i-1})-\CWstrut{i-1}^2}{\sqrt{\frac{1}{4} (\Lamblen{i}-\Lamblen{i-1})^2-\CWstrut{i-1}^2}}.
\end{equation}

In general, the deficit angle for any hinge would be given by \eqref{def_ang}.  But because all simplices are identical, the deficit angle on a trapezoidal hinge can be simplified to
\begin{equation}
\TrapDeficit{i} = 2 \pi - n \CWTrapdihedral{i},
\label{TrapDeficit}
\end{equation}
where $n$ is the number of faces meeting at the hinge and $\CWTrapdihedral{i}$ is the dihedral angle between any two adjacent faces.  Since each trapezoidal hinge corresponds to the world-sheet of a tetrahedral edge and each face on this hinge to the world-tube of a triangle at this edge, $n$ is equal to the number of triangles meeting at an edge; this number is listed in the last column of \tabref{tab:primary}.

Faces $ABCA^\prime B^\prime C^\prime$ and $ABDA^\prime B^\prime D^\prime$ meeting at hinge $ABA^\prime B^\prime$ will be separated by a dihedral angle of $\CWTrapdihedral{i}$; hence, $\CWTrapdihedral{i}$ can be determined from the scalar product of the two faces' unit normals.  Let $\Tensb{\hat{n}}_1$ denote the unit normal pointing into $ABCA^\prime B^\prime C^\prime$ and $\Tensb{\hat{n}}_2$ the unit normal out of $ABDA^\prime B^\prime D^\prime$; then in co-ordinate system \eqref{vertices}, they have components
\begin{IEEEeqnarray}{rCl}
\renewcommand{\arraystretch}{2.7}
\hat{n}_1^\mu &=& \displaystyle \frac{\left(0, 0, 1, -\imath \frac{1}{2\sqrt{6}}\, \Lamblendot{i} \right)}{\left(1-\frac{1}{24}\, \Lamblendot{i}^{\,2}\right)^{1/2}}\\
\shortintertext{and}
\hat{n}_2^\mu &=&\displaystyle \frac{\left(0, -2\sqrt{2}, 1, \imath \frac{\sqrt{3}}{2\sqrt{2}}\, \Lamblendot{i}\right)}{3\left(1-\frac{1}{24}\, \Lamblendot{i}^{\,2}\right)^{1/2}};
\end{IEEEeqnarray}
and therefore $\CWTrapdihedral{i}$ is given by
\begin{equation}
\cos \CWTrapdihedral{i} =\frac{1+\frac{1}{8}\, \Lamblendot{i}^{\,2}}{3-\frac{1}{8}\, \Lamblendot{i}^{\,2}}. \label{cosq}
\end{equation}

Four faces will meet at a triangular hinge in $\Cauchyt{i}$.  For hinge $ABC$ in \FigRef{fig:4block}, three of these faces are $ABCD$, $ABCA^\upharpoonleft B^\upharpoonleft C^\upharpoonleft$, and $ABCA^\downharpoonright B^\downharpoonright C^\downharpoonright$, where we use superscripts $\upharpoonleft$ and $\downharpoonright$ to denote the counterparts to vertices $A$, $B$, $C$ in $\Cauchyt{i+1}$ and $\Cauchyt{i-1}$, respectively.  The fourth face corresponds to the neighbouring tetrahedron, which we denote by $ABCE$.  By symmetry, $ABCD$ and $ABCE$ will form the same dihedral angle with $ABCA^\upharpoonleft B^\upharpoonleft C^\upharpoonleft$ and with $ABCA^\downharpoonright B^\downharpoonright C^\downharpoonright$; hence, there will only be two distinct dihedral angles surrounding this hinge.  We take $\CWTridihedral{\upharpoonleft}{i}$ to be the angle $ABCD$ and $ABCE$ form with $ABCA^\upharpoonleft B^\upharpoonleft C^\upharpoonleft$, and $\CWTridihedral{\downharpoonright}{i}$ to be the angle they form with $ABCA^\downharpoonright B^\downharpoonright C^\downharpoonright$.  Thus the deficit angle of $ABC$ is
\begin{equation}
\TriDeficit{i} = 2\pi - 2\CWTridihedral{\upharpoonleft}{i} - 2\CWTridihedral{\downharpoonright}{i}.
\end{equation}

\begin{figure}[htbp]
\input{hingeschem.pspdftex}
\caption{\label{fig:hingeschem}Two faces, separated by a dihedral angle of $\theta^{(j)}$, meet at a hinge.  The system has been projected onto the plane orthogonal to the hinge.  The deficit angle can be computed by taking the scalar product of the two vectors orthogonal to the faces, $\Tensb{n}_1$ and $\Tensb{n}_2$, or by taking the product of the two vectors tangent to the faces but orthogonal to the hinge, $\Tensb{u}_1$ and $\Tensb{u}_2$.  However, if we use the orthogonal vectors, we must be careful about their relative sign, otherwise we may end up taking the product with the incorrect vector, as exemplified by $-\Tensb{n}_2$.  Instead, we have no such uncertainty if we work with the tangent vectors.}
\end{figure}
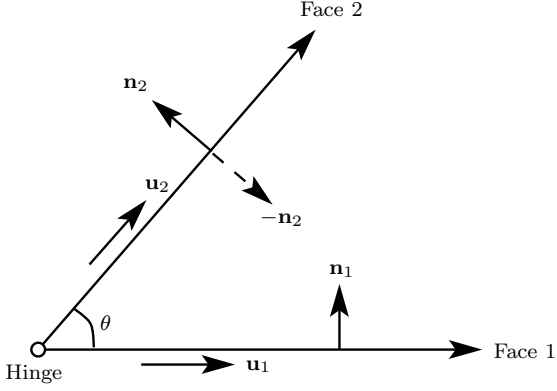

We just used vectors orthogonal to faces to calculate the previous dihedral angle, but with this approach, there may be uncertainties over the vectors' correct relative sign.  So we shall take a slightly different approach here; we shall instead use unit vectors tangent to the two faces but orthogonal to the hinge.  Since the hinge has co-dimension 2, there will always be a unique tangent vector satisfying these constraints for each face.  Our approach is depicted schematically in \FigRef{fig:hingeschem}.

We shall deduce $\CWTridihedral{\upharpoonleft}{i}$ first.  In the 4-block co-ordinates of \eqref{vertices}, the unit vector tangent to $ABCD$ but orthogonal to $ABC$ is simply
$$
\hat{u}_{ABCD}^\mu = (0,0,1,0),
$$
while the equivalent vector tangent to $ABCA^\upharpoonleft B^\upharpoonleft C^\upharpoonleft$ is
$$
\hat{u}_{ABCA^\upharpoonleft B^\upharpoonleft C^\upharpoonleft}^\mu = \frac{\left(0,\, 0,\, - \frac{1}{2\sqrt{6}}\Lamblendot{i}, \, \imath \right)}{\sqrt{\frac{1}{24}\Lamblendot{i}^2-1}}.
$$
Thus $\CWTridihedral{\upharpoonleft}{i}$ is given by
\begin{equation}
\cos \CWTridihedral{\upharpoonleft}{i} = -\frac{\frac{1}{2\sqrt{6}}\Lamblendot{i}}{\sqrt{\frac{1}{24}\Lamblendot{i}^2-1}}.
\end{equation}

By swapping $\Lamblen{i+1}$ for $\Lamblen{i-1}$, which appears implicitly in $\Lamblendot{i}$, we immediately obtain the corresponding expression for $\CWTridihedral{\downharpoonright}{i}$, that is,
\begin{equation}
\cos \CWTridihedral{\downharpoonright}{i} = \frac{\frac{1}{2\sqrt{6}}\Lamblendot{i-1}}{\sqrt{\frac{1}{24}\Lamblendot{i-1}^2-1}}.
\end{equation}

The final geometric quantities required for the Regge equations are the 4-block volumes.  The volume $\FourVol{i}$ of $ABCDA^\prime B^\prime C^\prime D^\prime$ is given by
\begin{equation}
\FourVol{i} = \frac{\imath}{24\sqrt{2}} ( \Lamblen{i+1}^2+\Lamblen{i}^2 ) ( \Lamblen{i+1}+\Lamblen{i} ) \left[\frac{3}{8}(\Lamblen{i+1}-\Lamblen{i})^2 -\CWstrut{i}^2 \right]^{1/2} \!\!.
\end{equation}
Varying this with respect to $\CWstrut{j}$ yields
\begin{equation}
\begin{split}
\frac{\partial \FourVol{i}}{\partial \CWstrut{j}} ={} & -\frac{\imath}{24\sqrt{2}} \, \CWstrut{i} \, (\Lamblen{i+1}^2+\Lamblen{i}^2 ) ( \Lamblen{i+1}+\Lamblen{i} ) \\
& {} \times \left[\frac{3}{8}(\Lamblen{i+1}-\Lamblen{i})^2 -\CWstrut{i}^2 \right]^{-1/2} \mkern-5mu \delta_{ij}.
\end{split}
\end{equation}
When entire 4-blocks are varied with respect to the tetrahedral edge-lengths, the situation is similar to the trapezoidal hinges: each edge $\Lamblen{j}$ is associated with a `past' 4-block between $\Cauchyt{j}$ and $\Cauchyt{j-1}$ and a `future' 4-block between $\Cauchyt{j}$ and $\Cauchyt{j+1}$.  Varying a `future' 4-block yields
\begin{equation}
\begin{split}
\frac{\partial \FourVol{i}}{\partial \Lamblen{j}} ={} & - \delta_{ij} \frac{\imath}{24\sqrt{2}} \\
& {} \times \! \left( \frac{\frac{3}{2}\, \Lamblen{i}^3 \, (\Lamblen{i+1} \! - \! \Lamblen{i}) \! + \! \CWstrut{i}^2 \, (\Lamblen{i+1}^2 \! + \! 2 \, \Lamblen{i+1}\Lamblen{i} \! + \! 3 \, \Lamblen{i}^2)}{\sqrt{\frac{3}{8}(\Lamblen{i+1} \! - \! \Lamblen{i})^2 \! - \! \CWstrut{i}^2}}\right) \! ,
\end{split}
\end{equation}
and varying a `past' 4-block yields
\begin{equation}
\begin{split}
\frac{\partial \FourVol{i-1}}{\partial \Lamblen{j}} ={} & \delta_{ij} \frac{\imath}{24\sqrt{2}} \\
& \! {} \times \!\! \left( \frac{\frac{3}{2}\, \Lamblen{i}^3 \, (\Lamblen{i} \! - \! \Lamblen{i-1}) \! - \! \CWstrut{i-1}^2 \, (\Lamblen{i-1}^2 \! + \! 2 \, \Lamblen{i-1}\Lamblen{i} \! + \! 3 \, \Lamblen{i}^2)}{\sqrt{\frac{3}{8}(\Lamblen{i} \! - \! \Lamblen{i-1})^2 \! -\! \CWstrut{i-1}^2}}\right) \! .
\end{split}
\end{equation}

We can now substitute these geometric quantities into the Regge equations above.  For the moment, we shall only do this for the Hamiltonian constraint \eqref{GlobConstraint0}, which yields
\begin{equation}
\Lamblen{i+1}^2+\Lamblen{i}^2 = 12\sqrt{2}\, \frac{\Nedge}{\Ntet \, \CosmoConst} \! \left(\frac{\frac{3}{8}(\Lamblen{i+1}-\Lamblen{i})^2 - \CWstrut{i}^2}{\frac{1}{4}(\Lamblen{i+1}-\Lamblen{i})^2 - \CWstrut{i}^2} \right)^{1/2} \!\!\!\!\! \left(2\pi - n\CWTrapdihedral{i}\right).
\label{GlobalParent}
\end{equation}
The other equation simplifies greatly in the continuum time limit, so we shall only present its continuum time form later on.

We now take the continuum time limit, where $\delta \RegTime{i}\to 0$, to obtain a differential equation for $\Lamblen{}(t)$.  In this limit, the tetrahedral edge-lengths and dihedral angles take the form
\begin{IEEEeqnarray*}{rCl}
\Lamblen{i} &\to& \Lamblen{}(t),\\
\Lamblen{i+1} &\to& \Lamblen{}(t) + \Lamblendot{}\, dt + \frac{1}{2}\, \Lamblenddot{}\, dt^2 + \Odt{3},\\
\Lamblen{i-1} &\to& \Lamblen{}(t) - \Lamblendot{}\, dt + \frac{1}{2}\, \Lamblenddot{}\, dt^2 + \Odt{3},\\
\CWTrapdihedral{i} &\to& \CWTrapdiS{\upharpoonleft}{}(t) + \CWTrapdiSdot{\upharpoonleft}{} \, dt + \Odt{2},\\
\CWTrapdihedral{i-1} &\to& \CWTrapdiS{\downharpoonright}{}(t) + \CWTrapdiSdot{\downharpoonright}{} \, dt + \Odt{2},\\
\CWTridihedral{\upharpoonleft}{i} &\to& \CWTridihedral{\upharpoonleft}{}(t) + \CWTridihedraldot{\upharpoonleft}{} \, dt + \Odt{2},\\
\CWTridihedral{\downharpoonright}{i} &\to& \CWTridihedral{\downharpoonright}{}(t) + \CWTridihedraldot{\downharpoonright}{} \, dt + \Odt{2},
\end{IEEEeqnarray*}
where in the context of continuum time, the overdot denotes a time-derivative.  The various geometric quantities now become
\begin{widetext}
\begin{IEEEeqnarray*}{rCl}
\Lamblendot{i} &\to& \Lamblendot{}\, dt + \frac{1}{2}\, \Lamblenddot{}\, dt^2 + \Odt{3},\\
\CWstrut{i}^2 &\to& \CWstrutdot{}^2 \, dt^2 + \Odt{3},\\
&& \textrm{where } \CWstrutdot{}^2 := \left(\frac{3}{8}\, \Lamblendot{}^2 - 1 \right),\\
\frac{\partial \TrapArea{i}}{\partial \CWstrut{i}} &\to& \Lamblen{} \, \CWstrutdot{} \left( \frac{1}{8}\, \Lamblendot{}^{\,2} - 1 \right)^{-1/2} + \Odtone,\\
\frac{\partial \TrapArea{i}}{\partial \Lamblen{i}} &\to& -\frac{\imath}{2}\frac{1}{\left(1-\frac{1}{8}\Lamblendot{}^2\right)^{1/2}} \left[\frac{1}{2} \Lamblen{} \Lamblendot{} \vphantom{\frac{dt}{1-\frac{1}{8}\Lamblendot{}^2}} + \frac{dt}{1-\frac{1}{8}\Lamblendot{}^2} \left(\frac{1}{4} \Lamblen{} \Lamblenddot{} + \frac{1}{2}\Lamblendot{}^2 - \frac{3}{64}\Lamblendot{}^4 -1 \right)\right] + \Odt{2},\\
\frac{\partial \TrapArea{i-1}}{\partial \Lamblen{i}} &\to& \frac{\imath}{2}\frac{1}{\left(1-\frac{1}{8}\Lamblendot{}^2\right)^{1/2}} \left[\frac{1}{2} \Lamblen{} \Lamblendot{} \vphantom{\frac{dt}{1-\frac{1}{8}\Lamblendot{}^2}} - \frac{dt}{1-\frac{1}{8}\Lamblendot{}^2} \left(\frac{1}{4} \Lamblen{} \Lamblenddot{} + \frac{1}{2}\Lamblendot{}^2 - \frac{3}{64}\Lamblendot{}^4 -1 \right)\right] + \Odt{2},\\
\frac{\partial \TriArea{i}}{\partial \Lamblen{i}} &\to& \frac{\sqrt{3}}{2} \Lamblen{},\\
\cos \CWTrapdihedral{i} &\to& \frac{1+\frac{1}{8}\Lamblendot{}^2}{3-\frac{1}{8}\Lamblendot{}^2} + \frac{\frac{1}{2}\Lamblendot{}\,\Lamblenddot{}}{\left(3-\frac{1}{8}\Lamblendot{}^2\right)^2}\, dt + \Odt{2},\\
\cos \CWTrapdihedral{i-1} &\to& \frac{1+\frac{1}{8}\Lamblendot{}^2}{3-\frac{1}{8}\Lamblendot{}^2} - \frac{\frac{1}{2}\Lamblendot{}\,\Lamblenddot{}}{\left(3-\frac{1}{8}\Lamblendot{}^2\right)^2}\, dt + \Odt{2},\\
\CWTrapdiS{\upharpoonleft}{} = \CWTrapdiS{\downharpoonright}{} &=& \arccos\left(\frac{1+\frac{1}{8}\Lamblendot{}^2}{3-\frac{1}{8}\Lamblendot{}^2} \right) + \Odtone,\\
\CWTrapdiSdot{\upharpoonleft}{} = -\CWTrapdiSdot{\downharpoonright}{} &=& -\frac{1}{4\sqrt{2}}\frac{1}{\left(1-\frac{1}{8}\Lamblendot{}^2\right)^{1/2}}\frac{\Lamblendot{}\,\Lamblenddot{}}{3-\frac{1}{8}\Lamblendot{}^2} \, dt + \Odt{2},\\
\cos \CWTridihedral{\upharpoonleft}{i} &\to& \frac{\imath}{2\sqrt{6}}\frac{1}{\left(1-\frac{1}{24}\Lamblendot{}^2\right)^{1/2}}\left(\Lamblendot{} + \frac{\frac{1}{2}\Lamblenddot{} dt}{1-\frac{1}{24}\Lamblendot{}^2} \right) + \Odt{2},\\
\cos \CWTridihedral{\downharpoonright}{i} &\to& -\frac{\imath}{2\sqrt{6}}\frac{1}{\left(1-\frac{1}{24}\Lamblendot{}^2\right)^{1/2}}\left(\Lamblendot{} - \frac{\frac{1}{2}\Lamblenddot{} dt}{1-\frac{1}{24}\Lamblendot{}^2} \right) + \Odt{2},\\
\CWTridihedral{\upharpoonleft}{} = \pi - \CWTridihedral{\downharpoonright}{} &=& \arccos\left(\frac{\frac{\imath}{2\sqrt{6}} \, \Lamblendot{}}{\left(1-\frac{1}{24}\Lamblendot{}^2\right)^{1/2}} \right) + \Odtone,\\
\CWTridihedraldot{\upharpoonleft}{} = \CWTridihedraldot{\downharpoonright}{} &=& -\frac{\imath}{4\sqrt{6}}\frac{\Lamblenddot{}}{1-\frac{1}{24}\Lamblendot{}^2} \, dt + \Odt{2},\\
\frac{\partial \FourVol{i}}{\partial \CWstrut{i}} &\to& -\frac{\imath}{6\sqrt{2}}\, \Lamblen{}^3\, \left(\frac{3}{8}\, \Lamblendot{}^2 - 1 \right)^{1/2} + \Odtone,\\
\frac{\partial \FourVol{i}}{\partial \Lamblen{i}} &\to& -\frac{\imath}{24\sqrt{2}} \left\{ \frac{3}{2}\Lamblen{}^3\Lamblendot{} + dt\left[\frac{3}{4} \Lamblen{}^3\Lamblenddot{} + 6 \Lamblen{}^2 \left(\frac{3}{8}\Lamblendot{}^2-1\right)\right] \right\} + \Odt{2},\\
\frac{\partial \FourVol{i-1}}{\partial \Lamblen{i}} &\to& \frac{\imath}{24\sqrt{2}} \left\{ \frac{3}{2}\Lamblen{}^3\Lamblendot{} - dt\left[\frac{3}{4} \Lamblen{}^3\Lamblenddot{} + 6 \Lamblen{}^2 \left(\frac{3}{8}\Lamblendot{}^2-1\right)\right] \right\} + \Odt{2}.
\end{IEEEeqnarray*}
\end{widetext}
Since $\CWTrapdiS{\upharpoonleft}{} = \CWTrapdiS{\downharpoonright}{}$, we shall henceforth denote this quantity simply by $\CWTrapdihedral{}$.  We can invert the continuum time expression for $\CWTrapdihedral{}$ above to parameterise $\Lamblendot{}^2$ in terms of $\CWTrapdihedral{}$, thus yielding
\begin{equation}
\Lamblendot{}^{\,2} = 8\left[1-2 \, \tan^2\left(\frac{1}{2}\, \CWTrapdihedral{}\right)\right].
\label{ldot}
\end{equation}

After substituting these results into the Regge equations \eqref{GlobalParent} and \eqref{GlobalEvol}, we obtain, to leading order in $dt$,
\begin{align}
\Lamblen{}^2 ={} & 6\sqrt{2}\, \frac{\Nedge}{\Ntet \, \CosmoConst} \frac{(2\pi - n\CWTrapdihedral{})}{\left(1-\frac{1}{8}\Lamblendot{}^2\right)^{1/2}},\label{GlobConstraint}\\
\begin{split}
0 ={} & \frac{\Nedge}{1-\frac{1}{8}\Lamblendot{}^2}\left[\frac{(2\pi-n\CWTrapdihedral{})}{\left(1-\frac{1}{8}\Lamblendot{}^2\right)^{1/2}}\left(\frac{1}{4} \Lamblen{} \Lamblenddot{}+\frac{1}{2}\Lamblendot{}^2-\frac{3}{64}\Lamblendot{}^4-1\right)\right.\\
& \left. \hphantom{\frac{\Nedge}{1-\frac{1}{8}\Lamblendot{}^2}\left[\right.} {}+ \frac{n}{8\sqrt{2}}\frac{\Lamblen{} \Lamblendot{}^2 \Lamblenddot{}}{3-\frac{1}{8}\Lamblendot{}^2} \vphantom{\frac{(2\pi-n\CWTrapdihedral{})}{\left[1-\frac{1}{8}\Lamblendot{}^2\right]^{1/2}}} \right] - \frac{\Ntri}{2\sqrt{2}}\frac{\Lamblen{} \Lamblenddot{}}{1-\frac{1}{24}\Lamblendot{}^2}\\
& {}- \frac{\Ntet \, \CosmoConst}{12\sqrt{2}} \, \Lamblen{}^2 \left[\frac{3}{4} \Lamblen{} \Lamblenddot{} + 6\left(\frac{3}{8}\Lamblendot{}^2-1\right)\right].
\end{split} \label{GlobEvolEqn}
\end{align}
Once again, the first equation is the Hamiltonian constraint while the second is the evolution equation.  From the Hamiltonian constraint, we see that one need only specify $l(t=0) = l_0$ as initial data on some initial Cauchy surface, and then $\dot{l}(t=0)$ will follow from the constraint.

It can be shown that the Hamiltonian constraint is actually a first integral of the evolution equation; a proof has been provided in \appendref{FirstIntegralProof}.  This implies that the Hamiltonian constraint is sufficient to determine the evolution of $\Lamblen{}(t)$, so we shall henceforth work with \eqref{GlobConstraint} only.

Using \eqref{cosq}, we can express $\Lamblen{}(t)$ and $\Lamblendot{}(t)$ parametrically in terms of $\theta$ to obtain
\begin{equation}
\Lamblen{}^2 = 6\, \frac{\Nedge}{\Ntet \, \CosmoConst} \frac{(2\pi - n\CWTrapdihedral{})}{\tan\left(\frac{1}{2}\CWTrapdihedral{}\right)} \label{parentl}.
\end{equation}
For the strut-length to be time-like, that is, for $\CWstrut{}(t)^2 < 0$, we require $\CWTrapdihedral{} > \pi / 3$, and for $\Lamblendot{}$ to be real, we require $\CWTrapdihedral{} \leq 2 \arctan (1/\sqrt{2} )$.  Hence, $\CWTrapdihedral{}$ must lie in the range
\begin{equation}
\frac{\pi}{3} < \CWTrapdihedral{} \leq 2 \arctan\left( \frac{1}{\sqrt{2}}\right).
\end{equation}

Now that we have $\Lamblen{}(t)$ and $\Lamblendot{}(t)$, we can determine $\Radhat{}(t)$.  Recall that we chose the integration constants in $\FriedScale(t)$ so that $\dot{\FriedScale}=0$ when $t=0$.  At $t=0$, we have that $\FriedScale^2=3 / \CosmoConst$.  On the other hand, $d\Radhat{} / dt$ will be zero when $\Lamblendot{}=0$, which happens when $\CWTrapdihedral{0} = 2 \, \arctan(1 / \sqrt{2} )$.  Inserting $\CWTrapdihedral{0}$ into \eqref{parentl}, we have that 
$$\Lamblen{0}^2 = \frac{3}{\CosmoConst}\frac{2\sqrt{2}\, \Nedge}{\Ntet} \left[2\pi - 2 n\arctan\left(\frac{1}{\sqrt{2}}\right)\right].$$
Therefore, we find that
\begin{equation}
\Radhat{}(t) = \left(\frac{\Ntet}{2\sqrt{2} \, \Nedge} \frac{1}{\left[2\pi - 2 n\arctan\left(\frac{1}{\sqrt{2}}\right)\right]}\right)^{1/2}\Lamblen{}(t).
\label{PRhat}
\end{equation}

\subsection{Local variation of the parent models}

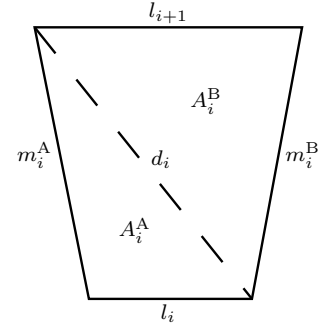
\begin{figure}[bht]
\input{traphinge.pspdftex}
\caption{\label{fig:hingefig}A diagonal $\CWdiag{i}$ divides the time-like hinge into a lower and upper triangular hinge, labelled A and B, respectively.  The two struts on the sides of the hinge are now considered independent quantities.}
\end{figure}

For comparison, we shall now derive the Regge equations by locally varying the action.  As mentioned earlier, this requires fully triangulating our skeleton; each trapezoidal hinge will now be divided by a diagonal into two triangular time-like hinges.  We shall label the lower triangular hinge by A and the upper by B, as depicted in \FigRef{fig:hingefig}.

After triangulation, the skeleton's hinges now comprise the time-like triangular hinges, the space-like triangular faces of the tetrahedra, and triangular hinges formed by two diagonal edges for its sides and one tetrahedral edge for its base.  This last set of hinges, an example being $ABC^\prime$, are new in that they have no counterpart in the global skeleton, unlike the first two sets.  When the lengths of the diagonals are set to be equal, these hinges reduce to a set of isosceles triangles.  It can be shown though that the corresponding deficit angles for these hinges are zero, so by virtue of the Schl\"afli identity, they will not contribute to the Regge equations.  Thus we can drop them from the action.

Therefore the Regge action for the triangulated skeleton can be written as
\begin{equation}
8 \pi\, \Action_{local} = \mkern-40mu \sum_{i \in \left\{\substack{\text{trapezoidal}\\\text{hinges}}\right\}} \mkern-40mu \left( \AreaA{i} \DeficitA{i} \! + \! \AreaB{i} \DeficitB{i} \right) + \mkern-35mu \sum_{i \in \left\{\substack{\text{triangular}\\\text{hinges}}\right\}} \mkern-35mu \TriArea{i} \TriDeficit{i} - \CosmoConst \mkern-30mu \sum_{i \in \left\{\vphantom{\substack{\text{trapezoidal}\\\text{hinges}}}\text{4-blocks}\right\}} \mkern-30mu \FourVol{i} \! ,\label{LocalAction}
\end{equation}
where, again, the superscripts A and B denote the lower and upper triangular subdivisions of the trapezoidal hinge as depicted in \FigRef{fig:allhinges}.  Here, we continue using $\FourVol{i}$ as a short-hand to denote the sum of the entire 4-block volume even though the 4-block is actually triangulated.

We shall focus on obtaining the Hamiltonian constraint first, that is, on varying with respect to the strut-lengths.  Locally varying with respect to an arbitrary strut-length $\CWstrut{j}$, we obtain an equation of the form
\begin{equation}
0 = \N{\text{A}} \frac{\partial \AreaA{i}}{\partial \CWstrut{j}} \DeficitA{i} + \N{\text{B}} \frac{\partial \AreaB{i}}{\partial \CWstrut{j}} \DeficitB{i} - \CosmoConst \frac{\partial \FourVolbar{i}}{\partial \CWstrut{j}},
\label{LocalConstraint}
\end{equation}
where
\begin{widetext}
\begin{equation*}
\frac{\partial \FourVolbar{i}}{\partial \CWstrut{j}} = \N{AA^\prime B^\prime C^\prime D^\prime} \frac{\partial \VolSub{AA^\prime B^\prime C^\prime D^\prime}{}}{\partial \CWstrut{j}} + \N{ABB^\prime C^\prime D^\prime} \frac{\partial \VolSub{ABB^\prime C^\prime D^\prime}{}}{\partial \CWstrut{j}} + \N{ABCC^\prime D^\prime} \frac{\partial \VolSub{ABCC^\prime D^\prime}{}}{\partial \CWstrut{j}} + \N{ABCDD^\prime} \frac{\partial \VolSub{ABCDD^\prime}{}}{\partial \CWstrut{j}},
\end{equation*}
\end{widetext}
and where $\VolSub{X}{}$ denotes the volume of 4-simplex $X$ situated between $\Cauchyt{i}$ and $\Cauchyt{i+1}$.  $\N{\text{A}}$ and $\N{\text{B}}$ are the numbers of lower and upper time-like triangular hinges meeting at $\CWstrut{j}$, while $\N{AA^\prime B^\prime C^\prime D^\prime}$, $\N{ABB^\prime C^\prime D^\prime}$, $\N{ABCC^\prime D^\prime}$, and $\N{ABCDD^\prime}$ are the numbers of 4-simplices corresponding, respectively, to $AA^\prime B^\prime C^\prime D^\prime$, $ABB^\prime C^\prime D^\prime$, $ABCC^\prime D^\prime$, and $ABCDD^\prime$ meeting at $\CWstrut{j}$ as well.  This equation only depends on geometric quantities involving the time-like hinges and the volumes of the triangulated 4-block.

We first consider the pair of triangular time-like hinges.  Using Heron of Alexandria's formula of classical antiquity, we can express their areas $\AreaA{i}$ and $\AreaB{i}$ in terms of the edge-lengths giving
\begin{align}
\begin{split}
\AreaA{i} ={} & \frac{1}{4} \left\{-\Lamblen{i}^{\,4} - \left(\CWstrut{i}^\text{A}\right)^4 - \CWdiag{i}^{\,4} \right.\\
& \left. \hphantom{\frac{1}{4} [} {} + 2\left[\Lamblen{i}^{\,2} \CWdiag{i}^{\,2} + \Lamblen{i}^{\,2} \left(\CWstrut{i}^\text{A}\right)^2  + \CWdiag{i}^{\,2} \left(\CWstrut{i}^\text{A}\right)^2 \right] \right\}^{1/2},
\end{split}\label{AreaA}\\
\begin{split}
\AreaB{i} ={} & \frac{1}{4} \left\{-\Lamblen{i+1}^{\,4} - \left(\CWstrut{i}^\text{B}\right)^4 - \CWdiag{i}^{\,4} \right.\\
& \left. \hphantom{\frac{1}{4} [} {}+ 2\left[\Lamblen{i+1}^{\,2} \CWdiag{i}^{\,2} + \Lamblen{i+1}^{\,2} \left(\CWstrut{i}^\text{B}\right)^2  + \CWdiag{i}^{\,2} \left(\CWstrut{i}^\text{B} \right)^2 \right] \right\}^{1/2}.
\end{split}\label{AreaB}
\end{align}
Differentiating with respect to their respective strut-lengths, we obtain
\begin{IEEEeqnarray}{rCl}
\frac{\partial \AreaA{i}}{\partial \CWstrut{j}^\text{A}} &=& \frac{\CWstrut{i}^\text{A}}{8 \AreaA{i}} \left[\Lamblen{i}^{\,2} + \CWdiag{i}^{\,2} - \left(\CWstrut{i}^\text{A}\right)^2 \right] \delta_{ij},\label{dAreaAdmA}\\
\frac{\partial \AreaB{i}}{\partial \CWstrut{j}^\text{B}} &=& \frac{\CWstrut{i}^\text{B}}{8 \AreaB{i}} \left[\Lamblen{i+1}^{\,2} + \CWdiag{i}^{\,2} - \left(\CWstrut{i}^\text{B}\right)^2 \right] \delta_{ij}\label{dAreaBdmB}.
\end{IEEEeqnarray}
If we now substitute in the strut and diagonal lengths given by \eqref{strut} and \eqref{diag}, and then take the continuum time limit, these derivatives simplify to
\begin{equation}
\frac{\partial \AreaA{i}}{\partial \CWstrut{j}^\text{A}} = \frac{\partial \AreaB{i}}{\partial \CWstrut{j}^\text{B}} = \frac{\Lamblen{}}{2} \frac{\CWstrutdot{}}{\left(\frac{1}{8}\Lamblendot{}^{\,2}-1\right)^{1/2}} \delta_{ij} + \Odtone,\label{dAreaTridmCont}
\end{equation}
where again, in the context of continuum time, the over-dot denotes a time-derivative.

After edge-lengths of the same type have been set equal, the deficit angles $\DeficitA{i}$ and $\DeficitB{i}$ of the time-like triangular hinges become identical to the deficit angle $\TrapDeficit{i}$ of the original trapezoidal hinge; that is, $\DeficitA{i} = \DeficitB{i} = \TrapDeficit{i}$.  Setting the edge-lengths equal makes the two triangular hinges be co-planar both with each other and with the original trapezoidal hinge; thus the 4-blocks meeting at the triangular hinges would be flat; the unit normals of the triangulated faces would be identical to the unit normals of the original faces; and the dihedral angles between the triangulated faces would be identical to the dihedral angles between the original faces.  Since the number of faces meeting at the triangulated hinge is the same as the number of faces at the original hinge, the deficit angles for the triangulated hinge and the original hinge are identical.  Additionally, since the dihedral angles are unchanged, then in the continuum time limit, $\Lamblendot{}$ would still be given by \eqref{ldot}.

We next consider the volumes of the 4-simplices.  To compute these volumes, we shall use the comparatively modern Cayley-Menger determinant instead, which generalises Heron's formula from areas of triangles to volumes of $n$-simplices.  Suppose we label the vertices of an $n$-simplex with numbers 0 to $n$; then the simplex's volume can be computed by the formula
\begin{equation}
\VolSub{(n)}{} = \left[\frac{(-1)^{n+1}}{2^n(n!)^2} \det (B) \right]^{1/2},
\label{4-vol}
\end{equation}
where $B$ is a symmetric matrix given by
\begin{equation}
B_{ij} =
\begin{dcases}
l_{i-2 \; j-2}^2 &\text{for } i,j > 1,\\
0 & \text{for } i=j=1,\\
1 & \text{otherwise},
\end{dcases}
\end{equation}
and where $l_{ij}=l_{ji}$ is the distance between vertices $i$ and $j$.  So for the volume of a 4-simplex, we have
\begin{equation}
\VolSub{(4)}{} = \frac{1}{96}\sqrt{-\det (B)},
\end{equation}
and
\begin{equation}
B = \left(
\begin{array}{cccccc}
0 & 1 & 1 & 1 & 1 & 1\\
1 & 0 & l_{01}^2 & l_{02}^2 & l_{03}^2 & l_{04}^2\\
1 & l_{01}^2 & 0 & l_{12}^2 & l_{13}^2 & l_{14}^2\\
1 & l_{02}^2 & l_{12}^2 & 0 & l_{23}^2 & l_{24}^2\\
1 & l_{03}^2 & l_{13}^2 & l_{23}^2& 0 & l_{34}^2\\
1 & l_{04}^2 & l_{14}^2 & l_{24}^2 & l_{34}^2 & 0
\end{array}
\right).
\label{4CayleyMenger}
\end{equation}

We now differentiate each 4-simplex volume with respect to its strut-length.  In our 4-block, each strut will be an edge of exactly one of the four 4-simplices, and each of the four 4-simplices will be attached to exactly one of the four struts.  When we differentiate each volume with respect to its associated strut-length, the resulting expression simplifies greatly if we then take the continuum time limit; therefore we shall present only these continuum time expressions here.  In the continuum time limit, we find that
\begin{widetext}
\begin{equation}
\begin{split}
\frac{\partial \VolSub{AA^\prime B^\prime C^\prime D^\prime}{}}{\partial \CWstrut{i}^{AA^\prime}} &= \frac{\partial \VolSub{ABB^\prime C^\prime D^\prime}{}}{\partial \CWstrut{i}^{BB^\prime}} = \frac{\partial \VolSub{ABCC^\prime D^\prime}{}}{\partial \CWstrut{i}^{CC^\prime}} = \frac{\partial \VolSub{ABCDD^\prime}{}}{\partial \CWstrut{i}^{DD^\prime}}\\
&= \frac{\Lamblen{}^3}{24\sqrt{2}} \left( 1 - \frac{3}{8} \Lamblendot{}^2 \right)^{1/2} + \Odtone.
\end{split}
\end{equation}
\end{widetext}

With the continuum time form of the relevant geometric quantities, we can now take the continuum time limit of the constraint equation \eqref{LocalConstraint} to obtain a differential equation for $\Lamblen{}(t)$.  Before doing this though, we can make some further simplifications to the continuum time form of \eqref{LocalConstraint}.  We saw in the continuum time limit that all derivatives of the time-like hinges' areas became identical, as did all derivatives of the 4-simplices' volumes.  Using this knowledge, we can significantly simplify the continuum time form of \eqref{LocalConstraint} to
$$
\N{\text{edges/vertex}} \frac{\partial A^\text{A}}{\partial m} \CWdeficit{}^\text{A} = \CosmoConst\, \N{\text{tetrahedra/vertex}} \frac{\partial V^{AA^\prime B^\prime C^\prime D^\prime}}{\partial m}.
$$
The constants $\N{\text{edges/vertex}}$ and $\N{\text{tetrahedra/vertex}}$ are the numbers of edges and tetrahedra, respectively, meeting at any single vertex; they follow from the fact that each strut corresponds to the world-line of a vertex, each time-like hinge at that strut to an edge at the vertex, and each 4-simplex at that strut to a tetrahedron at the vertex; therefore, the total number of triangular hinges meeting at the strut is identical to the number of edges at a vertex, $\N{\text{edges/vertex}}$, and the total number of 4-simplices at the strut is identical to the number of tetrahedra at a vertex, $\N{\text{tetrahedra/vertex}}$.

Substituting all geometric quantities into the above equation, we obtain
$$
\Lamblen{}^2 = \frac{12\sqrt{2}}{\CosmoConst} \frac{\N{\text{edges/vertex}}}{\N{\text{tetrahedra/vertex}}} \frac{(2\pi - n \CWTrapdihedral{})}{\left(1-\frac{1}{8}\Lamblendot{}^2\right)^{1/2}}.
$$
However, $\N{\text{edges/vertex}}$ and $\N{\text{tetrahedra/vertex}}$ are related to $\Nedge$ and $\Ntet$ by the relations
\begin{IEEEeqnarray*}{rCl}
\N{\text{edges/vertex}} &=& \frac{2 \Nedge}{\Nvert}\\
\shortintertext{and}
\N{\text{tetrahedra/vertex}} &=& \frac{4 \Ntet}{\Nvert},
\end{IEEEeqnarray*}
where $\Nvert$ is the number of vertices in a Cauchy surface and is also given by \tabref{tab:primary}.  We therefore recover equation \eqref{GlobConstraint}; thus in this case, local variation yields the same Hamiltonian constraint as global variation.

By locally varying action \eqref{LocalAction} with respect to $\Lamblen{i}$, we can also obtain an evolution equation identical to \eqref{GlobEvolEqn}.  To do this however, it turns out we must also make use of the Regge momentum constraints, that is, the equations obtained by locally varying the action with respect to each $\CWdiag{i}$.  It can be shown that such equations are of the form
\begin{equation}
0 = \CosmoConst \sum_j \frac{\partial \FourVol{j}}{\partial \CWdiag{i}},
\label{DiagRegEqn}
\end{equation}
where the summation is over all 4-blocks containing the diagonal being varied.  The area terms vanish from this equation, as it can be shown that $\partial \AreaA{i} / \partial \CWdiag{i} = -\partial \AreaB{i} / \partial \CWdiag{i}$ for any triangulated trapezoidal hinge.

To understand the relationship between the diagonal Regge equations and the evolution equations, let us first consider the variation of the 4-simplex volumes with respect to the diagonals and the tetrahedral edges.  These derivatives actually depend on which tetrahedral edge or diagonal is being varied.  Consider the triangulated 4-blocks lying between $\Cauchyt{i-1}$ and $\Cauchyt{i+1}$; if we vary the volumes with respect to the diagonals and then take the continuum time limit, we obtain
\begin{IEEEeqnarray*}{rCl}
\frac{\partial \FourVol{i}}{\partial \CWdiag{i}^{AB^\prime}} &\to& -\frac{\imath \delta t}{24\sqrt{2}}\Lamblen{}^2\left(\frac{1}{8}\Lamblendot{}^2+1\right) + \Odt{2},\\
\frac{\partial \FourVol{i}}{\partial \CWdiag{i}^{AC^\prime}} &\to& -\frac{\imath \delta t}{24\sqrt{2}}\Lamblen{}^2\left(\frac{1}{4}\Lamblendot{}^2\right) + \Odt{2},\\
\frac{\partial \FourVol{i}}{\partial \CWdiag{i}^{AD^\prime}} &\to& -\frac{\imath \delta t}{24\sqrt{2}}\Lamblen{}^2\left(\frac{3}{8}\Lamblendot{}^2-1\right) + \Odt{2},\\
\frac{\partial \FourVol{i}}{\partial \CWdiag{i}^{BC^\prime}} &\to& -\frac{\imath \delta t}{24\sqrt{2}}\Lamblen{}^2\left(\frac{1}{8}\Lamblendot{}^2+1\right) + \Odt{2},\\
\frac{\partial \FourVol{i}}{\partial \CWdiag{i}^{BD^\prime}} &\to& -\frac{\imath \delta t}{24\sqrt{2}}\Lamblen{}^2\left(\frac{1}{4}\Lamblendot{}^2\right) + \Odt{2},\\
\frac{\partial \FourVol{i}}{\partial \CWdiag{i}^{CD^\prime}} &\to& -\frac{\imath \delta t}{24\sqrt{2}}\Lamblen{}^2\left(\frac{1}{8}\Lamblendot{}^2+1\right) + \Odt{2}.
\end{IEEEeqnarray*}
Clearly the diagonals give different derivatives.  If we do a similar variation with respect to the tetrahedral edge-lengths, it can be shown that all six derivatives can be expressed in the form
$$
\frac{\partial \left(\FourVol{i}+\FourVol{i-1}\right)}{\partial \Lamblen{i}^{\,xy}} \to \frac{\imath \delta t}{24\sqrt{2}}\Lamblen{}^2 \! \left[2\left( \! 1 \! - \! \frac{3}{8}\Lamblendot{}^2 \! \right) \! - \! \frac{1}{4} \Lamblen{} \Lamblenddot{}\right] - \frac{\partial \FourVol{i}}{\partial \CWdiag{}^{\, xy^\prime}},
$$
where $\Lamblen{i}^{\,xy}$ is the length of the edge between vertices $x$ and $y$ and $\CWdiag{}^{\,xy^\prime}$ is the length of the diagonal triangulating the world-sheet between $\Cauchyt{i}$ and $\Cauchyt{i+1}$ of edge $\Lamblen{i}^{\,xy}$.  Thus each tetrahedral edge also gives a different derivative.

Now if we vary the action with respect to an arbitrary edge $\Lamblen{i}^{\,xy}$, we obtain an evolution equation that can be expressed in the form
\begin{equation}
0 = \Big( \text{RHS of \eqref{GlobEvolEqn}} \Big) + \CosmoConst \sum \frac{\partial V^{(4)}}{\partial \CWdiag{}^{\,xy^\prime}}, 
\label{LocalEvolEqn}
\end{equation}
where the summation is over all 4-blocks containing diagonal $\CWdiag{}^{\,xy^\prime}$; and by the diagonal Regge equation \eqref{DiagRegEqn}, this summation is equal to zero.  We thus recover the same evolution equation \eqref{GlobEvolEqn} as we obtained from globally varying the action.

Since the diagonal equation is just a sum of different volume derivatives, we deduce from the form of the volume derivatives above that the diagonal equation will necessarily have the form
$$
0 = -\CosmoConst \frac{\imath \delta t}{24\sqrt{2}}\Lamblen{}^2\left(\frac{1}{8} P \Lamblendot{}^2+Q\right),
$$
for some constants $P$ and $Q$.  Unfortunately, this equation does not give any physically meaningful solutions as it implies that either $\Lamblen{}=0$ or $\Lamblendot{}^2 = -8 \, Q/P$.  If $P$ and $Q$ have the same signs, then $\Lamblendot{}$ would be imaginary.  Even if $P$ and $Q$ have opposite signs, $\Lamblendot{}^2$ would still be a constant, and relation \eqref{ldot} for $\Lamblendot{}$ would only equal this constant at a single value of $\CWTrapdihedral{}$; this implies that there is only one moment in the evolution of the universe, as given parametrically by relations \eqref{parentl} and \eqref{ldot}, where such a diagonal equation could be satisfied.

Furthermore, without knowing how the triangulated 4-blocks fit together in the skeleton globally, we cannot determine the exact values for $P$ and $Q$.  The reason is as follows: a single diagonal will be shared by $n$ 4-blocks; the diagonal might behave like a $AB^\prime$-type diagonal in one 4-block but like a $AC^\prime$-type diagonal in a neighbouring 4-block, and these two 4-blocks will clearly have different contributions to the sum in \eqref{DiagRegEqn}; hence it is even conceivable that different diagonals may give different Regge equations.

We can however obtain one constraint equation by summing the diagonal Regge equations over all diagonals in the Cauchy surface $\Cauchyt{t}$, where $\Cauchyt{t}$ denotes the Cauchy surface at time $t$ in the continuum time model; it then follows that
\begin{IEEEeqnarray*}{rCl}
0 &=& \sum_{\substack{\CWdiag{}^{\,xy^\prime} \\[-1mm] \vphantom{\CWdiag{}^{\,xy^\prime}} \text{in } \Cauchyt{t}}} \mkern5mu  \sum_{\substack{ \vphantom{\CWdiag{}^{\,xy^\prime}} \FourVol{} \\[-1mm] \text{at } \CWdiag{}^{\,xy^\prime}}} \frac{\partial \FourVol{}}{\partial \CWdiag{}^{\,xy^\prime}}\\
&=& \sum_{\substack{ \vphantom{\CWdiag{}^{\,xy^\prime}} \FourVol{} \\[-1mm] \vphantom{\CWdiag{}^{\,xy^\prime}} \text{in } \Cauchyt{t}}} \mkern5mu  \sum_{\substack{ \CWdiag{}^{\,xy^\prime} \\[-1mm] \text{in } \vphantom{\CWdiag{}^{\,xy^\prime}} \FourVol{} }} \frac{\partial \FourVol{}}{\partial \CWdiag{}^{\,xy^\prime}}\\
&=& -\Ntet \frac{\imath \delta t}{24\sqrt{2}} \, \Lamblen{}^2\left(\frac{5}{4}\Lamblendot{}^2+2\right).
\end{IEEEeqnarray*}
If $\Lamblen{} \neq 0$, then this implies that $\Lamblendot{}^2 = -\frac{8}{5}$, which, as we have just remarked, is clearly non-physical.

We shall discuss possible reasons for the local model's unviability later on.

\subsection{Relationship between the global and local Regge equations}

As Brewin has pointed out, the global Regge equations can usually be related to the local Regge equations through a chain rule.  A necessary condition is that the global and local Regge actions be identical.  The summations over hinges in the global action \eqref{GlobAction} and in the local action \eqref{LocalAction} are identical for the following reasons: the diagonal hinges do not contribute to the local action; the area of each trapezoidal hinge would always equal the areas of its two constituent triangular hinges while the deficit angles would always be identical; and both the areas and the corresponding deficit angles of the triangular space-like hinges would always be identical in both the original and triangulated skeletons.  The volume components of the two actions are also identical since the volume of a CW 4-block should equal the sum of the volumes of its four constituent 4-simplices.  Therefore in this case, the global and local actions are indeed identical.

Then by use of the chain rule, we can relate the global Regge equations to the local Regge equations.  We begin with variation with respect to the struts.  We can express the global variation of the common action $\Action$ with respect to a global strut-length $\CWstrut{}$ as 
\begin{equation}
0 = \frac{\partial \Action}{\partial \CWstrut{}} = \sum_i \frac{\partial \Action}{\partial \CWstrutl{i}} \frac{\partial \CWstrutl{i}}{\partial \CWstrut{}} + \sum_i \frac{\partial \Action}{\partial \CWdiag{i}} \frac{\partial \CWdiag{i}}{\partial \CWstrut{}}, \label{strut-chain}
\end{equation}
where the summations are constrained to the region between a single pair of Cauchy surfaces $\Cauchyt{j}$ and $\Cauchyt{j+1}$, and where $\CWstrutl{i}$ denotes the length of a local strut and $\CWdiag{i}$ the length of a local diagonal.  Since we shall be setting all strut-lengths equal, then $\partial \CWstrutl{i} / \partial \CWstrut{} = 1$ for all $i$.  Additionally, as we saw above, $\partial \Action / \partial \CWstrutl{i}$ is $\Oconst$ to leading order in $dt$.  Thus the first summation has an overall leading order of $\Oconst$.  On the other hand, it can be shown that
$$
\frac{\partial \CWdiag{i}}{\partial \CWstrut{}} = \frac{m}{d},
$$
which becomes $\CWstrut{} / \Lamblen{}$ in the continuum time limit.  However $\CWstrut{}$ has a leading order of $\Odt{}$ in this limit, and as we saw above, $\partial \Action / \partial \CWdiag{i}$ has a leading order of $\Odtone$.  Thus the second summation has an overall leading order of $\Odt{2}$, which is higher than that of the first summation.  It therefore does not contribute to the Regge equation at leading order, and we can consequently simplify \eqref{strut-chain} to just
\begin{equation}
0 = \frac{\partial \Action}{\partial \CWstrut{}} = \sum_i \frac{\partial \Action}{\partial \CWstrutl{i}},
\end{equation}
clearly indicating that any solution of the local Regge equation will automatically be a solution of the global Regge equation as well.  And as we saw above, the two solutions are in fact identical.

Using the chain rule, we can also express the global variation of $\Action$ with respect to the tetrahedral edge-lengths $\Lamblen{}$ as
\begin{equation}
0 = \frac{\partial \Action}{\partial \Lamblen{}} = \sum_i \frac{\partial \Action}{\partial \Lamblenl{i}} \frac{\partial \Lamblenl{i}}{\partial \Lamblen{}} + \sum_i \frac{\partial \Action}{\partial \CWdiag{i}} \frac{\partial \CWdiag{i}}{\partial \Lamblen{}} + \sum_i \frac{\partial \Action}{\partial \CWdiag{i-1}} \frac{\partial \CWdiag{i-1}}{\partial \Lamblen{}},
\end{equation}
where the summations are constrained to a single Cauchy surface $\Cauchyt{j}$ as well as the regions between $\Cauchyt{j}$ and its neighbours $\Cauchyt{j-1}$ and $\Cauchyt{j+1}$, and where $\Lamblenl{i}$ denotes the length of a local tetrahedral edge.  Because the tetrahedral edge-lengths will all be set equal, then $\partial \Lamblenl{i} / \partial \Lamblen{} = 1$ for all $i$.  Additionally, it can be shown that $\partial \CWdiag{i} / \partial \Lamblen{} = \partial \CWdiag{i-1} / \partial \Lamblen{} = 1 / 2$ in the continuum time limit.  Thus, we can simplify the above equation to
\begin{equation}
0 = \frac{\partial \Action}{\partial \Lamblen{}} = \sum_i \frac{\partial \Action}{\partial \Lamblenl{i}} + \sum_i \frac{\partial \Action}{\partial \CWdiag{i}},
\end{equation}
where we have made use of the fact that $\sum_i \partial \Action / \partial \CWdiag{i} = \sum_i \partial \Action / \partial \CWdiag{i-1}$ in the continuum time limit.  However, it can be shown that in this case, both $\partial \Action / \partial \Lamblenl{i}$ and $\partial \Action / \partial \CWdiag{i}$ are $\Odtone$ at leading order.  Thus in contrast to the situation with the struts, a solution to $0 = \partial \Action / \partial \Lamblenl{i}$ by itself is not sufficient to be a solution to $0=\partial \Action / \partial \Lamblen{}$; we must also satisfy $0 = \partial \Action / \partial \CWdiag{i}$.  This is what we saw above, where we were able to recover the global evolution equation \eqref{GlobEvolEqn} from its local counterpart \eqref{LocalEvolEqn} only when we also made use of the diagonal Regge equation \eqref{DiagRegEqn}.

\subsection{Initial value equation for the parent models}

In (3+1)-formulations of general relativity, one would customarily determine the entire space-time by specifying a set of data on some initial Cauchy surface $\Cauchyt{0}$ and then evolving that data forwards in time to determine the rest of the space-time.  Naturally, the evolution equation would be derived from the Einstein field equations.  It has been shown \cite{HawkingEllis} that the required initial data consists of the first and second fundamental forms, $\Tensb{\firstform}$ and $\Tensb{\secondform}$; the former corresponds to the projection of the metric $\Tensb{\metric}$ into $\Cauchyt{0}$ and effectively determines the 3-dimensional intrinsic curvature of $\Cauchyt{0}$; the latter effectively determines the extrinsic curvature of $\Cauchyt{0}$ within the overall space-time.  

There is a set of constraint equations that the initial data must satisfy to be consistent with the Einstein field equations.  Let us express the Einstein field equation in the form
$$\Tensb{\EinsTens} = 8 \pi \Tensb{\StressEnergy},$$
where $\Tensb{\EinsTens}$ is the Einstein tensor and $\Tensb{\StressEnergy}$ the stress-energy tensor.  Let $\Tensb{n}$ denote a field of normalised one-forms everywhere orthogonal to $\Cauchyt{0}$.  By making use of the Gauss equation
\begin{equation}
{}^{(3)}\Riemann_{\mu\nu\sigma\rho} = \Riemann_{\alpha\beta\gamma\delta} \, \firstprojection{\alpha}{\mu} \, \firstprojection{\beta}{\nu} \, \firstprojection{\gamma}{\sigma} \, \firstprojection{\delta}{\rho} - \secondform_{\mu\sigma} \, \secondform_{\nu\rho} + \secondform_{\mu\rho} \, \secondform_{\nu\sigma},
\label{Gauss}
\end{equation}
which relates the 3-dimensional intrinsic curvature ${}^{(3)}\Riemann_{\mu\nu\sigma\rho}$ of $\Cauchyt{0}$ to its extrinsic curvature $\Tensb{\secondform}$ and its 4-dimensional intrinsic curvature $\Riemann_{\alpha\beta\gamma\delta}$, we can express the relation $\Tensb{\EinsTens}\left(\Tensb{n}, \Tensb{n} \right) = 8\pi \Tensb{\StressEnergy}\left(\Tensb{n}, \Tensb{n} \right)$ as
\begin{equation}
{}^{(3)}\RicScal + \left(\secondform^{\mu\nu} \, \firstform_{\mu\nu} \right)^2 - \secondform^{\mu\nu} \, \secondform^{\rho\sigma} \, \firstform_{\mu\rho} \, \firstform_{\nu\sigma}  = 16 \pi \rho,
\label{Hamiltonian}
\end{equation}
where ${}^{(3)}\RicScal$ is the 3-dimensional Ricci scalar of $\Cauchyt{0}$ and $\rho$ is the energy density of the matter source as measured by an observer co-moving with respect to $\Cauchyt{0}$.  Equation \eqref{Hamiltonian} gives the first constraint equation; it is actually the Hamiltonian constraint of the ADM formalism, where it is customarily derived by extremising the ADM action with respect to the lapse function \cite{MTW}.  Let $\left\{\Tensb{u}_i\right\}$, for $i=1,2,3$, denote a set of normalised basis vectors tangent to $\Cauchyt{0}$, and let $\vert$ denote covariant differentiation with respect to the metric connection implied by $\Tensb{\firstform}$.  By making use of the Gauss-Codazzi equation
\begin{equation}
\RicTens_{\sigma\rho} \, n^\sigma \firstprojection{\rho}{\mu} = \secondform^\sigma_{\phantom{\sigma} \mu \vert \sigma} - \secondform^\sigma_{\phantom{\sigma} \sigma \vert \mu},
\label{Gauss-Codazzi}
\end{equation}
which relates the extrinsic curvature $\Tensb{\secondform}$ of $\Cauchyt{0}$ to its 4-dimensional intrinsic curvature in the form of the Ricci tensor $\RicTens_{\sigma\rho}$, we can express the relation $\Tensb{\EinsTens}\left(\Tensb{n}, \Tensb{u}_i \right) = 8\pi \Tensb{\StressEnergy}\left(\Tensb{n}, \Tensb{u}_i \right)$ as
\begin{equation}
\left( \secondform^{\sigma\mu}_{\phantom{\sigma\mu} \vert \mu} \, \firstform_{\sigma\nu} - \secondform^{\sigma\mu}_{\phantom{\sigma\mu} \vert \nu} \, \firstform_{\sigma\mu} \right) u^\nu_{\phantom{\nu}i} = 8 \pi \, \StressEnergy_{\mu\nu} \, n^\mu \, u^\nu_{\phantom{\nu}i},
\label{Momentum}
\end{equation}
which is actually a set of three equations, one for each $i$.  This gives the rest of the constraint equations; these are the momentum constraints of the ADM formalism, where they are customarily derived by extremising the ADM action with respect to the shift functions \cite{MTW}.

Quite often, the initial surface $\Cauchyt{0}$ is chosen to be the surface at a moment of time symmetry, that is, the moment when the surface's extrinsic curvature, as given by the second fundamental form  $\Tensb{\secondform}$, vanishes.  In this case, the momentum constraints would vanish while the Hamiltonian constraint would simplify to
\begin{equation}
^{(3)}\RicScal = 16 \pi \rho;
\label{GenInitVal}
\end{equation}
this is known as the initial value equation at the moment of time symmetry.

When there is a cosmological constant, the Einstein field equations take the form
$$
\Tensb{\EinsTens} + \CosmoConst \Tensb{\metric} = 8 \pi \Tensb{\StressEnergy}.
$$
However we can always absorb the cosmological constant term $\CosmoConst \Tensb{\metric}$ into $\Tensb{\StressEnergy}$; it effectively acts as a perfect fluid source where $\rho_\CosmoConst = -p_\CosmoConst = \CosmoConst / 8\pi$; then the initial value equation for a vacuum $\CosmoConst$-universe can be expressed as
\begin{equation}
^{(3)}\RicScal = 2 \CosmoConst. \label{PInitVal}
\end{equation}

We shall now demonstrate that when this equation is applied to the time-symmetric Cauchy surface of our $\CosmoConst$-FLRW Regge model, the equation is satisfied.  However, the initial value equation \eqref{GenInitVal} and the Einstein field equations, from which it is derived, will only be satisfied in an average manner on a Regge Cauchy surface.  Curvature in the surface is concentrated at the hinges only, yet matter can be distributed away from the hinges where the skeleton is flat; thus the two sides of the equation will not agree in a point-wise manner.  This contradiction arises because the Einstein field equations actually apply to smooth manifolds rather than Regge skeletons; they come about by varying the Einstein-Hilbert action when the underlying manifold is smooth rather than discrete.  Thus by using the Einstein equations in this manner, we are effectively varying the Einstein-Hilbert action on a smooth manifold first and then applying the resulting field equations on a discrete manifold afterwards.  The standard approach in Regge calculus is to use a discrete manifold from the very beginning, with the field equations obtained being different as a result.  Clearly, the two approaches are not equivalent.

The parent models have a moment of time symmetry at the point of minimum expansion.  This happens when $\Lamblendot{} = 0$, corresponding to a dihedral angle of
$$\CWTrapdihedral{0} = 2 \arctan \left(\left.\left.\frac{1}{\sqrt{2}}\right) = \arccos \, \right(\frac{1}{3}\right),$$
and from the Regge equation \eqref{parentl}, the edge-lengths would then be
\begin{equation}
\Lamblen{0}^2 = 6\sqrt{2}\, \frac{\Nedge}{\Ntet \, \CosmoConst} (2\pi - n\CWTrapdihedral{0}). \label{PInitTLen}
\end{equation}
We shall show that this specifically is consistent with the initial value equation.

If a Regge Cauchy surface is sufficiently uniform such that there is a well-defined `volume per vertex', then as Wheeler noted \cite{Houches}, an average for ${}^{(3)}\RicScal$ can be given by
\begin{equation}
^{(3)}\RicScal = \frac{\sum_i \Lamblen{i} \CWdeficit{i}}{\text{`volume per vertex'}}, \label{RegRicScal}
\end{equation}
where the summation is over all edges radiating from a single vertex.  In this 3-dimensional Regge Cauchy surface, the tetrahedral edges are now the hinges, and $\CWdeficit{i}$ is the 3-dimensional deficit angle corresponding to edge $\Lamblen{i}$.  As the parent CW Cauchy surfaces are clearly very uniform, there is a well-defined `volume per vertex' given by
\begin{IEEEeqnarray*}{rCl}
\text{`volume per vertex'} &=& \frac{\Ntet}{\Nvert} (\text{volume of one tetrahedron})\\
&=& \frac{\Ntet}{6\sqrt{2} \, \Nvert}\, \Lamblen{0}^3.
\end{IEEEeqnarray*}
Since there are $2 \Nedge / \Nvert$ edges radiating out from any single vertex in the parent Cauchy surfaces, the summation in \eqref{RegRicScal} can be expressed as $\sum_i \Lamblen{i} \CWdeficit{i} = 2 (\Nedge / \Nvert) \Lamblen{0}\, \CWdeficit{}$, where $\CWdeficit{}$ is the common deficit angle of all edges.  Therefore, the initial value equation \eqref{PInitVal}, when applied to our model, becomes
\begin{equation}
\Lamblen{0}^2 = 6\sqrt{2}\, \frac{\Nedge}{\Ntet \, \CosmoConst}\, \CWdeficit{},
\end{equation}
which is exactly identical to \eqref{PInitTLen} provided $\CWdeficit{} = (2\pi - n\CWTrapdihedral{0})$.

Since $n$ triangular faces do meet at any single edge, then the deficit angle $\CWdeficit{}$ will have the form $(2\pi - n\CWTrapdiSbar{0})$, where $\CWTrapdiSbar{0}$ is the 3-dimensional dihedral angle between triangles in the Cauchy surface.  Thus to complete our proof, we must show that $\CWTrapdiSbar{0} = \CWTrapdihedral{0}$.  Consider a typical tetrahedron in $\mathbf{E}^3$ with vertices $A, B, C, D$, and assign the vertices to have co-ordinates identical to the spatial co-ordinates of their counterparts in \eqref{vertices}.  Edge $AB$ has faces $ABC$ and $ABD$ meeting at it; the unit normal to $ABC$ is
$$
\hat{n}^a_{ABC} = (0,0,1),
$$
and the unit normal to $ABD$ is
$$
\hat{n}^a_{ABD} = \left( 0,-\frac{2\sqrt{2}}{3},\frac{1}{3} \right);
$$
thus the dihedral angle between the two faces is given by
\begin{IEEEeqnarray*}{rCl}
\cos \CWTrapdiSbar{0} &=& \Tensb{\hat{n}}_{ABC} \cdot \Tensb{\hat{n}}_{ABD}\\
&=& \frac{1}{3}.
\end{IEEEeqnarray*}
We therefore see that $\CWTrapdiSbar{0} = \CWTrapdihedral{0} = \arccos ( 1 / 3 )$, and hence our models do satisfy the initial value equation at the moment of time symmetry.

\subsection{Discussion of the parent models}

Before examining the behaviour of our global Regge models, we shall first postulate on the reasons for the local models' failure.  After we have obtained the Regge equations from local variation, the standard approach in Regge calculus would be to specify a set of initial data on a single Cauchy surface and then determine the edge-lengths on all subsequent surfaces using the Regge equations alone.  Rather than doing just this, we have additionally constrained edges on each subsequent surface to be identical, in accordance with the CW geometric constraints.  As a result, the only variable left for the Regge equations to determine is the overall scaling of the edge-length on each surface.  The extra constraints were motivated by the Copernican principle, as we expected each Cauchy surface to be homogeneous and isotropic like the FLRW Cauchy surfaces they are intended to approximate.  The constraints implicitly assume that if we evolve from a Cauchy surface with identical edges, our subsequent surfaces will continue having identical edges because of the Copernican principle; we had assumed this would be the outcome even if we did not explicitly impose the constraints, so the constraints were really expected only to simplify calculations.  The Regge equations presented in \eqref{GlobConstraint} and \eqref{LocalEvolEqn} were obtained after the constraints were imposed.  

\begin{figure}[tb]
{\fontsize{8pt}{9.6pt}\input{ReggeParentVols}}
\caption{\label{fig:ReggeParentVols}The rate of expansion of the universe's volume $dU / dt$ versus the volume $U$ itself for the FLRW universe and the three different Regge models.  The Regge universe volume is given by the sum of the volumes of the Cauchy surface's constituent tetrahedra.}
\vspace{8pt}
{\fontsize{8pt}{9.6pt}\input{ReggeParentMatchedVols}}
\caption{\label{fig:ReggeParentMatchedVols}The rate of expansion of 3-sphere volumes $dU / dt$ versus the volume $U$ itself for the FLRW universe and the three different Regge models.  We have used $\Radhat{}(t)$ as defined in \eqref{PRhat} to be the Regge universes' 3-sphere radii.  As $\Radhat{}(t)$ was defined for all models to equal the FLRW scale factor $\FriedScale(t)$ when $d\Radhat{} / dt = \dot{\FriedScale} = 0$, then the volumes for all Regge models should equal the volume for the FLRW universe when $dU / dt =0$; hence all graphs above coincide at $dU / dt =0$.}
\end{figure}
\begin{figure*}[p]
\subfloat[\hspace{-0.79cm} \label{fig:ParentMatchRadGraph}]{\fontsize{8pt}{9.6pt}\input{ParentMatchRadGraph}}
\subfloat[\hspace{-0.71cm} ]{\fontsize{8pt}{9.6pt}\input{ParentAvgRadGraph}}\\
\subfloat[\hspace{-0.71cm} ]{\fontsize{8pt}{9.6pt}\input{ParentVertRadGraph}}
\subfloat[\hspace{-0.71cm} ]{\fontsize{8pt}{9.6pt}\input{ParentEdgeRadGraph}}\\
\subfloat[\hspace{-0.79cm} \label{fig:ParentFaceRadGraph}]{\fontsize{8pt}{9.6pt}\input{ParentFaceRadGraph}}
\subfloat[\hspace{-0.71cm} ]{\fontsize{8pt}{9.6pt}\input{ParentCentrRadGraph}}
\caption{\label{fig:ParentRadGraphs}The expansion rate of the universe's radius versus the radius itself for the FLRW model and the three different Regge models, where the Regge radius is taken to be (a) $\Radhat{}(t)$, as given by \eqref{PRhat}, (b) the average radius $\Radbar{}(t)$, as given by \eqref{AvgParentRad}, (c) the radius $\Rad{}(t)$ to vertices, as given by \eqref{Z0}, (d) the radius $\Rad{1}(t)$ to the centres of edges, as given by \eqref{PEdgeRad}, (e) the radius $\Rad{2}(t)$ to the centres of triangles, as given by \eqref{PFaceRad}, and (f) the radius $\Rad{3}(t)$ to tetrahedral centres, as given by \eqref{PCentrRad}.}
\end{figure*}
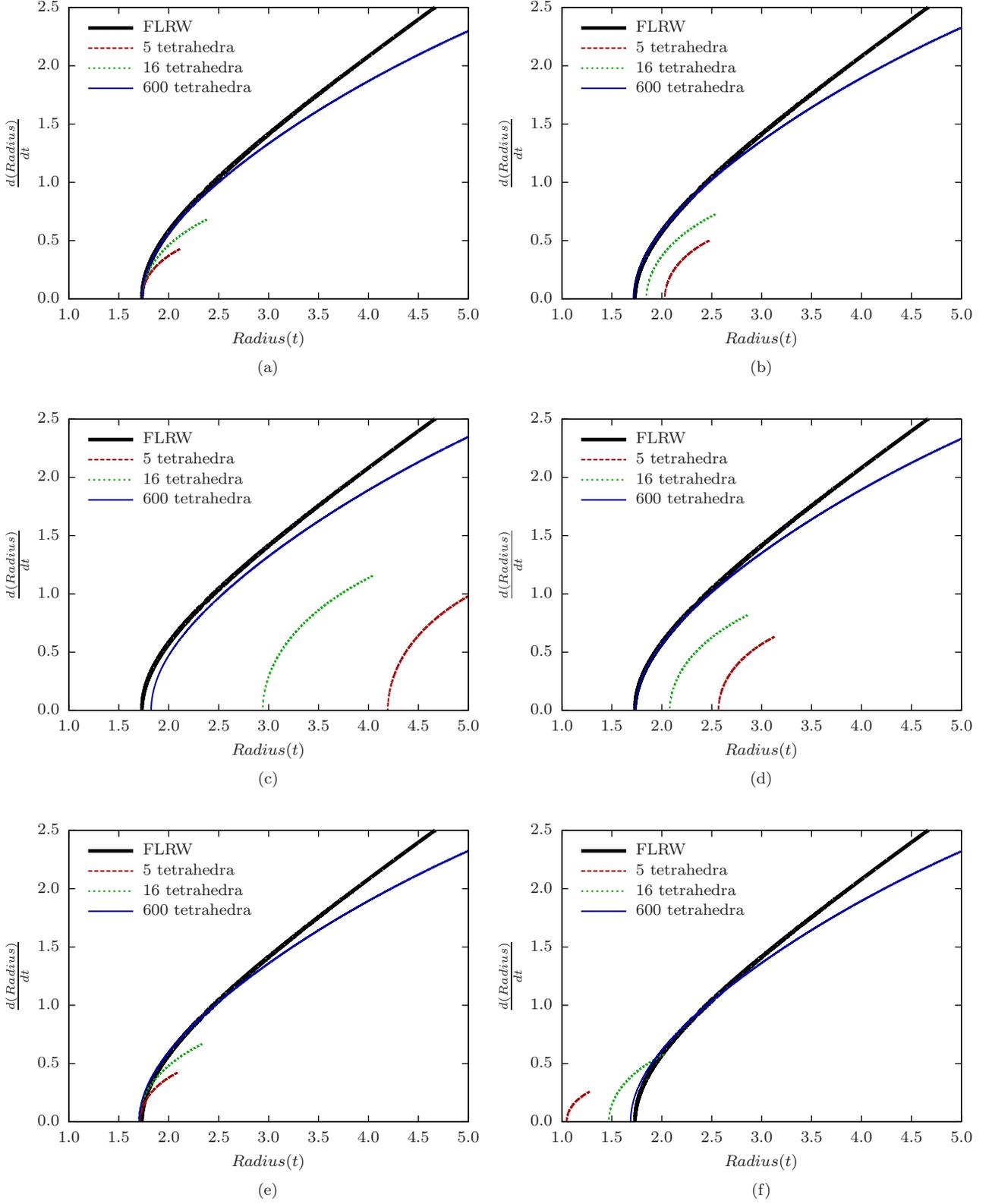

In the global model, all Cauchy surfaces did possess Copernican symmetries as we could readily swap any pair of vertices, tetrahedral edges, or struts on a surface without really changing the surface itself.  However this was no longer the case in the local model when we triangulated the Cauchy surfaces, as the diagonals seemed to disrupt homogeneity.  Not all vertices, for example, were connected to diagonals in the same way; thus they could not be swapped without non-trivially altering the surface itself: in a 4-block, like the one depicted in \FigRef{fig:4block} and \FigRef{fig:allhinges}, vertices $D$ and $A^\prime$ would not be assigned any diagonals while $A$ and $D^\prime$ would each be assigned three.  Because we no longer have perfect homogeneity, our expectation for edges to remain identical under evolution was no longer well founded.  Rather, we should allow the initial surface to evolve according to the unconstrained Regge equations; without the additional constraints, we should expect to have a different equation for each edge on the surface.

There may perhaps be a third method of varying the Regge skeleton, lying somewhere between a completely global variation and a completely local one, such that each edge of the original CW skeleton could be varied individually without having to break the symmetries inherent in the Cauchy surfaces.  We shall refer to this third approach as \emph{semi-local variation}.  When a single edge of a 4-block is being varied, it is possible to impose constraints on the 4-block's internal geometry, without needing to introduce extra independent edges, such that the geometry would still be well-defined under variation.  For instance, one could constrain each block's internal diagonals to be specific functions of the external edges; then when an external edge gets varied explicitly, the diagonals would get varied implicitly in accordance with the constraints.  The idea behind semi-local variation is to impose such constraints on the 4-blocks attached to the edge being varied; after variation, one would then impose whatever further constraints on the skeleton that are necessary to make all other 4-blocks consistent with CW 4-blocks.  We shall refer to the first set of constraints, the ones imposed on the 4-blocks around the varied edge, as \emph{local constraints} and the remaining constraints as \emph{global constraints}.  The local constraints, however, must be chosen in a way that keeps all tetrahedral edges identical, otherwise the model may still not be viable.  We leave to future consideration whether such a choice of constraints is possible.

We now turn to comparing the global Regge models against the continuum model.  We shall consider both the 3-sphere radii of the different models as well as the total volume of the universe.  In \FigRef{fig:ReggeParentVols}, we have plotted the rate of expansion of the universe's volume against the volume itself for each of the models.  The volumes of Regge universes were given by \eqref{ParentVol}, while the volume of the FLRW universe was given by \eqref{FLRW:closedU}.  We see that the Regge models have produced the correct qualitative dynamics of the space-time; for instance, they do not show a universe that eventually collapses back in on itself, nor do they show any instabilities in the evolution; rather, they show a universe that continues expanding indefinitely, like the continuum space-time being approximated.  In terms of accuracy, we see that as the number of tetrahedra increases, the model's accuracy improves, with the 600 tetrahedra model matching the FLRW space-time especially well when the universe is small.

For comparison, we have also plotted analogous graphs in \FigRef{fig:ReggeParentMatchedVols} where the Regge universes' volumes were taken to be volumes of 3-spheres of radius $\Radhat{}(t)$, as given by \eqref{PRhat}.  In these graphs, the Regge models also very closely approximate the FLRW universe at low volumes before diverging as the universe gets larger; the approximation again improves as the number of tetrahedra increases.

In \FigRef{fig:ParentRadGraphs}, we have plotted the expansion rate of the universe's radius against the radius itself, with each subfigure using a different measure of 3-sphere radius for the Regge models.  In all cases, the approximation to FLRW again improves as the number of tetrahedra increases but gradually diverges as the universe expands.  The figures also reveal that radius $\Radhat{}(t)$ gives the best approximation to the FLRW model.  This was somewhat expected given that $\Radhat{}(t)$ was deliberately defined so that it would match $\FriedScale(t)$ exactly at the point of minimum expansion, the point which corresponds to the beginning of all the graphs.  However, the radius $\Rad{2}(t)$ to the centres of triangles also gives a very good measure, as \FigRef{fig:ParentFaceRadGraph} reveals; this clearly indicates that centres of the triangles lie very close to 3-spheres of radius $\Radhat{}$, but it is not clear if there exists any underlying reason for this.

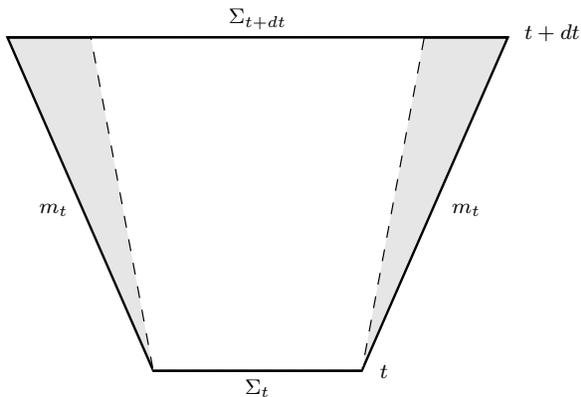
\begin{figure}[tb]
\centering
{\fontsize{8pt}{9.6pt}\input{4-block-spacelike.pspdftex}}
\caption{A schematic diagram, projected onto a 1+1 plane, of a 4-block with space-like struts.  The dashed lines correspond to null curves that originate from the 4-block vertices in $\Cauchyt{t}$.  These lines separate out the region of the 4-block interior that can be reached by causal curves from $\Cauchyt{t}$, the white region, from those that cannot, the shaded region.  Because of this causal structure, no past-directed causal curve from any vertex in $\Cauchyt{t+dt}$ will intercept $\Cauchyt{t}$, and therefore, $\Cauchyt{t+dt}$ lies outside the future domain of dependence of $\Cauchyt{t}$.  For Brewin's case, the above figure should be vertically inverted, since for a contracting universe, $\Cauchyt{t+dt}$ would be smaller than $\Cauchyt{t}$.}
\label{fig:4-block-spacelike}
\end{figure}

All of our graphs terminated at an end-point on the right; this corresponds to the moment when the time-like struts turn null.  Brewin remarked on a similar feature in his dust-filled FLRW models \cite{Brewin}: he obtained analogous graphs that also terminated at an end-point when the struts turned null.  However, Brewin was considering a closed universe, and the end-points appeared while the universe was contracting, whereas we are considering an open universe, and the end-points appear while it is expanding.  Brewin noted, in his case, that at the point when the struts turn null, the surface $\Cauchyt{t}$ would no longer lie in the past domain of dependence of the surface $\Cauchyt{t+dt}$; there would be points in $\Cauchyt{t+dt}$ which cannot be reached by any causal curves from any point in $\Cauchyt{t}$.  Since our Cauchy surfaces are expanding instead, it is the reverse that happens; that is, $\Cauchyt{t+dt}$ no longer lies in the \emph{future} domain of dependence of $\Cauchyt{t}$, as illustrated by \FigRef{fig:4-block-spacelike}.  Brewin suspected, in his case, that the end-point signalled the local curvature had become too large for the approximation to handle.  Nevertheless, in both Brewin's models and ours, increasing the number of tetrahedra does postpone the appearance of the end-point.  In his case, he was able to reach smaller volumes with a larger number of tetrahedra, while we are able to reach larger volumes, as \FigRef{fig:ReggeParentMatchedVols} shows.

\begin{figure}[tb]
\centering
{\fontsize{8pt}{9.6pt}\input{HubbleRadiusRatio}}
\caption{The ratio of the Hubble radius $1/H_0$ to the tetrahedral edge-length $\Lamblen{}(t)$ versus the edge-length for each of the three Regge models; the graphs terminate when the struts turn null.  The Regge Hubble parameter $H_0$ is given by $\Lamblendot{}(t) / \Lamblen{}(t)$.}
\label{fig:HubbleRadiusGraphs}
\end{figure}
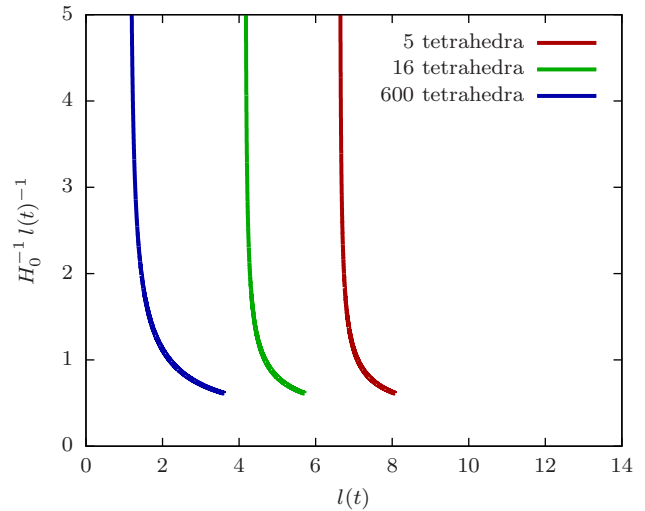

To investigate whether there is any relationship between the struts' turning null and the ratio of the Hubble radius $1/H_0$ to the tetrahedral edge-length $\Lamblen{}(t)$, we have plotted this ratio against $\Lamblen{}(t)$ in \FigRef{fig:HubbleRadiusGraphs}.  The Regge Hubble parameter $H_0$ is defined to be $\Lamblendot{}(t) / \Lamblen{}(t)$.\footnote{In the FLRW universe, Hubble parameter $H_0$ is defined to be $\dot{\FriedScale}/\FriedScale$.  We can define the Regge parameter analogously, using 3-sphere embedding radii rather than the FLRW scale factor $\FriedScale(t)$; however, regardless of which embedding radius we use, the quantity will always reduce to $\Lamblendot{}(t) / \Lamblen{}(t)$.}  The figure shows that the Hubble radius always falls below the edge-length before the struts turn null; thus there does not appear to be any relationship between the Hubble radius and the edge-length at the moment the struts turn null.

In all of our graphs, we noticed that the models diverge increasingly from FLRW as the universe gets larger.  We believe this is due to the finite resolution of our models trying to approximate an ever-expanding universe.  Regardless of the universe's size, the number of tetrahedra in any given approximation is always kept fixed; therefore the resolution will degrade as the universe gets larger.  This is consistent with our observation that the graph diverges more slowly from FLRW as the number of tetrahedra is increased, as the resolution of the models with a higher number of tetrahedra are able to `keep up' longer with FLRW.  This also suggests that the CW formalism would be more accurate when approximating closed cosmologies that do not expand indefinitely but rather attain a finite maximum size before collapsing in on themselves; in this case, the approximation's resolution would be better controlled.  Nevertheless, as we saw above, even for an infinitely expanding universe, the approximation was still able to yield qualitatively correct dynamics; this suggests the formalism should still be helpful in studying the dynamics of other space-times in general, and the information obtained would be especially invaluable if the exact space-time solution is unknown.

\section{Children models of closed vacuum $\CosmoConst$-FLRW universes}

From any parent model, one can always generate a secondary model by triangulating each parent tetrahedron into a set of smaller tetrahedra.  Brewin has devised a method that can subdivide any tetrahedra such that not only can it be applied to parent tetrahedra to generate secondary models, but it can also be applied to each child tetrahedron afterwards to generate even finer-grained models.  Thus, Brewin's scheme can in principle be applied indefinitely.  However, we shall only consider the first generation of children models.

\begin{figure}[tbh]
\input{subdiv-tet.pspdftex}
\caption{\label{fig:subdiv-tet}A partially subdivided parent tetrahedron.  Three of the mid-point vertices, $\midpt{AB}$, $\midpt{AD}$, and $\midpt{BD}$, as well as new edges of length $\BrewinBase{i}$ connecting them have been shown.  The parent edges $AB$, $AD$, and $BD$, have now been subdivided into two edges of length $\BrewinParent{i}$ each: $A\midpt{AB}$ and $\midpt{AB}B$, $A\midpt{AD}$ and $\midpt{AD}D$, and $B\midpt{BD}$ and $\midpt{BD}D$.  Also depicted is the central vertex $\central$.}
\end{figure}
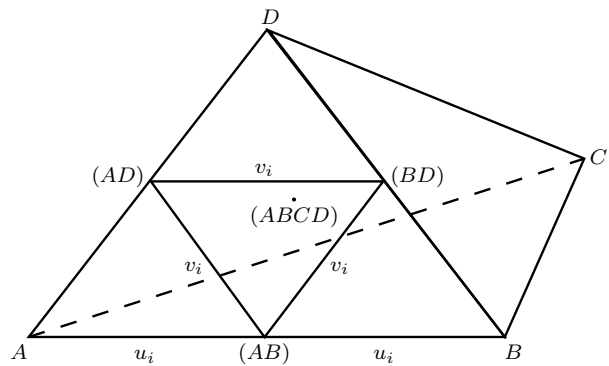

Under Brewin's scheme, each parent tetrahedron gets divided into 12 children tetrahedra.  Seven new vertices are introduced, six at the parent edges' mid-points, the seventh at the tetrahedral centre.  If the parent vertices are labelled $A$, $B$, $C$, $D$, then $\midpt{XY}$ will denote the mid-point vertex between any pair of parent vertices $X, Y \in \left\{A,B,C,D \right\}$, while $\central$ will denote the central vertex.  A partially subdivided parent tetrahedron is shown in \FigRef{fig:subdiv-tet}.

The scheme also introduces three distinct types of edges.  Edges connecting parent vertices to mid-point vertices, such as $A\midpt{AB}$, have length $\BrewinParent{i}$.  As there are six parent edges in a parent tetrahedron, so there are 12 length-$\BrewinParent{i}$ edges.  Edges connecting mid-points to mid-points, such as $\midpt{AB}\midpt{AC}$, have length $\BrewinBase{i}$.  Each face has three of these edges, and as there are four faces per parent tetrahedron, so there are 12 of these edges per parent tetrahedron as well.  Finally, edges connecting mid-points to $\central$, such as $\midpt{AB}\central$ have length $\BrewinCentral{i}$.  Each mid-point contributes one such edge and each parent edge contributes one mid-point, so there are six length-$\BrewinCentral{i}$ edges per parent tetrahedron.  No edges connect parent vertices to $\central$.

As a result, there will also be three distinct types of subdivided tetrahedra, each with four members per parent tetrahedron, thus giving the total of 12 children tetrahedra.  Each Type I tetrahedron consists of an equilateral base formed by three mid-point vertices and an apex at a parent vertex; an example would be $A\midpt{AB}\midpt{AC}\midpt{AD}$.  There are four of these tetrahedra in a parent tetrahedron, one per parent vertex.  Type II tetrahedra share the same equilateral base as Type I but have their apexes at $\central$ instead.  Because they share the same base, there is a 1-1 correspondence between Type I and Type II tetrahedra, so there are four Type II tetrahedra per parent as well.  Finally, a Type III tetrahedron consists of an apex at $\central$ as well and an equilateral base formed by the three mid-point vertices on a single parent face; an example would be $\midpt{AB}\midpt{AC}\midpt{BC}\central$.  Since each parent face is associated with one Type III tetrahedron, there are four Type III tetrahedra per parent tetrahedron.  We note that in terms of their edge-lengths, Type II and Type III tetrahedra are identical.  The different vertices, edges, and tetrahedra in the subdivided parent tetrahedron have been summarised in \tabref{tab:subdiv}.
\begin{table} [htb]
\renewcommand{\arraystretch}{1.2}
\caption{The vertices, edges, and tetrahedra of a subdivided parent tetrahedron.}
\begin{tabular}{>{\centering\arraybackslash}m{2.8cm} >{\centering\arraybackslash}m{3.7cm} >{\centering\arraybackslash}m{1.7cm}}
\hline\hline
Simplex type & Example & Number per parent tetrahedron\\
\hline
Parent vertices & $A$, $B$, $C$, $D$ & 4 \\
Mid-point vertices & $\midpt{AB}$, $\midpt{AC}$, $\midpt{AD}$, $\midpt{BC}$, $\midpt{BD}$, $\midpt{CD}$ & 6 \\
Central vertex & $\central$ & 1\\
&&\\
Edge $\BrewinParent{i}$ & $A\midpt{AB}$ & 12\\
Edge $\BrewinBase{i}$ & $\midpt{AB}\midpt{AC}$ & 12\\
Edge $\BrewinCentral{i}$ & $\midpt{AB}\central$ & 6\\
&&\\
Type I tetrahedra & $A\midpt{AB}\midpt{AC}\midpt{AD}$ & 4\\
Type II tetrahedra & $\midpt{AB}\midpt{AC}\midpt{AD}\central$ & 4\\
Type III tetrahedra & $\midpt{AB}\midpt{AC}\midpt{BC}\central$ & 4\\
\hline\hline
\end{tabular}
\label{tab:subdiv}
\end{table}

Finally, each vertex's world-line will generate a strut connecting one Cauchy surface to the next.  As there are three sets of vertices, there will also be three sets of struts, with all struts in the same set sharing the same length.  Following Brewin, we shall restrict our consideration to only models where all three sets are constrained to have the same length.

Since all Cauchy surfaces in continuum FLRW space-time are identical to each other apart from a time-dependent scale factor $\FriedScale(t)$, by analogy, we shall require the subdivided Cauchy surfaces to be identical to each other as well apart from an overall time-dependent scale factor; that is, we shall require
\begin{equation}
\renewcommand{\arraystretch}{2}
\begin{aligned}
\frac{\BrewinParent{i}}{\BrewinBase{i}} = \alpha && \text{and} && \frac{\BrewinCentral{i}}{\BrewinBase{i}} = \beta \quad \forall i,
\end{aligned}
\label{scaling}
\end{equation}
for some constants $\alpha$ and $\beta$ independent of the Cauchy surface.  This requirement is natural because our CW Cauchy surfaces are intended to approximate FLRW Cauchy surfaces; so if the only difference between two FLRW surfaces is an overall scaling $\FriedScale_0 \to \lambda\, \FriedScale_0$, then the only difference between the two CW surfaces approximating them should be a re-scaling of all lengths by $\lambda$ as well.  This is assuming that the two CW surfaces triangulate their respective FLRW surfaces in the same way; that is, no extra vertices, edges, or tetrahedra appear in one Regge surface but not the other.

\subsection{The 3-sphere embedding of children Cauchy surfaces}
\label{child-3sphere-embedding}

The subdivided Cauchy surface will have a slightly different embedding from that of the parent Cauchy surface.  The main difference is that each of the three sets of vertices can lie on its own 3-sphere, as the three sets are independent of each other.  Nonetheless, the three 3-spheres will share a common centre in $\mathbf{E}^4$.  As the Cauchy surface expands or contracts, the radii of the three 3-spheres will increase or decrease correspondingly.  Each 3-sphere can be parametrised using polar co-ordinates as given by \eqref{3sphere} though with different radii.  We shall denote the three radii by $\RadS{1}{i}$ for the parent vertices, $\RadS{2}{i}$ for the mid-points, and $\RadS{3}{i}$ for the central vertices.  As mentioned in the Introduction, our approach to this embedding will be completely different from that of Brewin \cite{Brewin}, who instead constrained all three sets of vertices to lie on the same 3-sphere.

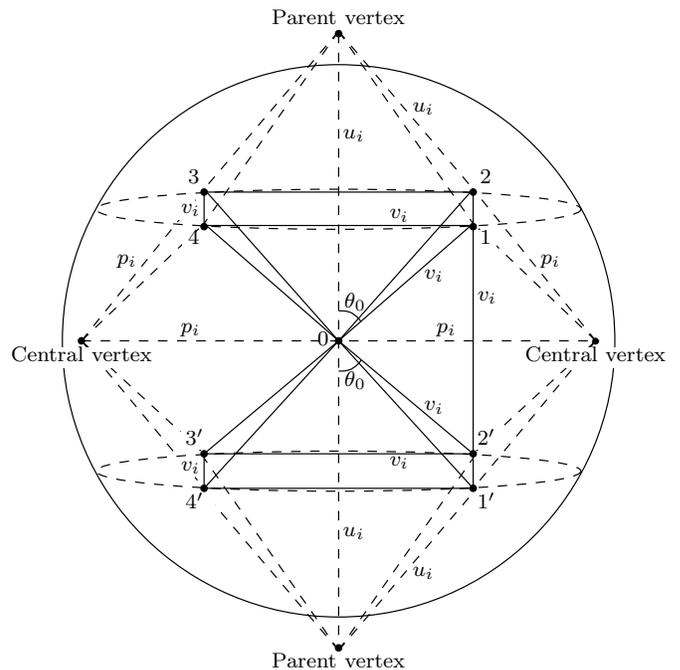
\begin{figure}[thbp]
\input{vertex-spoke.pspdftex}
\caption{\label{fig:vertex-spoke}The upper and lower spokes of mid-point vertices, shown in solid lines, surrounding mid-point vertex $0$ for the case of $n=4$.  One dimension has been suppressed.  All vertices in the two spokes are situated on a sphere of radius $\BrewinBase{i}$.  Each $m$, $m^\prime$ pair of vertices are also nearest neighbours to each other, being located on the same parent triangle; they are hence separated by $\BrewinBase{i}$.  Each vertex is also separated from its two neighbours in the same spoke by a distance of $\BrewinBase{i}$.  Also depicted are the two nearest parent vertices and two of the nearest central vertices.  The upper parent vertex is equidistant to each vertex of the upper spoke and to vertex 0, while the lower parent vertex is equidistant to each vertex of the lower spoke and to vertex 0.  The left central vertex is equidistant to vertices $0, 3, 4, 3^\prime, 4^\prime$, and the right central vertex to $0, 1, 2, 1^\prime, 2^\prime$.}
\end{figure}

We shall first embed a representative subset of the mid-point vertices.  As with the parent model, we can always choose our co-ordinates such that one of these vertices lies at $(0,0,0)$.  This vertex will have $n+n$ nearest mid-point neighbours to it, where $n$ is given by the final column of \tabref{tab:primary}.  To help understand the positions of these nearest neighbours, we shall refer to \FigRef{fig:subdiv-tet}.  Suppose our vertex at $(0,0,0)$ corresponds to the mid-point vertex $\midpt{AB}$ in \FigRef{fig:subdiv-tet}; then on each parent triangle containing this vertex, there will be two nearest mid-point neighbours, for instance $\midpt{AD}$ and $\midpt{BD}$ for triangle $ABD$ in the figure.  Since $n$ parent triangles share a parent edge, there will be $n$ parent triangles containing vertex $\midpt{AB}$; thus we shall have a total of $n$ nearest neighbours from $\midpt{AD}$ and its analogues and another $n$ nearest neighbours from $\midpt{BD}$ and its analogues.  This gives a combined set of $n+n$ nearest mid-point neighbours.  We shall refer to one of the two sets as the `upper' set and the other as the `lower' set.  As these neighbours are all located at a common distance of $\BrewinBase{i}$ from the first vertex, they are situated on a common 2-sphere centred on this vertex.  Hence, we can choose the $\chi$ co-ordinate of these neighbours to be determined by just the radius of this 2-sphere.  We denote this co-ordinate by $\chi=\chi_0$.

The upper and lower sets each form a spoke around the first vertex.  An example of such spokes is illustrated in \FigRef{fig:vertex-spoke} for the case of $n=4$.  We shall label the upper vertices by $1\ldots n$ and the lower ones by $1^\prime\ldots n^\prime$ such that if $m \in \left[1, n \right]$ is an $\midpt{AD}$-type vertex then $m^\prime \in \left[1^\prime, n^\prime \right]$ would be the $\midpt{BD}$-type counterpart located on the same parent triangle.  With this choice of labelling, we note that each $m$ and $m^\prime$ pair are nearest neighbours as well.  We can choose the $\theta$ co-ordinate to be such that the upper vertices are all located at $\theta = \theta_0$.  Then if we choose vertex $1$ to be at $\phi=0$, upper vertex $m$ will be located at $\phi=2(m-1)\pi/n$.  Having fixed the upper vertices, the lower vertices are constrained to be at $\theta=\pi-\theta_0$ with vertex $m^\prime$ having the same $\phi$ co-ordinate as its upper counterpart $m$.  These results are summarised in \tabref{tab:midvertexcoord}.
\begin{table} [htb]
\renewcommand{\arraystretch}{1.2}
\caption{\label{tab:midvertexcoord}The polar co-ordinates of a mid-point vertex and its $2n$ nearest mid-point neighbours.}
\begin{tabular}{>{\centering\arraybackslash}m{2cm} >{\centering\arraybackslash}m{2cm} >{\centering\arraybackslash}m{2cm} >{\centering\arraybackslash}m{2cm}}
\hline\hline
Vertex & $\chi$ & $\theta$ & $\phi$\\
\hline
0 & 0 & 0 & 0\\
1 & $\chi_0$ & $\theta_0$ & 0\\
2 & $\chi_0$ & $\theta_0$ & $\frac{2\pi}{n}$\\
$\vdots$ & $\vdots$ & $\vdots$ & $\vdots$\\
$n$ & $\chi_0$ & $\theta_0$ & $\frac{2(n-1)\pi}{n}$\\
$1^\prime$ & $\chi_0$ & $\pi-\theta_0$ & 0\\
$2^\prime$ & $\chi_0$ & $\pi-\theta_0$ & $\frac{2\pi}{n}$\\
$\vdots$ & $\vdots$ & $\vdots$ & $\vdots$\\
$n^\prime$ & $\chi_0$ & $\pi-\theta_0$ & $\frac{2(n-1)\pi}{n}$\\
\hline\hline
\end{tabular}
\end{table}

The parameters $\chi_0$, $\theta_0$, $\RadS{2}{i}$ can be related to the edge-length $\BrewinBase{i}$. As mentioned above, the $\chi_0$ co-ordinate is fixed by the distance between vertex $0$ and its nearest neighbours; this yields the relation
\begin{equation}
\BrewinBase{i}^2 = 2 \left(\RadS{2}{i}\right)^2(1-\cos\chi_0).
\end{equation}
From the distance between vertices $1$ and $2$, we have the relation
\begin{equation}
\BrewinBase{i}^2 = 2 \left(\RadS{2}{i}\right)^2\sin^2 \chi_0 \sin^2 \theta_0 \left(1-\cos \frac{2\pi}{n}\right),
\end{equation}
and from the distance between $1$ and $1^\prime$, we have the relation
\begin{equation}
\BrewinBase{i}^2 = 4 \left(\RadS{2}{i}\right)^2\sin^2\theta_0\cos^2\chi_0.
\end{equation}
Solving these equations yields
\begin{IEEEeqnarray}{rCl}
\cos \chi_0 &=& \frac{1+\cos \frac{2\pi}{n}}{2\,(1-\cos \frac{2\pi}{n})},\\
\cos \theta_0 &=& \left(\frac{1-\cos \frac{2\pi}{n}}{3-\cos \frac{2\pi}{n}}\right)^{1/2},\\
\SphRatioS{2} := \frac{\BrewinBase{i}}{\RadS{2}{i}} &=& \left(\frac{1-3\cos \frac{2\pi}{n}}{1-\cos \frac{2\pi}{n}}\right)^{1/2}.\label{Z2}
\end{IEEEeqnarray}
We see that once again, the ratio between the edge-length $\BrewinBase{i}$ and the 3-sphere radius $\RadS{2}{i}$ is independent of the Cauchy surface label $i$ and hence of time.  We see as well that the angular co-ordinates of the vertices are also independent of $i$, and therefore, as the edge-length $\BrewinBase{i}$ expands or contracts, the 3-sphere simply expands and contracts about its centre.

We now consider the embedding of the nearest parent vertices.  There will be two nearest parent neighbours to vertex $0$.  One of them will also be a nearest neighbour of all $n$ upper vertices and the other of all $n$ lower vertices, as illustrated in \FigRef{fig:vertex-spoke}.  By symmetry, the upper parent vertex will be located on the 2-dimensional plane containing vertex $0$ but orthogonal to the plane containing the upper $n$ vertices.  We can obtain one vector $\Tensb{p^{(1)}}$ in this plane by taking the average of the $n$ vertices and subtracting the position of vertex $0$.  This gives the vector
\begin{equation}
\left(p^{(1)}\right)^\mu = \left(-\SphRatioS{2},1,0,0\right).
\end{equation}
A second basis $\Tensb{q^{(1)}}$ for the plane can be obtained by requiring orthogonality to both $\Tensb{p^{(1)}}$ and any vector joining any two of the upper $n$ vertices.  This gives
\begin{equation}
\left(q^{(1)}\right)^\mu = \left(1, \SphRatioS{2},0,0\right).
\end{equation}
In terms of these two vectors, the location $\Tensb{r^{(1)}}$ of the parent vertex is then
$$
\left( r^{(1)} \right)^\mu = \RadS{2}{i}(1,0,0,0) + \lambda^{(1)} \left(p^{(1)}\right)^\mu +  \mu^{(1)} \left(q^{(1)} \right)^\mu,
$$
since both vectors $\Tensb{p^{(1)}}$ and $\Tensb{q^{(1)}}$ are given with respect to the position of vertex $0$.  We also require $\Tensb{r^{(1)}}$ be equidistant to vertex $0$ and all upper $n$ vertices; this constraint fixes $\lambda^{(1)}$ to be
\begin{equation}
\lambda^{(1)} = \frac{\BrewinBase{i}}{1+(\SphRatioS{2})^2},
\end{equation}
while $\mu^{(1)}$ remains a free parameter.

It can be shown that $\Tensb{q^{(1)}}$ is actually parallel to the radial vector connecting the parent vertex to the centre of the 3-sphere, that is,
\begin{equation}
\Tensb{r^{(1)}} = \RadS{1}{i} \Tensb{\hat{q}^{(1)}},
\end{equation}
where $\Tensb{\hat{q}^{(1)}}$ is simply $\Tensb{q^{(1)}}$ normalised.  We can therefore express $\Tensb{r^{(1)}}$ more simply as
\begin{equation}
\left(r^{(1)}\right)^\mu = \frac{\RadS{1}{i}}{\sqrt{1+(\SphRatioS{2})^2}} (1,\SphRatioS{2},0,0),
\end{equation}
and radius $\RadS{1}{i}$ is related to parameter $\mu^{(1)}$ by
\begin{equation}
\left(\RadS{1}{i}\right)^2 = \frac{\left(\SphRatioS{2} \frac{\mu^{(1)}}{\lambda^{(1)}}+1\right)^2}{1+(\SphRatioS{2})^2} \left(\RadS{2}{i}\right)^2.
\label{R1}
\end{equation}
Since the free parameter $\mu^{(1)}$ only changes $\Tensb{r^{(1)}}$ along the direction of $\Tensb{q^{(1)}}$, changing $\mu^{(1)}$ will only change the radius of the 3-sphere.

We can deduce the embedding of the nearest central vertices in a similar manner.  As there are $n$ parent tetrahedra hinging on a parent edge, there will be a total of $n$ central vertices that are nearest neighbours to vertex $0$.  Each of these $n$ vertices will be located along the central axis to each vertex quadruplet of the form $m$, $m^\prime$, $(m+1)$, $(m+1)^\prime$ for all $m \in [1,n]$, with vertices $(n+1)$ and $(n+1)^\prime$ being identified with vertices $1$ and $1^\prime$.  Two examples of central vertices are shown in \FigRef{fig:vertex-spoke}.  We shall focus on the central vertex equidistant to vertices $0$, $1$, $1^\prime$, $2$, $2^\prime$.  Similar to the situation with the parent vertex, the central vertex will be located on the 2-dimensional plane containing vertex $0$ but orthogonal to the plane containing vertices $1$, $1^\prime$, $2$, $2^\prime$.  A vector $\Tensb{p^{(2)}}$ in this plane can be found by taking the average of vertices $1$, $1^\prime$, $2$, $2^\prime$ and then subtracting the position of vertex $0$; this gives
\begin{equation}
\left(p^{(2)}\right)^\mu \! = \! \left( \! -\SphRatioS{2},0,\frac{2-\left(\SphRatioS{2}\right)^2}{\sqrt{3-\left(\SphRatioS{2}\right)^2}}, \! \left(\frac{2-\left(\SphRatioS{2}\right)^2}{3-\left(\SphRatioS{2}\right)^2}\right)^{1/2} \right) \! .
\end{equation}
The second basis $\Tensb{q^{(2)}}$ can be obtained by requiring orthogonality with $\Tensb{p^{(2)}}$ and with any vector connecting any of vertices $1$, $1^\prime$, $2$, $2^\prime$; this yields
\begin{widetext}
\begin{equation}
\left(q^{(2)}\right)^\mu = \left(\sqrt{2-\left(\SphRatioS{2}\right)^2},0,\SphRatioS{2} \left(\frac{2-\left(\SphRatioS{2}\right)^2}{3-\left(\SphRatioS{2}\right)^2}\right)^{1/2},\frac{\SphRatioS{2}}{\sqrt{3-\left(\SphRatioS{2}\right)^2}} \right).
\end{equation}
\end{widetext}
Then in terms of these two vectors, the position $\Tensb{r^{(2)}}$ of the vertex is
$$
\left( r^{(2)} \right)^\mu = \RadS{2}{i}(1,0,0,0) + \lambda^{(2)}\left( p^{(2)} \right)^\mu +  \mu^{(2)} \left( q^{(2)} \right)^\mu.
$$
Requiring $\Tensb{r^{(2)}}$ to be equidistant to vertices $0$, $1$, $1^\prime$, $2$, $2^\prime$ yields
\begin{equation}
\lambda^{(2)} = \frac{\BrewinBase{i}}{2},
\end{equation}
while $\mu^{(2)}$ remains a free parameter as well.

As with the parent vertex, it can be shown that $\Tensb{q^{(2)}}$ is actually parallel to the radial vector pointing from the centre of the 3-sphere to the central vertex, that is,
\begin{equation}
\Tensb{r^{(3)}} = \RadS{3}{i} \Tensb{\hat{q}^{(2)}},
\end{equation}
where $\Tensb{\hat{q}^{(2)}}$ is simply $\Tensb{q^{(2)}}$ normalised.  Therefore, we can also express $\Tensb{r^{(2)}}$ as
\begin{widetext}
\begin{equation}
\left(r^{(2)}\right)^\mu = \frac{\RadS{3}{i}}{\sqrt{2}}\left(\sqrt{2-\left(\SphRatioS{2}\right)^2},0,\SphRatioS{2} \left(\frac{2-\left(\SphRatioS{2}\right)^2}{3-\left(\SphRatioS{2}\right)^2}\right)^{1/2},\frac{\SphRatioS{2}}{\sqrt{3-\left(\SphRatioS{2}\right)^2}} \right),
\end{equation}
\end{widetext}
where $\RadS{3}{i}$ is related to $\mu^{(2)}$ by
\begin{equation}
\left(\RadS{3}{i}\right)^2 =\frac{1}{2}\left(\RadS{2}{i}\right)^2 \left(\frac{\BrewinBase{i}}{2}\SphRatioS{2}\mu^{(2)} + \sqrt{2-\left(\SphRatioS{2}\right)^2} \right)^2.
\label{R3}
\end{equation}
Since the free parameter $\mu^{(2)}$ only changes $\Tensb{r^{(2)}}$ along the direction of $\Tensb{q^{(2)}}$, changing $\mu^{(2)}$ will only change the radius of the 3-sphere.

We have previously related $\BrewinBase{i}$ to $\RadS{2}{i}$ in \eqref{Z2}.  We can also relate edge-lengths $\BrewinParent{i}$ and $\BrewinCentral{i}$ to the 3-sphere radii.  The relationship between $\BrewinParent{i}$ and $\RadS{1}{i}$ is
\begin{equation}
\BrewinParent{i}^2 = \left(\RadS{1}{i}\right)^2 - \frac{2 \RadS{1}{i}\RadS{2}{i}}{\sqrt{1+(\SphRatioS{2})^2}} + \left(\RadS{2}{i}\right)^2, \label{uR-relation}
\end{equation}
while that between $\BrewinCentral{i}$ and $\RadS{3}{i}$ is
\begin{equation}
\BrewinCentral{i}^2 = \left(\RadS{3}{i}\right)^2 - 2 \RadS{3}{i}\RadS{2}{i} \sqrt{2-(\SphRatioS{2})^2} + \left(\RadS{2}{i}\right)^2. \label{pR-relation}
\end{equation}

Since $\mu^{(1)}$ and $\mu^{(2)}$ are free parameters that determine the 3-sphere radii $\RadS{1}{i}$ and $\RadS{3}{i}$, respectively, we can re-express \eqref{R1} and \eqref{R3} in the form
\begin{equation}
\renewcommand{\arraystretch}{2}
\begin{aligned}
\RadS{1}{i} &= \bar{\alpha}\,\RadS{2}{i},\\
\RadS{3}{i} &= \bar{\beta}\,\RadS{2}{i},
\end{aligned}
\end{equation}
where the scaling factors $\bar{\alpha} > 0$ and $\bar{\beta} > 0$ now become the free parameters.  This effectively means that $\RadS{1}{i}$ and $\RadS{3}{i}$ can be freely chosen for some initial Cauchy surface, and from \eqref{uR-relation} and \eqref{pR-relation}, this choice would effectively determine $\BrewinParent{i}$ and $\BrewinCentral{i}$.  Therefore, the freedom to choose $\bar{\alpha}$ and $\bar{\beta}$ is equivalent to a freedom to choose $\alpha$ and $\beta$ in \eqref{scaling} for some initial Cauchy surface.

By altering $\bar{\alpha}$ or $\bar{\beta}$, we would expand or contract one of the 3-spheres relative to the others.  The vertices on the altered 3-sphere would simply shift radially inwards or outwards, but not angularly.  Because the shift is purely radial, a vertex on this 3-sphere would still remain equidistant to its nearest neighbours on the same 3-sphere, although that distance would change.  Similarly, the vertex would remain equidistant to its nearest neighbours on the $\RadS{2}{i}$ 3-sphere but with the distance altered as well.

As mentioned at the start of this section, all FLRW 3-spheres are identical to each other apart from an overall scaling $\FriedScale(t)$, and we approximate this symmetry by requiring ratios $\alpha$ and $\beta$ as defined in \eqref{scaling} to be constant; then the evolution of two sets of the Cauchy surface edge-lengths can be determined by the third set alone, which we shall take to be $\BrewinBase{i}$.  Since $\alpha$ and $\beta$ are equivalent to $\bar{\alpha}$ and $\bar{\beta}$, respectively, we can equivalently require that all of our Regge 3-spheres be identical to each other apart from an overall scaling; we would freely specify $\bar{\alpha}$ and $\bar{\beta}$ for some initial Cauchy surface, but our requirement would constrain $\bar{\alpha}$ and $\bar{\beta}$ to be the same for all subsequent surfaces.  As a result, the evolution of two of our 3-sphere radii can be determined by the third radius alone, and we shall choose $\RadS{2}{i}$ to be that sole dynamical radius.  The analogy between CW Cauchy surfaces and FLRW Cauchy surfaces is much clearer when working with these embedding 3-spheres and 3-sphere radii rather than with tetrahedral edge-lengths.  However regardless of whether we work with radii or edge-lengths, there is only one dynamical length parametrising the entire system.

\subsection{Child 4-block co-ordinates}

To facilitate the calculation of geometric quantities in the children models, we shall introduce a co-ordinate system similar to \eqref{vertices}.  Our approach is a modification of Collins and Williams' original approach and differs from that followed by Brewin, who uses non-Cartesian co-ordinates.  Let us consider a typical tetrahedron of the model in Cauchy surface $\Cauchyt{i}$.  It will always have an equilateral base of edge-length $\BrewinBase{i}$ regardless of the tetrahedron's type.  We label the base's vertices by $A$, $B$, $C$ and the apex by $D$.  The three edges meeting at the apex will all be of identical length, either $u_i$ or $p_i$, and without loss of generality, we shall work with $u_i$.  Then we have the freedom to set the co-ordinates for $A$, $B$, $C$, $D$ to be\footnote{We note that because of the underlying 3-sphere geometry, the apex of the neighbouring tetrahedron sharing base $ABC$ will in general be Lorentz-boosted relative to vertices $A$, $B$, $C$, $D$ in co-ordinate system \eqref{SLvertices}; its co-ordinates would in general take the form $\left(0, 0, -\axislentilde{i}\,\cosh\alpha_i, \, \imath \RegTime{i} + \imath \axislentilde{i} \,\sinh\alpha_i \right)$, where $\axislentilde{i}$ is the height of the neighbouring tetrahedron and $\alpha_i$ is a boost parameter.}
\begin{equation}
\renewcommand{\arraystretch}{2}
\begin{aligned}
A &= \displaystyle{\left(-\frac{\BrewinBase{i}}{2}, -\frac{\BrewinBase{i}}{2\sqrt{3}}, 0, \, \imath \RegTime{i} \right)},\\
B &= \displaystyle{\left(\frac{\BrewinBase{i}}{2}, -\frac{\BrewinBase{i}}{2\sqrt{3}}, 0, \, \imath \RegTime{i} \right)},\\
C &= \displaystyle{\left(0, \frac{\BrewinBase{i}}{\sqrt{3}}, 0, \, \imath \RegTime{i} \right)},\\
D &= \displaystyle{\left(0, 0, \axislen{i}, \, \imath \RegTime{i} \right)},
\end{aligned}
\label{SLvertices}
\end{equation}
where $\axislen{i}$ is the height of the tetrahedron and is given by
\begin{equation}
\axislen{i} = \sqrt{\BrewinParent{i}^2-\frac{1}{3}\BrewinBase{i}^2}.
\label{axis}
\end{equation}

This tetrahedron evolves to another in surface $\Cauchyt{i+1}$ with vertices labelled $A^\prime$, $B^\prime$, $C^\prime$, $D^\prime$.  The vertices' co-ordinates are determined by constraints on the edge-lengths and the requirement that the 4-block have no twist or shear.  The edge-length constraints are that the lengths of struts $AA^\prime$, $BB^\prime$, $CC^\prime$ be equal, the lengths of edges $A^\prime B^\prime$, $A^\prime C^\prime$, $B^\prime C^\prime$ be $\BrewinBase{i+1}$, and the lengths of edges $A^\prime D^\prime$, $B^\prime D^\prime$, $C^\prime D^\prime$ be $\BrewinParent{i+1}$.  The requirement of no twist or shear was also imposed on the parent 4-block; thus, the parent and child 4-block geometries are constrained in similar ways.  Once again, the constraint on the strut-lengths and the requirement of no twist or shear can be considered analogous to a choice of shift and lapse functions in the ADM formalism.  

Evolution should preserve the equilateral symmetry of the tetrahedron's base; therefore, the base should simply expand or contract uniformly about its centre.  However, it may also undergo a vertical displacement $\delta z_i$, which is determined by the struts' lengths.  

The requirement that apex $D^\prime$ be located at distance $\BrewinParent{i}$ from each base vertex is equivalent to the two constraints that $D^\prime$ lie at distance $\axislen{i+1}$ from the base's centre and that the central axis connecting $D^\prime$ to the base's centre lie orthogonally to the base.  As base $A^\prime B^\prime C^\prime$ defines a 2-dimensional plane in a (3+1)-dimensional Minkowski space-time, the subspace orthogonal to it would be a (1+1)-dimensional plane, and the tetrahedron's central axis can be oriented along any direction in this plane.  Combined with the first constraint, the second constraint implies that $D^\prime$ will lie on a hyperbola in this (1+1)-dimensional plane; exactly where on the hyperbola it lies depends on the length of strut $DD^\prime$.  Therefore the axis of the upper tetrahedron may in general be Lorentz-boosted relative to that of the lower tetrahedron.

Thus the upper vertices' co-ordinates are given most generally by
\begin{equation*}
\renewcommand{\arraystretch}{2}
\begin{aligned}
A^\prime &= \displaystyle{\left(-\frac{\BrewinBase{i+1}}{2}, -\frac{\BrewinBase{i+1}}{2\sqrt{3}}, \delta z_i, \, \imath \RegTime{i+1} \right)},\\
B^\prime &= \displaystyle{\left(\frac{\BrewinBase{i+1}}{2}, -\frac{\BrewinBase{i+1}}{2\sqrt{3}}, \delta z_i, \, \imath \RegTime{i+1} \right)},\\
C^\prime &= \displaystyle{\left(0, \frac{\BrewinBase{i+1}}{\sqrt{3}}, \delta z_i, \, \imath \RegTime{i+1} \right)},\\
D^\prime &= \displaystyle{\left(0, 0, \axislen{i+1} \cosh \psi_i + \delta z_i, \, \imath \RegTime{i+1} + \imath \axislen{i+1} \sinh \psi_i\right)},
\end{aligned}
\end{equation*}
where $\psi_i$ is the relative boost between the upper and lower tetrahedral axes.

However if the 4-block is to have no twist or shear, then we also require that the world-sheet generated by each tetrahedral edge between $\Cauchyt{i}$ and $\Cauchyt{i+1}$ be flat; in other words, the four vectors parallel to the four sides of this world-sheet must be co-planar.  When this requirement is imposed on any world-sheet involving $D^\prime$, such as $ADA^\prime D^\prime$, and when the scaling relations \eqref{scaling} are imposed, it can be shown that $\psi_i$ must be zero.  Therefore, the upper tetrahedron's co-ordinates can be simplified to
\begin{equation}
\renewcommand{\arraystretch}{2}
\begin{aligned}
A^\prime &= \displaystyle{\left(-\frac{\BrewinBase{i+1}}{2}, -\frac{\BrewinBase{i+1}}{2\sqrt{3}}, \delta z_i, \, \imath \RegTime{i+1} \right)},\\
B^\prime &= \displaystyle{\left(\frac{\BrewinBase{i+1}}{2}, -\frac{\BrewinBase{i+1}}{2\sqrt{3}}, \delta z_i, \, \imath \RegTime{i+1} \right)},\\
C^\prime &= \displaystyle{\left(0, \frac{\BrewinBase{i+1}}{\sqrt{3}}, \delta z_i, \, \imath \RegTime{i+1} \right)},\\
D^\prime &= \displaystyle{\left(0, 0, \axislen{i+1} + \delta z_i, \, \imath \RegTime{i+1}\right)}.
\end{aligned}
\label{SUvertices}
\end{equation}


We can now deduce $\delta z_i$ from the requirement that struts $AA^\prime$ and $DD^\prime$ have equal length; this gives the equation
$$
\frac{1}{3} \, \delta \BrewinBase{i}^2 + \delta z_i^2 = \left[\left(\sqrt{\alpha^2-\frac{1}{3}}\right) \delta \BrewinBase{i} + \delta z_i \right]^2,
$$
where $\delta \BrewinBase{i}$ denotes $\delta \BrewinBase{i} := \BrewinBase{i+1} - \BrewinBase{i}$.  Solving this yields
\begin{equation}
\delta z_i = \frac{1}{2\sqrt{\alpha^2-\frac{1}{3}}}\left(\frac{2}{3}-\alpha^2\right)\delta \BrewinBase{i}.
\end{equation}

\subsection{Geometric quantities}

We shall now derive the geometric quantities relevant to the Regge equations.  However, we shall only present those quantities relevant to the Hamiltonian constraint equation, that is, to varying the action with respect to the strut-lengths, as our plan is to use this constraint to study the behaviour of the children models, much like how we studied the behaviour of the parent models from the parent Hamiltonian constraint.  Moreover, we shall only present the local variation of those quantities, since global variation can always be related to local variation by the chain rule.

Before proceeding, we shall first remark on an implication that the strut-length constraint has on the relationship between $\alpha$ and $\beta$.  Using co-ordinates \eqref{SLvertices} and \eqref{SUvertices}, we can express the length of a parent vertex's strut as
\begin{equation}
\CWstrut{i}^2 = \frac{\alpha^4}{4\left(\alpha^2-\frac{1}{3}\right)} \, \delta \BrewinBase{i}^2 - \delta \RegTime{i}^2,
\end{equation}
where again, $\delta \RegTime{i}$ denotes $\delta \RegTime{i} := \RegTime{i+1} - \RegTime{i}$.  By simply swapping $\alpha$ for $\beta$ in the above expression, we obtain an equivalent expression for the length of a central vertex's strut, that is,
$$
\CWstrut{i}^2 = \frac{\beta^4}{4\left(\beta^2-\frac{1}{3}\right)} \, \delta \BrewinBase{i}^2 - \delta \RegTime{i}^2.
$$
As we require all strut-lengths to be equal between any pair of consecutive vertices, it follows that
\begin{equation}
\frac{\alpha^4}{\left(\alpha^2-\frac{1}{3}\right)} = \frac{\beta^4}{\left(\beta^2-\frac{1}{3}\right)},
\label{equal-struts}
\end{equation}
which implies that
\begin{equation}
\beta = \alpha \qquad \text{or} \qquad \beta = \frac{\displaystyle\alpha}{\sqrt{3\left(\alpha^2-\frac{1}{3}\right)}}.
\label{alpha-beta}
\end{equation}
This relationship between $\alpha$ and $\beta$ implies that one of the three edge-lengths, and hence one of the three 3-sphere radii, can no longer be independent of the other two edge-lengths and radii.

We shall now turn to presenting all geometric quantities relevant to the children models' Hamiltonian constraint.  As with the parent models, the only relevant geometric quantities are the varied time-like hinges generated by the tetrahedral edges' world-sheets, the corresponding deficit angles, and the 4-blocks' varied volumes.

We begin with the variation of the time-like hinges.  As with the parent models, the time-like hinges can again be triangulated in the same manner as depicted in \FigRef{fig:hingefig}; and we can again use equations \eqref{AreaA} to \eqref{dAreaBdmB} to obtain the variation of the resulting triangular hinges with respect to their strut-lengths $\CWstrut{i}$.  For the hinges generated by an edge of length $\BrewinParent{i}$, we find that
\begin{equation}
\frac{\partial \AreaAi{1}{i}}{\partial \CWstrutA{i}} = \frac{\partial \AreaBi{1}{i}}{\partial \CWstrutB{i}} = \frac{\alpha\, \CWstrut{i}}{2\delta \RegTime{i}}\frac{\BrewinBase{i+1}+\BrewinBase{i}}{\sqrt{\frac{1}{3}\frac{\alpha^2}{\alpha^2-\frac{1}{3}}\BrewinBasedot{i}^2-4}},
\label{SubHingeu}
\end{equation}
where $\BrewinBasedot{i}$ denotes $\BrewinBasedot{i} := \delta \BrewinBase{i} / \delta \RegTime{i}$.  For the hinges generated by an edge of length $\BrewinCentral{i}$, we can swap $\alpha$ for $\beta$ to obtain
\begin{equation}
\frac{\partial \AreaAi{3}{i}}{\partial \CWstrutA{i}} = \frac{\partial \AreaBi{3}{i}}{\partial \CWstrutB{i}} = \frac{\beta\, \CWstrut{i}}{2\delta \RegTime{i}}\frac{\BrewinBase{i+1}+\BrewinBase{i}}{\sqrt{\frac{1}{3}\frac{\beta^2}{\beta^2-\frac{1}{3}}\BrewinBasedot{i}^2-4}}.
\label{SubHingep}
\end{equation}
Finally for the hinges generated by $\BrewinBase{i}$, we find that
\begin{equation}
\frac{\partial \AreaAi{2}{i}}{\partial \CWstrutA{i}} = \frac{\partial \AreaBi{2}{i}}{\partial \CWstrutB{i}} = \frac{\CWstrut{i}\, (\BrewinBase{i+1}+\BrewinBase{i})}{4\sqrt{\CWstrut{i}^2 - \frac{1}{4}\,\delta \BrewinBase{i}^2}}.
\label{SubHingev}
\end{equation}

As we shall ultimately take the continuum time limit of the Regge equations, we shall at this stage present the continuum time limit of the quantities above for later use.  As $\delta \RegTime{i} \to 0$, equation \eqref{SubHingeu} becomes
\begin{equation}
\frac{\partial \AreaAi{1}{i}}{\partial \CWstrutA{i}} = \frac{\partial \AreaBi{1}{i}}{\partial \CWstrutB{i}} \to \alpha v \left(\frac{\frac{1}{4}\frac{\alpha^4}{\alpha^2-\frac{1}{3}}\BrewinBasedot{}^2-1}{\frac{1}{3}\frac{\alpha^2}{\alpha^2-\frac{1}{3}}\BrewinBasedot{}^2-4}\right)^{1/2};
\end{equation}
equation \eqref{SubHingep} becomes
\begin{equation}
\frac{\partial \AreaAi{3}{i}}{\partial \CWstrutA{i}} = \frac{\partial \AreaBi{3}{i}}{\partial \CWstrutB{i}} \to \beta v \left(\frac{\frac{1}{4}\frac{\beta^4}{\beta^2-\frac{1}{3}}\BrewinBasedot{}^2-1}{\frac{1}{3}\frac{\beta^2}{\beta^2-\frac{1}{3}}\BrewinBasedot{}^2-4}\right)^{1/2};
\end{equation}
and equation \eqref{SubHingev} becomes
\begin{equation}
\frac{\partial \AreaAi{2}{i}}{\partial \CWstrutA{i}} = \frac{\partial \AreaBi{2}{i}}{\partial \CWstrutB{i}} \to v \left(\frac{\frac{1}{4}\frac{\alpha^4}{\alpha^2-\frac{1}{3}}\BrewinBasedot{}^2-1}{\left(\frac{\alpha^4}{\alpha^2-\frac{1}{3}}-1\right)\BrewinBasedot{}^2-4}\right)^{1/2}.
\end{equation}

We next turn to determining the deficit angles.  In the 4-block described by \eqref{SLvertices} and \eqref{SUvertices}, there will be two different dihedral angles, one at hinges generated by the edges of length $\BrewinBase{i}$ and the other at hinges generated by the edges of length $\BrewinParent{i}$.

Let us first consider the dihedral angle at trapezoidal hinge $ABA^\prime B^\prime$, which is generated by a length-$\BrewinBase{i}$ edge.  Because this hinge is co-planar, the two triangular hinges that subdivide it will have the same dihedral angle as that of the original hinge.  The two faces in the 4-block meeting at $ABA^\prime B^\prime$ are $ABCA^\prime B^\prime C^\prime$ and $ABDA^\prime B^\prime D^\prime$, and their unit normals are, respectively,
\begin{equation}
\hat{n}_1^\mu = \left(1-\frac{\left(\alpha^2-\frac{2}{3}\right)^2}{4\left(\alpha^2-\frac{1}{3}\right)}\BrewinBasedot{i}^2\right)^{-1/2} \left(0, 0, 1, -\imath \frac{\alpha^2-\frac{2}{3}}{2\sqrt{\alpha^2-\frac{1}{3}}}\BrewinBasedot{i} \right)
\end{equation}
and
\begin{equation}
\begin{split}
\hat{n}_2^\mu ={} & \left[3\left(4\alpha^2-1\right)-\frac{\alpha^4\, \BrewinBasedot{i}^2}{4\left(\alpha^2-\frac{1}{3}\right)}\right]^{-1/2}\\
& {} \times \left(0, -2\sqrt{3\left(\alpha^2-\frac{1}{3}\right)}, 1, \imath \frac{\alpha^2 \BrewinBasedot{i}}{2\sqrt{\alpha^2-\frac{1}{3}}} \right).
\end{split}
\end{equation}
\vspace{4mm}
Therefore the dihedral angle between the two faces is given by
\begin{equation}
\begin{aligned}
\cos \CWSubdivTrapdi{1}{i} &= \Tensb{\hat{n}}_1 \cdot \Tensb{\hat{n}}_2\\
&= \frac{1 + \frac{\alpha^2-\frac{2}{3}}{4\left(\alpha^2-\frac{1}{3}\right)}\alpha^2 \BrewinBasedot{i}^2}{\left(1 \! - \! \frac{\left(\alpha^2-\frac{2}{3}\right)^2}{4\left(\alpha^2-\frac{1}{3}\right)}\BrewinBasedot{i}^2\right)^{1/2} \!\left[3\left(4\alpha^2 \! - \! 1\right) \! - \! \frac{\alpha^4\, \BrewinBasedot{i}^2}{4\left(\alpha^2-\frac{1}{3}\right)}\right]^{1/2}}.
\label{theta1}
\end{aligned}
\end{equation}

Let us next consider the dihedral angle at hinge $ADA^\prime D^\prime$, which is generated by length-$\BrewinParent{i}$ edges.  This hinge is also co-planar, and so the two triangular hinges that subdivide it will also have the same dihedral angle.  The faces meeting at this hinge are $ABDA^\prime B^\prime D^\prime$ and $ACDA^\prime C^\prime D^\prime$.  Face $ABDA^\prime B^\prime D^\prime$ repeats from before and so has unit normal $\Tensb{\hat{n}}_2$, while face $ACDA^\prime C^\prime D^\prime$ has unit normal
\begin{equation}
\begin{split}
\hat{n}_3^\mu ={} & \left(\frac{3\left(\alpha^2-\frac{1}{3}\right)}{3\left(4\alpha^2-1\right)-\frac{\alpha^4\, \BrewinBasedot{i}^2}{4\left(\alpha^2-\frac{1}{3}\right)}}\right)^{1/2}\\
& {} \times \left(-\sqrt{3}, 1, \frac{1}{\sqrt{3(\alpha^2-\frac{1}{3})}}, \imath \frac{\alpha^2 \BrewinBasedot{i}}{2\sqrt{3}\left(\alpha^2-\frac{1}{3}\right)} \right).
\end{split}
\end{equation}
Thus the dihedral angle is given by
\begin{equation}
\begin{aligned}
\cos \CWSubdivTrapdi{2}{i} &= \Tensb{\hat{n}}_2 \cdot \Tensb{\hat{n}}_3\\
&= \frac{3\left(2\alpha^2-1\right)+\frac{\alpha^4\, \BrewinBasedot{i}^2}{4\left(\alpha^2-\frac{1}{3}\right)}}{3\left(4\alpha^2-1\right)-\frac{\alpha^4\, \BrewinBasedot{i}^2}{4\left(\alpha^2-\frac{1}{3}\right)}}.
\end{aligned}
\label{theta2}
\end{equation}

There will be another pair of dihedral angles, analogous to the two above, in the 4-block with edges of length $\BrewinCentral{i}$ instead of $\BrewinParent{i}$.  These angles can immediately be obtained by swapping $\alpha$ for $\beta$ in the two expressions above, thus giving
\begin{equation}
\cos \CWSubdivTrapdi{3}{i} = \frac{1 + \frac{\beta^2-\frac{2}{3}}{4\left(\beta^2-\frac{1}{3}\right)}\beta^2 \BrewinBasedot{i}^2}{\left(1 \! - \! \frac{\left(\beta^2-\frac{2}{3}\right)^2}{4\left(\beta^2-\frac{1}{3}\right)}\BrewinBasedot{i}^2\right)^{1/2} \! \left[3\left(4\beta^2 \! - \! 1\right) \! - \! \frac{\beta^4\, \BrewinBasedot{i}^2}{4\left(\beta^2-\frac{1}{3}\right)}\right]^{1/2}}
\label{theta3}
\end{equation}
and
\begin{equation}
\cos \CWSubdivTrapdi{4}{i} = \frac{3\left(2\beta^2-1\right)+\frac{\beta^4\, \BrewinBasedot{i}^2}{4\left(\beta^2-\frac{1}{3}\right)}}{3\left(4\beta^2-1\right)-\frac{\beta^4\, \BrewinBasedot{i}^2}{4\left(\beta^2-\frac{1}{3}\right)}}.
\label{theta4}
\end{equation}
Therefore the child model has a total of four distinct dihedral angles.

With these dihedral angles, we can now deduce the deficit angles at each hinge.  We begin with hinges generated by length-$\BrewinParent{i}$ edges.  There will be $n$ space-like triangular faces meeting at such an edge, with $n$ again given by the last column of \tabref{tab:primary}, and hence there will be $n$ faces meeting at the hinge generated by that edge.  Between each adjacent pair of faces is a dihedral angle of $\CWSubdivTrapdi{2}{i}$, and hence the hinge's deficit angle is
\begin{equation}
\DeficitSub{1}{i} = 2\pi - n \CWSubdivTrapdi{2}{i}.
\label{delta_1}
\end{equation}

We next consider the deficit angle at hinges generated by length-$\BrewinBase{i}$ edges.  These edges are attached to all three types of tetrahedra, and each type contributes a different dihedral angle to the corresponding hinges' overall deficit angle; each Type I tetrahedron contributes $\CWSubdivTrapdi{1}{i}$, while each Type II and Type III tetrahedron contributes $\CWSubdivTrapdi{3}{i}$.  The number of Type I tetrahedra at a length-$\BrewinBase{i}$ edge is given by
\begin{widetext}
\begin{IEEEeqnarray*}{rCl}
\N{\text{Type I}} &=& \frac{(\text{No. of Type I})(\text{3 $\BrewinBase{i}$-edges per Type I})}{(\text{No. of $\BrewinBase{i}$-edges})}\\
&=&\frac{(\text{4 Type I per parent})(\text{No. of parents})(\text{3 $\BrewinBase{i}$-edges per Type I})}{(\text{3 $\BrewinBase{i}$-edges per parent triangle})(\text{No. of parent triangles})}\\
&=& 2,
\end{IEEEeqnarray*}
\end{widetext}
where `parent' refers to a parent tetrahedron, and where we have used the fact that, regardless of the model,
\begin{equation}
\frac{(\text{No. of parent triangles})}{(\text{No. of parents})} = 2;
\label{FaceTetRatio}
\end{equation}
this last identity follows from the fact that there are always four triangles per tetrahedron, but every triangle is always shared by two tetrahedra, so there are only half as many triangles per tetrahedron in a skeleton.  By similar combinatorics, we can show that the number of Type II tetrahedra is given by
\begin{IEEEeqnarray*}{rCl}
\N{\text{Type II}} &=& \frac{(\text{No. of Type II})(\text{3 $\BrewinBase{i}$-edges per Type II})}{(\text{No. of $\BrewinBase{i}$-edges})}\\
&=& 2,
\end{IEEEeqnarray*}
and the number of Type III by
\begin{IEEEeqnarray*}{rCl}
\N{\text{Type III}} &=& \frac{(\text{No. of Type III})(\text{3 $\BrewinBase{i}$-edges per Type III})}{(\text{No. of $\BrewinBase{i}$-edges})}\\
&=& 2.
\end{IEEEeqnarray*}
Therefore the deficit angle is
\begin{equation}
\DeficitSub{2}{i} = 2\pi - 2\CWSubdivTrapdi{1}{i} - 4\CWSubdivTrapdi{3}{i}.
\label{delta_2}
\end{equation}

Finally, we consider the deficit angle at hinges generated by length-$\BrewinCentral{i}$ edges.  Only Type II and III tetrahedra have such edges, and each will contribute a dihedral angle of $\CWSubdivTrapdi{4}{i}$.  The number of Type II and III tetrahedra at an edge is
\begin{widetext}
\begin{IEEEeqnarray*}{rCl}
\N{\text{Type II \& III}} &=& \frac{(\text{No. of Type II \& III per parent})(\text{3 $\BrewinCentral{i}$-edges per Type II or III})}{(\text{No. of $\BrewinCentral{i}$-edges per parent})}\\
&=& 4,
\end{IEEEeqnarray*}
\end{widetext}
and therefore the deficit angle is
\begin{equation}
\DeficitSub{3}{i} = 2\pi - 4\CWSubdivTrapdi{4}{i}.
\label{delta_3}
\end{equation}

The final geometric quantities to derive are the 4-simplices' varied volumes.  A 4-block generated by a tetrahedron can again be triangulated in the same manner as shown in \FigRef{fig:allhinges}, and the volumes of the four 4-simplices generated can again be calculated in the same manner as with the parent models, using equations \eqref{4-vol} to \eqref{4CayleyMenger}.  Again, we vary each of these volumes with respect to their associated strut-length.  The resulting expressions also simplify greatly after the continuum time limit is taken, so we therefore present only the continuum time expressions again.  For the 4-block generated by a Type I tetrahedron, the derivatives of all four 4-simplices' volumes simplify to
\begin{equation}
\frac{\partial V^{(4)}_\text{I}}{\partial \CWstrut{i}} \! \to \! \frac{v^3}{16\sqrt{3}}\sqrt{ \! \left(\alpha^2 \! - \! \frac{1}{3}\right) \!\! \left(1 \! - \! \frac{\alpha^4}{4\left(\alpha^2 \! - \! \frac{1}{3}\right)} \BrewinBasedot{}^2\right) \! } + \Odtone \! .
\end{equation}
The 4-block generated by Type II and Type III tetrahedra are identical, and the derivatives of all 4-simplices' volumes simplify to
\begin{equation}
\frac{\partial V^{(4)}_\text{II \& III}}{\partial \CWstrut{i}} \! \to \! \frac{v^3}{16\sqrt{3}}\sqrt{ \! \left(\beta^2 \! - \! \frac{1}{3}\right) \!\! \left(1 \! - \! \frac{\beta^4}{4\left(\beta^2 \! - \! \frac{1}{3}\right)} \BrewinBasedot{}^2\right) \!}  + \Odtone \! .
\end{equation}

\subsection{Varying the Regge action}\label{VaryChildAction}

We shall now derive the Regge equations for the children models.  We initially attempted to locally vary the Regge action with respect to the strut lengths, however this resulted in a mutually inconsistent set of equations.  Because each child model has three distinct sets of struts, local variation would yield a total of three distinct constraint equations, one for each set.  Any of the constraint equations could then be used to determine the behaviour of our dynamical variable $\BrewinBase{i}$, but with three constraint equations determining a single $\BrewinBase{i}$, the system of equations risked being over-determined.  This indeed happened: we obtained three constraint equations that gave mutually inconsistent relations for $\BrewinBase{i}$ regardless of which choice we made for $\beta$ in \eqref{alpha-beta}.  Therefore, we can perform only a global variation of the action, and this is what we shall now present.

The Regge action for a child model can be expressed as
\begin{equation}
\begin{split}
\Action ={} & \frac{1}{8\pi} \sum_i \left[\Ns{\text{Area 1}}\left(\AreaAi{1}{i}+\AreaBi{1}{i}\right)\DeficitSub{1}{i} \vphantom{V^{(4)}_{i,\, \text{II \& III}}} \right. \\
& \hphantom{\frac{1}{8\pi} \sum_i \left[\right.} {} + \Ns{\text{Area 2}}\left(\AreaAi{2}{i}+\AreaBi{2}{i}\right)\DeficitSub{2}{i}\\
& \hphantom{\frac{1}{8\pi} \sum_i \left[\right.} {} + \Ns{\text{Area 3}}\left(\AreaAi{3}{i}+\AreaBi{3}{i}\right)\DeficitSub{3}{i}\\
& \hphantom{\frac{1}{8\pi} \sum_i \left[\right.} {} - \left. 4 \Ntet\, \CosmoConst \left(V^{(4)}_{i,\, \text{I}} + 2 V^{(4)}_{i,\, \text{II \& III}} \right) \right],
\end{split}
\label{SAction}
\end{equation}
where the summation $i$ is over the Cauchy surfaces, $\Ntet$ is the number of parent tetrahedra, and $V_{i,\, \text{X}}^{(4)}$ is the volume of the entire 4-block of a Type X tetrahedron, that is the combined volume of all four constituent 4-simplices of the 4-block.  Each coefficient $\Ns{\text{Area X}}$, where $\text{X} = 1,2,3$, denotes the number of AX, BX triangular hinge pairs in a Cauchy surface and would equal the number of tetrahedral edges that generate the pairs.  So $\Ns{\text{Area 1}}$ is given by twice the number of parent edges $\Nedge$, that is,
\begin{equation}
\Ns{\text{Area 1}} = 2 \Nedge.\label{NArea1}
\end{equation}
As there are three $\BrewinBase{i}$-edges per parent face, then $N^{\text{Area 2}}$ is
\begin{equation}
\Ns{\text{Area 2}} = 3 \Ntri = 6 \Ntet,\label{NArea2}
\end{equation}
where $\Ntri$ is the number of parent faces, and the second equality follows from \eqref{FaceTetRatio}.  Finally, $\Ns{\text{Area 3}}$ is determined by the fact that each parent tetrahedron has six $\BrewinCentral{i}$-edges, none of which is shared with any other parent tetrahedra.  Therefore,
\begin{equation}
\Ns{\text{Area 3}} = 6 \Ntet.\label{NArea3}
\end{equation}

We can use the chain rule to express any globally varied quantity as a sum of locally varied quantities.  For a common strut-length $\CWstrut{i}$, the chain rule takes the form
\begin{equation*}
\begin{split}
\frac{\partial}{\partial \CWstrut{i}} ={} & \frac{\partial \Lambstrut{1}{j}}{\partial \CWstrut{i}}\frac{\partial}{\partial \Lambstrut{1}{j}} + \frac{\partial \Lambstrut{2}{j}}{\partial \CWstrut{i}}\frac{\partial}{\partial \Lambstrut{2}{j}} + \frac{\partial \Lambstrut{3}{j}}{\partial \CWstrut{i}}\frac{\partial}{\partial \Lambstrut{3}{j}}\\
& \quad {}+ \frac{\partial \Lambdiag{1}{j}}{\partial \CWstrut{i}}\frac{\partial}{\partial \Lambdiag{1}{j}} + \frac{\partial \Lambdiag{2}{j}}{\partial \CWstrut{i}}\frac{\partial}{\partial \Lambdiag{2}{j}} + \frac{\partial \Lambdiag{3}{j}}{\partial \CWstrut{i}}\frac{\partial}{\partial \Lambdiag{3}{j}}.
\end{split}
\end{equation*}
Since all struts have equal length, then
$$\frac{\partial m^{(k)}_j}{\partial \CWstrut{i}} = \delta_{ij} \qquad \forall k.$$
We also have that
$$\frac{\partial d^{(k)}_j}{\partial \CWstrut{i}} = \frac{\CWstrut{i}}{\Lambdiag{k}{i}} \, \delta_{ij} = \Odtone \qquad \forall k.$$
However we found that the leading order of the Regge equation $0= \partial \Action / \partial \Lambstrut{k}{i}$ was $\Oconst$ for all $k$, so once again, the diagonal derivatives do not contribute.  Hence to leading order, the chain rule can be simplified to the form
$$
\frac{\partial}{\partial \CWstrut{i}} = \frac{\partial}{\partial \Lambstrut{1}{i}} + \frac{\partial}{\partial \Lambstrut{2}{i}} + \frac{\partial}{\partial \Lambstrut{3}{i}}.
$$

We have used this chain rule to globally vary the Regge action with respect to $\CWstrut{i}$ to obtain the Hamiltonian constraint for the children models, and we have then taken the continuum time limit.  The resulting equation is
\begin{equation}
\begin{split}
\BrewinBase{}^2 ={} & \frac{4\sqrt{3}}{\CosmoConst}\frac{\alpha^2}{\beta^2}\left[\sqrt{\alpha^2 - \frac{1}{3}}\left(\frac{\alpha^2}{\beta^2} + 2 \right)\right]^{-1}\\
& {} \times \left(\alpha \frac{\Nedge}{\Ntet} \frac{2\pi - n \CWSubdivTrapdi{2}{}}{\sqrt{4-\frac{1}{3}\frac{\alpha^2}{\alpha^2-\frac{1}{3}}\BrewinBasedot{}^2}} \vphantom{\frac{2\pi - 2\CWSubdivTrapdi{1}{} - 4\CWSubdivTrapdi{3}{}}{\sqrt{4-\left(\frac{\alpha^4}{\alpha^2-\frac{1}{3}}-1\right)\BrewinBasedot{}^2}}} + 3 \frac{2\pi - 2\CWSubdivTrapdi{1}{} - 4\CWSubdivTrapdi{3}{}}{\sqrt{4-\left(\frac{\alpha^4}{\alpha^2-\frac{1}{3}}-1\right)\BrewinBasedot{}^2}}\right.\\
& \hphantom{{} \times (}{}+ \left. 3 \beta \frac{2\pi - 4\CWSubdivTrapdi{4}{}}{\sqrt{4-\frac{1}{3}\frac{\alpha^2}{\beta^2}\frac{\alpha^2}{\alpha^2-\frac{1}{3}}\BrewinBasedot{}^2}} \right),
\end{split}
\label{SReggeEqn}
\end{equation}
where relation \eqref{equal-struts} has been used to simplify the expression.  From \eqref{theta2}, we can parametrise $\BrewinBasedot{}$ in terms of $\cos\CWSubdivTrapdi{2}{}$ through the expression
\begin{equation}
\BrewinBasedot{}^2 = \frac{12\left(\alpha^2-\frac{1}{3}\right)\left[\left(4\alpha^2-1\right)\cos\CWSubdivTrapdi{2}{} - \left(2\alpha^2-1\right) \right]}{\alpha^4\left(1+\cos\CWSubdivTrapdi{2}{}\right)}.
\label{SReggedvdt}
\end{equation}

As with the parent model, the range of $\CWSubdivTrapdi{2}{}$ is bounded from above by the requirement that $\BrewinBasedot{}^2 \geq 0$ and from below by the requirement that the strut-lengths be time-like, that is, $\left(\CWstrut{i}\right)^2 < 0$; this leads to the range of
\begin{equation}
\frac{\pi}{3} < \CWSubdivTrapdi{2}{} \leq \arccos \left( \frac{2\alpha^2-1}{4\alpha^2-1} \right).
\end{equation}

We have also varied the action with respect to $\BrewinBase{i}$ to obtain the evolution equation for the children models.  The calculation, which we have not shown, is similar to that for the parent models.  In the continuum time limit, we obtained
\begin{widetext}
\begin{IEEEeqnarray}{rCl}
0 &=& \frac{\Nedge}{\Ntet}\frac{1}{\left(4-\frac{1}{3}\frac{\alpha^2\BrewinBasedot{}^2}{\alpha^2-\frac{1}{3}}\right)^{1/2}}\left[\frac{3n \alpha^6 \BrewinBase{} \BrewinBasedot{}^2\BrewinBaseddot{}}{2\sqrt{\alpha^2-\frac{1}{3}}\left[3(4\alpha^2-1)-\frac{\alpha^4\BrewinBasedot{}^2}{4\left(\alpha^2-\frac{1}{3}\right)}\right]\left(3-\frac{\alpha^2\BrewinBasedot{}^2}{4\left(\alpha^2-\frac{1}{3}\right)}\right)^{1/2}} \right.\nonumber\\
&& \hphantom{\frac{\Nedge}{\Ntet}\frac{1}{\left[4-\frac{1}{3}\frac{\alpha^2\BrewinBasedot{}^2}{\alpha^2-\frac{1}{3}}\right]^{1/2}}\left[\right.} {}+ \left. \frac{16\alpha \left(2\pi - n\CWSubdivTrapdi{2}{} \right)}{4-\frac{1}{3}\frac{\alpha^2\BrewinBasedot{}^2}{\alpha^2-\frac{1}{3}}}\left(\frac{\alpha^4}{4}\BrewinBase{}\BrewinBaseddot{} - \frac{\alpha^6\BrewinBasedot{}^4}{48\left(\alpha^2-\frac{1}{3}\right)^2} + \frac{\alpha^4\BrewinBasedot{}^2}{2\left(\alpha^2-\frac{1}{3}\right)} - \frac{\alpha^2\BrewinBasedot{}^2}{4} - 1 \right) \vphantom{\frac{3n \alpha^6 \BrewinBase{} \BrewinBasedot{}^2\BrewinBaseddot{}}{2\sqrt{\alpha^2-\frac{1}{3}}\left[3(4\alpha^2-1)-\frac{\alpha^4\BrewinBasedot{}^2}{4\left(\alpha^2-\frac{1}{3}\right)}\right]\left[3-\frac{\alpha^2\BrewinBasedot{}^2}{4\left(\alpha^2-\frac{1}{3}\right)}\right]^{1/2}}}  \right]\nonumber\\
&& {}+ \frac{1}{\left(4-\left(\frac{\alpha^4}{\alpha^2-\frac{1}{3}}-1\right)\BrewinBasedot{}^2\right)^{1/2}} \nonumber\\
&& {} \times \left[3\sqrt{3}\BrewinBase{} \BrewinBasedot{}^2\BrewinBaseddot{} \left(\frac{4\alpha^2\left(\alpha^2-\frac{2}{3}\right)+2\left(\alpha^2-\frac{1}{3}\right)\left(\frac{\alpha^4}{\alpha^2-\frac{1}{3}}-1\right)-\frac{1}{2}\alpha^2\left(\alpha^2-\frac{2}{3}\right)\left(\frac{\alpha^4}{\alpha^2-\frac{1}{3}}-1\right)\BrewinBasedot{}^2}{\sqrt{\alpha^2-\frac{1}{3}}\left[4-\left(\frac{\alpha^4}{\alpha^2-\frac{1}{3}}-1\right)\BrewinBasedot{}^2\right]^{1/2}\left(1- \frac{\left(\alpha^2-\frac{2}{3}\right)^2}{4\left(\alpha^2-\frac{1}{3}\right)}\BrewinBasedot{}^2\right) \left[3\left(4\alpha^2-1\right)- \frac{\alpha^4\BrewinBasedot{}^2}{4\left(\alpha^2-\frac{1}{3}\right)}\right] } \right.\right. \nonumber\\
&& \hphantom{\times \left[3\sqrt{3}\BrewinBase{} \BrewinBasedot{}^2\BrewinBaseddot{} \left[ \right. \right.} {}+ \left.\frac{8\beta^2\left(\beta^2-\frac{2}{3}\right)+4\left(\beta^2-\frac{1}{3}\right)\left(\frac{\beta^4}{\beta^2-\frac{1}{3}}-1\right)-\beta^2\left(\beta^2-\frac{2}{3}\right)\left(\frac{\beta^4}{\beta^2-\frac{1}{3}}-1\right)\BrewinBasedot{}^2}{\sqrt{\beta^2-\frac{1}{3}}\left[4-\left(\frac{\beta^4}{\beta^2-\frac{1}{3}}-1\right)\BrewinBasedot{}^2\right]^{1/2}\left(1- \frac{\left(\beta^2-\frac{2}{3}\right)^2}{4\left(\beta^2-\frac{1}{3}\right)}\BrewinBasedot{}^2\right) \left[3\left(4\beta^2-1\right)- \frac{\beta^4\BrewinBasedot{}^2}{4\left(\beta^2-\frac{1}{3}\right)}\right] } \right)\nonumber\\
&& \hphantom{{} \times [} {}+ \left.\frac{48 \left(2\pi - 2\CWSubdivTrapdi{1}{} - 4\CWSubdivTrapdi{3}{} \right)}{4-\left(\frac{\alpha^4}{\alpha^2-\frac{1}{3}}-1\right)\BrewinBasedot{}^2}\left(\frac{1}{4}\BrewinBase{}\BrewinBaseddot{} - \frac{\left(\frac{\alpha^4}{\alpha^2-\frac{1}{3}}-1\right)\alpha^4\BrewinBasedot{}^4}{16\left(\alpha^2-\frac{1}{3}\right)} + \frac{\alpha^4\BrewinBasedot{}^2}{2\left(\alpha^2-\frac{1}{3}\right)} - \frac{\BrewinBasedot{}^2}{4} - 1 \right) \vphantom{\frac{\frac{\alpha^4}{\alpha^2-\frac{1}{3}}}{\left[3-\frac{\alpha^2\BrewinBasedot{}^2}{4\left(\alpha^2-\frac{1}{3}\right)}\right]^{1/2}}} \right]\nonumber\\
&& {}+ \frac{6}{\left[4-\frac{1}{3}\frac{\beta^2\BrewinBasedot{}^2}{\beta^2-\frac{1}{3}}\right]^{1/2}}\left[\frac{3\beta^6 \BrewinBase{} \BrewinBasedot{}^2\BrewinBaseddot{}}{\sqrt{\beta^2-\frac{1}{3}}\left[3(4\beta^2-1)-\frac{\beta^4\BrewinBasedot{}^2}{4\left(\beta^2-\frac{1}{3}\right)}\right]\left(3-\frac{\beta^2\BrewinBasedot{}^2}{4\left(\beta^2-\frac{1}{3}\right)}\right)^{1/2}} \right.\nonumber\\
&& \hphantom{+ \frac{6}{\left[4-\frac{1}{3}\frac{\beta^2\BrewinBasedot{}^2}{\beta^2-\frac{1}{3}}\right]^{1/2}}\left[\right.} {}+ \left.\frac{8\beta \left(2\pi - 4\CWSubdivTrapdi{4}{} \right)}{4-\frac{1}{3}\frac{\beta^2\BrewinBasedot{}^2}{\beta^2-\frac{1}{3}}}\left(\frac{\beta^4}{4}\BrewinBase{}\BrewinBaseddot{} - \frac{\beta^6\BrewinBasedot{}^4}{48\left(\beta^2-\frac{1}{3}\right)^2} + \frac{\beta^4\BrewinBasedot{}^2}{2\left(\beta^2-\frac{1}{3}\right)} - \frac{\beta^2\BrewinBasedot{}^2}{4} - 1 \right) \vphantom{[\frac{3\beta^6 \BrewinBase{} \BrewinBasedot{}^2\BrewinBaseddot{}}{\left[\frac{\beta^2\BrewinBasedot{}^2}{\left(\beta^2-\frac{1}{3}\right)}\right]^{1/2}}} \right]\nonumber\\
&& {}+ \frac{12\sqrt{3}\left(\alpha^2-\frac{1}{4}\right)\alpha^2 \BrewinBase{} \BrewinBaseddot{}}{\sqrt{\alpha^2-\frac{1}{3}}\left[3\left(4\alpha^2-1\right)- \frac{\alpha^4\BrewinBasedot{}^2}{4\left(\alpha^2-\frac{1}{3}\right)}\right]} + \frac{24\sqrt{3}\left(\beta^2-\frac{1}{4}\right)\beta^2 \BrewinBase{} \BrewinBaseddot{}}{\sqrt{\beta^2-\frac{1}{3}}\left[3\left(4\beta^2-1\right)- \frac{\beta^4\BrewinBasedot{}^2}{4\left(\beta^2-\frac{1}{3}\right)}\right]}\nonumber\\
&& {}+ \frac{3\sqrt{3}}{2}\BrewinBase{} \BrewinBaseddot{}\left(\frac{\alpha^2-\frac{2}{3}}{\sqrt{\alpha^2-\frac{1}{3}}\left(1- \frac{\left(\alpha^2-\frac{2}{3}\right)^2}{4\left(\alpha^2-\frac{1}{3}\right)}\BrewinBasedot{}^2\right)} + \frac{\beta^2-\frac{2}{3}}{\sqrt{\beta^2-\frac{1}{3}}\left(1- \frac{\left(\beta^2-\frac{2}{3}\right)^2}{4\left(\beta^2-\frac{1}{3}\right)}\BrewinBasedot{}^2\right)}\right)\nonumber\\
&& {}- \frac{\CosmoConst}{4\sqrt{3}}\BrewinBase{}^2\left[\left(\frac{\alpha^4}{\sqrt{\alpha^2-\frac{1}{3}}} + 2\frac{\beta^4}{\sqrt{\beta^2-\frac{1}{3}}}\right)\left(\BrewinBase{} \BrewinBaseddot{} + 3\BrewinBasedot{}^2 \right) -  12\left(\sqrt{\alpha^2-\frac{1}{3}} + 2\sqrt{\beta^2-\frac{1}{3}} \right) \right] \label{SEvolEqn}.
\end{IEEEeqnarray}
\end{widetext}
If this evolution equation is to be consistent with the Hamiltonian constraint \eqref{SReggeEqn} and with relation \eqref{SReggedvdt}, then we require the Hamiltonian constraint to be its first integral.  In \appendref{FirstIntegralProof}, we have proven that this is possible if $\alpha = \beta = 1$, that is, if the children tetrahedra are all equilateral, and we have found indications that this requirement fails for any other values of $\alpha$ and $\beta$.

By taking $\alpha = \beta = 1$, we can further simplify our model.  The dihedral angles now become
\begin{equation}
\begin{aligned}
\cos \CWSubdivTrapdi{1}{} &= \cos \CWSubdivTrapdi{2}{} = \cos \CWSubdivTrapdi{3}{} = \cos \CWSubdivTrapdi{4}{} \\
& = \frac{1+\frac{1}{8}\BrewinBasedot{}^2}{3-\frac{1}{8}\BrewinBasedot{}^2},
\end{aligned}
\end{equation}
which is identical to its counterpart \eqref{cosq} for the parent models.  Since all dihedral angles are identical, we shall henceforth drop the superscript.  Then, the Hamiltonian constraint \eqref{SReggeEqn} simplifies to
\begin{equation}
\BrewinBase{}^2 = \frac{\sqrt{2}}{\CosmoConst} \frac{1}{\sqrt{1-\frac{1}{8}\BrewinBasedot{}^2}} \left[\frac{\Nedge}{\Ntet}\, \left(2\pi - n \CWTrapdihedral{}\right) +\: 6\, \left( 2\pi - 5\CWTrapdihedral{}\right) \right],
\label{SimplifiedSubdivConstr}
\end{equation}
and relation \eqref{SReggedvdt} for $\BrewinBasedot{}$ simplifies to
\begin{equation}
\BrewinBasedot{}^2 = 8\left[1-2\tan^2\left(\frac{1}{2}\CWTrapdihedral{}\right)\right],
\label{SimplifiedSubdivv-dot}
\end{equation}
which is identical to its parent model counterpart as given by \eqref{ldot}.  We can now simply use \eqref{SimplifiedSubdivConstr} and \eqref{SimplifiedSubdivv-dot} rather than the evolution equation to determine the behaviour of our models.

Finally, we present the volume of the subdivided Regge universe, which is again simply the sum of the volumes of all constituent tetrahedra in a Cauchy surface.  This volume is
\begin{equation}
\tilde{U}_\Ntet(\RegTime{i}) = \frac{1}{\sqrt{3}} \Ntet \left(1 + 2\, \frac{\beta^2}{\alpha^2} \right) \sqrt{\alpha^2-\frac{1}{3}}\, \BrewinBase{}(\RegTime{i})^3,
\label{subdiv-tet-vol}
\end{equation}
which reduces to
\begin{equation}
\tilde{U}_\Ntet(\RegTime{i}) = \sqrt{2}\, \Ntet \, \BrewinBase{}(\RegTime{i})^3,
\label{subdiv-tet-vol-simplified}
\end{equation}
when $\alpha = \beta = 1$.

\subsection{Initial value equation for the children models}

Like the parent models, the children models also admit a moment of time symmetry at the moment of minimum expansion, when $\BrewinBasedot{} = 0$.  At this moment, it follows from \eqref{SReggeEqn} that 
\begin{equation}
\begin{split}
\BrewinBase{0}^2 ={} & \frac{2\sqrt{3}}{\CosmoConst}\frac{\alpha^2}{\beta^2}\left[\sqrt{\alpha^2 - \frac{1}{3}}\left(\frac{\alpha^2}{\beta^2} + 2 \right)\right]^{-1} \\
& {} \times \left[\alpha \frac{\Nedge}{\Ntet} \left( 2\pi - n \CWSubdivTrapdi{2}{0} \right) + 3 \left( 2\pi - 2\CWSubdivTrapdi{1}{0} - 4\CWSubdivTrapdi{3}{0} \right) \right. \\
& \hphantom{{} \times [ }{} +  \left. 3\, \beta \left( 2\pi - 4\CWSubdivTrapdi{4}{0} \right) \vphantom{\frac{\Nedge}{\Ntet}} \right],
\end{split}
\label{SInitLen}
\end{equation}
where
\begin{IEEEeqnarray*}{rCl}
\cos \CWSubdivTrapdi{1}{0} &=& \frac{1}{\sqrt{3(4\alpha^2-1)}},\\
\cos \CWSubdivTrapdi{2}{0} &=& \frac{2\alpha^2-1}{4\alpha^2-1},\\
\cos \CWSubdivTrapdi{3}{0} &=& \frac{1}{\sqrt{3(4\beta^2-1)}},\\
\cos \CWSubdivTrapdi{4}{0} &=& \frac{2\beta^2-1}{4\beta^2-1}.
\end{IEEEeqnarray*}
The three deficit angles $\DeficitSub{1}{i}$, $\DeficitSub{2}{i}$, $\DeficitSub{3}{i}$, given, respectively, by \eqref{delta_1}, \eqref{delta_2}, and \eqref{delta_3}, now become
\begin{equation}
\begin{aligned}
\CWdeficit{0}^{(1)} &= 2\pi - n \CWSubdivTrapdi{2}{0}\\
\CWdeficit{0}^{(2)} &= 2\pi - 2\CWSubdivTrapdi{1}{0} - 4\CWSubdivTrapdi{3}{0}\\
\CWdeficit{0}^{(3)} &= 2\pi - 4\CWSubdivTrapdi{4}{0}.
\end{aligned}
\label{4D_time-sym_angs}
\end{equation}

We shall now demonstrate that \eqref{SInitLen} is also consistent with the initial value equation.  Unlike the parent models however, vertices in children Cauchy surfaces are no longer identical, so the notion of a `volume per vertex' becomes harder to define; thus we cannot apply Wheeler's definition of ${}^{(3)}\RicScal$, as given by \eqref{RegRicScal}, to the initial value equation \eqref{GenInitVal}.  Instead, we shall integrate \eqref{GenInitVal} over the entire Cauchy surface $\Cauchyt{0}$.  The left-hand side becomes \cite{Regge}
$$
\int_{\Cauchyt{0}} {}^{(3)}\RicScal \; d^3 x = 2 \mkern-20mu \sum_{i \,\in\, \left\{ \text{hinges}\right\}} \mkern-20mu \ell_i \CWdeficit{i},
$$
where the integration measure is unity because the Regge tetrahedra are flat, the summation is over all edges in $\Cauchyt{0}$ because these are the hinges of a 3-dimensional skeleton, $\CWdeficit{i}$ is the 3-dimensional deficit angle of an edge, and $\ell_i$ is the edge's length.  The initial value equation \eqref{GenInitVal} therefore becomes
\begin{equation}
\displaystyle{\sum_{i \,\in\, \left\{ \text{hinges}\right\}} \mkern-9mu \ell_i\, \deficit{i}} = \displaystyle{8 \pi \int_{\Cauchyt{0}} \rho \; d^3 x}.
\label{IntegratedReggeInitVal}
\end{equation}
To the best of our knowledge, this form of the Regge initial value equation is novel and applies generally to any Regge Cauchy surface at a moment of time symmetry.  For our particular model where the matter content is simply the cosmological constant $\CosmoConst$, once again we have that $\rho_\CosmoConst = \CosmoConst / 8\pi$, and hence, the integration on the right-hand side simplifies to $\CosmoConst \, \tilde{U}_\Ntet(\RegTime{i})$.  Making use of \eqref{subdiv-tet-vol}, we can therefore express the initial value equation for the children models as
\begin{equation}
\sum_{i \in \left\{ \text{edges} \right\}} \ell_i \CWdeficit{i} = \frac{1}{\sqrt{3}} \Ntet\, \CosmoConst \left(1 + 2\, \frac{\beta^2}{\alpha^2} \right) \sqrt{\alpha^2-\frac{1}{3}}\, \BrewinBase{0}^3.
\label{SInitVal}
\end{equation}

Recall that a child Cauchy surface has only three distinct edge-lengths, which for surface $\Cauchyt{0}$ are $\BrewinParent{0}$, $\BrewinBase{0}$, and $\BrewinCentral{0}$, and that all edges with the same length are identical, implying that they would be associated with the same deficit angle in the 3-dimensional skeleton of $\Cauchyt{0}$.  We shall denote the 3-dimensional deficit angle associated with $\BrewinParent{0}$ by $\CWdeficitbar{0}^{(1)}$, with $\BrewinBase{0}$ by $\CWdeficitbar{0}^{(2)}$, and with $\BrewinCentral{0}$ by $\CWdeficitbar{0}^{(3)}$.  Then the summation on the left-hand side of \eqref{SInitVal} can be expressed as
\begin{equation*}
\begin{split}
\sum_{i \in \left\{ \text{edges} \right\}} \ell_i \CWdeficit{i} ={} & \left(\Ns{\text{Edge $\BrewinParent{0}$}} \right)\, \BrewinParent{0}\, \CWdeficitbar{0}^{(1)} + \left(\Ns{\text{Edge $\BrewinBase{0}$}} \right)\, \BrewinBase{0}\, \CWdeficitbar{0}^{(2)}\\
& {} + \left(\Ns{\text{Edge $\BrewinCentral{0}$}} \right)\, \BrewinCentral{0}\, \CWdeficitbar{0}^{(3)},
\end{split}
\end{equation*}
where $\Ns{\text{Edge $\BrewinParent{0}$}}$, $\Ns{\text{Edge $\BrewinCentral{0}$}}$, and $\Ns{\text{Edge $\BrewinBase{0}$}}$ denote the numbers of edges with lengths $\BrewinParent{0}$, $\BrewinCentral{0}$, and $\BrewinBase{0}$, respectively, in $\Cauchyt{0}$.  Making use of the scaling relations \eqref{scaling} and the fact that
\begin{align*}
\Ns{\text{Edge $\BrewinParent{0}$}} &= \Ns{\text{Area 1}}, \\
\Ns{\text{Edge $\BrewinBase{0}$}} &= \Ns{\text{Area 2}}, \\
\Ns{\text{Edge $\BrewinCentral{0}$}} &= \Ns{\text{Area 3}},
\end{align*}
we further simplify the summation to
$$
\sum_{i \in \left\{ \text{edges} \right\}} \ell_i \CWdeficit{i} = 2 \left( \Nedge\, \alpha\, \CWdeficitbar{0}^{(1)} + 3 \Ntet\, \CWdeficitbar{0}^{(2)} + 3 \Ntet\, \beta\, \CWdeficitbar{0}^{(3)} \right) \BrewinBase{0},
$$
where we have used relations \eqref{NArea1} to \eqref{NArea3} to substitute for $\Ns{\text{Area 1}}$, $\Ns{\text{Area 2}}$, and $\Ns{\text{Area 3}}$.

Substituting back into \eqref{SInitVal}, we solve for $\BrewinBase{0}$ to obtain
\begin{equation}
\begin{split}
\BrewinBase{0}^2 = {} & \frac{2\sqrt{3}}{\CosmoConst}\frac{\alpha^2}{\beta^2}\left[\sqrt{\alpha^2 - \frac{1}{3}}\left(\frac{\alpha^2}{\beta^2} + 2 \right)\right]^{-1} \\
& {} \times \left(\alpha \frac{\Nedge}{\Ntet} \CWdeficitbar{0}^{(1)} + 3\, \CWdeficitbar{0}^{(2)} + 3\, \beta\, \CWdeficitbar{0}^{(3)} \right).
\end{split}
\end{equation}
This expression is identical to \eqref{SInitLen} provided the 3-dimensional deficit angles $\CWdeficitbar{0}^{(i)}$ equal their 4-dimensional counterparts $\CWdeficit{0}^{(i)}$, as given by \eqref{4D_time-sym_angs}, for $i=1,2,3$.

The deficit angles $\CWdeficit{0}^{(i)}$ and $\CWdeficitbar{0}^{(i)}$ have identical forms though.  Recall that each edge in the hypersurface generates a trapezoidal hinge, and each triangle meeting at the edge generates a 3-dimensional face meeting at the hinge.  Thus from each of deficit angles $\CWdeficit{0}^{(i)}$, we can immediately deduce the form of its counterpart $\CWdeficitbar{0}^{(i)}$; it follows that
\begin{IEEEeqnarray*}{rCl}
\CWdeficitbar{i}^{(1)} &=& 2\pi - n\CWSubdivTrapdiSbar{1}{0},\\
\CWdeficitbar{i}^{(2)} &=& 2\pi - 2\CWSubdivTrapdiSbar{1}{0} - 4\CWSubdivTrapdiSbar{3}{0},\\
\CWdeficitbar{i}^{(3)} &=& 2\pi - 4\CWSubdivTrapdiSbar{4}{0},
\end{IEEEeqnarray*}
where $\CWSubdivTrapdiSbar{i}{0}$ is the 3-dimensional dihedral angle between the two triangles that generate the two 3-dimensional faces separated by dihedral angle $\CWSubdivTrapdi{i}{0}$.  To complete our proof then, we need only demonstrate $\CWSubdivTrapdiSbar{i}{0} = \CWSubdivTrapdi{i}{0}$ for all $i$.

Consider a typical tetrahedron in $\mathbf{E}^3$ with vertices $A, B, C, D$.  Let the vertices' co-ordinates be identical to the spatial co-ordinates of their counterparts in \eqref{SLvertices}.  This tetrahedron has two distinct hinges, $AB$ and $AD$.  We first consider the dihedral angle at $AB$.  Triangles $ABC$ and $ABD$ meet at this hinge and are separated by the dihedral angle $\CWSubdivTrapdiSbar{1}{0}$.  The unit normal to $ABC$ is $(0,0,1)$, and the unit normal to $ABD$ is
\begin{equation*}
\begin{split}
&\displaystyle\frac{1}{\sqrt{\BrewinParent{0}^2-\frac{1}{4}\BrewinBase{0}^2}} \left(0,\,-\sqrt{\BrewinParent{0}^2-\frac{1}{3}\BrewinBase{0}^2},\,\frac{\BrewinBase{0}}{2\sqrt{3}}\right)\\
&\qquad \displaystyle =  \frac{1}{\sqrt{\alpha^2-\frac{1}{4}}} \left(0,\, -\sqrt{\alpha^2-\frac{1}{3}},\, \frac{1}{2\sqrt{3}}\right).
\end{split}
\end{equation*}
From their scalar product, we find that
$$
\cos \CWSubdivTrapdiSbar{1}{0} = \frac{1}{\sqrt{3(4\alpha^2-1)}},
$$
and thus we see that $\CWSubdivTrapdiSbar{1}{0} = \CWSubdivTrapdi{1}{0}$.

Next, we consider the dihedral angle at hinge $AD$.  Triangles $ABD$ and $ACD$ meet at this hinge and are separated by the dihedral angle $\CWSubdivTrapdiSbar{2}{0}$.  The unit normal to $ABD$ is the same as before, while the unit normal to $ACD$ is
\begin{equation*}
\begin{split}
& \displaystyle\frac{1}{\sqrt{3\left (4 \BrewinParent{0}^2 - \BrewinBase{0}^2 \right)}} \left(3 \sqrt{\BrewinParent{0}^2 - \frac{1}{3} \BrewinBase{0}^2},\, -\sqrt{3 \BrewinParent{0}^2 - \BrewinBase{0}^2},\, -\BrewinBase{0} \right)\\
&\qquad \displaystyle = \frac{1}{\sqrt{3\left (4 \alpha^2 - 1 \right)}} \left(3 \sqrt{\alpha^2 - \frac{1}{3}},\, -\sqrt{3 \alpha^2 - 1},\, -1 \right).
\end{split}
\end{equation*}
The scalar product gives
$$
\cos \CWSubdivTrapdiSbar{2}{0} = \frac{2\alpha^2-1}{4\alpha^2-1},
$$
and we can similarly conclude that $\CWSubdivTrapdiSbar{2}{0} = \CWSubdivTrapdi{2}{0}$.

Finally, we can readily obtain analogous expressions for $\CWSubdivTrapdiSbar{3}{0}$ and $\CWSubdivTrapdiSbar{4}{0}$ by swapping $\BrewinParent{0}$ for $\BrewinCentral{0}$ or equivalently $\alpha$ for $\beta$ in all of the above expressions.  From $\CWSubdivTrapdiSbar{1}{0}$, we immediately obtain
$$
\cos \CWSubdivTrapdiSbar{3}{0} = \frac{1}{\sqrt{3(4\beta^2-1)}},
$$
and from $\CWSubdivTrapdiSbar{4}{0}$, we obtain
$$
\cos \CWSubdivTrapdiSbar{4}{0} = \frac{2\beta^2-1}{4\beta^2-1}.
$$
Therefore $\CWSubdivTrapdiSbar{3}{0} = \CWSubdivTrapdi{3}{0}$ and $\CWSubdivTrapdiSbar{4}{0} = \CWSubdivTrapdi{4}{0}$, and we can now conclude that the Regge equation \eqref{SReggeEqn} for the children models does indeed satisfy the initial value equation.

\subsection{Discussion of the children models}

We now compare the behaviour of the children models against their corresponding parent models as well as against the FLRW universe.  We shall only consider the models where the children tetrahedra are all equilateral, as these are the only models for which the evolution and constraint equations are consistent.

\begin{figure}[bht]
{\fontsize{8pt}{9.6pt}\input{SubdivVol-CombinedGraphs}}
\caption{The expansion rate of the universe's volume against the volume itself for all parent models and their children models.} \label{fig:SubdivVol-CombinedGraphs}
\end{figure}
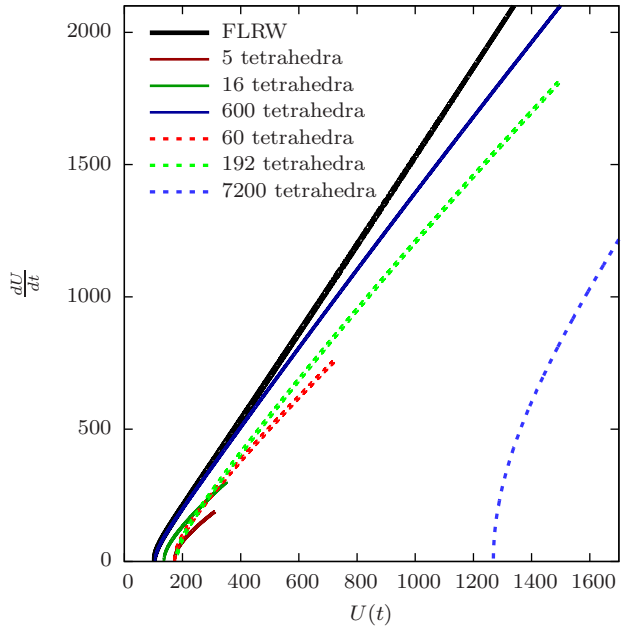

\begin{table}[bht]
\renewcommand{\arraystretch}{1.2}
\caption{\label{tab:MinVols}The minimum volume and the fractional difference from the FLRW minimum for each Regge model and for the FLRW model.}
\begin{tabular}{>{\centering\arraybackslash}m{2.5cm} >{\centering\arraybackslash}m{2.5cm} >{\centering\arraybackslash}m{2.5cm}}
\hline\hline
Model & Minimum volume & Fractional difference from FLRW \\
\hline
FLRW & 102.567937639753 & 0\\
5-tetrahedral parent & 171.741398309775 & 0.67442\\
16-tetrahedral parent & 135.70186007972 & 0.32304\\
600-tetrahedral parent & 105.692461545881 & 0.03046\\
60-tetrahedral child & 172.637934789289 & 0.68316\\
192-tetrahedral child & 179.191098180344 & 0.74705\\
7200-tetrahedral child & 1268.30953855058 & 11.36556\\
\hline\hline
\end{tabular}
\end{table}

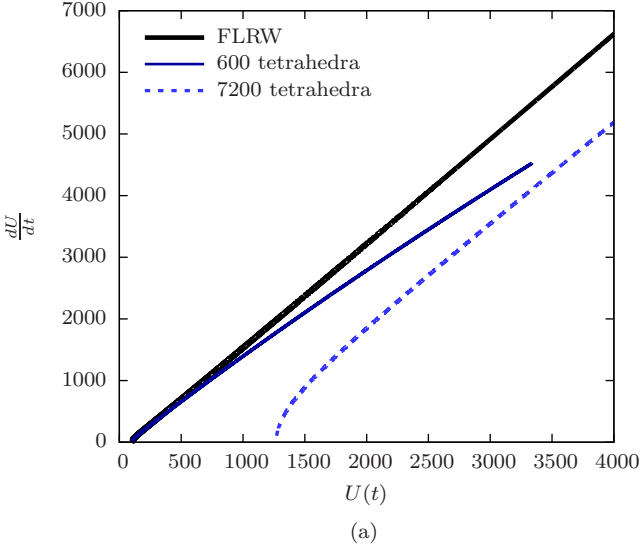
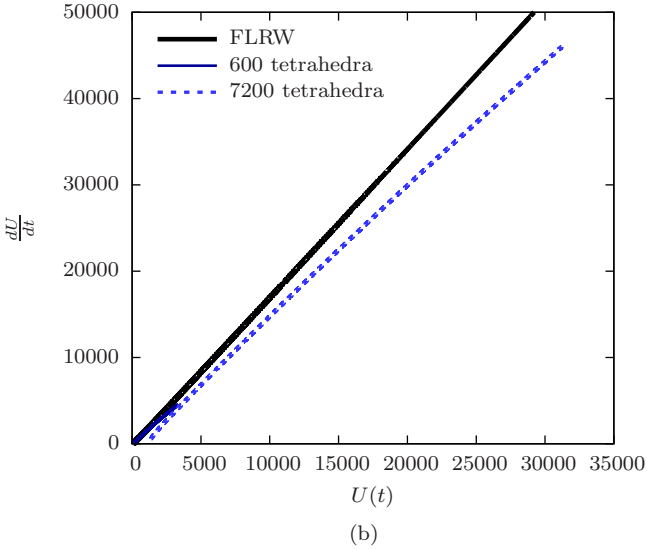
\begin{figure}[htbp]
\subfloat[\hspace{-1.07cm} ]{\fontsize{8pt}{9.6pt}\input{SubdivVol-CombinedGraphs3}}\\
\subfloat[\hspace{-1.07cm} ]{\fontsize{8pt}{9.6pt}\input{SubdivVol-CombinedGraphs2}}
\caption{\label{fig:SubdivVol-CombinedGraphs2}The expansion rate of the universe's volume against the volume itself for the 600-tetrahedral parent and its 7200-tetrahedral child model; (a) focuses on the region around the origin while (b) shows both Regge graphs in their entirety.}
\end{figure}

\Figuref{fig:SubdivVol-CombinedGraphs} and \Figuref{fig:SubdivVol-CombinedGraphs2} compare the graphs of $dU / dt$ against $U$ for each of our models, with $U$ given by \eqref{subdiv-tet-vol-simplified} for the subdivided Regge models and by \eqref{ParentVol} for the parent models.  The graphs in \FigRef{fig:SubdivVol-CombinedGraphs2} focus on the 600-tetrahedral parent and its 7200-tetrahedral child, with the bottom plot extended so as to reveal the graphs' endpoints.  Once again, the models reproduce the correct qualitative dynamics of the universe being approximated.  However, at low volumes, subdividing the tetrahedra actually made the accuracy of the Regge approximation worse.  We can see this more concretely in \tabref{tab:MinVols}, where we list the minimum volumes for each model and their fractional difference from the FLRW minimum.  While increasing the number of tetrahedra in the parent models brought the minimum volume closer to the FLRW value, increasing the number in the children models actually brought it further away.  In fact, the worst parent model was still more accurate than the best child model.

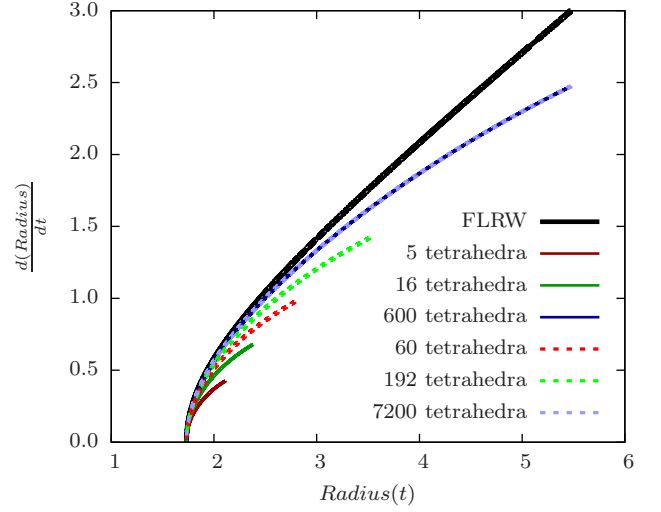
\begin{figure}[tbh]
{\fontsize{8pt}{9.6pt}\input{SubdivRhat-CombinedGraphs}}
\caption{\label{fig:SubdivRhatCombinedRad}A combined graph of the radius' expansion rate against the radius itself, using $\Radhat{}(t)$ as the Regge radius.}
\end{figure}

We also see that all models again diverge from the FLRW model as the universe expanded; however increasing the number of tetrahedra reduces the rate of divergence, and in this sense, increasing the number of tetrahedra improves the Regge approximation.  We again believe this lower rate of divergence is the result of more tetrahedra providing a higher resolution approximation that can `keep up' longer with the FLRW model.  We also note that each model terminates at large volumes whenever the strut becomes null and that this endpoint gets increasingly delayed as the number of tetrahedra is increased.  Thus these figures reveal that each child model provides an approximation that starts off worse than its parent but is later much better by virtue of its more robust resolution.  Indeed, if one were to extrapolate all graphs past their end-points to very large volumes, the 7200-tetrahedral model would ultimately provide the best performance.

As with the parent models, we can define a 3-sphere radius $\Radhat{}(t)$ analogous to \eqref{R-hat} for children Cauchy surfaces such that $\Radhat{}(0) = \FriedScale(0)$; that is, $\Radhat{}(t)$ and $\FriedScale(t)$ match at the moment of minimum expansion; thus, we define $\Radhat{}(t)$ to be
\begin{equation}
\Radhat{}(t) = \frac{\FriedScale(0)}{\BrewinBase{min}}\, \BrewinBase{}(t),
\label{child-R-hat}
\end{equation}
where $\BrewinBase{min}$ is the minimum value of $\BrewinBase{}$ and is given by \eqref{SReggeEqn} when $\CWSubdivTrapdi{2}{} = \arccos [ (2\alpha^2 \! - \! 1)/(4\alpha^2 \! - \! 1) ] = \arccos (1/3)$.  Such a definition is possible because, as discussed at the end of \secref{child-3sphere-embedding}, all dynamical length-scales in the system are related to each other by time-independent scalings such that there is really only one independent dynamical length-scale describing the entire model; therefore any dynamical length-scale, when re-scaled by an appropriate constant, will yield the same $\Radhat{}(t)$.  Note that when $n=5$, then $\Lamblen{}(t)$ as given by \eqref{GlobConstraint} and $\BrewinBase{}(t)$ as given by \eqref{SimplifiedSubdivConstr} will be identical apart from an overall constant factor.  When this happens, the two corresponding models will have the same $\Radhat{}(t)$.  This happens for the 600-tetrahedral parent model and its 7200-tetrahedral child.

\begin{figure}[tbh]
\subfloat[\hspace{-1.07cm} ]{\fontsize{8pt}{9.6pt}\input{SubdivRhatVol-CombinedGraphs2}}\\
\subfloat[\hspace{-1.07cm} ]{\fontsize{8pt}{9.6pt}\input{SubdivRhatVol-CombinedGraphs}}
\caption{\label{fig:SubdivRhatCombinedVol}A combined graph of the 3-sphere volume's expansion rate against the volume itself, using $\Radhat{}(t)$ as the Regge 3-sphere radius; (a) focuses on the region around the origin while (b) shows all Regge graphs in their entirety.}
\end{figure}
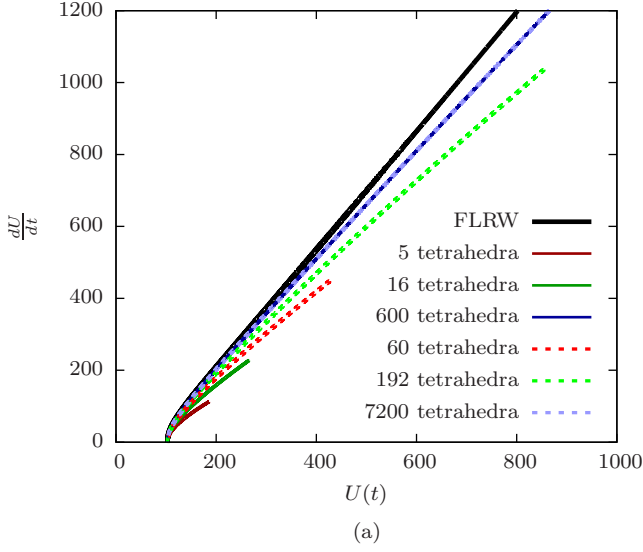
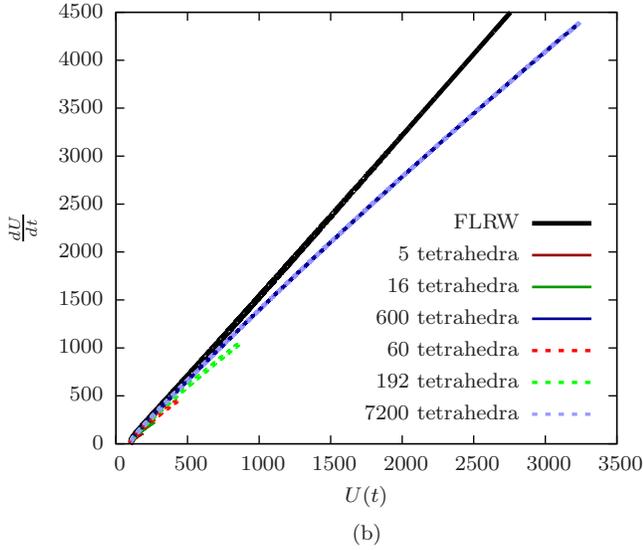
We now examine the behaviours of $d\Radhat{} / dt$ versus $\Radhat{}(t)$ and of the corresponding 3-sphere volumes.  \Figuref{fig:SubdivRhatCombinedRad} shows the relationship between $d\Radhat{} / dt$ and $\Radhat{}(t)$ for each of the Regge models and the relationship between $d\FriedScale / dt$ and $\FriedScale(t)$ for the FLRW model.  \Figuref{fig:SubdivRhatCombinedVol} shows the relationship between the corresponding 3-sphere volumes and their rates of expansion.  We note that in both figures, the graphs for the 600-tetrahedral model and its 7200-tetrahedral child coincide, as expected.  The definition of $\Radhat{}(t)$ has clearly removed any variability in the initial performance of the Regge models; now, the Regge models with more tetrahedra consistently outperform those with fewer.  The rate of divergence from FLRW is again reduced and the graphs' end-points further delayed as the number of tetrahedra is increased.  Thus in these graphs, increasing the number of tetrahedra clearly improves the Regge approximation.

As mentioned at the start of \secref{VaryChildAction}, we were unable to obtain a consistent set of Regge equations when we locally varied the children models' action.  We suspect that we may not actually have complete freedom to specify the lengths of all three sets of struts, contrary to Brewin's claim.  Instead, two of the strut-lengths may depend on the third.  The dependence may actually be determined by two of the three constraint equations obtained from varying the struts locally; they may perhaps act as evolution equations for the two non-independent sets of struts, one equation for one set; the third constraint equation would still determine the evolution of the tetrahedral edges.  We note though that so long as the strut-lengths remain interdependent, there will also be some constraint between the tetrahedral edge-length ratios $\alpha$ and $\beta$ similar to \eqref{alpha-beta}; this would follow from a similar reasoning to that which led to \eqref{alpha-beta}.  After determining the lengths of the two non-independent sets of struts, one could then deduce from them the constraint between $\alpha$ and $\beta$.

Nevertheless, we may still run into the same problem encountered when locally varying the parent model.  There, we discovered that for the evolution equation \eqref{LocalEvolEqn} to be a first integral of the Hamiltonian constraint \eqref{parentl}, the momentum constraints \eqref{DiagRegEqn} must also be satisfied.  However these latter constraints required the model to behave in an unphysical manner, and thus, the model broke down.  We suspected the underlying cause to be the breaking of Copernican symmetries from introducing the diagonals.  With the children models, we should also expect the diagonals to break Copernican symmetries, possibly rendering, for example, the mid-point vertices inequivalent to each other.  Nevertheless, even if the local model were not viable, it would, through a chain-rule relationship analogous to \eqref{strut-chain}, produce an alternative but viable global model, and the properties of this new model would be worthy of further investigation.  Indeed, it may possess some desirable advantages over our current child model, such as, for instance, having a Hamiltonian constraint that is unconditionally a first integral of the evolution equation.  However, we shall leave for now a more thorough examination into the viability and properties of such models, local and global, to future study.

\vspace{5mm}
\begin{acknowledgments}
The authors would like to thank Leo Brewin and Adrian Gentle for much helpful discussion.  RGL acknowledges partial financial support from the Cambridge Commonwealth Trust.
\end{acknowledgments}

\appendix

\setcounter{section}{0}
\renewcommand{\thesection}{\Alph{section}}
\titleformat{\section}[block]
  {\normalfont\bfseries}{APPENDIX \thesection:}{1em}{\centering \MakeUppercase{#1}}

\section{Average radius of a CW Cauchy surface}
\label{AvgRad}
In this appendix, we shall present the computation of the numerical factors in \eqref{AvgParentRad} for the average radius of a CW Cauchy surface.  To compute these factors, we must first determine the radius to any arbitrary point $\Tensb{v}$ in the Cauchy surface.  In terms of the position of vertices $A$, $B$, $C$, $D$ of our representative tetrahedron, the position of $\Tensb{v}$ in the tetrahedron can be expressed as
\begin{equation}
\Tensb{v} = \alpha\, \mathbf{A} + \beta\, \mathbf{B} + \gamma\, \mathbf{C} + \delta\, \mathbf{D},
\end{equation}
where $\mathbf{A}$, $\mathbf{B}$, $\mathbf{C}$, $\mathbf{D}$ are the position vectors of vertices $A$, $B$, $C$, $D$ relative to the embedding 3-sphere's centre, and where constants $\alpha$, $\beta$, $\gamma$, $\delta$ satisfy $0 \leq \alpha, \beta, \gamma, \delta \leq 1$ and $\alpha+\beta+\gamma+\delta = 1$.  Working in the $\mathbf{E}^4$ co-ordinate system of \eqref{3sphere}, we can take vertices $A$, $B$, $C$, $D$ to have co-ordinates given by vertices 1 to 4, respectively, in \tabref{tab:3sphvertices}.  Then the radius of $\Tensb{v}$ is given by
\begin{widetext}
\begin{equation}
\Rad{v}(\RegTime{i}) = \frac{\Rad{}(\RegTime{i})}{\sqrt{1-2 \cos \frac{2\pi}{n}}} \left[\left(1 \! - \! 2 \cos \frac{2\pi}{n}\right)(\alpha^2 \! + \! \beta^2 \! + \! \gamma^2 \! + \! \delta^2) + 2\cos \frac{2\pi}{n}(\alpha\beta \! + \! \alpha\gamma \! + \! \alpha\delta \! + \! \beta\gamma \! + \! \beta\delta \! + \! \gamma\delta) \right]^{1/2} \!\! ,
\end{equation}
\end{widetext}
where $\Rad{}(\RegTime{i})$ is the radius to any of the vertices and is determined from the edge-lengths $\Lamblen{}(\RegTime{i})$ using \eqref{Z0}.

To get the average radius, we therefore need to compute the multiple integral
\begin{equation}
\Radbar{}(\RegTime{i}) = \frac{1}{\mathcal{N}}\displaystyle\int_{\alpha=0}^1 \int_{\beta=0}^{1-\alpha} \int_{\gamma=0}^{1-\alpha-\beta} \Rad{v}\, d\alpha\, d\beta\, d\gamma,
\label{ParentAvgRadIntegral}
\end{equation}
where the normalisation $\mathcal{N}$ is given by
\begin{equation}
\mathcal{N} = \displaystyle\int_{\alpha=0}^1 \int_{\beta=0}^{1-\alpha} \int_{\gamma=0}^{1-\alpha-\beta} d\alpha\, d\beta\, d\gamma = \frac{1}{6}.
\end{equation}
For each of the three models, the integral \eqref{ParentAvgRadIntegral} was evaluated numerically to obtain the factors in \eqref{AvgParentRad}.

\section{Proof of constraint equation being a first integral of the evolution equation}
\label{FirstIntegralProof}

In this appendix, we shall first prove that the Hamiltonian constraint \eqref{GlobConstraint} is a first integral of the evolution equation \eqref{GlobEvolEqn} for the parent models.  We shall next provide a partial proof that the Hamiltonian constraint \eqref{SReggeEqn} is a first integral of the evolution equation \eqref{SEvolEqn} for the children models.

To facilitate comparison with the constraint equation, we shall first simplify the parent model's evolution equation.  Note that the number of edges $\Nedge$ in a Cauchy surface is related to the number of triangles $\Ntri$ by the relation $\Nedge = 3 \Ntri / n$, where $n$ is the number of triangles meeting at an edge; this relation follows because a triangle has three edges, but an edge joins $n$ triangles together.  By using this relation to substitute for $\Ntri$ and by substituting \eqref{GlobConstraint} into the $\Lamblen{}^2$ factor in the last term of \eqref{GlobEvolEqn}, we can simplify \eqref{GlobEvolEqn} to get
\begin{equation}
\begin{split}
0 = {} & \frac{1}{1-\frac{1}{8}\Lamblendot{}^2} \\
& {}\times \! \left[\frac{1}{8}\frac{(2\pi-n\CWTrapdihedral{})}{\left(1 \! - \! \frac{1}{8}\Lamblendot{}^2\right)^{1/2}} \! \left(\frac{3}{8}\Lamblendot{}^2 \! - \! 1\right) \! \left(\Lamblen{} \Lamblenddot{} \! + \! 2\Lamblendot{}^2 \! - \! 16\right) \! + \! \frac{n}{2\sqrt{2}} \Lamblen{} \Lamblenddot{}\right] \! .
\end{split}
\label{AppProof0}
\end{equation}

Now we consider the constraint equation \eqref{GlobConstraint} itself.  To demonstrate that this is a first integral of \eqref{GlobEvolEqn}, we first differentiate it; this gives
\begin{equation}
\Lamblen{}\Lamblendot{} = 3\sqrt{2}\,\frac{\Nedge}{\Ntet \, \CosmoConst}\frac{1}{\left(1-\frac{1}{8}\Lamblendot{}^2\right)^{1/2}}\left(-n\CWTrapdihedraldot{} +\frac{1}{8} \, \Lamblendot{}\,\Lamblenddot{} \, \frac{(2\pi-n\CWTrapdihedral{})}{1-\frac{1}{8}\Lamblendot{}^2} \right).
\label{AppProof1}
\end{equation}
The derivative $\CWTrapdihedraldot{}$ can be determined simply by differentiating $\CWTrapdihedral{} = \arccos\left(\frac{1+\frac{1}{8}\Lamblendot{}^2}{3-\frac{1}{8}\Lamblendot{}^2} \right)$, yielding
$$\CWTrapdihedraldot{} = -\frac{1}{2\sqrt{2}}\frac{1}{\left(1-\frac{1}{8}\Lamblendot{}^2\right)^{1/2}}\frac{\Lamblendot{}\,\Lamblenddot{}}{3-\frac{1}{8}\Lamblendot{}^2},$$
and \eqref{AppProof1} now becomes
\begin{widetext}
\begin{equation}
\Lamblen{} = 3\sqrt{2}\,\frac{\Nedge}{\Ntet \, \CosmoConst}\frac{1}{\left(1-\frac{1}{8}\Lamblendot{}^2\right)^{1/2}} \left(\frac{n}{2\sqrt{2}}\frac{1}{\left(1-\frac{1}{8}\Lamblendot{}^2\right)^{1/2}}\frac{\Lamblenddot{}}{3-\frac{1}{8}\Lamblendot{}^2} +\frac{1}{8} \, \Lamblenddot{} \, \frac{(2\pi-n\CWTrapdihedral{})}{1-\frac{1}{8}\Lamblendot{}^2} \right) \! .
\label{AppProof2}
\end{equation}
\end{widetext}
We next multiply through by $\Lamblen{}$ and replace the $\Lamblen{}^2$ on the left-hand side by \eqref{GlobConstraint}.  After further simplification, we arrive at
$$
0 = \frac{1}{8}\frac{(2\pi-n\CWTrapdihedral{})}{\left(1-\frac{1}{8}\Lamblendot{}^2\right)^{1/2}}\left(\Lamblen{} \Lamblenddot{}+2\Lamblendot{}^2-16\right) + \frac{n}{2\sqrt{2}} \frac{\Lamblen{} \Lamblenddot{}}{\left(\frac{3}{8}\Lamblendot{}^2-1\right)},
$$
which is clearly identical to \eqref{AppProof0} apart from an irrelevant overall factor of $(\frac{3}{8}\Lamblendot{}^2-1) / (1-\frac{1}{8}\Lamblendot{}^2)$.  The numerator of this factor is simply the square of the strut-length and is therefore constrained to be negative.  This in turn implies that the denominator must be strictly positive.  Hence, the factor is never singular nor zero.  Therefore, we can conclude that the constraint equation is indeed a first integral of the evolution equation for the parent models.

We now turn to the children models.  Following the example of the parent model, we similarly differentiate constraint equation \eqref{SReggeEqn} with respect to time and simplify the resulting expression to get
\begin{widetext}
\begin{IEEEeqnarray}{rCl}
0 &=& \frac{\Nedge}{\Ntet}\frac{1}{\left(4-\frac{1}{3}\frac{\alpha^2\BrewinBasedot{}^2}{\alpha^2-\frac{1}{3}}\right)^{1/2}} \nonumber\\
&& {} \times \left[\frac{3n \alpha^4 \BrewinBase{} \BrewinBaseddot{}}{2\sqrt{\alpha^2-\frac{1}{3}}\left[3(4\alpha^2-1)-\frac{\alpha^4\BrewinBasedot{}^2}{4\left(\alpha^2-\frac{1}{3}\right)}\right]\left(3-\frac{\alpha^2\BrewinBasedot{}^2}{4\left(\alpha^2-\frac{1}{3}\right)}\right)^{1/2}} + \frac{\alpha \left(2\pi - n\CWSubdivTrapdi{2}{} \right)}{4-\frac{1}{3}\frac{\alpha^2\BrewinBasedot{}^2}{\alpha^2-\frac{1}{3}}}\left(\frac{1}{3}\frac{\alpha^2}{\alpha^2-\frac{1}{3}}\BrewinBase{}\BrewinBaseddot{} + \frac{2}{3}\frac{\alpha^2\BrewinBasedot{}^2}{\alpha^2-\frac{1}{3}} - 8 \right) \right]\nonumber\\
&& {} + \frac{1}{\left[4-\left(\frac{\alpha^4}{\alpha^2-\frac{1}{3}}-1\right)\BrewinBasedot{}^2\right]^{1/2}} \nonumber\\
&& {} \times \left\{6\sqrt{3}\BrewinBase{} \BrewinBaseddot{} \left(\frac{3\left(\alpha^2-\frac{1}{3}\right)^2\left(1-\frac{\alpha^2}{9\left(\alpha^2-\frac{1}{3}\right)^2}\right) - \frac{1}{4}\left(\alpha^2-\frac{2}{3}\right)\left(\frac{\alpha^4}{\alpha^2-\frac{1}{3}}-1\right)\alpha^2\BrewinBasedot{}^2}{\sqrt{\alpha^2-\frac{1}{3}}\left[4-\left(\frac{\alpha^4}{\alpha^2-\frac{1}{3}}-1\right)\BrewinBasedot{}^2\right]^{1/2}\left(1- \frac{\left(\alpha^2-\frac{2}{3}\right)^2}{4\left(\alpha^2-\frac{1}{3}\right)}\BrewinBasedot{}^2\right) \left[3\left(4\alpha^2-1\right)- \frac{\alpha^4\BrewinBasedot{}^2}{4\left(\alpha^2-\frac{1}{3}\right)}\right] } \right. \right. \nonumber\\
&& \hphantom{{} \times [6\sqrt{3}\BrewinBase{} \BrewinBaseddot{} [} {}+ \left.\frac{6\left(\beta^2-\frac{1}{3}\right)^2\left(1-\frac{\beta^2}{9\left(\beta^2-\frac{1}{3}\right)^2}\right) - \frac{1}{2}\left(\beta^2-\frac{2}{3}\right)\left(\frac{\beta^4}{\beta^2-\frac{1}{3}}-1\right)\beta^2\BrewinBasedot{}^2}{\sqrt{\beta^2-\frac{1}{3}}\left[4-\left(\frac{\beta^4}{\beta^2-\frac{1}{3}}-1\right)\BrewinBasedot{}^2\right]^{1/2}\left(1- \frac{\left(\beta^2-\frac{2}{3}\right)^2}{4\left(\beta^2-\frac{1}{3}\right)}\BrewinBasedot{}^2\right) \left[3\left(4\beta^2-1\right)- \frac{\beta^4\BrewinBasedot{}^2}{4\left(\beta^2-\frac{1}{3}\right)}\right] } \right)\nonumber\\
&& \hphantom{{} \times [} {}+ \left.\frac{\left(2\pi - 2\CWSubdivTrapdi{1}{} - 4\CWSubdivTrapdi{3}{} \right)}{4-\left(\frac{\alpha^4}{\alpha^2-\frac{1}{3}}-1\right)\BrewinBasedot{}^2}\left[3\left(\frac{\alpha^4}{\alpha^2-\frac{1}{3}}-1\right)\BrewinBase{}\BrewinBaseddot{} + 6\left(\frac{\alpha^4}{\alpha^2-\frac{1}{3}}-1\right)\BrewinBasedot{}^2 - 24 \right] \vphantom{\left[\frac{\left(\frac{\alpha^2}{9\left(\alpha^2-\frac{1}{3}\right)^2}\right)}{\left[\left(\frac{\beta^4}{\beta^2-\frac{1}{3}}\right)\right]^{1/2}}\right]} \right\}\nonumber\\
&& {}+ \frac{1}{\left(4-\frac{1}{3}\frac{\beta^2\BrewinBasedot{}^2}{\beta^2-\frac{1}{3}}\right)^{1/2}} \nonumber\\
&& {} \times \left[\frac{18\beta^4 \BrewinBase{} \BrewinBaseddot{}}{\sqrt{\beta^2-\frac{1}{3}}\left[3(4\beta^2-1)-\frac{\beta^4\BrewinBasedot{}^2}{4\left(\beta^2-\frac{1}{3}\right)}\right]\left(3-\frac{\beta^2\BrewinBasedot{}^2}{4\left(\beta^2-\frac{1}{3}\right)}\right)^{1/2}} + \frac{\beta \left(2\pi - 4\CWSubdivTrapdi{4}{} \right)}{4-\frac{1}{3}\frac{\beta^2\BrewinBasedot{}^2}{\beta^2-\frac{1}{3}}}\left(\frac{\beta^2}{\beta^2-\frac{1}{3}}\BrewinBase{}\BrewinBaseddot{} + \frac{2\beta^2\BrewinBasedot{}^2}{\beta^2-\frac{1}{3}} - 24 \right) \right]
\label{ExpandedSConstraint}
\end{IEEEeqnarray}
\end{widetext}
If we use \eqref{SReggeEqn} to expand out the $\BrewinBase{}^2$ term at the end of \eqref{SEvolEqn}, we hope to find that this expanded form of \eqref{SEvolEqn} will be identical to \eqref{ExpandedSConstraint} apart from some overall factor.  If this is the case, then we shall have demonstrated that \eqref{SReggeEqn} is a first integral of \eqref{SEvolEqn}.

Note that \eqref{SEvolEqn} and \eqref{ExpandedSConstraint} can be separated into four distinct components corresponding to a $(\Nedge / \Ntet) \left(2\pi - n\CWSubdivTrapdi{2}{} \right)$ term, a $\left(2\pi - 2\CWSubdivTrapdi{1}{} - 4\CWSubdivTrapdi{3}{} \right)$ term, a $\left(2\pi - 4\CWSubdivTrapdi{4}{} \right)$ term, and everything else.  To identify the factor, our approach will be to match each component independently of the rest.  We begin with the $(\Nedge / \Ntet) \left(2\pi - n\CWSubdivTrapdi{2}{} \right)$ component which requires the following equality to hold:
\begin{widetext}
\begin{equation*}
16\left(\frac{\alpha^4}{4} \frac{\alpha^2-\frac{4}{3}}{\alpha^2-\frac{1}{3}}\BrewinBase{} \BrewinBaseddot{} + \frac{1}{24} \frac{\alpha^6\BrewinBasedot{}^4}{\left(\alpha^2-\frac{1}{3}\right)^2} - \frac{\alpha^2\left(\alpha^2+\frac{1}{3}\right)\BrewinBasedot{}^2}{2\left(\alpha^2-\frac{1}{3}\right)} + \frac{\alpha^6 \BrewinBase{} \BrewinBasedot{}^2 \BrewinBaseddot{}}{48 \left(\alpha^2-\frac{1}{3}\right)^2} + 2 \right) = factor \times \left(\frac{1}{3} \frac{\alpha^2}{\alpha^2-\frac{1}{3}}\BrewinBase{} \BrewinBaseddot{} + \frac{2}{3}\frac{\alpha^2\BrewinBasedot{}^2}{\alpha^2-\frac{1}{3}} -8 \right).
\end{equation*}
\end{widetext}
The highest order terms on the left-hand side are the $\BrewinBasedot{}^4$ and $\BrewinBase{} \BrewinBasedot{}^2 \BrewinBaseddot{}$ terms, which are of order $\BrewinBasedot{}^2$ higher than the right-hand side.  Thus we make the ansatz that the factor is of the form
$$
factor = A + B\, \BrewinBasedot{}^2
$$
for some constants $A$ and $B$ to be determined.  By matching the numerical constants on the two sides, we find that $A = -4$.  However if we now match the $\BrewinBase{} \BrewinBaseddot{}$ terms, we obtain the expression
$$
4\alpha^4 \frac{\alpha^2-\frac{4}{3}}{\alpha^2-\frac{1}{3}} = -\frac{4}{3} \frac{\alpha^2}{\alpha^2-\frac{1}{3}}.
$$
This equality can only be true if $\alpha = \pm1 \text{ or } \pm 1/\sqrt{3}$.  However since $\alpha$ is the ratio between the lengths of two edges, we require $\alpha > 0$, and for the strut-length to be finite, we also require $\alpha > 1/\sqrt{3}$.  Therefore the only acceptable solution is $\alpha = 1$.  After making this choice, we can continue matching the remaining terms to find that $B = 3/2$.  Thus the factor is given by
$$
factor = -4\left(1 - \frac{3}{8} \BrewinBasedot{}^2 \right).
$$

By matching the $\left(2\pi - 4\CWSubdivTrapdi{4}{} \right)$ component in a similar manner, we also find that $\beta = 1$ and that the terms from the two equations also differ by the same factor.

If we now make the substitution $\alpha = \beta = 1$, we find that the evolution equation \eqref{SEvolEqn} is indeed identical to the time-derivative of the constraint equation \eqref{ExpandedSConstraint} multiplied by the overall factor of $-4[1 - (3/8) \BrewinBasedot{}^2 ]$.  Therefore when $\alpha = \beta = 1$, the constraint equation is a first integral of the evolution equation.

\bibliography{Lambda_Regge_model}

\end{document}

%% file: 4-block.pspdftex
\begin{picture}(0,0)%
\includegraphics{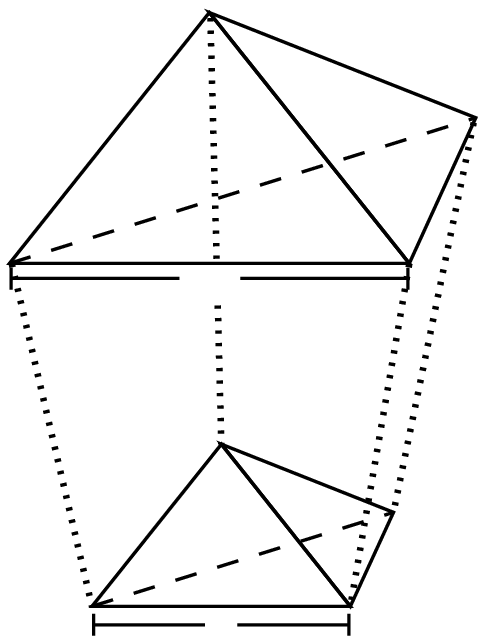}%
\end{picture}%
\setlength{\unitlength}{3947sp}%
\begingroup\makeatletter\ifx\SetFigFont\undefined%
\gdef\SetFigFont#1#2#3#4#5{%
  \reset@font\fontsize{#1}{#2pt}%
  \fontfamily{#3}\fontseries{#4}\fontshape{#5}%
  \selectfont}%
\fi\endgroup%
\begin{picture}(2528,3230)(1754,662)
\put(2764,2361){\makebox(0,0)[b]{\smash{{\SetFigFont{8}{9.6}{\rmdefault}{\mddefault}{\updefault}{\color[rgb]{0,0,0}$\Lamblen{i+1}$}%
}}}}
\put(2790,3745){\makebox(0,0)[b]{\smash{{\SetFigFont{8}{9.6}{\rmdefault}{\mddefault}{\updefault}{\color[rgb]{0,0,0}$D^\prime$}%
}}}}
\put(4058,3184){\makebox(0,0)[lb]{\smash{{\SetFigFont{8}{9.6}{\rmdefault}{\mddefault}{\updefault}{\color[rgb]{0,0,0}$C^\prime$}%
}}}}
\put(3737,2447){\makebox(0,0)[lb]{\smash{{\SetFigFont{8}{9.6}{\rmdefault}{\mddefault}{\updefault}{\color[rgb]{0,0,0}$B^\prime$}%
}}}}
\put(3822,1924){\makebox(0,0)[lb]{\smash{{\SetFigFont{8}{9.6}{\rmdefault}{\mddefault}{\updefault}{\color[rgb]{0,0,0}$\CWstrut{i}$}%
}}}}
\put(3472,801){\makebox(0,0)[lb]{\smash{{\SetFigFont{8}{9.6}{\rmdefault}{\mddefault}{\updefault}{\color[rgb]{0,0,0}$B$}%
}}}}
\put(3677,1275){\makebox(0,0)[lb]{\smash{{\SetFigFont{8}{9.6}{\rmdefault}{\mddefault}{\updefault}{\color[rgb]{0,0,0}$C$}%
}}}}
\put(2891,1658){\makebox(0,0)[b]{\smash{{\SetFigFont{8}{9.6}{\rmdefault}{\mddefault}{\updefault}{\color[rgb]{0,0,0}$D$}%
}}}}
\put(2154,801){\makebox(0,0)[rb]{\smash{{\SetFigFont{8}{9.6}{\rmdefault}{\mddefault}{\updefault}{\color[rgb]{0,0,0}$A$}%
}}}}
\put(1769,2447){\makebox(0,0)[rb]{\smash{{\SetFigFont{8}{9.6}{\rmdefault}{\mddefault}{\updefault}{\color[rgb]{0,0,0}$A^\prime$}%
}}}}
\put(2817,726){\makebox(0,0)[b]{\smash{{\SetFigFont{8}{9.6}{\rmdefault}{\mddefault}{\updefault}{\color[rgb]{0,0,0}$\Lamblen{i}$}%
}}}}
\put(3925,1009){\makebox(0,0)[lb]{\smash{{\SetFigFont{8}{9.6}{\rmdefault}{\mddefault}{\updefault}{\color[rgb]{0,0,0}$\RegTime{i}$}%
}}}}
\put(4267,2672){\makebox(0,0)[lb]{\smash{{\SetFigFont{8}{9.6}{\rmdefault}{\mddefault}{\updefault}{\color[rgb]{0,0,0}$\RegTime{i+1}$}%
}}}}
\end{picture}%

%% file: allhinges.pspdftex
\begin{picture}(0,0)%
\includegraphics{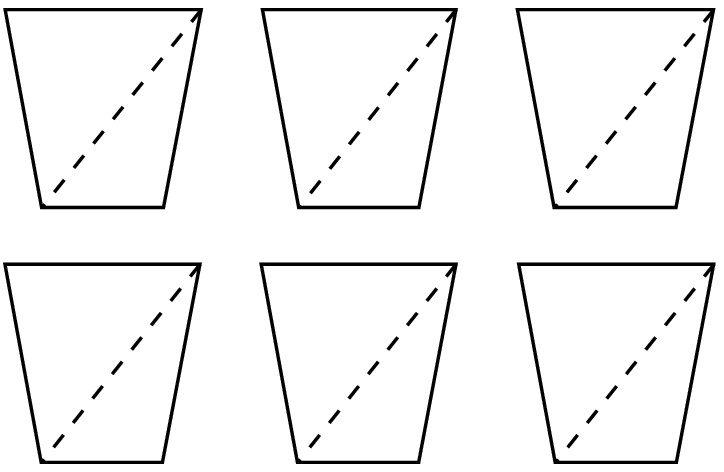}%
\end{picture}%
\setlength{\unitlength}{3947sp}%
\begingroup\makeatletter\ifx\SetFigFont\undefined%
\gdef\SetFigFont#1#2#3#4#5{%
  \reset@font\fontsize{#1}{#2pt}%
  \fontfamily{#3}\fontseries{#4}\fontshape{#5}%
  \selectfont}%
\fi\endgroup%
\begin{picture}(3468,2478)(-3391,-2943)
\put(-1343,-1675){\makebox(0,0)[b]{\smash{{\SetFigFont{8}{9.6}{\rmdefault}{\mddefault}{\updefault}{\color[rgb]{0,0,0}$C$}%
}}}}
\put(-2399,-576){\makebox(0,0)[b]{\smash{{\SetFigFont{8}{9.6}{\rmdefault}{\mddefault}{\updefault}{\color[rgb]{0,0,0}$B^\prime$}%
}}}}
\put(-2568,-1675){\makebox(0,0)[b]{\smash{{\SetFigFont{8}{9.6}{\rmdefault}{\mddefault}{\updefault}{\color[rgb]{0,0,0}$B$}%
}}}}
\put(-3376,-576){\makebox(0,0)[b]{\smash{{\SetFigFont{8}{9.6}{\rmdefault}{\mddefault}{\updefault}{\color[rgb]{0,0,0}$A^\prime$}%
}}}}
\put(-3230,-1675){\makebox(0,0)[b]{\smash{{\SetFigFont{8}{9.6}{\rmdefault}{\mddefault}{\updefault}{\color[rgb]{0,0,0}$A$}%
}}}}
\put(-3376,-1799){\makebox(0,0)[b]{\smash{{\SetFigFont{8}{9.6}{\rmdefault}{\mddefault}{\updefault}{\color[rgb]{0,0,0}$A^\prime$}%
}}}}
\put(-2574,-2897){\makebox(0,0)[b]{\smash{{\SetFigFont{8}{9.6}{\rmdefault}{\mddefault}{\updefault}{\color[rgb]{0,0,0}$D$}%
}}}}
\put(-3231,-2897){\makebox(0,0)[b]{\smash{{\SetFigFont{8}{9.6}{\rmdefault}{\mddefault}{\updefault}{\color[rgb]{0,0,0}$A$}%
}}}}
\put(-2406,-1799){\makebox(0,0)[b]{\smash{{\SetFigFont{8}{9.6}{\rmdefault}{\mddefault}{\updefault}{\color[rgb]{0,0,0}$D^\prime$}%
}}}}
\put(-107,-2897){\makebox(0,0)[b]{\smash{{\SetFigFont{8}{9.6}{\rmdefault}{\mddefault}{\updefault}{\color[rgb]{0,0,0}$D$}%
}}}}
\put(-764,-2897){\makebox(0,0)[b]{\smash{{\SetFigFont{8}{9.6}{\rmdefault}{\mddefault}{\updefault}{\color[rgb]{0,0,0}$C$}%
}}}}
\put(-907,-1799){\makebox(0,0)[b]{\smash{{\SetFigFont{8}{9.6}{\rmdefault}{\mddefault}{\updefault}{\color[rgb]{0,0,0}$C^\prime$}%
}}}}
\put( 62,-1799){\makebox(0,0)[b]{\smash{{\SetFigFont{8}{9.6}{\rmdefault}{\mddefault}{\updefault}{\color[rgb]{0,0,0}$D^\prime$}%
}}}}
\put(-2145,-1799){\makebox(0,0)[b]{\smash{{\SetFigFont{8}{9.6}{\rmdefault}{\mddefault}{\updefault}{\color[rgb]{0,0,0}$B^\prime$}%
}}}}
\put(-1344,-2897){\makebox(0,0)[b]{\smash{{\SetFigFont{8}{9.6}{\rmdefault}{\mddefault}{\updefault}{\color[rgb]{0,0,0}$D$}%
}}}}
\put(-2002,-2897){\makebox(0,0)[b]{\smash{{\SetFigFont{8}{9.6}{\rmdefault}{\mddefault}{\updefault}{\color[rgb]{0,0,0}$B$}%
}}}}
\put(-1175,-1799){\makebox(0,0)[b]{\smash{{\SetFigFont{8}{9.6}{\rmdefault}{\mddefault}{\updefault}{\color[rgb]{0,0,0}$D^\prime$}%
}}}}
\put(-912,-576){\makebox(0,0)[b]{\smash{{\SetFigFont{8}{9.6}{\rmdefault}{\mddefault}{\updefault}{\color[rgb]{0,0,0}$B^\prime$}%
}}}}
\put( 61,-576){\makebox(0,0)[b]{\smash{{\SetFigFont{8}{9.6}{\rmdefault}{\mddefault}{\updefault}{\color[rgb]{0,0,0}$C^\prime$}%
}}}}
\put(-769,-1675){\makebox(0,0)[b]{\smash{{\SetFigFont{8}{9.6}{\rmdefault}{\mddefault}{\updefault}{\color[rgb]{0,0,0}$B$}%
}}}}
\put(-109,-1675){\makebox(0,0)[b]{\smash{{\SetFigFont{8}{9.6}{\rmdefault}{\mddefault}{\updefault}{\color[rgb]{0,0,0}$C$}%
}}}}
\put(-1175,-576){\makebox(0,0)[b]{\smash{{\SetFigFont{8}{9.6}{\rmdefault}{\mddefault}{\updefault}{\color[rgb]{0,0,0}$C^\prime$}%
}}}}
\put(-1995,-1675){\makebox(0,0)[b]{\smash{{\SetFigFont{8}{9.6}{\rmdefault}{\mddefault}{\updefault}{\color[rgb]{0,0,0}$A$}%
}}}}
\put(-2138,-576){\makebox(0,0)[b]{\smash{{\SetFigFont{8}{9.6}{\rmdefault}{\mddefault}{\updefault}{\color[rgb]{0,0,0}$A^\prime$}%
}}}}
\end{picture}%

%% file: 3sphere.pspdftex
\begin{picture}(0,0)%
\includegraphics{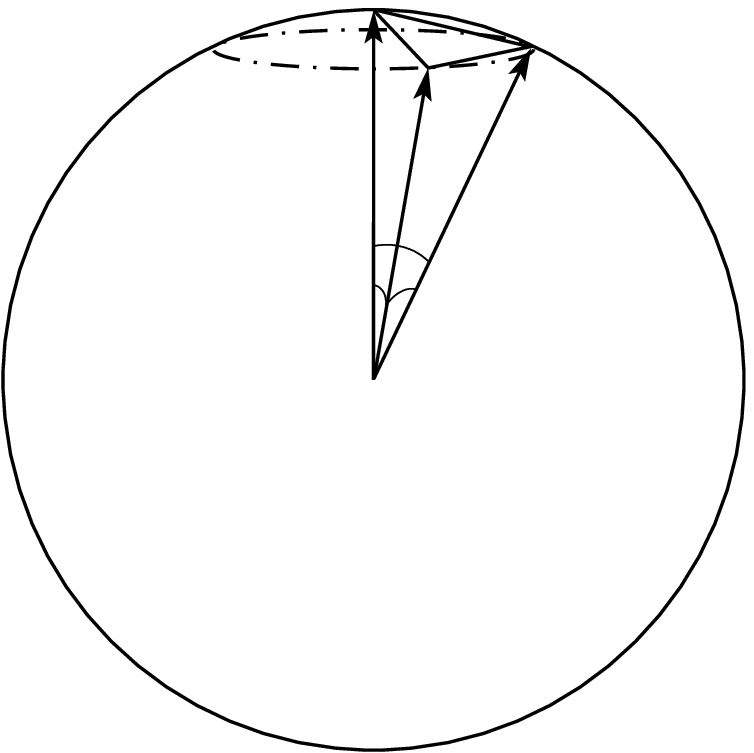}%
\end{picture}%
\setlength{\unitlength}{3947sp}%
\begingroup\makeatletter\ifx\SetFigFont\undefined%
\gdef\SetFigFont#1#2#3#4#5{%
  \reset@font\fontsize{#1}{#2pt}%
  \fontfamily{#3}\fontseries{#4}\fontshape{#5}%
  \selectfont}%
\fi\endgroup%
\begin{picture}(3586,3591)(2939,-2451)
\put(4776,-187){\makebox(0,0)[b]{\smash{{\SetFigFont{8}{9.6}{\rmdefault}{\mddefault}{\updefault}{\color[rgb]{0,0,0}$\theta$}%
}}}}
\put(5258,752){\makebox(0,0)[b]{\smash{{\SetFigFont{8}{9.6}{\rmdefault}{\mddefault}{\updefault}{\color[rgb]{0,0,0}$\Lamblen{i}$}%
}}}}
\put(4890,-199){\makebox(0,0)[b]{\smash{{\SetFigFont{8}{9.6}{\rmdefault}{\mddefault}{\updefault}{\color[rgb]{0,0,0}$\phi$}%
}}}}
\put(4890,860){\makebox(0,0)[rb]{\smash{{\SetFigFont{8}{9.6}{\rmdefault}{\mddefault}{\updefault}{\color[rgb]{0,0,0}$\Lamblen{i}$}%
}}}}
\put(5182,1041){\makebox(0,0)[b]{\smash{{\SetFigFont{8}{9.6}{\rmdefault}{\mddefault}{\updefault}{\color[rgb]{0,0,0}$\Lamblen{i}$}%
}}}}
\put(4717,247){\makebox(0,0)[rb]{\smash{{\SetFigFont{8}{9.6}{\rmdefault}{\mddefault}{\updefault}{\color[rgb]{0,0,0}$\Rad{i}$}%
}}}}
\put(4915, -7){\makebox(0,0)[b]{\smash{{\SetFigFont{8}{9.6}{\rmdefault}{\mddefault}{\updefault}{\color[rgb]{0,0,0}$\theta$}%
}}}}
\end{picture}%

%% file: hingeschem.pspdftex
\begin{picture}(0,0)%
\includegraphics{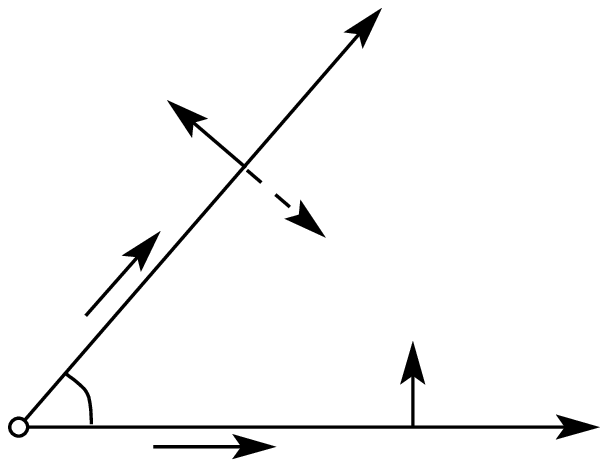}%
\end{picture}%
\setlength{\unitlength}{3947sp}%
\begingroup\makeatletter\ifx\SetFigFont\undefined%
\gdef\SetFigFont#1#2#3#4#5{%
  \reset@font\fontsize{#1}{#2pt}%
  \fontfamily{#3}\fontseries{#4}\fontshape{#5}%
  \selectfont}%
\fi\endgroup%
\begin{picture}(3410,2414)(3792,-2123)
\put(4819,-884){\makebox(0,0)[rb]{\smash{{\SetFigFont{8}{9.6}{\rmdefault}{\mddefault}{\updefault}{\color[rgb]{0,0,0}$\Tensb{u}_2$}%
}}}}
\put(5293,-2019){\makebox(0,0)[lb]{\smash{{\SetFigFont{8}{9.6}{\rmdefault}{\mddefault}{\updefault}{\color[rgb]{0,0,0}$\Tensb{u}_1$}%
}}}}
\put(5829,201){\makebox(0,0)[b]{\smash{{\SetFigFont{8}{9.6}{\rmdefault}{\mddefault}{\updefault}{\color[rgb]{0,0,0}Face 2}%
}}}}
\put(3961,-2077){\makebox(0,0)[b]{\smash{{\SetFigFont{8}{9.6}{\rmdefault}{\mddefault}{\updefault}{\color[rgb]{0,0,0}Hinge}%
}}}}
\put(4377,-1772){\makebox(0,0)[lb]{\smash{{\SetFigFont{8}{9.6}{\rmdefault}{\mddefault}{\updefault}{\color[rgb]{0,0,0}$\theta$}%
}}}}
\put(5518,-1098){\makebox(0,0)[b]{\smash{{\SetFigFont{8}{9.6}{\rmdefault}{\mddefault}{\updefault}{\color[rgb]{0,0,0}$-\Tensb{n}_2$}%
}}}}
\put(6849,-1946){\makebox(0,0)[lb]{\smash{{\SetFigFont{8}{9.6}{\rmdefault}{\mddefault}{\updefault}{\color[rgb]{0,0,0}Face 1}%
}}}}
\put(5894,-1420){\makebox(0,0)[b]{\smash{{\SetFigFont{8}{9.6}{\rmdefault}{\mddefault}{\updefault}{\color[rgb]{0,0,0}$\Tensb{n}_1$}%
}}}}
\put(4600,-250){\makebox(0,0)[b]{\smash{{\SetFigFont{8}{9.6}{\rmdefault}{\mddefault}{\updefault}{\color[rgb]{0,0,0}$\Tensb{n}_2$}%
}}}}
\end{picture}%

%% file: traphinge.pspdftex
\begin{picture}(0,0)%
\includegraphics{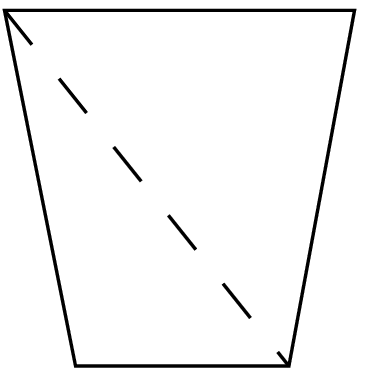}%
\end{picture}%
\setlength{\unitlength}{3947sp}%
\begingroup\makeatletter\ifx\SetFigFont\undefined%
\gdef\SetFigFont#1#2#3#4#5{%
  \reset@font\fontsize{#1}{#2pt}%
  \fontfamily{#3}\fontseries{#4}\fontshape{#5}%
  \selectfont}%
\fi\endgroup%
\begin{picture}(1724,2041)(2354,-2523)
\put(3108,-1510){\makebox(0,0)[lb]{\smash{{\SetFigFont{8}{9.6}{\rmdefault}{\mddefault}{\updefault}{\color[rgb]{0,0,0}$\CWdiag{i}$}%
}}}}
\put(3006,-1956){\makebox(0,0)[b]{\smash{{\SetFigFont{8}{9.6}{\rmdefault}{\mddefault}{\updefault}{\color[rgb]{0,0,0}$\Area{A}{i}$}%
}}}}
\put(3952,-1483){\makebox(0,0)[lb]{\smash{{\SetFigFont{8}{9.6}{\rmdefault}{\mddefault}{\updefault}{\color[rgb]{0,0,0}$\CWstrut{i}^\textrm{B}$}%
}}}}
\put(3216,-581){\makebox(0,0)[b]{\smash{{\SetFigFont{8}{9.6}{\rmdefault}{\mddefault}{\updefault}{\color[rgb]{0,0,0}$\Lamblen{i+1}$}%
}}}}
\put(3216,-2477){\makebox(0,0)[b]{\smash{{\SetFigFont{8}{9.6}{\rmdefault}{\mddefault}{\updefault}{\color[rgb]{0,0,0}$\Lamblen{i}$}%
}}}}
\put(2488,-1483){\makebox(0,0)[rb]{\smash{{\SetFigFont{8}{9.6}{\rmdefault}{\mddefault}{\updefault}{\color[rgb]{0,0,0}$\CWstrut{i}^\textrm{A}$}%
}}}}
\put(3453,-1142){\makebox(0,0)[b]{\smash{{\SetFigFont{8}{9.6}{\rmdefault}{\mddefault}{\updefault}{\color[rgb]{0,0,0}$\Area{B}{i}$}%
}}}}
\end{picture}%

%% file: ReggeParentVols.tex
\begingroup
  \fontfamily{Verdana}%
  \selectfont
  \makeatletter
  \providecommand\color[2][]{%
    \GenericError{(gnuplot) \space\space\space\@spaces}{%
      Package color not loaded in conjunction with
      terminal option `colourtext'%
    }{See the gnuplot documentation for explanation.%
    }{Either use 'blacktext' in gnuplot or load the package
      color.sty in LaTeX.}%
    \renewcommand\color[2][]{}%
  }%
  \providecommand\includegraphics[2][]{%
    \GenericError{(gnuplot) \space\space\space\@spaces}{%
      Package graphicx or graphics not loaded%
    }{See the gnuplot documentation for explanation.%
    }{The gnuplot epslatex terminal needs graphicx.sty or graphics.sty.}%
    \renewcommand\includegraphics[2][]{}%
  }%
  \providecommand\rotatebox[2]{#2}%
  \@ifundefined{ifGPcolor}{%
    \newif\ifGPcolor
    \GPcolorfalse
  }{}%
  \@ifundefined{ifGPblacktext}{%
    \newif\ifGPblacktext
    \GPblacktexttrue
  }{}%
  \let\gplgaddtomacro\g@addto@macro
  \gdef\gplbacktext{}%
  \gdef\gplfronttext{}%
  \makeatother
  \ifGPblacktext
    \def\colorrgb#1{}%
    \def\colorgray#1{}%
  \else
    \ifGPcolor
      \def\colorrgb#1{\color[rgb]{#1}}%
      \def\colorgray#1{\color[gray]{#1}}%
      \expandafter\def\csname LTw\endcsname{\color{white}}%
      \expandafter\def\csname LTb\endcsname{\color{black}}%
      \expandafter\def\csname LTa\endcsname{\color{black}}%
      \expandafter\def\csname LT0\endcsname{\color[rgb]{1,0,0}}%
      \expandafter\def\csname LT1\endcsname{\color[rgb]{0,1,0}}%
      \expandafter\def\csname LT2\endcsname{\color[rgb]{0,0,1}}%
      \expandafter\def\csname LT3\endcsname{\color[rgb]{1,0,1}}%
      \expandafter\def\csname LT4\endcsname{\color[rgb]{0,1,1}}%
      \expandafter\def\csname LT5\endcsname{\color[rgb]{1,1,0}}%
      \expandafter\def\csname LT6\endcsname{\color[rgb]{0,0,0}}%
      \expandafter\def\csname LT7\endcsname{\color[rgb]{1,0.3,0}}%
      \expandafter\def\csname LT8\endcsname{\color[rgb]{0.5,0.5,0.5}}%
    \else
      \def\colorrgb#1{\color{black}}%
      \def\colorgray#1{\color[gray]{#1}}%
      \expandafter\def\csname LTw\endcsname{\color{white}}%
      \expandafter\def\csname LTb\endcsname{\color{black}}%
      \expandafter\def\csname LTa\endcsname{\color{black}}%
      \expandafter\def\csname LT0\endcsname{\color{black}}%
      \expandafter\def\csname LT1\endcsname{\color{black}}%
      \expandafter\def\csname LT2\endcsname{\color{black}}%
      \expandafter\def\csname LT3\endcsname{\color{black}}%
      \expandafter\def\csname LT4\endcsname{\color{black}}%
      \expandafter\def\csname LT5\endcsname{\color{black}}%
      \expandafter\def\csname LT6\endcsname{\color{black}}%
      \expandafter\def\csname LT7\endcsname{\color{black}}%
      \expandafter\def\csname LT8\endcsname{\color{black}}%
    \fi
  \fi
  \setlength{\unitlength}{0.0500bp}%
  \begin{picture}(4896.00,3960.00)%
    \gplgaddtomacro\gplbacktext{%
      \csname LTb\endcsname%
      \put(688,512){\makebox(0,0)[r]{\strut{} 0}}%
      \put(688,919){\makebox(0,0)[r]{\strut{} 100}}%
      \put(688,1326){\makebox(0,0)[r]{\strut{} 200}}%
      \put(688,1733){\makebox(0,0)[r]{\strut{} 300}}%
      \put(688,2140){\makebox(0,0)[r]{\strut{} 400}}%
      \put(688,2546){\makebox(0,0)[r]{\strut{} 500}}%
      \put(688,2953){\makebox(0,0)[r]{\strut{} 600}}%
      \put(688,3360){\makebox(0,0)[r]{\strut{} 700}}%
      \put(688,3767){\makebox(0,0)[r]{\strut{} 800}}%
      \put(784,352){\makebox(0,0){\strut{} 100}}%
      \put(1549,352){\makebox(0,0){\strut{} 200}}%
      \put(2313,352){\makebox(0,0){\strut{} 300}}%
      \put(3078,352){\makebox(0,0){\strut{} 400}}%
      \put(3842,352){\makebox(0,0){\strut{} 500}}%
      \put(4607,352){\makebox(0,0){\strut{} 600}}%
      \put(128,2139){\rotatebox{-270}{\makebox(0,0){\strut{}$\frac{dU}{dt}$}}}%
      \put(2695,112){\makebox(0,0){\strut{}$U(t)$}}%
    }%
    \gplgaddtomacro\gplfronttext{%
      \csname LTb\endcsname%
      \put(1519,3564){\makebox(0,0)[l]{\strut{}$\text{FLRW}$}}%
      \csname LTb\endcsname%
      \put(1519,3364){\makebox(0,0)[l]{\strut{}$\text{5 tetrahedra}$}}%
      \csname LTb\endcsname%
      \put(1519,3164){\makebox(0,0)[l]{\strut{}$\text{16 tetrahedra}$}}%
      \csname LTb\endcsname%
      \put(1519,2964){\makebox(0,0)[l]{\strut{}$\text{600 tetrahedra}$}}%
    }%
    \gplbacktext
    \put(0,0){\includegraphics{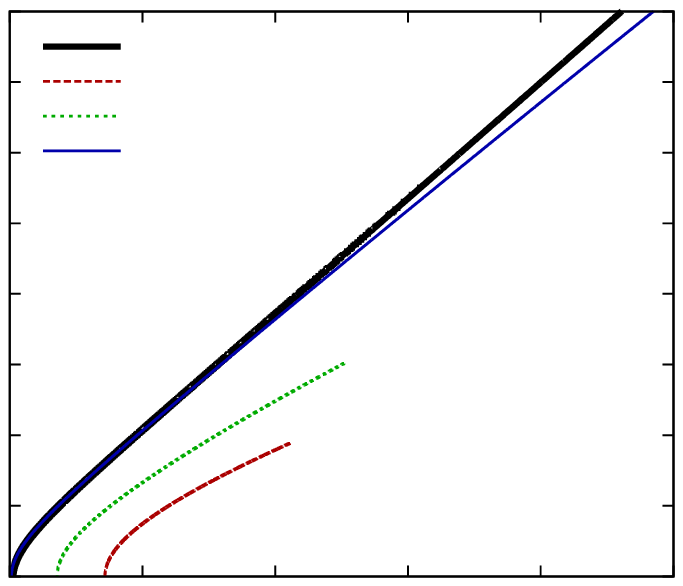}}%
    \gplfronttext
  \end{picture}%
\endgroup

%% file: ReggeParentMatchedVols.tex
\begingroup
  \fontfamily{Verdana}%
  \selectfont
  \makeatletter
  \providecommand\color[2][]{%
    \GenericError{(gnuplot) \space\space\space\@spaces}{%
      Package color not loaded in conjunction with
      terminal option `colourtext'%
    }{See the gnuplot documentation for explanation.%
    }{Either use 'blacktext' in gnuplot or load the package
      color.sty in LaTeX.}%
    \renewcommand\color[2][]{}%
  }%
  \providecommand\includegraphics[2][]{%
    \GenericError{(gnuplot) \space\space\space\@spaces}{%
      Package graphicx or graphics not loaded%
    }{See the gnuplot documentation for explanation.%
    }{The gnuplot epslatex terminal needs graphicx.sty or graphics.sty.}%
    \renewcommand\includegraphics[2][]{}%
  }%
  \providecommand\rotatebox[2]{#2}%
  \@ifundefined{ifGPcolor}{%
    \newif\ifGPcolor
    \GPcolorfalse
  }{}%
  \@ifundefined{ifGPblacktext}{%
    \newif\ifGPblacktext
    \GPblacktexttrue
  }{}%
  \let\gplgaddtomacro\g@addto@macro
  \gdef\gplbacktext{}%
  \gdef\gplfronttext{}%
  \makeatother
  \ifGPblacktext
    \def\colorrgb#1{}%
    \def\colorgray#1{}%
  \else
    \ifGPcolor
      \def\colorrgb#1{\color[rgb]{#1}}%
      \def\colorgray#1{\color[gray]{#1}}%
      \expandafter\def\csname LTw\endcsname{\color{white}}%
      \expandafter\def\csname LTb\endcsname{\color{black}}%
      \expandafter\def\csname LTa\endcsname{\color{black}}%
      \expandafter\def\csname LT0\endcsname{\color[rgb]{1,0,0}}%
      \expandafter\def\csname LT1\endcsname{\color[rgb]{0,1,0}}%
      \expandafter\def\csname LT2\endcsname{\color[rgb]{0,0,1}}%
      \expandafter\def\csname LT3\endcsname{\color[rgb]{1,0,1}}%
      \expandafter\def\csname LT4\endcsname{\color[rgb]{0,1,1}}%
      \expandafter\def\csname LT5\endcsname{\color[rgb]{1,1,0}}%
      \expandafter\def\csname LT6\endcsname{\color[rgb]{0,0,0}}%
      \expandafter\def\csname LT7\endcsname{\color[rgb]{1,0.3,0}}%
      \expandafter\def\csname LT8\endcsname{\color[rgb]{0.5,0.5,0.5}}%
    \else
      \def\colorrgb#1{\color{black}}%
      \def\colorgray#1{\color[gray]{#1}}%
      \expandafter\def\csname LTw\endcsname{\color{white}}%
      \expandafter\def\csname LTb\endcsname{\color{black}}%
      \expandafter\def\csname LTa\endcsname{\color{black}}%
      \expandafter\def\csname LT0\endcsname{\color{black}}%
      \expandafter\def\csname LT1\endcsname{\color{black}}%
      \expandafter\def\csname LT2\endcsname{\color{black}}%
      \expandafter\def\csname LT3\endcsname{\color{black}}%
      \expandafter\def\csname LT4\endcsname{\color{black}}%
      \expandafter\def\csname LT5\endcsname{\color{black}}%
      \expandafter\def\csname LT6\endcsname{\color{black}}%
      \expandafter\def\csname LT7\endcsname{\color{black}}%
      \expandafter\def\csname LT8\endcsname{\color{black}}%
    \fi
  \fi
  \setlength{\unitlength}{0.0500bp}%
  \begin{picture}(4896.00,3960.00)%
    \gplgaddtomacro\gplbacktext{%
      \csname LTb\endcsname%
      \put(688,512){\makebox(0,0)[r]{\strut{} 0}}%
      \put(688,919){\makebox(0,0)[r]{\strut{} 100}}%
      \put(688,1326){\makebox(0,0)[r]{\strut{} 200}}%
      \put(688,1733){\makebox(0,0)[r]{\strut{} 300}}%
      \put(688,2140){\makebox(0,0)[r]{\strut{} 400}}%
      \put(688,2546){\makebox(0,0)[r]{\strut{} 500}}%
      \put(688,2953){\makebox(0,0)[r]{\strut{} 600}}%
      \put(688,3360){\makebox(0,0)[r]{\strut{} 700}}%
      \put(688,3767){\makebox(0,0)[r]{\strut{} 800}}%
      \put(784,352){\makebox(0,0){\strut{} 100}}%
      \put(1549,352){\makebox(0,0){\strut{} 200}}%
      \put(2313,352){\makebox(0,0){\strut{} 300}}%
      \put(3078,352){\makebox(0,0){\strut{} 400}}%
      \put(3842,352){\makebox(0,0){\strut{} 500}}%
      \put(4607,352){\makebox(0,0){\strut{} 600}}%
      \put(128,2139){\rotatebox{-270}{\makebox(0,0){\strut{}$\frac{dU}{dt}$}}}%
      \put(2695,112){\makebox(0,0){\strut{}$U(t)$}}%
    }%
    \gplgaddtomacro\gplfronttext{%
      \csname LTb\endcsname%
      \put(1519,3564){\makebox(0,0)[l]{\strut{}\text{FLRW}}}%
      \csname LTb\endcsname%
      \put(1519,3364){\makebox(0,0)[l]{\strut{}$\text{5 tetrahedra}$}}%
      \csname LTb\endcsname%
      \put(1519,3164){\makebox(0,0)[l]{\strut{}$\text{16 tetrahedra}$}}%
      \csname LTb\endcsname%
      \put(1519,2964){\makebox(0,0)[l]{\strut{}$\text{600 tetrahedra}$}}%
    }%
    \gplbacktext
    \put(0,0){\includegraphics{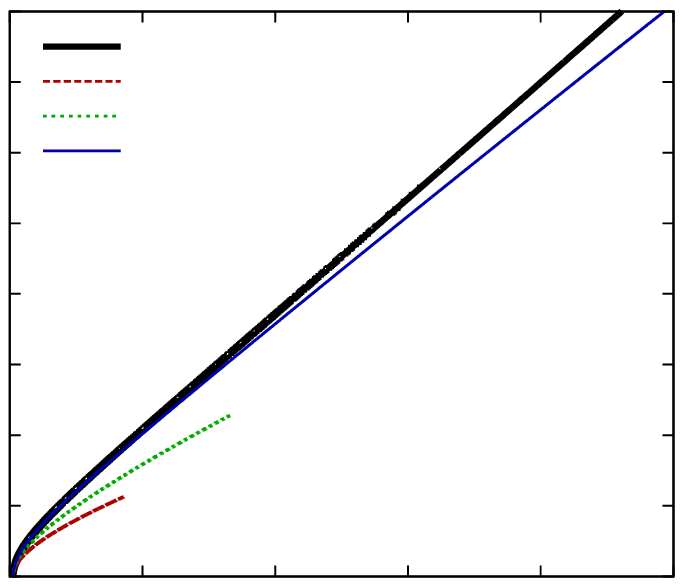}}%
    \gplfronttext
  \end{picture}%
\endgroup

%% file: ParentMatchRadGraph.tex
\begingroup
  \fontfamily{Verdana}%
  \selectfont
  \makeatletter
  \providecommand\color[2][]{%
    \GenericError{(gnuplot) \space\space\space\@spaces}{%
      Package color not loaded in conjunction with
      terminal option `colourtext'%
    }{See the gnuplot documentation for explanation.%
    }{Either use 'blacktext' in gnuplot or load the package
      color.sty in LaTeX.}%
    \renewcommand\color[2][]{}%
  }%
  \providecommand\includegraphics[2][]{%
    \GenericError{(gnuplot) \space\space\space\@spaces}{%
      Package graphicx or graphics not loaded%
    }{See the gnuplot documentation for explanation.%
    }{The gnuplot epslatex terminal needs graphicx.sty or graphics.sty.}%
    \renewcommand\includegraphics[2][]{}%
  }%
  \providecommand\rotatebox[2]{#2}%
  \@ifundefined{ifGPcolor}{%
    \newif\ifGPcolor
    \GPcolorfalse
  }{}%
  \@ifundefined{ifGPblacktext}{%
    \newif\ifGPblacktext
    \GPblacktexttrue
  }{}%
  \let\gplgaddtomacro\g@addto@macro
  \gdef\gplbacktext{}%
  \gdef\gplfronttext{}%
  \makeatother
  \ifGPblacktext
    \def\colorrgb#1{}%
    \def\colorgray#1{}%
  \else
    \ifGPcolor
      \def\colorrgb#1{\color[rgb]{#1}}%
      \def\colorgray#1{\color[gray]{#1}}%
      \expandafter\def\csname LTw\endcsname{\color{white}}%
      \expandafter\def\csname LTb\endcsname{\color{black}}%
      \expandafter\def\csname LTa\endcsname{\color{black}}%
      \expandafter\def\csname LT0\endcsname{\color[rgb]{1,0,0}}%
      \expandafter\def\csname LT1\endcsname{\color[rgb]{0,1,0}}%
      \expandafter\def\csname LT2\endcsname{\color[rgb]{0,0,1}}%
      \expandafter\def\csname LT3\endcsname{\color[rgb]{1,0,1}}%
      \expandafter\def\csname LT4\endcsname{\color[rgb]{0,1,1}}%
      \expandafter\def\csname LT5\endcsname{\color[rgb]{1,1,0}}%
      \expandafter\def\csname LT6\endcsname{\color[rgb]{0,0,0}}%
      \expandafter\def\csname LT7\endcsname{\color[rgb]{1,0.3,0}}%
      \expandafter\def\csname LT8\endcsname{\color[rgb]{0.5,0.5,0.5}}%
    \else
      \def\colorrgb#1{\color{black}}%
      \def\colorgray#1{\color[gray]{#1}}%
      \expandafter\def\csname LTw\endcsname{\color{white}}%
      \expandafter\def\csname LTb\endcsname{\color{black}}%
      \expandafter\def\csname LTa\endcsname{\color{black}}%
      \expandafter\def\csname LT0\endcsname{\color{black}}%
      \expandafter\def\csname LT1\endcsname{\color{black}}%
      \expandafter\def\csname LT2\endcsname{\color{black}}%
      \expandafter\def\csname LT3\endcsname{\color{black}}%
      \expandafter\def\csname LT4\endcsname{\color{black}}%
      \expandafter\def\csname LT5\endcsname{\color{black}}%
      \expandafter\def\csname LT6\endcsname{\color{black}}%
      \expandafter\def\csname LT7\endcsname{\color{black}}%
      \expandafter\def\csname LT8\endcsname{\color{black}}%
    \fi
  \fi
  \setlength{\unitlength}{0.0500bp}%
  \begin{picture}(4896.00,3600.00)%
    \gplgaddtomacro\gplbacktext{%
      \csname LTb\endcsname%
      \put(544,512){\makebox(0,0)[r]{\strut{}$0.0$}}%
      \put(544,1091){\makebox(0,0)[r]{\strut{}$0.5$}}%
      \put(544,1670){\makebox(0,0)[r]{\strut{}$1.0$}}%
      \put(544,2249){\makebox(0,0)[r]{\strut{}$1.5$}}%
      \put(544,2828){\makebox(0,0)[r]{\strut{}$2.0$}}%
      \put(544,3407){\makebox(0,0)[r]{\strut{}$2.5$}}%
      \put(640,352){\makebox(0,0){\strut{}$  1.0$}}%
      \put(1136,352){\makebox(0,0){\strut{}$  1.5$}}%
      \put(1632,352){\makebox(0,0){\strut{}$  2.0$}}%
      \put(2128,352){\makebox(0,0){\strut{}$  2.5$}}%
      \put(2624,352){\makebox(0,0){\strut{}$  3.0$}}%
      \put(3119,352){\makebox(0,0){\strut{}$  3.5$}}%
      \put(3615,352){\makebox(0,0){\strut{}$  4.0$}}%
      \put(4111,352){\makebox(0,0){\strut{}$  4.5$}}%
      \put(4607,352){\makebox(0,0){\strut{}$  5.0$}}%
      \put(128,1959){\rotatebox{-270}{\makebox(0,0){\strut{}$\frac{d(Radius)}{dt}$}}}%
      \put(2623,112){\makebox(0,0){\strut{}$Radius(t)$}}%
    }%
    \gplgaddtomacro\gplfronttext{%
      \csname LTb\endcsname%
      \put(1375,3204){\makebox(0,0)[l]{\strut{}$\text{FLRW}$}}%
      \csname LTb\endcsname%
      \put(1375,3004){\makebox(0,0)[l]{\strut{}$\text{5 tetrahedra}$}}%
      \csname LTb\endcsname%
      \put(1375,2804){\makebox(0,0)[l]{\strut{}$\text{16 tetrahedra}$}}%
      \csname LTb\endcsname%
      \put(1375,2604){\makebox(0,0)[l]{\strut{}$\text{600 tetrahedra}$}}%
    }%
    \gplbacktext
    \put(0,0){\includegraphics{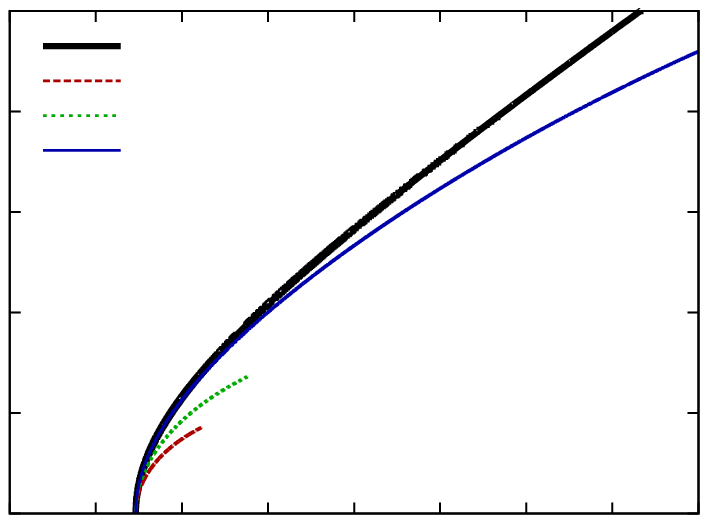}}%
    \gplfronttext
  \end{picture}%
\endgroup

%% file: ParentAvgRadGraph.tex
\begingroup
  \fontfamily{Verdana}%
  \selectfont
  \makeatletter
  \providecommand\color[2][]{%
    \GenericError{(gnuplot) \space\space\space\@spaces}{%
      Package color not loaded in conjunction with
      terminal option `colourtext'%
    }{See the gnuplot documentation for explanation.%
    }{Either use 'blacktext' in gnuplot or load the package
      color.sty in LaTeX.}%
    \renewcommand\color[2][]{}%
  }%
  \providecommand\includegraphics[2][]{%
    \GenericError{(gnuplot) \space\space\space\@spaces}{%
      Package graphicx or graphics not loaded%
    }{See the gnuplot documentation for explanation.%
    }{The gnuplot epslatex terminal needs graphicx.sty or graphics.sty.}%
    \renewcommand\includegraphics[2][]{}%
  }%
  \providecommand\rotatebox[2]{#2}%
  \@ifundefined{ifGPcolor}{%
    \newif\ifGPcolor
    \GPcolorfalse
  }{}%
  \@ifundefined{ifGPblacktext}{%
    \newif\ifGPblacktext
    \GPblacktexttrue
  }{}%
  \let\gplgaddtomacro\g@addto@macro
  \gdef\gplbacktext{}%
  \gdef\gplfronttext{}%
  \makeatother
  \ifGPblacktext
    \def\colorrgb#1{}%
    \def\colorgray#1{}%
  \else
    \ifGPcolor
      \def\colorrgb#1{\color[rgb]{#1}}%
      \def\colorgray#1{\color[gray]{#1}}%
      \expandafter\def\csname LTw\endcsname{\color{white}}%
      \expandafter\def\csname LTb\endcsname{\color{black}}%
      \expandafter\def\csname LTa\endcsname{\color{black}}%
      \expandafter\def\csname LT0\endcsname{\color[rgb]{1,0,0}}%
      \expandafter\def\csname LT1\endcsname{\color[rgb]{0,1,0}}%
      \expandafter\def\csname LT2\endcsname{\color[rgb]{0,0,1}}%
      \expandafter\def\csname LT3\endcsname{\color[rgb]{1,0,1}}%
      \expandafter\def\csname LT4\endcsname{\color[rgb]{0,1,1}}%
      \expandafter\def\csname LT5\endcsname{\color[rgb]{1,1,0}}%
      \expandafter\def\csname LT6\endcsname{\color[rgb]{0,0,0}}%
      \expandafter\def\csname LT7\endcsname{\color[rgb]{1,0.3,0}}%
      \expandafter\def\csname LT8\endcsname{\color[rgb]{0.5,0.5,0.5}}%
    \else
      \def\colorrgb#1{\color{black}}%
      \def\colorgray#1{\color[gray]{#1}}%
      \expandafter\def\csname LTw\endcsname{\color{white}}%
      \expandafter\def\csname LTb\endcsname{\color{black}}%
      \expandafter\def\csname LTa\endcsname{\color{black}}%
      \expandafter\def\csname LT0\endcsname{\color{black}}%
      \expandafter\def\csname LT1\endcsname{\color{black}}%
      \expandafter\def\csname LT2\endcsname{\color{black}}%
      \expandafter\def\csname LT3\endcsname{\color{black}}%
      \expandafter\def\csname LT4\endcsname{\color{black}}%
      \expandafter\def\csname LT5\endcsname{\color{black}}%
      \expandafter\def\csname LT6\endcsname{\color{black}}%
      \expandafter\def\csname LT7\endcsname{\color{black}}%
      \expandafter\def\csname LT8\endcsname{\color{black}}%
    \fi
  \fi
  \setlength{\unitlength}{0.0500bp}%
  \begin{picture}(4896.00,3600.00)%
    \gplgaddtomacro\gplbacktext{%
      \csname LTb\endcsname%
      \put(544,512){\makebox(0,0)[r]{\strut{}$0.0$}}%
      \put(544,1091){\makebox(0,0)[r]{\strut{}$0.5$}}%
      \put(544,1670){\makebox(0,0)[r]{\strut{}$1.0$}}%
      \put(544,2249){\makebox(0,0)[r]{\strut{}$1.5$}}%
      \put(544,2828){\makebox(0,0)[r]{\strut{}$2.0$}}%
      \put(544,3407){\makebox(0,0)[r]{\strut{}$2.5$}}%
      \put(640,352){\makebox(0,0){\strut{}$  1.0$}}%
      \put(1136,352){\makebox(0,0){\strut{}$  1.5$}}%
      \put(1632,352){\makebox(0,0){\strut{}$  2.0$}}%
      \put(2128,352){\makebox(0,0){\strut{}$  2.5$}}%
      \put(2624,352){\makebox(0,0){\strut{}$  3.0$}}%
      \put(3119,352){\makebox(0,0){\strut{}$  3.5$}}%
      \put(3615,352){\makebox(0,0){\strut{}$  4.0$}}%
      \put(4111,352){\makebox(0,0){\strut{}$  4.5$}}%
      \put(4607,352){\makebox(0,0){\strut{}$  5.0$}}%
      \put(128,1959){\rotatebox{-270}{\makebox(0,0){\strut{}$\frac{d(Radius)}{dt}$}}}%
      \put(2623,112){\makebox(0,0){\strut{}$Radius(t)$}}%
    }%
    \gplgaddtomacro\gplfronttext{%
      \csname LTb\endcsname%
      \put(1375,3204){\makebox(0,0)[l]{\strut{}$\text{FLRW}$}}%
      \csname LTb\endcsname%
      \put(1375,3004){\makebox(0,0)[l]{\strut{}$\text{5 tetrahedra}$}}%
      \csname LTb\endcsname%
      \put(1375,2804){\makebox(0,0)[l]{\strut{}$\text{16 tetrahedra}$}}%
      \csname LTb\endcsname%
      \put(1375,2604){\makebox(0,0)[l]{\strut{}$\text{600 tetrahedra}$}}%
    }%
    \gplbacktext
    \put(0,0){\includegraphics{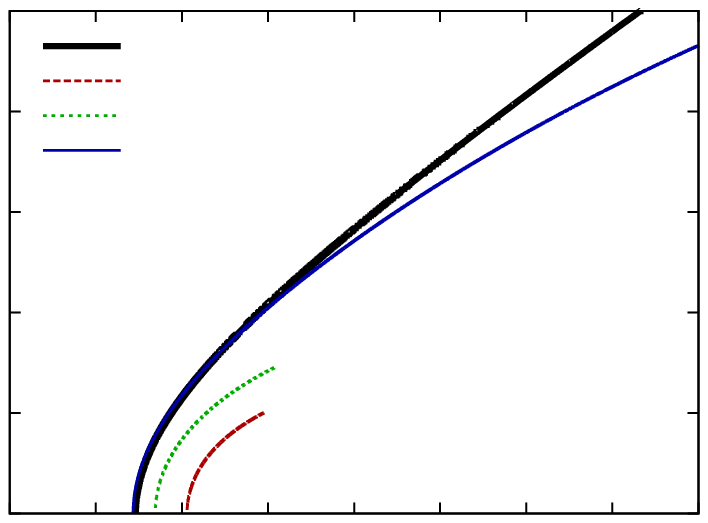}}%
    \gplfronttext
  \end{picture}%
\endgroup

%% file: ParentVertRadGraph.tex
\begingroup
  \fontfamily{Verdana}%
  \selectfont
  \makeatletter
  \providecommand\color[2][]{%
    \GenericError{(gnuplot) \space\space\space\@spaces}{%
      Package color not loaded in conjunction with
      terminal option `colourtext'%
    }{See the gnuplot documentation for explanation.%
    }{Either use 'blacktext' in gnuplot or load the package
      color.sty in LaTeX.}%
    \renewcommand\color[2][]{}%
  }%
  \providecommand\includegraphics[2][]{%
    \GenericError{(gnuplot) \space\space\space\@spaces}{%
      Package graphicx or graphics not loaded%
    }{See the gnuplot documentation for explanation.%
    }{The gnuplot epslatex terminal needs graphicx.sty or graphics.sty.}%
    \renewcommand\includegraphics[2][]{}%
  }%
  \providecommand\rotatebox[2]{#2}%
  \@ifundefined{ifGPcolor}{%
    \newif\ifGPcolor
    \GPcolorfalse
  }{}%
  \@ifundefined{ifGPblacktext}{%
    \newif\ifGPblacktext
    \GPblacktexttrue
  }{}%
  \let\gplgaddtomacro\g@addto@macro
  \gdef\gplbacktext{}%
  \gdef\gplfronttext{}%
  \makeatother
  \ifGPblacktext
    \def\colorrgb#1{}%
    \def\colorgray#1{}%
  \else
    \ifGPcolor
      \def\colorrgb#1{\color[rgb]{#1}}%
      \def\colorgray#1{\color[gray]{#1}}%
      \expandafter\def\csname LTw\endcsname{\color{white}}%
      \expandafter\def\csname LTb\endcsname{\color{black}}%
      \expandafter\def\csname LTa\endcsname{\color{black}}%
      \expandafter\def\csname LT0\endcsname{\color[rgb]{1,0,0}}%
      \expandafter\def\csname LT1\endcsname{\color[rgb]{0,1,0}}%
      \expandafter\def\csname LT2\endcsname{\color[rgb]{0,0,1}}%
      \expandafter\def\csname LT3\endcsname{\color[rgb]{1,0,1}}%
      \expandafter\def\csname LT4\endcsname{\color[rgb]{0,1,1}}%
      \expandafter\def\csname LT5\endcsname{\color[rgb]{1,1,0}}%
      \expandafter\def\csname LT6\endcsname{\color[rgb]{0,0,0}}%
      \expandafter\def\csname LT7\endcsname{\color[rgb]{1,0.3,0}}%
      \expandafter\def\csname LT8\endcsname{\color[rgb]{0.5,0.5,0.5}}%
    \else
      \def\colorrgb#1{\color{black}}%
      \def\colorgray#1{\color[gray]{#1}}%
      \expandafter\def\csname LTw\endcsname{\color{white}}%
      \expandafter\def\csname LTb\endcsname{\color{black}}%
      \expandafter\def\csname LTa\endcsname{\color{black}}%
      \expandafter\def\csname LT0\endcsname{\color{black}}%
      \expandafter\def\csname LT1\endcsname{\color{black}}%
      \expandafter\def\csname LT2\endcsname{\color{black}}%
      \expandafter\def\csname LT3\endcsname{\color{black}}%
      \expandafter\def\csname LT4\endcsname{\color{black}}%
      \expandafter\def\csname LT5\endcsname{\color{black}}%
      \expandafter\def\csname LT6\endcsname{\color{black}}%
      \expandafter\def\csname LT7\endcsname{\color{black}}%
      \expandafter\def\csname LT8\endcsname{\color{black}}%
    \fi
  \fi
  \setlength{\unitlength}{0.0500bp}%
  \begin{picture}(4896.00,3600.00)%
    \gplgaddtomacro\gplbacktext{%
      \csname LTb\endcsname%
      \put(544,512){\makebox(0,0)[r]{\strut{}$0.0$}}%
      \put(544,1091){\makebox(0,0)[r]{\strut{}$0.5$}}%
      \put(544,1670){\makebox(0,0)[r]{\strut{}$1.0$}}%
      \put(544,2249){\makebox(0,0)[r]{\strut{}$1.5$}}%
      \put(544,2828){\makebox(0,0)[r]{\strut{}$2.0$}}%
      \put(544,3407){\makebox(0,0)[r]{\strut{}$2.5$}}%
      \put(640,352){\makebox(0,0){\strut{}$  1.0$}}%
      \put(1136,352){\makebox(0,0){\strut{}$  1.5$}}%
      \put(1632,352){\makebox(0,0){\strut{}$  2.0$}}%
      \put(2128,352){\makebox(0,0){\strut{}$  2.5$}}%
      \put(2624,352){\makebox(0,0){\strut{}$  3.0$}}%
      \put(3119,352){\makebox(0,0){\strut{}$  3.5$}}%
      \put(3615,352){\makebox(0,0){\strut{}$  4.0$}}%
      \put(4111,352){\makebox(0,0){\strut{}$  4.5$}}%
      \put(4607,352){\makebox(0,0){\strut{}$  5.0$}}%
      \put(128,1959){\rotatebox{-270}{\makebox(0,0){\strut{}$\frac{d(Radius)}{dt}$}}}%
      \put(2623,112){\makebox(0,0){\strut{}$Radius(t)$}}%
    }%
    \gplgaddtomacro\gplfronttext{%
      \csname LTb\endcsname%
      \put(1375,3204){\makebox(0,0)[l]{\strut{}$\text{FLRW}$}}%
      \csname LTb\endcsname%
      \put(1375,3004){\makebox(0,0)[l]{\strut{}$\text{5 tetrahedra}$}}%
      \csname LTb\endcsname%
      \put(1375,2804){\makebox(0,0)[l]{\strut{}$\text{16 tetrahedra}$}}%
      \csname LTb\endcsname%
      \put(1375,2604){\makebox(0,0)[l]{\strut{}$\text{600 tetrahedra}$}}%
    }%
    \gplbacktext
    \put(0,0){\includegraphics{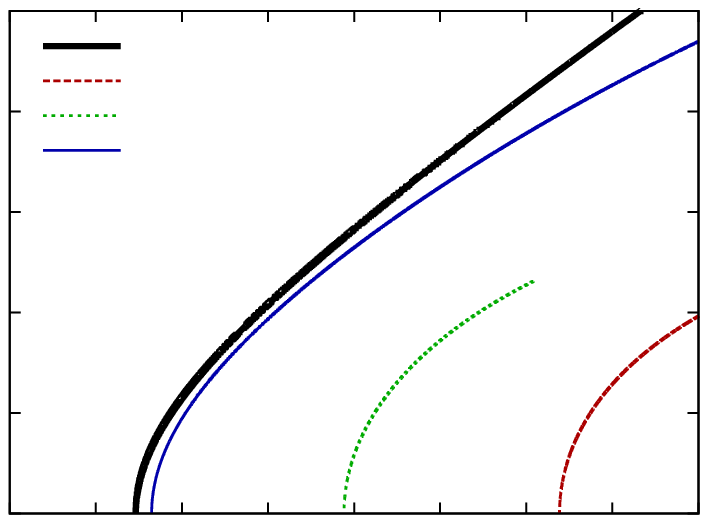}}%
    \gplfronttext
  \end{picture}%
\endgroup

%% file: ParentEdgeRadGraph.tex
\begingroup
  \fontfamily{Verdana}%
  \selectfont
  \makeatletter
  \providecommand\color[2][]{%
    \GenericError{(gnuplot) \space\space\space\@spaces}{%
      Package color not loaded in conjunction with
      terminal option `colourtext'%
    }{See the gnuplot documentation for explanation.%
    }{Either use 'blacktext' in gnuplot or load the package
      color.sty in LaTeX.}%
    \renewcommand\color[2][]{}%
  }%
  \providecommand\includegraphics[2][]{%
    \GenericError{(gnuplot) \space\space\space\@spaces}{%
      Package graphicx or graphics not loaded%
    }{See the gnuplot documentation for explanation.%
    }{The gnuplot epslatex terminal needs graphicx.sty or graphics.sty.}%
    \renewcommand\includegraphics[2][]{}%
  }%
  \providecommand\rotatebox[2]{#2}%
  \@ifundefined{ifGPcolor}{%
    \newif\ifGPcolor
    \GPcolorfalse
  }{}%
  \@ifundefined{ifGPblacktext}{%
    \newif\ifGPblacktext
    \GPblacktexttrue
  }{}%
  \let\gplgaddtomacro\g@addto@macro
  \gdef\gplbacktext{}%
  \gdef\gplfronttext{}%
  \makeatother
  \ifGPblacktext
    \def\colorrgb#1{}%
    \def\colorgray#1{}%
  \else
    \ifGPcolor
      \def\colorrgb#1{\color[rgb]{#1}}%
      \def\colorgray#1{\color[gray]{#1}}%
      \expandafter\def\csname LTw\endcsname{\color{white}}%
      \expandafter\def\csname LTb\endcsname{\color{black}}%
      \expandafter\def\csname LTa\endcsname{\color{black}}%
      \expandafter\def\csname LT0\endcsname{\color[rgb]{1,0,0}}%
      \expandafter\def\csname LT1\endcsname{\color[rgb]{0,1,0}}%
      \expandafter\def\csname LT2\endcsname{\color[rgb]{0,0,1}}%
      \expandafter\def\csname LT3\endcsname{\color[rgb]{1,0,1}}%
      \expandafter\def\csname LT4\endcsname{\color[rgb]{0,1,1}}%
      \expandafter\def\csname LT5\endcsname{\color[rgb]{1,1,0}}%
      \expandafter\def\csname LT6\endcsname{\color[rgb]{0,0,0}}%
      \expandafter\def\csname LT7\endcsname{\color[rgb]{1,0.3,0}}%
      \expandafter\def\csname LT8\endcsname{\color[rgb]{0.5,0.5,0.5}}%
    \else
      \def\colorrgb#1{\color{black}}%
      \def\colorgray#1{\color[gray]{#1}}%
      \expandafter\def\csname LTw\endcsname{\color{white}}%
      \expandafter\def\csname LTb\endcsname{\color{black}}%
      \expandafter\def\csname LTa\endcsname{\color{black}}%
      \expandafter\def\csname LT0\endcsname{\color{black}}%
      \expandafter\def\csname LT1\endcsname{\color{black}}%
      \expandafter\def\csname LT2\endcsname{\color{black}}%
      \expandafter\def\csname LT3\endcsname{\color{black}}%
      \expandafter\def\csname LT4\endcsname{\color{black}}%
      \expandafter\def\csname LT5\endcsname{\color{black}}%
      \expandafter\def\csname LT6\endcsname{\color{black}}%
      \expandafter\def\csname LT7\endcsname{\color{black}}%
      \expandafter\def\csname LT8\endcsname{\color{black}}%
    \fi
  \fi
  \setlength{\unitlength}{0.0500bp}%
  \begin{picture}(4896.00,3600.00)%
    \gplgaddtomacro\gplbacktext{%
      \csname LTb\endcsname%
      \put(544,512){\makebox(0,0)[r]{\strut{}$0.0$}}%
      \put(544,1091){\makebox(0,0)[r]{\strut{}$0.5$}}%
      \put(544,1670){\makebox(0,0)[r]{\strut{}$1.0$}}%
      \put(544,2249){\makebox(0,0)[r]{\strut{}$1.5$}}%
      \put(544,2828){\makebox(0,0)[r]{\strut{}$2.0$}}%
      \put(544,3407){\makebox(0,0)[r]{\strut{}$2.5$}}%
      \put(640,352){\makebox(0,0){\strut{}$  1.0$}}%
      \put(1136,352){\makebox(0,0){\strut{}$  1.5$}}%
      \put(1632,352){\makebox(0,0){\strut{}$  2.0$}}%
      \put(2128,352){\makebox(0,0){\strut{}$  2.5$}}%
      \put(2624,352){\makebox(0,0){\strut{}$  3.0$}}%
      \put(3119,352){\makebox(0,0){\strut{}$  3.5$}}%
      \put(3615,352){\makebox(0,0){\strut{}$  4.0$}}%
      \put(4111,352){\makebox(0,0){\strut{}$  4.5$}}%
      \put(4607,352){\makebox(0,0){\strut{}$  5.0$}}%
      \put(128,1959){\rotatebox{-270}{\makebox(0,0){\strut{}$\frac{d(Radius)}{dt}$}}}%
      \put(2623,112){\makebox(0,0){\strut{}$Radius(t)$}}%
    }%
    \gplgaddtomacro\gplfronttext{%
      \csname LTb\endcsname%
      \put(1375,3204){\makebox(0,0)[l]{\strut{}$\text{FLRW}$}}%
      \csname LTb\endcsname%
      \put(1375,3004){\makebox(0,0)[l]{\strut{}$\text{5 tetrahedra}$}}%
      \csname LTb\endcsname%
      \put(1375,2804){\makebox(0,0)[l]{\strut{}$\text{16 tetrahedra}$}}%
      \csname LTb\endcsname%
      \put(1375,2604){\makebox(0,0)[l]{\strut{}$\text{600 tetrahedra}$}}%
    }%
    \gplbacktext
    \put(0,0){\includegraphics{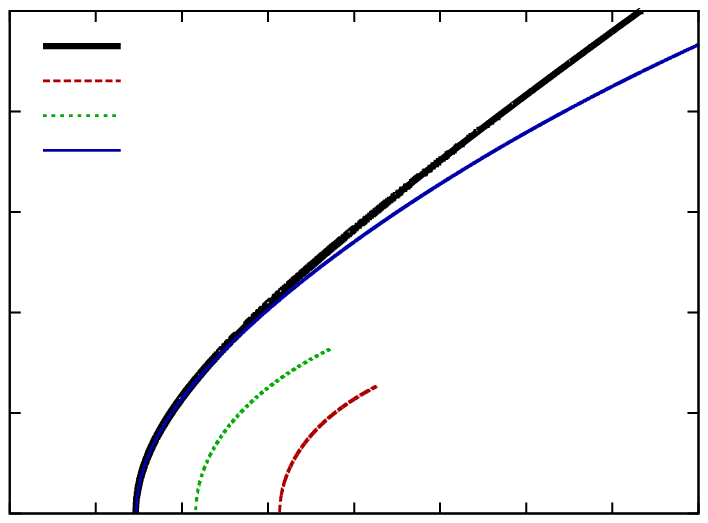}}%
    \gplfronttext
  \end{picture}%
\endgroup

%% file: ParentFaceRadGraph.tex
\begingroup
  \fontfamily{Verdana}%
  \selectfont
  \makeatletter
  \providecommand\color[2][]{%
    \GenericError{(gnuplot) \space\space\space\@spaces}{%
      Package color not loaded in conjunction with
      terminal option `colourtext'%
    }{See the gnuplot documentation for explanation.%
    }{Either use 'blacktext' in gnuplot or load the package
      color.sty in LaTeX.}%
    \renewcommand\color[2][]{}%
  }%
  \providecommand\includegraphics[2][]{%
    \GenericError{(gnuplot) \space\space\space\@spaces}{%
      Package graphicx or graphics not loaded%
    }{See the gnuplot documentation for explanation.%
    }{The gnuplot epslatex terminal needs graphicx.sty or graphics.sty.}%
    \renewcommand\includegraphics[2][]{}%
  }%
  \providecommand\rotatebox[2]{#2}%
  \@ifundefined{ifGPcolor}{%
    \newif\ifGPcolor
    \GPcolorfalse
  }{}%
  \@ifundefined{ifGPblacktext}{%
    \newif\ifGPblacktext
    \GPblacktexttrue
  }{}%
  \let\gplgaddtomacro\g@addto@macro
  \gdef\gplbacktext{}%
  \gdef\gplfronttext{}%
  \makeatother
  \ifGPblacktext
    \def\colorrgb#1{}%
    \def\colorgray#1{}%
  \else
    \ifGPcolor
      \def\colorrgb#1{\color[rgb]{#1}}%
      \def\colorgray#1{\color[gray]{#1}}%
      \expandafter\def\csname LTw\endcsname{\color{white}}%
      \expandafter\def\csname LTb\endcsname{\color{black}}%
      \expandafter\def\csname LTa\endcsname{\color{black}}%
      \expandafter\def\csname LT0\endcsname{\color[rgb]{1,0,0}}%
      \expandafter\def\csname LT1\endcsname{\color[rgb]{0,1,0}}%
      \expandafter\def\csname LT2\endcsname{\color[rgb]{0,0,1}}%
      \expandafter\def\csname LT3\endcsname{\color[rgb]{1,0,1}}%
      \expandafter\def\csname LT4\endcsname{\color[rgb]{0,1,1}}%
      \expandafter\def\csname LT5\endcsname{\color[rgb]{1,1,0}}%
      \expandafter\def\csname LT6\endcsname{\color[rgb]{0,0,0}}%
      \expandafter\def\csname LT7\endcsname{\color[rgb]{1,0.3,0}}%
      \expandafter\def\csname LT8\endcsname{\color[rgb]{0.5,0.5,0.5}}%
    \else
      \def\colorrgb#1{\color{black}}%
      \def\colorgray#1{\color[gray]{#1}}%
      \expandafter\def\csname LTw\endcsname{\color{white}}%
      \expandafter\def\csname LTb\endcsname{\color{black}}%
      \expandafter\def\csname LTa\endcsname{\color{black}}%
      \expandafter\def\csname LT0\endcsname{\color{black}}%
      \expandafter\def\csname LT1\endcsname{\color{black}}%
      \expandafter\def\csname LT2\endcsname{\color{black}}%
      \expandafter\def\csname LT3\endcsname{\color{black}}%
      \expandafter\def\csname LT4\endcsname{\color{black}}%
      \expandafter\def\csname LT5\endcsname{\color{black}}%
      \expandafter\def\csname LT6\endcsname{\color{black}}%
      \expandafter\def\csname LT7\endcsname{\color{black}}%
      \expandafter\def\csname LT8\endcsname{\color{black}}%
    \fi
  \fi
  \setlength{\unitlength}{0.0500bp}%
  \begin{picture}(4896.00,3600.00)%
    \gplgaddtomacro\gplbacktext{%
      \csname LTb\endcsname%
      \put(544,512){\makebox(0,0)[r]{\strut{}$0.0$}}%
      \put(544,1091){\makebox(0,0)[r]{\strut{}$0.5$}}%
      \put(544,1670){\makebox(0,0)[r]{\strut{}$1.0$}}%
      \put(544,2249){\makebox(0,0)[r]{\strut{}$1.5$}}%
      \put(544,2828){\makebox(0,0)[r]{\strut{}$2.0$}}%
      \put(544,3407){\makebox(0,0)[r]{\strut{}$2.5$}}%
      \put(640,352){\makebox(0,0){\strut{}$  1.0$}}%
      \put(1136,352){\makebox(0,0){\strut{}$  1.5$}}%
      \put(1632,352){\makebox(0,0){\strut{}$  2.0$}}%
      \put(2128,352){\makebox(0,0){\strut{}$  2.5$}}%
      \put(2624,352){\makebox(0,0){\strut{}$  3.0$}}%
      \put(3119,352){\makebox(0,0){\strut{}$  3.5$}}%
      \put(3615,352){\makebox(0,0){\strut{}$  4.0$}}%
      \put(4111,352){\makebox(0,0){\strut{}$  4.5$}}%
      \put(4607,352){\makebox(0,0){\strut{}$  5.0$}}%
      \put(128,1959){\rotatebox{-270}{\makebox(0,0){\strut{}$\frac{d(Radius)}{dt}$}}}%
      \put(2623,112){\makebox(0,0){\strut{}$Radius(t)$}}%
    }%
    \gplgaddtomacro\gplfronttext{%
      \csname LTb\endcsname%
      \put(1375,3204){\makebox(0,0)[l]{\strut{}$\text{FLRW}$}}%
      \csname LTb\endcsname%
      \put(1375,3004){\makebox(0,0)[l]{\strut{}$\text{5 tetrahedra}$}}%
      \csname LTb\endcsname%
      \put(1375,2804){\makebox(0,0)[l]{\strut{}$\text{16 tetrahedra}$}}%
      \csname LTb\endcsname%
      \put(1375,2604){\makebox(0,0)[l]{\strut{}$\text{600 tetrahedra}$}}%
    }%
    \gplbacktext
    \put(0,0){\includegraphics{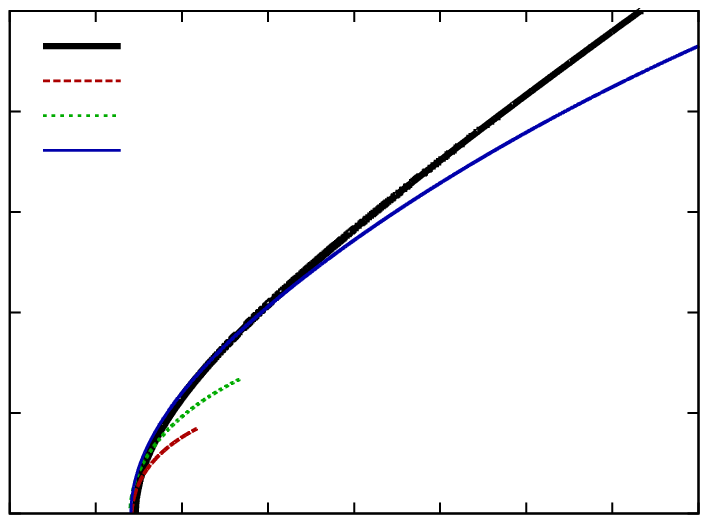}}%
    \gplfronttext
  \end{picture}%
\endgroup

%% file: ParentCentrRadGraph.tex
\begingroup
  \fontfamily{Verdana}%
  \selectfont
  \makeatletter
  \providecommand\color[2][]{%
    \GenericError{(gnuplot) \space\space\space\@spaces}{%
      Package color not loaded in conjunction with
      terminal option `colourtext'%
    }{See the gnuplot documentation for explanation.%
    }{Either use 'blacktext' in gnuplot or load the package
      color.sty in LaTeX.}%
    \renewcommand\color[2][]{}%
  }%
  \providecommand\includegraphics[2][]{%
    \GenericError{(gnuplot) \space\space\space\@spaces}{%
      Package graphicx or graphics not loaded%
    }{See the gnuplot documentation for explanation.%
    }{The gnuplot epslatex terminal needs graphicx.sty or graphics.sty.}%
    \renewcommand\includegraphics[2][]{}%
  }%
  \providecommand\rotatebox[2]{#2}%
  \@ifundefined{ifGPcolor}{%
    \newif\ifGPcolor
    \GPcolorfalse
  }{}%
  \@ifundefined{ifGPblacktext}{%
    \newif\ifGPblacktext
    \GPblacktexttrue
  }{}%
  \let\gplgaddtomacro\g@addto@macro
  \gdef\gplbacktext{}%
  \gdef\gplfronttext{}%
  \makeatother
  \ifGPblacktext
    \def\colorrgb#1{}%
    \def\colorgray#1{}%
  \else
    \ifGPcolor
      \def\colorrgb#1{\color[rgb]{#1}}%
      \def\colorgray#1{\color[gray]{#1}}%
      \expandafter\def\csname LTw\endcsname{\color{white}}%
      \expandafter\def\csname LTb\endcsname{\color{black}}%
      \expandafter\def\csname LTa\endcsname{\color{black}}%
      \expandafter\def\csname LT0\endcsname{\color[rgb]{1,0,0}}%
      \expandafter\def\csname LT1\endcsname{\color[rgb]{0,1,0}}%
      \expandafter\def\csname LT2\endcsname{\color[rgb]{0,0,1}}%
      \expandafter\def\csname LT3\endcsname{\color[rgb]{1,0,1}}%
      \expandafter\def\csname LT4\endcsname{\color[rgb]{0,1,1}}%
      \expandafter\def\csname LT5\endcsname{\color[rgb]{1,1,0}}%
      \expandafter\def\csname LT6\endcsname{\color[rgb]{0,0,0}}%
      \expandafter\def\csname LT7\endcsname{\color[rgb]{1,0.3,0}}%
      \expandafter\def\csname LT8\endcsname{\color[rgb]{0.5,0.5,0.5}}%
    \else
      \def\colorrgb#1{\color{black}}%
      \def\colorgray#1{\color[gray]{#1}}%
      \expandafter\def\csname LTw\endcsname{\color{white}}%
      \expandafter\def\csname LTb\endcsname{\color{black}}%
      \expandafter\def\csname LTa\endcsname{\color{black}}%
      \expandafter\def\csname LT0\endcsname{\color{black}}%
      \expandafter\def\csname LT1\endcsname{\color{black}}%
      \expandafter\def\csname LT2\endcsname{\color{black}}%
      \expandafter\def\csname LT3\endcsname{\color{black}}%
      \expandafter\def\csname LT4\endcsname{\color{black}}%
      \expandafter\def\csname LT5\endcsname{\color{black}}%
      \expandafter\def\csname LT6\endcsname{\color{black}}%
      \expandafter\def\csname LT7\endcsname{\color{black}}%
      \expandafter\def\csname LT8\endcsname{\color{black}}%
    \fi
  \fi
  \setlength{\unitlength}{0.0500bp}%
  \begin{picture}(4896.00,3600.00)%
    \gplgaddtomacro\gplbacktext{%
      \csname LTb\endcsname%
      \put(544,512){\makebox(0,0)[r]{\strut{}$0.0$}}%
      \put(544,1091){\makebox(0,0)[r]{\strut{}$0.5$}}%
      \put(544,1670){\makebox(0,0)[r]{\strut{}$1.0$}}%
      \put(544,2249){\makebox(0,0)[r]{\strut{}$1.5$}}%
      \put(544,2828){\makebox(0,0)[r]{\strut{}$2.0$}}%
      \put(544,3407){\makebox(0,0)[r]{\strut{}$2.5$}}%
      \put(640,352){\makebox(0,0){\strut{}$  1.0$}}%
      \put(1136,352){\makebox(0,0){\strut{}$  1.5$}}%
      \put(1632,352){\makebox(0,0){\strut{}$  2.0$}}%
      \put(2128,352){\makebox(0,0){\strut{}$  2.5$}}%
      \put(2624,352){\makebox(0,0){\strut{}$  3.0$}}%
      \put(3119,352){\makebox(0,0){\strut{}$  3.5$}}%
      \put(3615,352){\makebox(0,0){\strut{}$  4.0$}}%
      \put(4111,352){\makebox(0,0){\strut{}$  4.5$}}%
      \put(4607,352){\makebox(0,0){\strut{}$  5.0$}}%
      \put(128,1959){\rotatebox{-270}{\makebox(0,0){\strut{}$\frac{d(Radius)}{dt}$}}}%
      \put(2623,112){\makebox(0,0){\strut{}$Radius(t)$}}%
    }%
    \gplgaddtomacro\gplfronttext{%
      \csname LTb\endcsname%
      \put(1375,3204){\makebox(0,0)[l]{\strut{}$\text{FLRW}$}}%
      \csname LTb\endcsname%
      \put(1375,3004){\makebox(0,0)[l]{\strut{}$\text{5 tetrahedra}$}}%
      \csname LTb\endcsname%
      \put(1375,2804){\makebox(0,0)[l]{\strut{}$\text{16 tetrahedra}$}}%
      \csname LTb\endcsname%
      \put(1375,2604){\makebox(0,0)[l]{\strut{}$\text{600 tetrahedra}$}}%
    }%
    \gplbacktext
    \put(0,0){\includegraphics{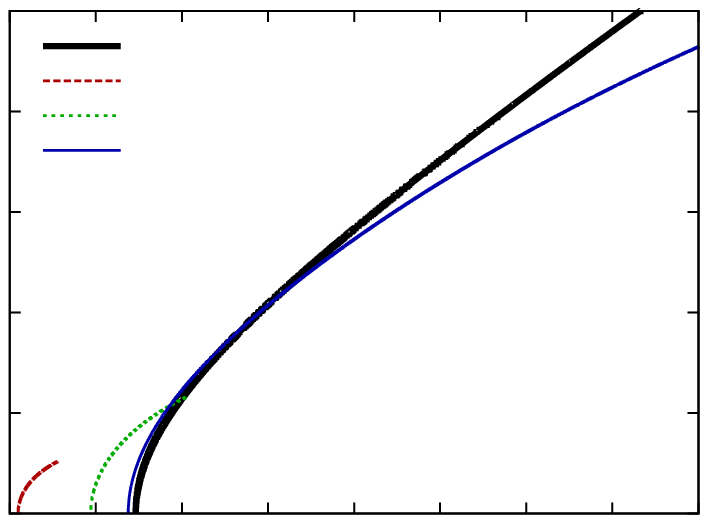}}%
    \gplfronttext
  \end{picture}%
\endgroup

%% file: 4-block-spacelike.pspdftex
\begin{picture}(0,0)%
\includegraphics{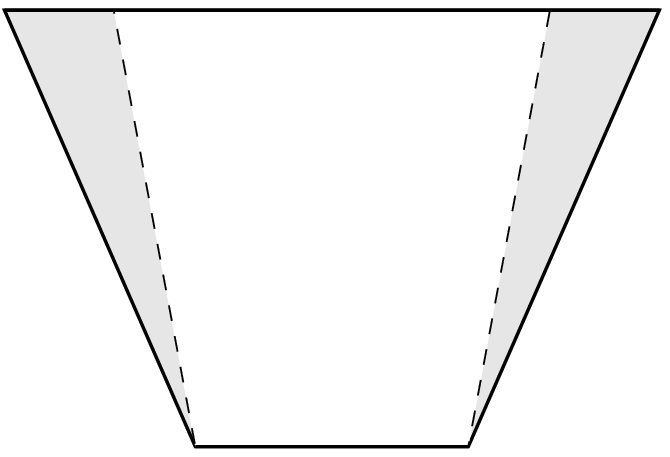}%
\end{picture}%
\setlength{\unitlength}{3947sp}%
\begingroup\makeatletter\ifx\SetFigFont\undefined%
\gdef\SetFigFont#1#2#3#4#5{%
  \reset@font\fontsize{#1}{#2pt}%
  \fontfamily{#3}\fontseries{#4}\fontshape{#5}%
  \selectfont}%
\fi\endgroup%
\begin{picture}(3285,2490)(2114,-3664)
\put(3707,-1285){\makebox(0,0)[b]{\smash{{\SetFigFont{8}{9.6}{\rmdefault}{\mddefault}{\updefault}{\color[rgb]{0,0,0}$\Cauchyt{t+dt}$}%
}}}}
\put(3707,-3613){\makebox(0,0)[b]{\smash{{\SetFigFont{8}{9.6}{\rmdefault}{\mddefault}{\updefault}{\color[rgb]{0,0,0}$\Cauchyt{t}$}%
}}}}
\put(4477,-3511){\makebox(0,0)[lb]{\smash{{\SetFigFont{8}{9.6}{\rmdefault}{\mddefault}{\updefault}{\color[rgb]{0,0,0}$t$}%
}}}}
\put(5384,-1384){\makebox(0,0)[lb]{\smash{{\SetFigFont{8}{9.6}{\rmdefault}{\mddefault}{\updefault}{\color[rgb]{0,0,0}$t+dt$}%
}}}}
\put(2516,-2476){\makebox(0,0)[rb]{\smash{{\SetFigFont{8}{9.6}{\rmdefault}{\mddefault}{\updefault}{\color[rgb]{0,0,0}$\CWstrut{t}$}%
}}}}
\put(4928,-2476){\makebox(0,0)[lb]{\smash{{\SetFigFont{8}{9.6}{\rmdefault}{\mddefault}{\updefault}{\color[rgb]{0,0,0}$\CWstrut{t}$}%
}}}}
\end{picture}%

%% file: HubbleRadiusRatio.tex
\begingroup
  \fontfamily{Verdana}%
  \selectfont
  \makeatletter
  \providecommand\color[2][]{%
    \GenericError{(gnuplot) \space\space\space\@spaces}{%
      Package color not loaded in conjunction with
      terminal option `colourtext'%
    }{See the gnuplot documentation for explanation.%
    }{Either use 'blacktext' in gnuplot or load the package
      color.sty in LaTeX.}%
    \renewcommand\color[2][]{}%
  }%
  \providecommand\includegraphics[2][]{%
    \GenericError{(gnuplot) \space\space\space\@spaces}{%
      Package graphicx or graphics not loaded%
    }{See the gnuplot documentation for explanation.%
    }{The gnuplot epslatex terminal needs graphicx.sty or graphics.sty.}%
    \renewcommand\includegraphics[2][]{}%
  }%
  \providecommand\rotatebox[2]{#2}%
  \@ifundefined{ifGPcolor}{%
    \newif\ifGPcolor
    \GPcolorfalse
  }{}%
  \@ifundefined{ifGPblacktext}{%
    \newif\ifGPblacktext
    \GPblacktexttrue
  }{}%
  \let\gplgaddtomacro\g@addto@macro
  \gdef\gplbacktext{}%
  \gdef\gplfronttext{}%
  \makeatother
  \ifGPblacktext
    \def\colorrgb#1{}%
    \def\colorgray#1{}%
  \else
    \ifGPcolor
      \def\colorrgb#1{\color[rgb]{#1}}%
      \def\colorgray#1{\color[gray]{#1}}%
      \expandafter\def\csname LTw\endcsname{\color{white}}%
      \expandafter\def\csname LTb\endcsname{\color{black}}%
      \expandafter\def\csname LTa\endcsname{\color{black}}%
      \expandafter\def\csname LT0\endcsname{\color[rgb]{1,0,0}}%
      \expandafter\def\csname LT1\endcsname{\color[rgb]{0,1,0}}%
      \expandafter\def\csname LT2\endcsname{\color[rgb]{0,0,1}}%
      \expandafter\def\csname LT3\endcsname{\color[rgb]{1,0,1}}%
      \expandafter\def\csname LT4\endcsname{\color[rgb]{0,1,1}}%
      \expandafter\def\csname LT5\endcsname{\color[rgb]{1,1,0}}%
      \expandafter\def\csname LT6\endcsname{\color[rgb]{0,0,0}}%
      \expandafter\def\csname LT7\endcsname{\color[rgb]{1,0.3,0}}%
      \expandafter\def\csname LT8\endcsname{\color[rgb]{0.5,0.5,0.5}}%
    \else
      \def\colorrgb#1{\color{black}}%
      \def\colorgray#1{\color[gray]{#1}}%
      \expandafter\def\csname LTw\endcsname{\color{white}}%
      \expandafter\def\csname LTb\endcsname{\color{black}}%
      \expandafter\def\csname LTa\endcsname{\color{black}}%
      \expandafter\def\csname LT0\endcsname{\color{black}}%
      \expandafter\def\csname LT1\endcsname{\color{black}}%
      \expandafter\def\csname LT2\endcsname{\color{black}}%
      \expandafter\def\csname LT3\endcsname{\color{black}}%
      \expandafter\def\csname LT4\endcsname{\color{black}}%
      \expandafter\def\csname LT5\endcsname{\color{black}}%
      \expandafter\def\csname LT6\endcsname{\color{black}}%
      \expandafter\def\csname LT7\endcsname{\color{black}}%
      \expandafter\def\csname LT8\endcsname{\color{black}}%
    \fi
  \fi
    \setlength{\unitlength}{0.0500bp}%
    \ifx\gptboxheight\undefined%
      \newlength{\gptboxheight}%
      \newlength{\gptboxwidth}%
      \newsavebox{\gptboxtext}%
    \fi%
    \setlength{\fboxrule}{0.5pt}%
    \setlength{\fboxsep}{1pt}%
\begin{picture}(4896.00,3960.00)%
    \gplgaddtomacro\gplbacktext{%
      \csname LTb\endcsname%
      \put(472,512){\makebox(0,0)[r]{\strut{}$0$}}%
      \put(472,1163){\makebox(0,0)[r]{\strut{}$1$}}%
      \put(472,1814){\makebox(0,0)[r]{\strut{}$2$}}%
      \put(472,2465){\makebox(0,0)[r]{\strut{}$3$}}%
      \put(472,3116){\makebox(0,0)[r]{\strut{}$4$}}%
      \put(472,3767){\makebox(0,0)[r]{\strut{}$5$}}%
      \put(568,352){\makebox(0,0){\strut{}$0$}}%
      \put(1145,352){\makebox(0,0){\strut{}$2$}}%
      \put(1722,352){\makebox(0,0){\strut{}$4$}}%
      \put(2299,352){\makebox(0,0){\strut{}$6$}}%
      \put(2876,352){\makebox(0,0){\strut{}$8$}}%
      \put(3453,352){\makebox(0,0){\strut{}$10$}}%
      \put(4030,352){\makebox(0,0){\strut{}$12$}}%
      \put(4607,352){\makebox(0,0){\strut{}$14$}}%
    }%
    \gplgaddtomacro\gplfronttext{%
      \csname LTb\endcsname%
      \put(128,2139){\rotatebox{-270}{\makebox(0,0){\strut{}$H_0^{-1} \, \Lamblen{}(t)^{-1}$}}}%
      \put(2587,112){\makebox(0,0){\strut{}$\Lamblen{}(t)$}}%
      \csname LTb\endcsname%
      \put(3872,3554){\makebox(0,0)[r]{\strut{}5 tetrahedra}}%
      \csname LTb\endcsname%
      \put(3872,3354){\makebox(0,0)[r]{\strut{}16 tetrahedra}}%
      \csname LTb\endcsname%
      \put(3872,3154){\makebox(0,0)[r]{\strut{}600 tetrahedra}}%
    }%
    \gplbacktext
    \put(0,0){\includegraphics{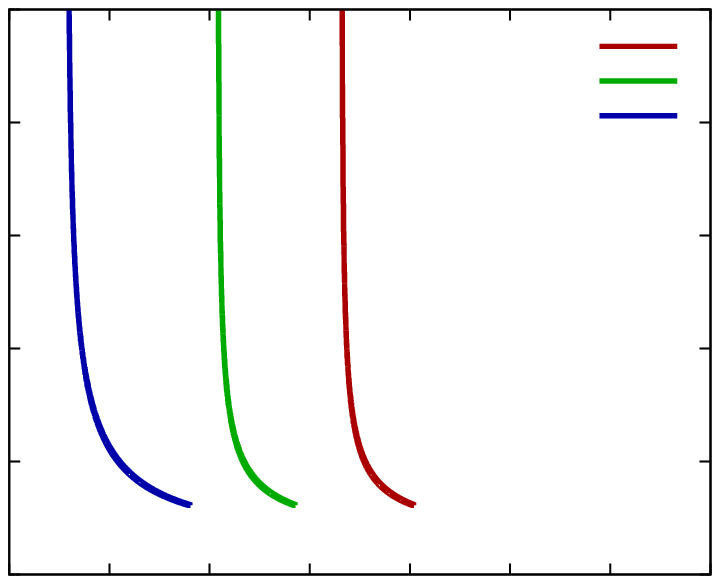}}%
    \gplfronttext
  \end{picture}%
\endgroup

%% file: subdiv-tet.pspdftex
\begin{picture}(0,0)%
\includegraphics{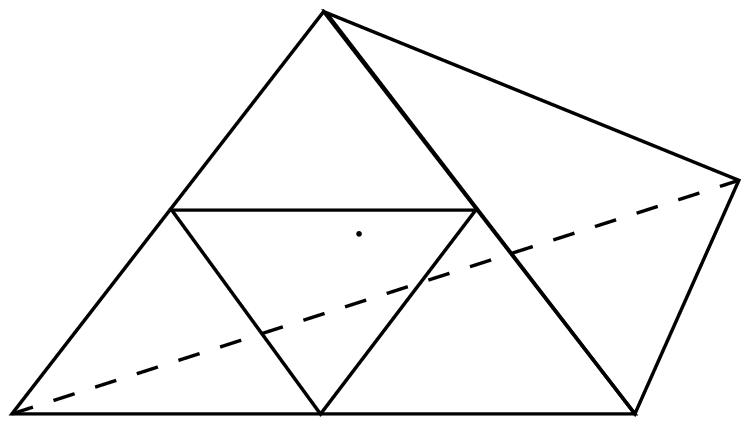}%
\end{picture}%
\setlength{\unitlength}{3947sp}%
\begingroup\makeatletter\ifx\SetFigFont\undefined%
\gdef\SetFigFont#1#2#3#4#5{%
  \reset@font\fontsize{#1}{#2pt}%
  \fontfamily{#3}\fontseries{#4}\fontshape{#5}%
  \selectfont}%
\fi\endgroup%
\begin{picture}(3616,2260)(2190,-3662)
\put(3935,-2744){\makebox(0,0)[b]{\smash{{\SetFigFont{8}{9.6}{\rmdefault}{\mddefault}{\updefault}{\color[rgb]{0,0,0}$\central$}%
}}}}
\put(5791,-2380){\makebox(0,0)[lb]{\smash{{\SetFigFont{8}{9.6}{\rmdefault}{\mddefault}{\updefault}{\color[rgb]{0,0,0}$C$}%
}}}}
\put(2205,-3611){\makebox(0,0)[b]{\smash{{\SetFigFont{8}{9.6}{\rmdefault}{\mddefault}{\updefault}{\color[rgb]{0,0,0}$A$}%
}}}}
\put(3747,-3611){\makebox(0,0)[b]{\smash{{\SetFigFont{8}{9.6}{\rmdefault}{\mddefault}{\updefault}{\color[rgb]{0,0,0}$\midpt{AB}$}%
}}}}
\put(4538,-2492){\makebox(0,0)[lb]{\smash{{\SetFigFont{8}{9.6}{\rmdefault}{\mddefault}{\updefault}{\color[rgb]{0,0,0}$\midpt{BD}$}%
}}}}
\put(2995,-2492){\makebox(0,0)[rb]{\smash{{\SetFigFont{8}{9.6}{\rmdefault}{\mddefault}{\updefault}{\color[rgb]{0,0,0}$\midpt{AD}$}%
}}}}
\put(3747,-2441){\makebox(0,0)[b]{\smash{{\SetFigFont{8}{9.6}{\rmdefault}{\mddefault}{\updefault}{\color[rgb]{0,0,0}$\BrewinBase{i}$}%
}}}}
\put(4161,-3057){\makebox(0,0)[lb]{\smash{{\SetFigFont{8}{9.6}{\rmdefault}{\mddefault}{\updefault}{\color[rgb]{0,0,0}$\BrewinBase{i}$}%
}}}}
\put(3366,-3057){\makebox(0,0)[rb]{\smash{{\SetFigFont{8}{9.6}{\rmdefault}{\mddefault}{\updefault}{\color[rgb]{0,0,0}$\BrewinBase{i}$}%
}}}}
\put(4500,-3611){\makebox(0,0)[b]{\smash{{\SetFigFont{8}{9.6}{\rmdefault}{\mddefault}{\updefault}{\color[rgb]{0,0,0}$\BrewinParent{i}$}%
}}}}
\put(2995,-3611){\makebox(0,0)[b]{\smash{{\SetFigFont{8}{9.6}{\rmdefault}{\mddefault}{\updefault}{\color[rgb]{0,0,0}$\BrewinParent{i}$}%
}}}}
\put(3784,-1501){\makebox(0,0)[b]{\smash{{\SetFigFont{8}{9.6}{\rmdefault}{\mddefault}{\updefault}{\color[rgb]{0,0,0}$D$}%
}}}}
\put(5310,-3611){\makebox(0,0)[b]{\smash{{\SetFigFont{8}{9.6}{\rmdefault}{\mddefault}{\updefault}{\color[rgb]{0,0,0}$B$}%
}}}}
\end{picture}%

%% file: vertex-spoke.pspdftex
\begin{picture}(0,0)%
\includegraphics{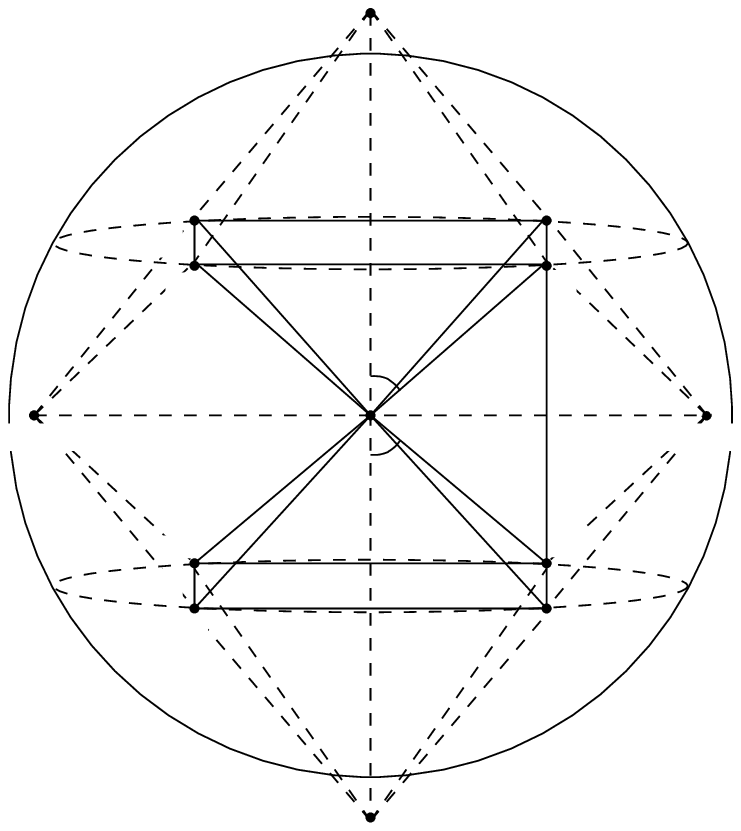}%
\end{picture}%
\setlength{\unitlength}{3947sp}%
\begingroup\makeatletter\ifx\SetFigFont\undefined%
\gdef\SetFigFont#1#2#3#4#5{%
  \reset@font\fontsize{#1}{#2pt}%
  \fontfamily{#3}\fontseries{#4}\fontshape{#5}%
  \selectfont}%
\fi\endgroup%
\begin{picture}(4043,4145)(1298,-3226)
\put(3425,-1444){\makebox(0,0)[b]{\smash{{\SetFigFont{8}{9.6}{\rmdefault}{\mddefault}{\updefault}{\color[rgb]{0,0,0}$\theta_0$}%
}}}}
\put(3860,-1597){\makebox(0,0)[lb]{\smash{{\SetFigFont{8}{9.6}{\rmdefault}{\mddefault}{\updefault}{\color[rgb]{0,0,0}$v_i$}%
}}}}
\put(3702,-1945){\makebox(0,0)[b]{\smash{{\SetFigFont{8}{9.6}{\rmdefault}{\mddefault}{\updefault}{\color[rgb]{0,0,0}$v_i$}%
}}}}
\put(3321,-3211){\makebox(0,0)[b]{\smash{{\SetFigFont{8}{9.6}{\rmdefault}{\mddefault}{\updefault}{\color[rgb]{0,0,0}Parent vertex}%
}}}}
\put(4250,-2214){\makebox(0,0)[b]{\smash{{\SetFigFont{8}{9.6}{\rmdefault}{\mddefault}{\updefault}{\color[rgb]{0,0,0}$1^\prime$}%
}}}}
\put(2417,-2214){\makebox(0,0)[b]{\smash{{\SetFigFont{8}{9.6}{\rmdefault}{\mddefault}{\updefault}{\color[rgb]{0,0,0}$4^\prime$}%
}}}}
\put(4193,-905){\makebox(0,0)[lb]{\smash{{\SetFigFont{8}{9.6}{\rmdefault}{\mddefault}{\updefault}{\color[rgb]{0,0,0}$v_i$}%
}}}}
\put(2053,-677){\makebox(0,0)[rb]{\smash{{\SetFigFont{8}{9.6}{\rmdefault}{\mddefault}{\updefault}{\color[rgb]{0,0,0}$p_i$}%
}}}}
\put(4589,-677){\makebox(0,0)[lb]{\smash{{\SetFigFont{8}{9.6}{\rmdefault}{\mddefault}{\updefault}{\color[rgb]{0,0,0}$p_i$}%
}}}}
\put(2453,-357){\makebox(0,0)[rb]{\smash{{\SetFigFont{8}{9.6}{\rmdefault}{\mddefault}{\updefault}{\color[rgb]{0,0,0}$v_i$}%
}}}}
\put(2417,-1815){\makebox(0,0)[b]{\smash{{\SetFigFont{8}{9.6}{\rmdefault}{\mddefault}{\updefault}{\color[rgb]{0,0,0}$3^\prime$}%
}}}}
\put(3425,-954){\makebox(0,0)[b]{\smash{{\SetFigFont{8}{9.6}{\rmdefault}{\mddefault}{\updefault}{\color[rgb]{0,0,0}$\theta_0$}%
}}}}
\put(3321,829){\makebox(0,0)[b]{\smash{{\SetFigFont{8}{9.6}{\rmdefault}{\mddefault}{\updefault}{\color[rgb]{0,0,0}Parent vertex}%
}}}}
\put(3997,-1107){\makebox(0,0)[b]{\smash{{\SetFigFont{8}{9.6}{\rmdefault}{\mddefault}{\updefault}{\color[rgb]{0,0,0}$p_i$}%
}}}}
\put(3350,-2385){\makebox(0,0)[lb]{\smash{{\SetFigFont{8}{9.6}{\rmdefault}{\mddefault}{\updefault}{\color[rgb]{0,0,0}$u_i$}%
}}}}
\put(2391,-1107){\makebox(0,0)[b]{\smash{{\SetFigFont{8}{9.6}{\rmdefault}{\mddefault}{\updefault}{\color[rgb]{0,0,0}$p_i$}%
}}}}
\put(3702,-391){\makebox(0,0)[b]{\smash{{\SetFigFont{8}{9.6}{\rmdefault}{\mddefault}{\updefault}{\color[rgb]{0,0,0}$v_i$}%
}}}}
\put(3350,110){\makebox(0,0)[lb]{\smash{{\SetFigFont{8}{9.6}{\rmdefault}{\mddefault}{\updefault}{\color[rgb]{0,0,0}$u_i$}%
}}}}
\put(3860,-771){\makebox(0,0)[lb]{\smash{{\SetFigFont{8}{9.6}{\rmdefault}{\mddefault}{\updefault}{\color[rgb]{0,0,0}$v_i$}%
}}}}
\put(3258,-1192){\makebox(0,0)[rb]{\smash{{\SetFigFont{8}{9.6}{\rmdefault}{\mddefault}{\updefault}{\color[rgb]{0,0,0}0}%
}}}}
\put(3786,-2639){\makebox(0,0)[lb]{\smash{{\SetFigFont{8}{9.6}{\rmdefault}{\mddefault}{\updefault}{\color[rgb]{0,0,0}$u_i$}%
}}}}
\put(3786,279){\makebox(0,0)[lb]{\smash{{\SetFigFont{8}{9.6}{\rmdefault}{\mddefault}{\updefault}{\color[rgb]{0,0,0}$u_i$}%
}}}}
\put(4244,-172){\makebox(0,0)[b]{\smash{{\SetFigFont{8}{9.6}{\rmdefault}{\mddefault}{\updefault}{\color[rgb]{0,0,0}2}%
}}}}
\put(2410,-553){\makebox(0,0)[b]{\smash{{\SetFigFont{8}{9.6}{\rmdefault}{\mddefault}{\updefault}{\color[rgb]{0,0,0}4}%
}}}}
\put(4250,-1815){\makebox(0,0)[b]{\smash{{\SetFigFont{8}{9.6}{\rmdefault}{\mddefault}{\updefault}{\color[rgb]{0,0,0}$2^\prime$}%
}}}}
\put(4934,-1285){\makebox(0,0)[b]{\smash{{\SetFigFont{8}{9.6}{\rmdefault}{\mddefault}{\updefault}{\color[rgb]{0,0,0}Central vertex}%
}}}}
\put(2410,-172){\makebox(0,0)[b]{\smash{{\SetFigFont{8}{9.6}{\rmdefault}{\mddefault}{\updefault}{\color[rgb]{0,0,0}3}%
}}}}
\put(1706,-1285){\makebox(0,0)[b]{\smash{{\SetFigFont{8}{9.6}{\rmdefault}{\mddefault}{\updefault}{\color[rgb]{0,0,0}Central vertex}%
}}}}
\put(2453,-1975){\makebox(0,0)[rb]{\smash{{\SetFigFont{8}{9.6}{\rmdefault}{\mddefault}{\updefault}{\color[rgb]{0,0,0}$v_i$}%
}}}}
\put(4244,-553){\makebox(0,0)[b]{\smash{{\SetFigFont{8}{9.6}{\rmdefault}{\mddefault}{\updefault}{\color[rgb]{0,0,0}1}%
}}}}
\end{picture}%

%% file: SubdivVol-CombinedGraphs.tex
\begingroup
  \fontfamily{Verdana}%
  \selectfont
  \makeatletter
  \providecommand\color[2][]{%
    \GenericError{(gnuplot) \space\space\space\@spaces}{%
      Package color not loaded in conjunction with
      terminal option `colourtext'%
    }{See the gnuplot documentation for explanation.%
    }{Either use 'blacktext' in gnuplot or load the package
      color.sty in LaTeX.}%
    \renewcommand\color[2][]{}%
  }%
  \providecommand\includegraphics[2][]{%
    \GenericError{(gnuplot) \space\space\space\@spaces}{%
      Package graphicx or graphics not loaded%
    }{See the gnuplot documentation for explanation.%
    }{The gnuplot epslatex terminal needs graphicx.sty or graphics.sty.}%
    \renewcommand\includegraphics[2][]{}%
  }%
  \providecommand\rotatebox[2]{#2}%
  \@ifundefined{ifGPcolor}{%
    \newif\ifGPcolor
    \GPcolorfalse
  }{}%
  \@ifundefined{ifGPblacktext}{%
    \newif\ifGPblacktext
    \GPblacktexttrue
  }{}%
  \let\gplgaddtomacro\g@addto@macro
  \gdef\gplbacktext{}%
  \gdef\gplfronttext{}%
  \makeatother
  \ifGPblacktext
    \def\colorrgb#1{}%
    \def\colorgray#1{}%
  \else
    \ifGPcolor
      \def\colorrgb#1{\color[rgb]{#1}}%
      \def\colorgray#1{\color[gray]{#1}}%
      \expandafter\def\csname LTw\endcsname{\color{white}}%
      \expandafter\def\csname LTb\endcsname{\color{black}}%
      \expandafter\def\csname LTa\endcsname{\color{black}}%
      \expandafter\def\csname LT0\endcsname{\color[rgb]{1,0,0}}%
      \expandafter\def\csname LT1\endcsname{\color[rgb]{0,1,0}}%
      \expandafter\def\csname LT2\endcsname{\color[rgb]{0,0,1}}%
      \expandafter\def\csname LT3\endcsname{\color[rgb]{1,0,1}}%
      \expandafter\def\csname LT4\endcsname{\color[rgb]{0,1,1}}%
      \expandafter\def\csname LT5\endcsname{\color[rgb]{1,1,0}}%
      \expandafter\def\csname LT6\endcsname{\color[rgb]{0,0,0}}%
      \expandafter\def\csname LT7\endcsname{\color[rgb]{1,0.3,0}}%
      \expandafter\def\csname LT8\endcsname{\color[rgb]{0.5,0.5,0.5}}%
    \else
      \def\colorrgb#1{\color{black}}%
      \def\colorgray#1{\color[gray]{#1}}%
      \expandafter\def\csname LTw\endcsname{\color{white}}%
      \expandafter\def\csname LTb\endcsname{\color{black}}%
      \expandafter\def\csname LTa\endcsname{\color{black}}%
      \expandafter\def\csname LT0\endcsname{\color{black}}%
      \expandafter\def\csname LT1\endcsname{\color{black}}%
      \expandafter\def\csname LT2\endcsname{\color{black}}%
      \expandafter\def\csname LT3\endcsname{\color{black}}%
      \expandafter\def\csname LT4\endcsname{\color{black}}%
      \expandafter\def\csname LT5\endcsname{\color{black}}%
      \expandafter\def\csname LT6\endcsname{\color{black}}%
      \expandafter\def\csname LT7\endcsname{\color{black}}%
      \expandafter\def\csname LT8\endcsname{\color{black}}%
    \fi
  \fi
  \setlength{\unitlength}{0.0500bp}%
  \begin{picture}(4896.00,4896.00)%
    \gplgaddtomacro\gplbacktext{%
      \csname LTb\endcsname%
      \put(784,512){\makebox(0,0)[r]{\strut{} 0}}%
      \put(784,1510){\makebox(0,0)[r]{\strut{} 500}}%
      \put(784,2508){\makebox(0,0)[r]{\strut{} 1000}}%
      \put(784,3506){\makebox(0,0)[r]{\strut{} 1500}}%
      \put(784,4503){\makebox(0,0)[r]{\strut{} 2000}}%
      \put(880,352){\makebox(0,0){\strut{} 0}}%
      \put(1318,352){\makebox(0,0){\strut{} 200}}%
      \put(1757,352){\makebox(0,0){\strut{} 400}}%
      \put(2195,352){\makebox(0,0){\strut{} 600}}%
      \put(2634,352){\makebox(0,0){\strut{} 800}}%
      \put(3072,352){\makebox(0,0){\strut{} 1000}}%
      \put(3511,352){\makebox(0,0){\strut{} 1200}}%
      \put(3949,352){\makebox(0,0){\strut{} 1400}}%
      \put(4388,352){\makebox(0,0){\strut{} 1600}}%
      \put(128,2607){\rotatebox{-270}{\makebox(0,0){\strut{}$\frac{dU}{dt}$}}}%
      \put(2743,112){\makebox(0,0){\strut{}$U(t)$}}%
    }%
    \gplgaddtomacro\gplfronttext{%
      \csname LTb\endcsname%
      \put(1615,4500){\makebox(0,0)[l]{\strut{}$\text{FLRW}$}}%
      \csname LTb\endcsname%
      \put(1615,4300){\makebox(0,0)[l]{\strut{}$\text{5 tetrahedra}$}}%
      \csname LTb\endcsname%
      \put(1615,4100){\makebox(0,0)[l]{\strut{}$\text{16 tetrahedra}$}}%
      \csname LTb\endcsname%
      \put(1615,3900){\makebox(0,0)[l]{\strut{}$\text{600 tetrahedra}$}}%
      \csname LTb\endcsname%
      \put(1615,3700){\makebox(0,0)[l]{\strut{}$\text{60 tetrahedra}$}}%
      \csname LTb\endcsname%
      \put(1615,3500){\makebox(0,0)[l]{\strut{}$\text{192 tetrahedra}$}}%
      \csname LTb\endcsname%
      \put(1615,3300){\makebox(0,0)[l]{\strut{}$\text{7200 tetrahedra}$}}%
    }%
    \gplbacktext
    \put(0,0){\includegraphics{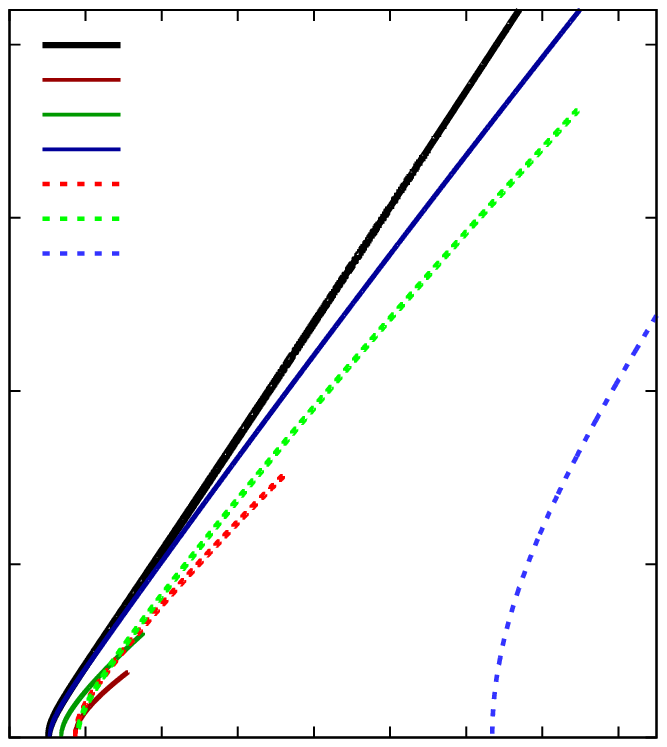}}%
    \gplfronttext
  \end{picture}%
\endgroup

%% file: SubdivVol-CombinedGraphs3.tex
\begingroup
  \fontfamily{Verdana}%
  \selectfont
  \makeatletter
  \providecommand\color[2][]{%
    \GenericError{(gnuplot) \space\space\space\@spaces}{%
      Package color not loaded in conjunction with
      terminal option `colourtext'%
    }{See the gnuplot documentation for explanation.%
    }{Either use 'blacktext' in gnuplot or load the package
      color.sty in LaTeX.}%
    \renewcommand\color[2][]{}%
  }%
  \providecommand\includegraphics[2][]{%
    \GenericError{(gnuplot) \space\space\space\@spaces}{%
      Package graphicx or graphics not loaded%
    }{See the gnuplot documentation for explanation.%
    }{The gnuplot epslatex terminal needs graphicx.sty or graphics.sty.}%
    \renewcommand\includegraphics[2][]{}%
  }%
  \providecommand\rotatebox[2]{#2}%
  \@ifundefined{ifGPcolor}{%
    \newif\ifGPcolor
    \GPcolorfalse
  }{}%
  \@ifundefined{ifGPblacktext}{%
    \newif\ifGPblacktext
    \GPblacktexttrue
  }{}%
  \let\gplgaddtomacro\g@addto@macro
  \gdef\gplbacktext{}%
  \gdef\gplfronttext{}%
  \makeatother
  \ifGPblacktext
    \def\colorrgb#1{}%
    \def\colorgray#1{}%
  \else
    \ifGPcolor
      \def\colorrgb#1{\color[rgb]{#1}}%
      \def\colorgray#1{\color[gray]{#1}}%
      \expandafter\def\csname LTw\endcsname{\color{white}}%
      \expandafter\def\csname LTb\endcsname{\color{black}}%
      \expandafter\def\csname LTa\endcsname{\color{black}}%
      \expandafter\def\csname LT0\endcsname{\color[rgb]{1,0,0}}%
      \expandafter\def\csname LT1\endcsname{\color[rgb]{0,1,0}}%
      \expandafter\def\csname LT2\endcsname{\color[rgb]{0,0,1}}%
      \expandafter\def\csname LT3\endcsname{\color[rgb]{1,0,1}}%
      \expandafter\def\csname LT4\endcsname{\color[rgb]{0,1,1}}%
      \expandafter\def\csname LT5\endcsname{\color[rgb]{1,1,0}}%
      \expandafter\def\csname LT6\endcsname{\color[rgb]{0,0,0}}%
      \expandafter\def\csname LT7\endcsname{\color[rgb]{1,0.3,0}}%
      \expandafter\def\csname LT8\endcsname{\color[rgb]{0.5,0.5,0.5}}%
    \else
      \def\colorrgb#1{\color{black}}%
      \def\colorgray#1{\color[gray]{#1}}%
      \expandafter\def\csname LTw\endcsname{\color{white}}%
      \expandafter\def\csname LTb\endcsname{\color{black}}%
      \expandafter\def\csname LTa\endcsname{\color{black}}%
      \expandafter\def\csname LT0\endcsname{\color{black}}%
      \expandafter\def\csname LT1\endcsname{\color{black}}%
      \expandafter\def\csname LT2\endcsname{\color{black}}%
      \expandafter\def\csname LT3\endcsname{\color{black}}%
      \expandafter\def\csname LT4\endcsname{\color{black}}%
      \expandafter\def\csname LT5\endcsname{\color{black}}%
      \expandafter\def\csname LT6\endcsname{\color{black}}%
      \expandafter\def\csname LT7\endcsname{\color{black}}%
      \expandafter\def\csname LT8\endcsname{\color{black}}%
    \fi
  \fi
  \setlength{\unitlength}{0.0500bp}%
  \begin{picture}(4896.00,3960.00)%
    \gplgaddtomacro\gplbacktext{%
      \csname LTb\endcsname%
      \put(784,512){\makebox(0,0)[r]{\strut{} 0}}%
      \put(784,977){\makebox(0,0)[r]{\strut{} 1000}}%
      \put(784,1442){\makebox(0,0)[r]{\strut{} 2000}}%
      \put(784,1907){\makebox(0,0)[r]{\strut{} 3000}}%
      \put(784,2372){\makebox(0,0)[r]{\strut{} 4000}}%
      \put(784,2837){\makebox(0,0)[r]{\strut{} 5000}}%
      \put(784,3302){\makebox(0,0)[r]{\strut{} 6000}}%
      \put(784,3767){\makebox(0,0)[r]{\strut{} 7000}}%
      \put(880,352){\makebox(0,0){\strut{} 0}}%
      \put(1346,352){\makebox(0,0){\strut{} 500}}%
      \put(1812,352){\makebox(0,0){\strut{} 1000}}%
      \put(2278,352){\makebox(0,0){\strut{} 1500}}%
      \put(2744,352){\makebox(0,0){\strut{} 2000}}%
      \put(3209,352){\makebox(0,0){\strut{} 2500}}%
      \put(3675,352){\makebox(0,0){\strut{} 3000}}%
      \put(4141,352){\makebox(0,0){\strut{} 3500}}%
      \put(4607,352){\makebox(0,0){\strut{} 4000}}%
      \put(128,2139){\rotatebox{-270}{\makebox(0,0){\strut{}$\frac{dU}{dt}$}}}%
      \put(2743,112){\makebox(0,0){\strut{}$U(t)$}}%
    }%
    \gplgaddtomacro\gplfronttext{%
      \csname LTb\endcsname%
      \put(1615,3564){\makebox(0,0)[l]{\strut{}$\text{FLRW}$}}%
      \csname LTb\endcsname%
      \put(1615,3364){\makebox(0,0)[l]{\strut{}$\text{600 tetrahedra}$}}%
      \csname LTb\endcsname%
      \put(1615,3164){\makebox(0,0)[l]{\strut{}$\text{7200 tetrahedra}$}}%
    }%
    \gplbacktext
    \put(0,0){\includegraphics{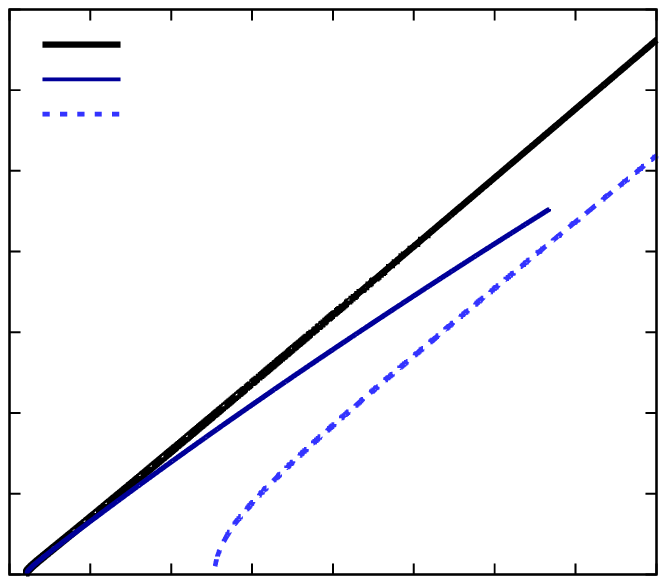}}%
    \gplfronttext
  \end{picture}%
\endgroup

%% file: SubdivVol-CombinedGraphs2.tex
\begingroup
  \fontfamily{Verdana}%
  \selectfont
  \makeatletter
  \providecommand\color[2][]{%
    \GenericError{(gnuplot) \space\space\space\@spaces}{%
      Package color not loaded in conjunction with
      terminal option `colourtext'%
    }{See the gnuplot documentation for explanation.%
    }{Either use 'blacktext' in gnuplot or load the package
      color.sty in LaTeX.}%
    \renewcommand\color[2][]{}%
  }%
  \providecommand\includegraphics[2][]{%
    \GenericError{(gnuplot) \space\space\space\@spaces}{%
      Package graphicx or graphics not loaded%
    }{See the gnuplot documentation for explanation.%
    }{The gnuplot epslatex terminal needs graphicx.sty or graphics.sty.}%
    \renewcommand\includegraphics[2][]{}%
  }%
  \providecommand\rotatebox[2]{#2}%
  \@ifundefined{ifGPcolor}{%
    \newif\ifGPcolor
    \GPcolorfalse
  }{}%
  \@ifundefined{ifGPblacktext}{%
    \newif\ifGPblacktext
    \GPblacktexttrue
  }{}%
  \let\gplgaddtomacro\g@addto@macro
  \gdef\gplbacktext{}%
  \gdef\gplfronttext{}%
  \makeatother
  \ifGPblacktext
    \def\colorrgb#1{}%
    \def\colorgray#1{}%
  \else
    \ifGPcolor
      \def\colorrgb#1{\color[rgb]{#1}}%
      \def\colorgray#1{\color[gray]{#1}}%
      \expandafter\def\csname LTw\endcsname{\color{white}}%
      \expandafter\def\csname LTb\endcsname{\color{black}}%
      \expandafter\def\csname LTa\endcsname{\color{black}}%
      \expandafter\def\csname LT0\endcsname{\color[rgb]{1,0,0}}%
      \expandafter\def\csname LT1\endcsname{\color[rgb]{0,1,0}}%
      \expandafter\def\csname LT2\endcsname{\color[rgb]{0,0,1}}%
      \expandafter\def\csname LT3\endcsname{\color[rgb]{1,0,1}}%
      \expandafter\def\csname LT4\endcsname{\color[rgb]{0,1,1}}%
      \expandafter\def\csname LT5\endcsname{\color[rgb]{1,1,0}}%
      \expandafter\def\csname LT6\endcsname{\color[rgb]{0,0,0}}%
      \expandafter\def\csname LT7\endcsname{\color[rgb]{1,0.3,0}}%
      \expandafter\def\csname LT8\endcsname{\color[rgb]{0.5,0.5,0.5}}%
    \else
      \def\colorrgb#1{\color{black}}%
      \def\colorgray#1{\color[gray]{#1}}%
      \expandafter\def\csname LTw\endcsname{\color{white}}%
      \expandafter\def\csname LTb\endcsname{\color{black}}%
      \expandafter\def\csname LTa\endcsname{\color{black}}%
      \expandafter\def\csname LT0\endcsname{\color{black}}%
      \expandafter\def\csname LT1\endcsname{\color{black}}%
      \expandafter\def\csname LT2\endcsname{\color{black}}%
      \expandafter\def\csname LT3\endcsname{\color{black}}%
      \expandafter\def\csname LT4\endcsname{\color{black}}%
      \expandafter\def\csname LT5\endcsname{\color{black}}%
      \expandafter\def\csname LT6\endcsname{\color{black}}%
      \expandafter\def\csname LT7\endcsname{\color{black}}%
      \expandafter\def\csname LT8\endcsname{\color{black}}%
    \fi
  \fi
  \setlength{\unitlength}{0.0500bp}%
  \begin{picture}(4896.00,3960.00)%
    \gplgaddtomacro\gplbacktext{%
      \csname LTb\endcsname%
      \put(880,512){\makebox(0,0)[r]{\strut{} 0}}%
      \put(880,1163){\makebox(0,0)[r]{\strut{} 10000}}%
      \put(880,1814){\makebox(0,0)[r]{\strut{} 20000}}%
      \put(880,2465){\makebox(0,0)[r]{\strut{} 30000}}%
      \put(880,3116){\makebox(0,0)[r]{\strut{} 40000}}%
      \put(880,3767){\makebox(0,0)[r]{\strut{} 50000}}%
      \put(976,352){\makebox(0,0){\strut{} 0}}%
      \put(1495,352){\makebox(0,0){\strut{} 5000}}%
      \put(2013,352){\makebox(0,0){\strut{} 10000}}%
      \put(2532,352){\makebox(0,0){\strut{} 15000}}%
      \put(3051,352){\makebox(0,0){\strut{} 20000}}%
      \put(3570,352){\makebox(0,0){\strut{} 25000}}%
      \put(4088,352){\makebox(0,0){\strut{} 30000}}%
      \put(4607,352){\makebox(0,0){\strut{} 35000}}%
      \put(128,2139){\rotatebox{-270}{\makebox(0,0){\strut{}$\frac{dU}{dt}$}}}%
      \put(2791,112){\makebox(0,0){\strut{}$U(t)$}}%
    }%
    \gplgaddtomacro\gplfronttext{%
      \csname LTb\endcsname%
      \put(1711,3564){\makebox(0,0)[l]{\strut{}$\text{FLRW}$}}%
      \csname LTb\endcsname%
      \put(1711,3364){\makebox(0,0)[l]{\strut{}$\text{600 tetrahedra}$}}%
      \csname LTb\endcsname%
      \put(1711,3164){\makebox(0,0)[l]{\strut{}$\text{7200 tetrahedra}$}}%
    }%
    \gplbacktext
    \put(0,0){\includegraphics{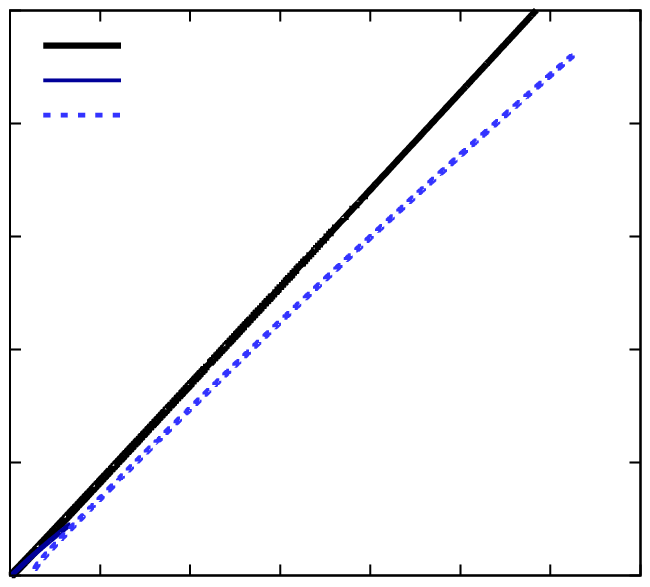}}%
    \gplfronttext
  \end{picture}%
\endgroup

%% file: SubdivRhat-CombinedGraphs.tex
\begingroup
  \fontfamily{Verdana}%
  \selectfont
  \makeatletter
  \providecommand\color[2][]{%
    \GenericError{(gnuplot) \space\space\space\@spaces}{%
      Package color not loaded in conjunction with
      terminal option `colourtext'%
    }{See the gnuplot documentation for explanation.%
    }{Either use 'blacktext' in gnuplot or load the package
      color.sty in LaTeX.}%
    \renewcommand\color[2][]{}%
  }%
  \providecommand\includegraphics[2][]{%
    \GenericError{(gnuplot) \space\space\space\@spaces}{%
      Package graphicx or graphics not loaded%
    }{See the gnuplot documentation for explanation.%
    }{The gnuplot epslatex terminal needs graphicx.sty or graphics.sty.}%
    \renewcommand\includegraphics[2][]{}%
  }%
  \providecommand\rotatebox[2]{#2}%
  \@ifundefined{ifGPcolor}{%
    \newif\ifGPcolor
    \GPcolorfalse
  }{}%
  \@ifundefined{ifGPblacktext}{%
    \newif\ifGPblacktext
    \GPblacktexttrue
  }{}%
  \let\gplgaddtomacro\g@addto@macro
  \gdef\gplbacktext{}%
  \gdef\gplfronttext{}%
  \makeatother
  \ifGPblacktext
    \def\colorrgb#1{}%
    \def\colorgray#1{}%
  \else
    \ifGPcolor
      \def\colorrgb#1{\color[rgb]{#1}}%
      \def\colorgray#1{\color[gray]{#1}}%
      \expandafter\def\csname LTw\endcsname{\color{white}}%
      \expandafter\def\csname LTb\endcsname{\color{black}}%
      \expandafter\def\csname LTa\endcsname{\color{black}}%
      \expandafter\def\csname LT0\endcsname{\color[rgb]{1,0,0}}%
      \expandafter\def\csname LT1\endcsname{\color[rgb]{0,1,0}}%
      \expandafter\def\csname LT2\endcsname{\color[rgb]{0,0,1}}%
      \expandafter\def\csname LT3\endcsname{\color[rgb]{1,0,1}}%
      \expandafter\def\csname LT4\endcsname{\color[rgb]{0,1,1}}%
      \expandafter\def\csname LT5\endcsname{\color[rgb]{1,1,0}}%
      \expandafter\def\csname LT6\endcsname{\color[rgb]{0,0,0}}%
      \expandafter\def\csname LT7\endcsname{\color[rgb]{1,0.3,0}}%
      \expandafter\def\csname LT8\endcsname{\color[rgb]{0.5,0.5,0.5}}%
    \else
      \def\colorrgb#1{\color{black}}%
      \def\colorgray#1{\color[gray]{#1}}%
      \expandafter\def\csname LTw\endcsname{\color{white}}%
      \expandafter\def\csname LTb\endcsname{\color{black}}%
      \expandafter\def\csname LTa\endcsname{\color{black}}%
      \expandafter\def\csname LT0\endcsname{\color{black}}%
      \expandafter\def\csname LT1\endcsname{\color{black}}%
      \expandafter\def\csname LT2\endcsname{\color{black}}%
      \expandafter\def\csname LT3\endcsname{\color{black}}%
      \expandafter\def\csname LT4\endcsname{\color{black}}%
      \expandafter\def\csname LT5\endcsname{\color{black}}%
      \expandafter\def\csname LT6\endcsname{\color{black}}%
      \expandafter\def\csname LT7\endcsname{\color{black}}%
      \expandafter\def\csname LT8\endcsname{\color{black}}%
    \fi
  \fi
  \setlength{\unitlength}{0.0500bp}%
  \begin{picture}(4896.00,3960.00)%
    \gplgaddtomacro\gplbacktext{%
      \csname LTb\endcsname%
      \put(640,512){\makebox(0,0)[r]{\strut{}$  0.0$}}%
      \put(640,1055){\makebox(0,0)[r]{\strut{}$  0.5$}}%
      \put(640,1597){\makebox(0,0)[r]{\strut{}$  1.0$}}%
      \put(640,2140){\makebox(0,0)[r]{\strut{}$  1.5$}}%
      \put(640,2682){\makebox(0,0)[r]{\strut{}$  2.0$}}%
      \put(640,3225){\makebox(0,0)[r]{\strut{}$  2.5$}}%
      \put(640,3767){\makebox(0,0)[r]{\strut{}$  3.0$}}%
      \put(736,352){\makebox(0,0){\strut{} 1}}%
      \put(1510,352){\makebox(0,0){\strut{} 2}}%
      \put(2284,352){\makebox(0,0){\strut{} 3}}%
      \put(3059,352){\makebox(0,0){\strut{} 4}}%
      \put(3833,352){\makebox(0,0){\strut{} 5}}%
      \put(4607,352){\makebox(0,0){\strut{} 6}}%
      \put(128,2139){\rotatebox{-270}{\makebox(0,0){\strut{}$\frac{d(Radius)}{dt}$}}}%
      \put(2671,112){\makebox(0,0){\strut{}$Radius(t)$}}%
    }%
    \gplgaddtomacro\gplfronttext{%
      \csname LTb\endcsname%
      \put(3872,2175){\makebox(0,0)[r]{\strut{}$\text{FLRW}$}}%
      \csname LTb\endcsname%
      \put(3872,1935){\makebox(0,0)[r]{\strut{}$\text{5 tetrahedra}$}}%
      \csname LTb\endcsname%
      \put(3872,1695){\makebox(0,0)[r]{\strut{}$\text{16 tetrahedra}$}}%
      \csname LTb\endcsname%
      \put(3872,1455){\makebox(0,0)[r]{\strut{}$\text{600 tetrahedra}$}}%
      \csname LTb\endcsname%
      \put(3872,1215){\makebox(0,0)[r]{\strut{}$\text{60 tetrahedra}$}}%
      \csname LTb\endcsname%
      \put(3872,975){\makebox(0,0)[r]{\strut{}$\text{192 tetrahedra}$}}%
      \csname LTb\endcsname%
      \put(3872,735){\makebox(0,0)[r]{\strut{}$\text{7200 tetrahedra}$}}%
    }%
    \gplbacktext
    \put(0,0){\includegraphics{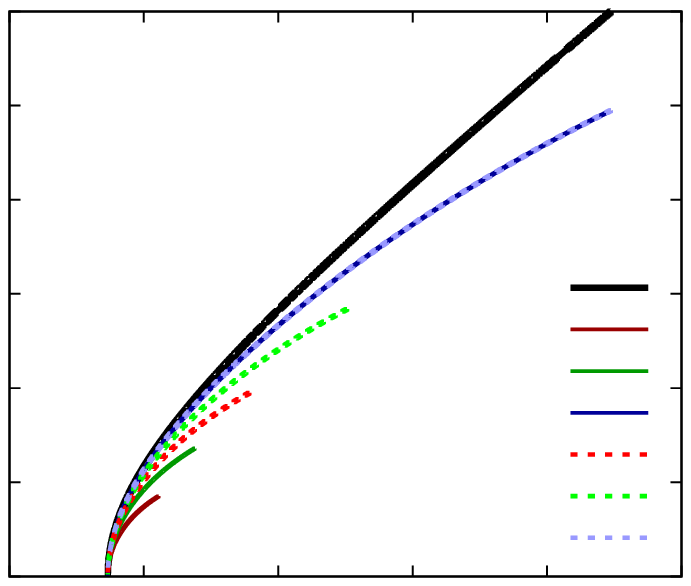}}%
    \gplfronttext
  \end{picture}%
\endgroup

%% file: SubdivRhatVol-CombinedGraphs2.tex
\begingroup
  \fontfamily{Verdana}%
  \selectfont
  \makeatletter
  \providecommand\color[2][]{%
    \GenericError{(gnuplot) \space\space\space\@spaces}{%
      Package color not loaded in conjunction with
      terminal option `colourtext'%
    }{See the gnuplot documentation for explanation.%
    }{Either use 'blacktext' in gnuplot or load the package
      color.sty in LaTeX.}%
    \renewcommand\color[2][]{}%
  }%
  \providecommand\includegraphics[2][]{%
    \GenericError{(gnuplot) \space\space\space\@spaces}{%
      Package graphicx or graphics not loaded%
    }{See the gnuplot documentation for explanation.%
    }{The gnuplot epslatex terminal needs graphicx.sty or graphics.sty.}%
    \renewcommand\includegraphics[2][]{}%
  }%
  \providecommand\rotatebox[2]{#2}%
  \@ifundefined{ifGPcolor}{%
    \newif\ifGPcolor
    \GPcolorfalse
  }{}%
  \@ifundefined{ifGPblacktext}{%
    \newif\ifGPblacktext
    \GPblacktexttrue
  }{}%
  \let\gplgaddtomacro\g@addto@macro
  \gdef\gplbacktext{}%
  \gdef\gplfronttext{}%
  \makeatother
  \ifGPblacktext
    \def\colorrgb#1{}%
    \def\colorgray#1{}%
  \else
    \ifGPcolor
      \def\colorrgb#1{\color[rgb]{#1}}%
      \def\colorgray#1{\color[gray]{#1}}%
      \expandafter\def\csname LTw\endcsname{\color{white}}%
      \expandafter\def\csname LTb\endcsname{\color{black}}%
      \expandafter\def\csname LTa\endcsname{\color{black}}%
      \expandafter\def\csname LT0\endcsname{\color[rgb]{1,0,0}}%
      \expandafter\def\csname LT1\endcsname{\color[rgb]{0,1,0}}%
      \expandafter\def\csname LT2\endcsname{\color[rgb]{0,0,1}}%
      \expandafter\def\csname LT3\endcsname{\color[rgb]{1,0,1}}%
      \expandafter\def\csname LT4\endcsname{\color[rgb]{0,1,1}}%
      \expandafter\def\csname LT5\endcsname{\color[rgb]{1,1,0}}%
      \expandafter\def\csname LT6\endcsname{\color[rgb]{0,0,0}}%
      \expandafter\def\csname LT7\endcsname{\color[rgb]{1,0.3,0}}%
      \expandafter\def\csname LT8\endcsname{\color[rgb]{0.5,0.5,0.5}}%
    \else
      \def\colorrgb#1{\color{black}}%
      \def\colorgray#1{\color[gray]{#1}}%
      \expandafter\def\csname LTw\endcsname{\color{white}}%
      \expandafter\def\csname LTb\endcsname{\color{black}}%
      \expandafter\def\csname LTa\endcsname{\color{black}}%
      \expandafter\def\csname LT0\endcsname{\color{black}}%
      \expandafter\def\csname LT1\endcsname{\color{black}}%
      \expandafter\def\csname LT2\endcsname{\color{black}}%
      \expandafter\def\csname LT3\endcsname{\color{black}}%
      \expandafter\def\csname LT4\endcsname{\color{black}}%
      \expandafter\def\csname LT5\endcsname{\color{black}}%
      \expandafter\def\csname LT6\endcsname{\color{black}}%
      \expandafter\def\csname LT7\endcsname{\color{black}}%
      \expandafter\def\csname LT8\endcsname{\color{black}}%
    \fi
  \fi
  \setlength{\unitlength}{0.0500bp}%
  \begin{picture}(4896.00,3960.00)%
    \gplgaddtomacro\gplbacktext{%
      \csname LTb\endcsname%
      \put(736,512){\makebox(0,0)[r]{\strut{}$0$}}%
      \put(736,1055){\makebox(0,0)[r]{\strut{}$200$}}%
      \put(736,1597){\makebox(0,0)[r]{\strut{}$400$}}%
      \put(736,2140){\makebox(0,0)[r]{\strut{}$600$}}%
      \put(736,2682){\makebox(0,0)[r]{\strut{}$800$}}%
      \put(736,3225){\makebox(0,0)[r]{\strut{}$1000$}}%
      \put(736,3767){\makebox(0,0)[r]{\strut{}$1200$}}%
      \put(832,352){\makebox(0,0){\strut{} 0}}%
      \put(1587,352){\makebox(0,0){\strut{} 200}}%
      \put(2342,352){\makebox(0,0){\strut{} 400}}%
      \put(3097,352){\makebox(0,0){\strut{} 600}}%
      \put(3852,352){\makebox(0,0){\strut{} 800}}%
      \put(4607,352){\makebox(0,0){\strut{} 1000}}%
      \put(128,2139){\rotatebox{-270}{\makebox(0,0){\strut{}$\frac{dU}{dt}$}}}%
      \put(2719,112){\makebox(0,0){\strut{}$U(t)$}}%
    }%
    \gplgaddtomacro\gplfronttext{%
      \csname LTb\endcsname%
      \put(3872,2175){\makebox(0,0)[r]{\strut{}$\text{FLRW}$}}%
      \csname LTb\endcsname%
      \put(3872,1935){\makebox(0,0)[r]{\strut{}$\text{5 tetrahedra}$}}%
      \csname LTb\endcsname%
      \put(3872,1695){\makebox(0,0)[r]{\strut{}$\text{16 tetrahedra}$}}%
      \csname LTb\endcsname%
      \put(3872,1455){\makebox(0,0)[r]{\strut{}$\text{600 tetrahedra}$}}%
      \csname LTb\endcsname%
      \put(3872,1215){\makebox(0,0)[r]{\strut{}$\text{60 tetrahedra}$}}%
      \csname LTb\endcsname%
      \put(3872,975){\makebox(0,0)[r]{\strut{}$\text{192 tetrahedra}$}}%
      \csname LTb\endcsname%
      \put(3872,735){\makebox(0,0)[r]{\strut{}$\text{7200 tetrahedra}$}}%
    }%
    \gplbacktext
    \put(0,0){\includegraphics{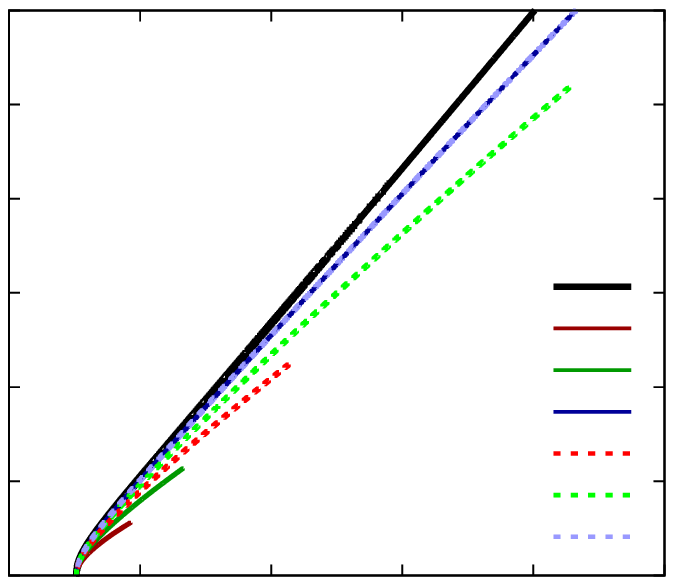}}%
    \gplfronttext
  \end{picture}%
\endgroup

%% file: SubdivRhatVol-CombinedGraphs.tex
\begingroup
  \fontfamily{Verdana}%
  \selectfont
  \makeatletter
  \providecommand\color[2][]{%
    \GenericError{(gnuplot) \space\space\space\@spaces}{%
      Package color not loaded in conjunction with
      terminal option `colourtext'%
    }{See the gnuplot documentation for explanation.%
    }{Either use 'blacktext' in gnuplot or load the package
      color.sty in LaTeX.}%
    \renewcommand\color[2][]{}%
  }%
  \providecommand\includegraphics[2][]{%
    \GenericError{(gnuplot) \space\space\space\@spaces}{%
      Package graphicx or graphics not loaded%
    }{See the gnuplot documentation for explanation.%
    }{The gnuplot epslatex terminal needs graphicx.sty or graphics.sty.}%
    \renewcommand\includegraphics[2][]{}%
  }%
  \providecommand\rotatebox[2]{#2}%
  \@ifundefined{ifGPcolor}{%
    \newif\ifGPcolor
    \GPcolorfalse
  }{}%
  \@ifundefined{ifGPblacktext}{%
    \newif\ifGPblacktext
    \GPblacktexttrue
  }{}%
  \let\gplgaddtomacro\g@addto@macro
  \gdef\gplbacktext{}%
  \gdef\gplfronttext{}%
  \makeatother
  \ifGPblacktext
    \def\colorrgb#1{}%
    \def\colorgray#1{}%
  \else
    \ifGPcolor
      \def\colorrgb#1{\color[rgb]{#1}}%
      \def\colorgray#1{\color[gray]{#1}}%
      \expandafter\def\csname LTw\endcsname{\color{white}}%
      \expandafter\def\csname LTb\endcsname{\color{black}}%
      \expandafter\def\csname LTa\endcsname{\color{black}}%
      \expandafter\def\csname LT0\endcsname{\color[rgb]{1,0,0}}%
      \expandafter\def\csname LT1\endcsname{\color[rgb]{0,1,0}}%
      \expandafter\def\csname LT2\endcsname{\color[rgb]{0,0,1}}%
      \expandafter\def\csname LT3\endcsname{\color[rgb]{1,0,1}}%
      \expandafter\def\csname LT4\endcsname{\color[rgb]{0,1,1}}%
      \expandafter\def\csname LT5\endcsname{\color[rgb]{1,1,0}}%
      \expandafter\def\csname LT6\endcsname{\color[rgb]{0,0,0}}%
      \expandafter\def\csname LT7\endcsname{\color[rgb]{1,0.3,0}}%
      \expandafter\def\csname LT8\endcsname{\color[rgb]{0.5,0.5,0.5}}%
    \else
      \def\colorrgb#1{\color{black}}%
      \def\colorgray#1{\color[gray]{#1}}%
      \expandafter\def\csname LTw\endcsname{\color{white}}%
      \expandafter\def\csname LTb\endcsname{\color{black}}%
      \expandafter\def\csname LTa\endcsname{\color{black}}%
      \expandafter\def\csname LT0\endcsname{\color{black}}%
      \expandafter\def\csname LT1\endcsname{\color{black}}%
      \expandafter\def\csname LT2\endcsname{\color{black}}%
      \expandafter\def\csname LT3\endcsname{\color{black}}%
      \expandafter\def\csname LT4\endcsname{\color{black}}%
      \expandafter\def\csname LT5\endcsname{\color{black}}%
      \expandafter\def\csname LT6\endcsname{\color{black}}%
      \expandafter\def\csname LT7\endcsname{\color{black}}%
      \expandafter\def\csname LT8\endcsname{\color{black}}%
    \fi
  \fi
  \setlength{\unitlength}{0.0500bp}%
  \begin{picture}(4896.00,3960.00)%
    \gplgaddtomacro\gplbacktext{%
      \csname LTb\endcsname%
      \put(736,512){\makebox(0,0)[r]{\strut{}$0$}}%
      \put(736,874){\makebox(0,0)[r]{\strut{}$500$}}%
      \put(736,1235){\makebox(0,0)[r]{\strut{}$1000$}}%
      \put(736,1597){\makebox(0,0)[r]{\strut{}$1500$}}%
      \put(736,1959){\makebox(0,0)[r]{\strut{}$2000$}}%
      \put(736,2320){\makebox(0,0)[r]{\strut{}$2500$}}%
      \put(736,2682){\makebox(0,0)[r]{\strut{}$3000$}}%
      \put(736,3044){\makebox(0,0)[r]{\strut{}$3500$}}%
      \put(736,3405){\makebox(0,0)[r]{\strut{}$4000$}}%
      \put(736,3767){\makebox(0,0)[r]{\strut{}$4500$}}%
      \put(832,352){\makebox(0,0){\strut{} 0}}%
      \put(1371,352){\makebox(0,0){\strut{} 500}}%
      \put(1911,352){\makebox(0,0){\strut{} 1000}}%
      \put(2450,352){\makebox(0,0){\strut{} 1500}}%
      \put(2989,352){\makebox(0,0){\strut{} 2000}}%
      \put(3528,352){\makebox(0,0){\strut{} 2500}}%
      \put(4068,352){\makebox(0,0){\strut{} 3000}}%
      \put(4607,352){\makebox(0,0){\strut{} 3500}}%
      \put(128,2139){\rotatebox{-270}{\makebox(0,0){\strut{}$\frac{dU}{dt}$}}}%
      \put(2719,112){\makebox(0,0){\strut{}$U(t)$}}%
    }%
    \gplgaddtomacro\gplfronttext{%
      \csname LTb\endcsname%
      \put(3872,2175){\makebox(0,0)[r]{\strut{}$\text{FLRW}$}}%
      \csname LTb\endcsname%
      \put(3872,1935){\makebox(0,0)[r]{\strut{}$\text{5 tetrahedra}$}}%
      \csname LTb\endcsname%
      \put(3872,1695){\makebox(0,0)[r]{\strut{}$\text{16 tetrahedra}$}}%
      \csname LTb\endcsname%
      \put(3872,1455){\makebox(0,0)[r]{\strut{}$\text{600 tetrahedra}$}}%
      \csname LTb\endcsname%
      \put(3872,1215){\makebox(0,0)[r]{\strut{}$\text{60 tetrahedra}$}}%
      \csname LTb\endcsname%
      \put(3872,975){\makebox(0,0)[r]{\strut{}$\text{192 tetrahedra}$}}%
      \csname LTb\endcsname%
      \put(3872,735){\makebox(0,0)[r]{\strut{}$\text{7200 tetrahedra}$}}%
    }%
    \gplbacktext
    \put(0,0){\includegraphics{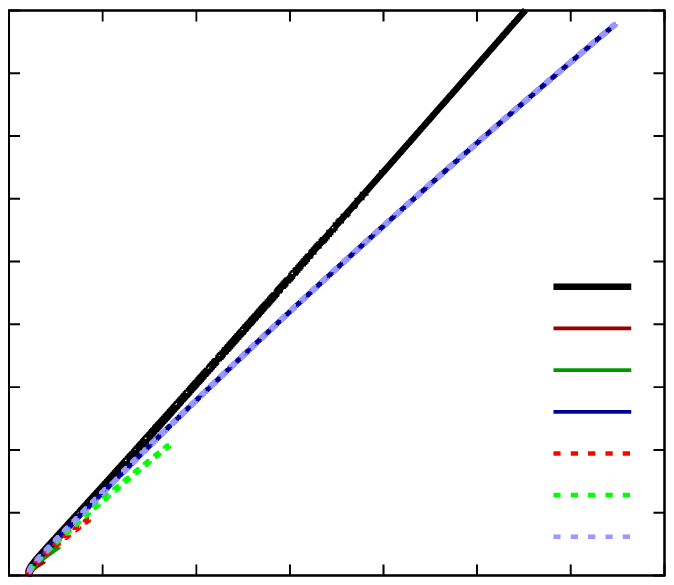}}%
    \gplfronttext
  \end{picture}%
\endgroup